\documentclass[twoside,11pt]{article}

\usepackage{amsfonts}
\usepackage{amsmath}
\usepackage{amssymb}
\usepackage{graphicx}
\usepackage{subfigure}

\usepackage{algorithm}
\usepackage{algorithmic}

\usepackage{paralist}

\newtheorem{definition}{Definition}

\newtheorem{theorem}{Theorem}
\newtheorem{lemma}{Lemma}
\newtheorem{claim}{Claim}
\newtheorem{fact}{Fact}
\newtheorem{corollary}{Corollary}
\newenvironment{Proof}{\noindent {\em Proof:}}{\\\hspace*{\fill}\mbox{$\diamond$}}

\newcommand{\Probab}[1]{\mbox{}{\bf{Pr}}\left[#1\right]}
\newcommand{\Expect}[1]{\mbox{}{\bf{E}}\left[#1\right]}
\newcommand{\Varnce}[1]{\mbox{}{\bf{Var}}\left[#1\right]}
\newcommand{\Trace }[1]{\mbox{}{\bf{Tr}}\left(#1\right)}
\newcommand{\Sqrt  }[1]{\mbox{}\left(#1\right)^{1/2}}

\newcommand{\FNorm }[1]{\mbox{}\left\|#1\right\|_F  }
\newcommand{\FNormS}[1]{\mbox{}\left\|#1\right\|_F^2}

\newcommand{\TNorm }[1]{\mbox{}\left\|#1\right\|_2  }
\newcommand{\TNormS}[1]{\mbox{}\left\|#1\right\|_2^2}

\newcommand{\XNorm }[1]{\mbox{}\left\|#1\right\|_{\xi}  }
\newcommand{\XNormS}[1]{\mbox{}\left\|#1\right\|_{\xi}^2}

\newcommand{\VTTNorm }[1]{\mbox{}\left\|#1\right\|_2  }
\newcommand{\VTTNormS}[1]{\mbox{}\left\|#1\right\|_2^2}

\newcommand{\VINorm }[1]{\mbox{}\left\|#1\right\|_{\infty}  }

\newcommand{\abs }[1]{\left|#1\right|}

\renewcommand{\sectionmark}[1]{\markboth{\textsc{M. W. Mahoney}}
{\textsc{Lecture Notes on Randomized Linear Algebra}}}
\renewcommand{\subsectionmark}[1]{\markboth{\textsc{M. W. Mahoney}}
{\textsc{Lecture Notes on Randomized Linear Algebra}}}

\newlength{\defbaselineskip}
\begin{document}

\title{
Lecture Notes on Randomized Linear Algebra
}

\author{
Michael W. Mahoney%
\thanks{
International Computer Science Institute 
and 
Department of Statistics, 
University of California at Berkeley, 
Berkeley, CA. 
E-mail: mmahoney@stat.berkeley.edu.
}
}

\date{}


\clearpage\maketitle
\thispagestyle{empty}


\newpage
\thispagestyle{empty}

\begin{abstract}

\noindent
These are lecture notes that are based on the lectures from a class I taught on the topic of Randomized Linear Algebra (RLA) at UC Berkeley during the Fall 2013 semester.
These notes are unchanged, relative to the notes that have been available on my web page since then; but, in response to a number of requests, I decided to put them all together as a single file and post them on the arXiv.  
In particular, RLA is a timely topic that is receiving a lot of interdisciplinary attention, and there has been a lot of development in RLA during the last few years, but these notes do not reflect any of these recent developments.  
They do, however, represent the state-of-the-art in terms of a general overview of the area of RLA, as of~then.  

More recent overviews include the following: 
the overview for a general audience by Petros Drineas and me (``RandNLA: Randomized Numerical Linear Algebra,'' in \emph{Communications of the ACM}, 59(6), 80-90 (2016));  
the overview on sketching techniques from a theoretical computer science perspective by David Woodruff (``Sketching as a Tool for Numerical Linear Algebra,'' \emph{FnTTCS}, 10(1-2), 1-157 (2014));  
and
the overview on implementational aspects in Matlab by Shusen Wang (``A Practical Guide to Randomized Matrix Computations with MATLAB Implementations,'' arXiv:1505.07570).  
These more recent references complement well the older overview of randomized low-rank approximation methods from a numerical perspective by Halko et al. (``Finding Structure with Randomness: Probabilistic Algorithms for Constructing Approximate Matrix Decompositions,'' in \emph{SIAM Review}, 53(2), 217-288 (2011))  
and the general overview monograph by me (``Randomized Algorithms for Matrices and Data,'' \emph{FnTML}, 3(2), 123-224 (2011)).  
Readers interested in more recent developments in RLA than provided by these lecture notes should consult these more recent references or wait for v2 of these notes, which will be posted in due course.

Finally, these notes are still very rough, and they likely contain typos and errors.
Moreover, the topic of RLA is still under rapid development.
For both of these reasons, feedback and comments---both in terms of specific technical issues as well as general scope---are most welcome.

\vspace{5mm}

\hfill \hfill Michael W. Mahoney

\hfill \hfill August 2016
 
\end{abstract}

\newpage
\rule{6in}{0.1mm}
\tableofcontents
\rule{6in}{0.1mm}
\newpage


\section{%
(09/04/2013):  Overview of topics and class}

This course will provide an overview of the general area of Randomized Linear Algebra.
Today we will start with some general discussion, and next time we will start to get into the details.
There is no particular reading for today, but here is an overview of the area where some of these themes are discussed and that will be a useful reference throughout the semester.
\begin{itemize}
\item
Mahoney, ``Randomized Algorithms for Matrices and Data,'' FnTML 2011.
\end{itemize}

\subsection{Initial thoughts on randomized matrix algorithms}

This course will cover recent developments in randomized matrix algorithms of interest in large-scale machine learning and statistical data analysis applications.
By this, we will mean basic algorithms for fundamental matrix problems---such as matrix multiplication, least-squares regression, low-rank matrix approximation, and so on---that use randomization in some nontrivial way.
This area goes by the name RandNLA (Randomized Numerical Linear Algebra) or RLA (Randomized Linear Algebra).
It has led to several rather remarkable theoretical, implementation, and empirical successes so far, and a lot more is currently being developed by researchers.

Although very elementary forms of randomization are commonly used in linear algebra, e.g., the starting vector in Lanczos algorithms are typically random vectors, randomization has historically been anathema in matrix algorithms and numerical linear algebra.
This stands in stark contrast to its widespread use in RandNLA in recent years, where it has proven to be a powerful resource for improved computation, and to its widespread acceptance in various forms
in machine learning, statistical data analysis, etc.
The recently-developed randomized matrix algorithms that we will cover in this class typically use randomness to perform random sampling, i.e., choosing, typically in a judicious manner to highlight structural properties of interest, a small number of columns or rows or elements from the matrix, or performing a random projection, i.e., projecting in a data-agnostic manner the original data to a much lower dimensional space.
In either case, one typically hopes that the ``sketch'' of the original data that is thereby constructed is ``similar'' to the original full data set, so that if one runs relatively more expensive computations of interest on the sketch, then one gets a good approximation to the output of computations on the full data set.  

While there has been a lot of interest in these randomized matrix algorithms,  in and of themselves, much of the interest arises since many additional machine learning and data analysis methods either directly call these algorithms as black boxes or indirectly use very similar ideas within their analysis.
Indeed, having been motivated by large-scale statistical data analysis problems, the area of RandNLA has received attention by and been developed by researchers from theoretical computer science, statistics, numerical linear algebra, optimization, scientific computing, data analysis, machine learning, as well as domain sciences such as astronomy, genetics, and internet data analysis.
Given this diversity of approaches, the challenge is to distill out common algorithmic and statistical principles responsible for the success of these methods, to highlight commonalities and differences, so that these methods may be applied more generally.
Thus, in additional to explaining the basic ideas underlying these methods and what is going on ``under the hood'' that makes these methods work, the course will also make connections with how these and similar ideas appear in other related machine learning and data analysis problems.

Today, we will start with a high-level description of some background ideas, mainly to set some context, and then we will describe a simple algorithmic primitive for sampling---uniformly or non-uniformly---in a not-immediately-obvious way from a large matrix.
Next time, we will get into details of our first non-trivial RandNLA algorithm for a matrix problem.

\subsection{Data, models of data, and approaches to computing on data}

Before proceeding, it is worth remembering that, although we will often 
refer to the data as a matrix, matrices are just matrices, and data are 
whatever data are. 
That is, if you are already thinking of the data as consisting of $m$ things, 
each of which is described by $n$ features, or as the correlations between 
$n$ pairs of things, then you have already made very strong assumptions 
about the data---that may or may not be appropriate in particular 
applications.
Examples of data include discretized images, base pair information read out 
of a genetic sequencing machine, click logs at an internet site, records of 
consumer transactions, call records to a 911 site of criminal activity in a 
city, and so on; and in none of those examples are matrices explicitly 
mentioned.
Matrices appear in these and other applications since they are a useful way 
to \emph{model} the data.
By this, we mean a useful way to encode data into a mathematical object such 
that we can run computations of interest on that object in a reasonable 
time and get answers that are useful in some sense in the application domain 
that generated the data.

Of course, matrices are not the only way to model data.
Here are several other common ways data can be modeled.
\begin{itemize}
\item
\textbf{Turing machine.} 
This a tape with $\{0,1\}$ entries, upon which Turing machine operations are performed.
This is a popular way to model data if we are interested in characterizing the algorithmic complexity of problems, especially in the sense of polynomial-time equivalence.
\item
\textbf{Database table.} 
This a table of $(key,value)$ pairs, upon which database/logical operations are performed.
This is a popular way to model data in the the field of databases, and it is of interest to us since really large-scale data are typically held in some sort of database.
\item
\textbf{Strings.}
This is a sequence of characters, e.g., an array of bytes/words that stores a sequence of characters or more general arrays/lists.  
This is popular in certain theoretical areas, e.g., formal language theory, as well as in certain data analysis areas, e.g., bioinformatics. 
\item
\textbf{Graphs.} 
This is a set of things and pairs of things, i.e., $G=(V,E)$, where $V$ is a set of nodes, and $E \subset V \times V$ is a set of edges.
This is a popular way to model data since data often consist of $n$ things and some sort of the pairwise relationship between these things.
\end{itemize}
Of course, these models are not inconsistent, and it is often helpful to 
model the data in different ways, depending on what one is interested in 
doing.
For example, we may want to model the data as a real-valued matrix, since we 
are interested in performing matrix computations on the data, but if we are
going to implement our matrix algorithms on a computer, then we need to 
represent those real numbers in terms of a fixed number of bits, in which 
case we would need to implement algorithms in a manner that is well-behaved
with respect to this discretization.
Alternatively, we may want to model the data as a matrix, but the data might
be large enough that they are stored in a database, in which case we would 
have to implement those computations in a manner that respected the 
constraints imposed by the database query language. 

One of the main points is that for each of these ways to model the data, 
certain types of operations tend to be relatively easy, and other types of 
operations tend to be relatively difficult.
For example, modeling the data in a discrete way such as with a graph or 
a Turing machine tape makes it relatively easy to make statements about 
worst-case algorithmic complexity; but this is often not robust to the 
addition of a bit of noise.
Indeed, one of the benefits of modeling the data as real-valued matrices is 
that the geometry of Euclidean spaces provides a robustness that, in 
addition to leading to relatively fast algorithms, underlies good 
statistical inferential properties that are often of interest.
That being said, while viewing real-valued matrices as consisting of real 
numbers, rather than fixed precision approximations to real numbers, as they 
are actually stored on a computer, is convenient, it is often not robust in 
the presence of even very low-order bit roundoff error.
To the extent that most modern researchers who use matrix algorithms have 
the luxury of ignoring such issues, it is because a large body of numerical 
analysts and scientific computers have worried about these issues for them 
and have black-boxed these issues from them.

As an example of one of the many challenges in providing matrix algorithms 
for a wide range of very large-scale statistical data analysis applications, 
modern variants of these problems can re-arise either when researchers who 
have never thought about these issues implement nontrivial matrix algorithms 
in large-scale settings or also when researchers apply traditional algorithms 
to data matrices that are structured very differently than matrices that have
arisen in the past.
Different fields parameterize problems in different ways, and 
seemingly-minor differences can have fundamental consequences for the 
appropriateness and applicability of different algorithms.
One of the major challenges in the area, which will be a theme that arises
throughout the course, is to develop algorithmic principles that allow 
researchers to draw strength from the experiences of matrix computations in 
the past, while addressing the novel and peculiar features of matrix-based 
data that arise in modern massive data set applications.

This is a good point to say what is ``large'' in large-scale data.
There is a lot of hype about large-scale and massive and big data that 
wasn't present even a few years ago, and so it is useful to keep particular
examples and categorizations in mind as we proceed, since large means 
different things to different people in different contexts.
One of the most useful categorizations is the following:
\begin{itemize}
\item
\textbf{Small.} 
A data set is small if you can look at the data and fairly-easily find solutions to problems of interest with any one of several algorithmic tools, e.g., one's favorite method.
This is common, especially in areas that focus on the development of methods qua methods.
\item
\textbf{Medium.}
A data set is medium-sized if it fits into RAM and one can fairly-easily run computations of interest in a reasonable length of time and get answers to questions of interest.
This is common, especially in areas that use methods for downstream goals of primary interest.
\item
\textbf{Large.} 
A data set is large if it doesn't easily fit into RAM and/or one can't relatively-easily run computations of interest.
This is increasingly common, and although it is the domain of a relatively-small set of users of data, it provides an important forcing function in general but for the development of algorithmic and statistical methods in particular.
\end{itemize}
The main point of this informal categorization is that as one goes from 
medium-sized to large-scale one does not have random access to the data, 
and so details of memory access become increasingly important.
A related issue is that communication is often a more precious resource than
computation, and so this must be taken into account at the start of designing
algorithms.
The details of the memory access issues can vary in different application
areas---e.g., streaming settings, moderate-sized databases, MapReduce or 
Hadoop-style environments, multi-core settings, etc.
From the perspective of this class, however, we will see that often similar 
algorithmic ideas or algorithmic principles hold in multiple settings, 
basically since those ideas exploit basic structural properties of vector 
spaces, but that those ideas have to be instantiated in somewhat different 
ways depending on the particular setting.
For example, a random projection algorithm might solve a given problem in 
an idealized sense, but one should work with versions of random projections 
that optimize what matters most, e.g., one should consider an Hadamard-based 
or Gaussian-based projection depending on whether one is interested in 
optimizing FLOPs or communication.
The coarse algorithmic ideas are similar in both of those settings, and 
it's often not so useful to ``over-optimize'' to the details of an idealized 
computational model, since it can hide the breadth of applicability of a
basic algorithmic idea.

Randomization can be thought about in different ways:
\begin{itemize}
\item
Vague philosophical hope that you will find something useful if you randomly 
sample.
\item
Statistical approach: observed data are a noisy/random version of ground truth.
\item
Algorithmic approach: randomness is a computational resource for faster 
algorithms on a given observed data set.
\end{itemize}

This parallels two major perspectives on the data that will be a common theme:
\begin{itemize}
\item
Algorithmic perspective (common in computer science theory, databases, etc.):
in this case,
the data/algorithms are typically discrete;
do worst-case analysis on a given data set;
interested in optimizing running time and other resources;
models for computation and data access.
\item
Statistical perspective (common in statistics, machine learning, natural 
sciences, etc.):
in this case,
the data/algorithms are often continuous;
make reasonable niceness assumptions on the data;
ultimately interested in inferences about the world and not data per se;
models for the data to help inference. 
\end{itemize}

Randomization can be useful in several ways:
\begin{itemize}
\item
Faster algorithms: worst-case theory, numerical implementation, clock time
\item
Simpler algorithms: to state, implement, and analyze
\item
More interpretable algorithms and output: select actual columns 
\item
Implicit regularization: randomness in the algorithms helps to avoid 
overfitting to a given data set
\item
Organize algorithms to modern computational architectures better.
\end{itemize}

\subsection{Examples of matrix-based data and matrix computations}

If we are going to develop algorithms for data modeled as a matrix, then 
here are several examples of matrix-based data to keep in mind.
\begin{itemize}
\item
\textbf{Object-feature data matrix.} 
This is perhaps the most common way matrices arise, in which case one has $m$ objects or things, each of which are described by $n$ features, e.g., term-document data, people-SNPs data, etc.
\item
\textbf{Correlation matrices.} 
This is basically $X^TX$, where $X$ or $X^T$ is an object-feature data matrix, perhaps appropriately normalized.
\item
\textbf{Kernels and similarity matrices.}
These are popular in machine learning.
The former are basically SPSD matrices, while the latter are basically entry-wise non-negative matrices, and there are several common procedures to go from one to the other.
\item
\textbf{Laplacians or Adjacency matrices of graphs.}
These are central to spectral graph theory where one considers eigenvectors and eigenvalues and related quantities of matrices associated with a graph.
\item
\textbf{PDEs and discretization of continuum operators.}
Matrices that arise here can come in one of several forms, and implementations of (low-rank, in particular) RandNLA algorithms are often developed for these matrices, but they often come with relatively strong domain-specific niceness assumptions.
\end{itemize}

And here is a motivating application (one of many, but one to which I am 
partial) to keep in mind as we develop and analyze algorithms over the semester.
Recall that \emph{the} human genome consists of roughly $3B$ base pairs, 
but every individual is distinct, and so there are differences.
Of the many types of differences, perhaps the most amenable to large-scale
data analysis are SNPs, which are single locations in the genome where a
non-negligible fraction of the population has one base pair and a 
non-negligible fraction has another different base pair.
Very roughly, e.g., depending on how one defined the minor-allele frequency, 
these SNPs occur in $1/1000$ base paris, and so there are roughly $3M$ SNPs.
HapMap considered roughly $400$ people, and subsequent studies considered 
$1000s$ or $10,000s$ of people.
So, we can easily get matrices of size roughly, say, $10^4 \times 10^6$.
There are typically one of two goals of interest: either do population 
genetics, or do some sort of personalized medicine.

Among the many algorithmic/statistical challenges---for this particular 
motivating application, as well as much more generally---here are two 
prototypical examples.
\begin{itemize}
\item
\textbf{Low-rank matrix approximation.}
Do PCA or SVD to get a good low-rank approximation, and do stuff.
That is, compute a full/partial SVD/PCA/QR to get a small number of 
eigenvectors; appeal to a model selection rule to determine the number of 
eigenvectors to keep; and use those eigenvectors to cluster or classify.
This is no problem if the matrices are of size $10^2 \times 10^4$; it is 
challenging but possible if the matrices are of size $10^4 \times 10^6$; 
and it is essentially impossible if the matrices are of size 
$10^4 \times 10^8$. 
\item
\textbf{Column subset selection problem.}
Select a small number of representative SNPs, and do stuff.
That is, compute a full/partial SVD/PCA/QR to get a small number of 
eigenvectors; appeal to a model selection rule to determine the number of 
eigenvectors to keep; and ``interpret'' those eigenvectors i.t.o. processes
generating the data or use them to select actual SNPs.
This is no problem if the data really are generated from a Gaussian process, 
since in that case the eigenvectors mean something in terms of the data, 
but otherwise the reification is typically no good.
\end{itemize}
There are many variants of these basic methods, e.g. ``nonlinear'' kernel 
methods, ``sparse'' SVD, other feature selection methods, etc.; and nearly 
all, either structurally or in terms of the algorithms to implement them 
have similar challenges to those two just outlined.

\subsection{A simple model for accessing large-scale matrix data}

Now, we informally define the Pass-Efficient Model, which is a computational 
model in which the computational resources are the number of passes over the 
data and the additional space and additional time required, and we use it 
to present several technical sampling lemmas that illustrate how to draw 
uniform and nonuniform samples when the data are not in RAM.

Such data streaming models have been widely-studied in recent years, 
especially in the theory of algorithms, and the basic idea is that the data 
are so large that they stream by and one must compute on the stream without 
storing the entire data set.
The Pass-Efficient Model, in particular, is motivated by the observation 
that in many applications one has often the ability to generate and store 
very large amounts of data, but one often does not have random access to 
that data.  
For example, the data may be stored on external storage such as a tape, or 
the data may be stored in a distributed data center.
To model this phenomenon, consider a situation in which the three scarce 
computational resources of interest are number of passes over the data and 
the additional space and time required.
Although this model is quite idealized, understanding how matrix algorithms 
behave with respect to it will help us get beyond a vanilla RAM model in 
which any element of a data matrix can be accessed at random.

\begin{definition}
In the \emph{Pass-Efficient Model}, the only access an algorithm has to the 
data is via a pass, where a \emph{pass} over the data is a sequential read 
of the entire input data set.
In addition to the external storage space to store the data and to a small 
\emph{number of passes} over the data, an algorithm in the Pass-Efficient 
Model is permitted to use \emph{additional RAM space} and \emph{additional 
computation time}.
An algorithm is considered \emph{pass-efficient} if it requires a small 
constant number of passes and additional space and time which are sublinear 
in the length of the data stream in order to compute the solution (or a 
``description'' of the solution).
\end{definition}

Recall that, if the data are represented by a $m \times n$ matrix, e.g., $n$ 
vectors $a_i \in \mathbb{R}^{m}$, $i = 1, \ldots, n$, then the data can be 
presented column-wise or row-wise or in some other arbitrary order. 
The \emph{sparse-unordered representation} of data is a form of data 
representation in which each element of the data stream consists of a pair 
$((i,j),A_{ij})$ where the elements in the data stream may be unordered with 
respect to the indices $(i,j)$ and only the nonzero elements of the matrix $A$ 
need to be presented.  
This very general form is suited to applications where, e.g., multiple agents 
may write parts of a matrix to a central database and where one cannot make 
assumptions about the rules for write-conflict resolution.  
In the simplest form, the data stream read by algorithms in the 
Pass-Efficient Model is assumed to be presented in the 
\emph{sparse-unordered representation}, but in many cases stronger results 
can be obtained when the data matrix is assumed to be presented, e.g., 
column-wise or row-wise.

Next, we present two related sampling lemmas that will be used by our 
subsequent algorithms. 
Since many of our subsequent algorithms will involve constructing random 
samples, from either uniform or non-uniform distributions, we would like to 
be able to select random samples in a pass-efficient manner.
To this end, consider the \textsc{Select} algorithm, described below, which 
does just this.  
The algorithm reads a stream, assumed to consist of non-negative entries, and
in constant additional space and time (where we assume that we are 
representing the real numbers in the stream in constant size) it selects and
returns an element from that stream with a probability proportional to its 
size.

\begin{algorithm}
\caption{The \textsc{Select} Algorithm.}
\label{alg:AlgorithmSelect}
\begin{algorithmic}[1]
    \REQUIRE $\{a_1, \ldots, a_n\}$, $a_i \ge 0$, read in one pass, i.e., 
             one sequential read, over the data.
    \ENSURE  $i^{*}, a_{i^{*}}$.
    \STATE $D=0$.
    \FOR{$i = 1$ to $n$}
       \STATE $D = D + a_i$.
       \STATE With probability $a_i/D$, let $i^{*}=i$ and $a_{i^{*}}=a_i$.
    \ENDFOR
    \STATE Return $i^{*}, a_{i^{*}}$.
\end{algorithmic}
\end{algorithm}

The following lemma establishes that in one pass over the data one can sample
an element according to certain probability distributions.

\begin{lemma}
\label{selectlemma1}
Suppose that $\{a_1, \ldots, a_n\}$, $a_i \ge 0$, are read in one pass, i.e., 
one sequential read over the data, by the \textsc{Select} algorithm.
Then the \textsc{Select} algorithm requires $O(1)$ additional storage space 
and returns $i^{*}$ such that
$ \Probab{i^{*}=i} = a_i/\sum_{i^{\prime}=1}^{n}a_{i^{\prime}} $.
\end{lemma}
\begin{Proof}
First, note that retaining the selected value and the running sum requires 
$O(1)$ additional space.
The remainder of the proof is by induction.
After reading the first element $a_1$, $i^{*}=1$ with probability $a_1/a_1=1$.
Let $D_{\ell  } = \sum_{i^{\prime}=1}^{{\ell  }}a_{i^{\prime}}$ and suppose 
that the algorithm has read $a_1, \ldots, a_{\ell}$ thus far and has retained 
the running sum $D_{\ell}$ and a sample $i^{*}$ such that 
$ \Probab{i^{*} = i} = a_i/D_{\ell} $.
Upon reading $a_{\ell+1}$ the algorithm lets $i^{*}=\ell+1$ with probability 
$a_{\ell+1}/D_{\ell+1}$ and retains $i^{*}$ at its previous value otherwise.
At that point, clearly $\Probab{i^{*}=\ell+1} = a_{\ell+1}/D_{\ell+1}$; 
furthermore for $i = 1, \ldots, \ell$, 
$\Probab{i^{*}=i} 
   = \frac{a_i}{D_{\ell}}\left(1-\frac{a_{\ell+1}}{D_{\ell+1}}\right) 
   = \frac{a_{i}}{D_{\ell+1}}$. 
By induction this results holds when $\ell+1 = n$ and the lemma follows.
\end{Proof}

\noindent
Clearly, in a single pass over the data this algorithm can be run in parallel 
with $O(s)$ total memory units to return $s$ independent samples 
$i^{*}_1, \ldots, i^{*}_s$ such that for each $i^{*}_{t}$, 
$t = 1, \ldots, s$, we have
$ \Probab{i_{t}^{*}=i} = a_i/\sum_{i^{\prime}=1}^{n}a_{i^{\prime}} $.
Also, one can clearly use this algorithm to sample with respect to other 
distributions that depend (or don't depend) on the $a_i$, e.g., the 
uniform distribution, probabilities proportional to $a_i^2$, etc.

The next lemma is a modification of the previous lemma to deal with the case 
where a matrix is read in the sparse-unordered representation and one wants 
to choose a row label with a certain probability.  
Note that a trivial modification would permit choosing a column label.

\begin{lemma}
\label{selectlemma2}
Suppose that $A \in \mathbb{R}^{m \times n}$, is presented in the 
sparse-unordered representation and is read in one pass, i.e., one sequential 
read over the data, by the \textsc{Select} algorithm.
Then the algorithm requires $O(1)$ additional storage space and returns 
$i^{*}$, $j^{*}$ such that
$ \Probab{i^{*}=i \wedge j^{*}=j} = A_{i^{*}j^{*}}^{2}/\FNormS{A} $
and thus
$ \Probab{i^{*}=i} = \VTTNormS{A_{(i^{*})}}/\FNormS{A} $.
\end{lemma}
\begin{Proof}
Since $A_{i^{*}j^{*}}^{2} > 0$ the first claim follows from 
Lemma \ref{selectlemma1}; the second follows since
$$
\Probab{i^{*}=i} = \sum_{j=1}^{n}\Probab{i^{*}=i \wedge j^{*}=j} 
                   = \sum_{j=1}^{n}\frac{A_{i^{*}j^{*}}^{2}}{\FNormS{A}} 
                   = \frac{\VTTNormS{A_{(i^{*})}}}{\FNormS{A}}     .
$$
\end{Proof}

Note that, in particular, this lemma implies that we can select columns and 
rows from a matrix according to a probability distribution that is 
proportional to the squared Euclidean norms, and that we can do it in two 
``passes'' over the data. 
More precisely, in one pass and $O(1)$ additional space and time, we can 
choose choose the index of a column with a probability proportional to the 
Euclidean norm squared of that column (and with $O(s)$ additional space and 
time, we can choose the indices of $s$ columns), and in the second pass we
can pull out that column---provided, of course, that we have $O(n)$ 
additional space.

\subsection{Quick overview of ideas and topics to be covered in the class}

The ideas to be discussed in this class arose and were developed in several related research areas, and they have been applied in various forms in a wide range of theoretical and practical applications.
Not surprisingly, then, there has been some reinvention of the wheel, and it can be difficult for even experts in one area to understand the contributions and developments from other areas.
That being said, the ideas have already proven remarkably fruitful: they have led to qualitatively improved worst-case bounds for fundamental matrix problems, they have led to numerical implementations that beat state-of-the-art solvers, and they have led to improved machine learning and data analysis that have been used in a wide range of scientific and internet applications. 
In this course, we will try to distill out the basic algorithmic and statistical ideas that make these methods work.
We will do so by focusing on a few very basic linear algebraic problems that underlie all or nearly all of the extensions and applications.

To illustrate this, a fundamental primitive and the first matrix problem that we will consider will be that of approximating the product of two matrices.
Say that we have an $m \times n$ matrix $A$ and an $n \times p$ matrix $B$, and assume that we are interested in computing the product $AB$.
\begin{itemize}
\item
\textbf{Traditional perspective on matrix multiplication.}
The obvious well-known way to compute the product $AB$ is with the usual three-loop algorithm.
In this case, one views an element of $AB$ as an \emph{inner product between a row of $A$ and an column of $B$}.
\item
\textbf{RandNLA perspective on matrix multiplication.}
A less obvious way is to view the product $AB$ is as a sum of $n$ terms, each of which is an \emph{outer product between a column of $A$ and a row of $B$}.
Viewed this latter way, we can try to  construct some sort of ``sketch'' of the columns of $A$ and the rows of $B$---let's represent those sketches as matrices $C$ and $R$, respectively---and approximate the product $AB$ by the product $CR$.
\end{itemize}
If the sketches are linear, as they almost always are and as they will be in this class, then we can represent the sketches themselves as a matrix.
Let's say that the $n \times c$ matrix $S$ represents the sketching operation.
Then, observe that $C=AS$ and $R=S^TB$, in which case $AB \approx CR = ASS^TB$.
We will quantify the quality of $CR$ by bounding the norm of the error, i.e., by providing an upper bound for $\XNorm{AB-ASS^TB}$, where $\xi$ represents some matrix norm such as the spectral or Frobenius or trace norm.
This fundamental matrix multiplication primitive will appear again and again in many different guises.
Of course, described this way, the sketching matrix $S$ can be anything---deterministic or randomized, efficient or intractable to compute, etc. 
It turns out that if the sketches are randomized---basically consisting of random sampling or random projection operations---then in many cases we can obtain results that are ``better'' than with deterministic~methods. 
 
(This comment about ``better'' requires some clarification and comes with a number of caveats.
First, the randomized methods might simply fail and return an answer that
is extremely bad---after all one could flip a fair coin heads $100$ times in 
a row---and so we will have to control for that.
Second, lower bounds in general are hard to come by, and since they often 
say more about the computational model being considered than about the 
problem being considered, and they are often not robust to minor variations 
in problem statement or minor variations in models of data access.
There are lower bounds in this area, but because of this non-robustness and 
because these lower bounds have yet to have impact on numerical, statistical, 
machine learning, or downstream scientific and internet applications of 
these algorithms, we will not focus on them in this class.
Third, ``better'' might mean faster algorithms in worst-case theory, or
numerical implementations that are faster in terms or wall-clock time, or 
implementations in parallel and distributed environments where traditional 
methods fail to run, or applications that are more useful to downstream 
scientists who tend to view these methods as black boxes.
We will consider all of these notions of better throughout the class.)

The randomized sketches we will consider will come in one of two flavors:
random sampling sketches, in which each column of $S$ has one nonzero 
entry, which defines which (rescaled) columns of $A$ (remember, here we are
post-multiplying $A$ by $S$---we won't use this convention throughout---which
means that $S$ is working on the columns of $A$, which also means that $S^T$
is working on the rows of $B$); and random projection sketches, in which 
case $S$ is dense or nearly dense and consists of i.i.d. random variables
drawn from, e.g., a Gaussian distribution.
The random projections are data-agnostic, in the sense that they can be 
constructed without looking at the data, while the random sampling typically 
needs to be performed in a way that identifies and extracts relevant 
structure in the data.
Vanilla versions of both of these procedures can often be shown to perform 
well, in the sense that they return relatively-good answers, but they are 
often not faster than solving the original problem.
Thus, more complex versions also exist and will be considered.
For example, $S$ could consist of the composition of a Hadamard-like
rotation and a uniform sampling or sparse projection matrix.
In this case, the projection or sampling quality is nearly as good as the
vanilla version, while being much faster.

Thus, the main themes in RandNLA algorithms will be that we need to construct a sketch in one of two complementary ways.
\begin{itemize}
\item
\textbf{Randomly sample.}
Here, one identifies some sort of uniformity structure and use that to construct an importance sampling distribution with which to construct a random sample.
\item
\textbf{Randomly project.}
Here, one performs a random projection which rotates to a random basis where the nonuniformity structure is uniformized and so where we can sample u.a.r.  
\end{itemize}
Then, in either case, RandNLA algorithms do one of two things.
\begin{itemize}
\item
\textbf{Solve the subproblem with a black box.}
Here, the smaller subproblem is solved with a traditional black box solver.
This is most appropriate for theoretical analysis and when one is not concerned with the complexity of the algorithm with respect to the $\epsilon$ error parameter. 
\item
\textbf{Solve a preconditioned version of the original problem.}
Here, one uses the sample to construct a preconditioner for the original problem and solves the original problem with a traditional preconditioned iterative method.
This is most appropriate when one is interested in high precision solutions where the dependence on the $\epsilon$ error parameter is important.
\end{itemize}
The quality of the bounds will boil down to certain structural results (i.e., depending on the linear algebraic structure) and then an additional error from the random sampling process, and we will use matrix perturbation bounds to establish that those latter errors are small.
The most obvious example, where we will discuss this in greatest detail, is for the least-squares problem, but the same idea holds for the low-rank approximation problem, etc.
(In addition, although we will not have the time to go into it in quite as much detail, we should note that the same ideas hold not only in RAM, but also in streaming and parallel and distributed environments; we won't spend as much time on those extensions, but the basic idea will be to refrain from optimizing FLOPs and instead worry about other things like communication.)

\subsection{Course announcement}

For completeness, here is the description of the course from the initial coarse announcement.

Matrices are a popular way to model data (e.g., term-document data,
people-SNP data, social network data, machine learning kernels, and so on),
but the size-scale, noise properties, and diversity of modern data presents
serious challenges for many traditional deterministic matrix algorithms.
The course will cover the theory and practice of randomized algorithms for
large-scale matrix problems 
arising in modern massive data set analysis.
Topics to be covered include: underlying theory, including the
Johnson-Lindenstrauss lemma, random sampling and projection algorithms, and
connections between representative problems such as matrix multiplication,
least-squares regression, least-absolute deviations regression, low-rank
matrix approximation, etc.; numerical and computational issues that arise in
practice in implementing algorithms in different computational environments;
machine learning and statistical issues, as they arise in modern large-scale
data applications; and extensions/connections to related problems as well as
recent work that builds on the basic methods.
Appropriate for graduate students in computer science, statistics, and
mathematics, as well as computationally-inclined students from application
domains.
Here are several representative topics:
\begin{itemize}
\item
Introduction and Overview
\item
Approximate Matrix Multiplication: Building Blocks and Establishing Concentration
\item
Random Projections: Slow, Fast, and Subspace
\item
Least-squares Regression: Sampling versus Projections, Low versus High Precision
\item
Low-rank Matrix Approximation: Additive-error, Relative-error, and Fewer Samples
\item
Element-wise Sampling and Applications
\item
Solving Square and Non-square Linear Equations
\item
Preserving Sparsity in Theory and in Practice
\item
Extensions and Applications:
Kernel-based Learning; Matrix Completion; Graph Sparsification;
$\ell_p$ Regression and Convex Optimization; Parallel and Distributed Environments;
Etc.
\end{itemize}

\newpage

\section{%
(09/09/2013):  Approximating Matrix Multiplication}

Approximating the product of two matrices with random sampling or random projection methods is a fundamental operation that is of interest in and of itself as well as since it is used in a critical way as a primitive for many RandNLA algorithms.
In this class, we will introduce a basic algorithm; and in this and the next few classes, we will discusses several related methods.
Here is the reading for today.
\begin{itemize}
\item
Drineas, Kannan, and Mahoney, ``Fast Monte Carlo Algorithms for Matrices I: Approximating Matrix Multiplication''
\end{itemize}

\subsection{Some notation}

Before proceeding, here is some notation that we will use.
For a vector $x \in \mathbb{R}^n$ we let 
$\VTTNorm{x} = \left(\sum_{i=1}^n |x_i|^2\right)^{1/2}$ denote its Euclidean 
length.
For a matrix $A \in \mathbb{R}^{m \times n}$ we let 
$A^{(j)}$, $j=1,\ldots,n$, denote the $j$-th column of $A$ as a 
column vector and $A_{(i)}$, $i=1,\ldots,m$, denote the $i$-th 
row of $A$ as a row vector.
We denote matrix norms by $\|A\|_{\xi}$, using subscripts to distinguish 
between various norms.  
Of particular interest will be the Frobenius norm which is defined by
\begin{equation}
\FNorm{A} = \sqrt{\sum_{i=1}^m \sum_{j=1}^n A_{ij}^2}     ,
\end{equation}
and the spectral norm which is defined by
\begin{equation}
\TNorm{A} = \sup_{x\in \mathbb{R}^n,\ x\neq 0} \frac{\VTTNorm{Ax}}{\VTTNorm{x}} .
\end{equation}
These norms are related to each other as:
$ \TNorm{A} \leq \FNorm{A} \leq \sqrt{n} \TNorm{A} $.
Both of these norms provide a measure of the ``size'' of the matrix $A$.
There are several situations in which we will be interested in measuring the 
size of a matrix---e.g., the error that is incurred by a random sampling or
random projection process can be viewed as a matrix, and we will be 
interested in showing that it is small in an appropriate norm.

\subsection{Approximating matrix multiplication by random sampling}

We will start by considering a very simple randomized algorithm to 
approximate the product of two matrices.  
Matrix multiplication is a fundamental linear algebraic problem, and this 
randomized algorithm for it is of interest in its own right.
In addition, this algorithm is of interest since matrix multiplication is a 
primitive that is used---often ``under the hood'' or within the 
analysis---for many many other randomized algorithms for many many other 
matrix problems, and thus this algorithm will appear---either explicitly or 
implicitly, e.g., within the analysis---throughout the course.

The problem is the following: given an arbitrary $m \times n$ matrix $A$ and
an arbitrary $n \times p$ matrix $B$, compute, exactly or approximately, the
product $AB$.
As a starting point, the well-known three-loop algorithm to solve this 
problem is the following.

\begin{algorithm}
\caption{Vanilla three-look matrix multiplication algorithm.}
\label{alg:three-loop}
\begin{algorithmic}[1]
    \REQUIRE An $m \times n$ matrix $A$ and an $n \times p$ matrix $B$
    \ENSURE  The product $AB$
    \FOR{$i = 1$ to $m$}
       \FOR{$j = 1$ to $p$}
          \STATE $(AB)_{ij}=0$
          \FOR{$k = 1$ to $n$}
             \STATE $(AB)_{ik} \, +\!= \, A_{ij} B_{jk}$     
          \ENDFOR
       \ENDFOR
    \ENDFOR
    \STATE Return $AB$
\end{algorithmic}
\end{algorithm}

The running time of this algorithm is $O(mnp)$ time, which is $O(n^3)$ 
time if $m=n=p$.
Note in particular that this algorithm loops over all pairs of elements 
in the product matrix and computes that element as a dot product or inner
product between the $i$-th row of $A$ and the $j$-th column of $B$.

The question of interest here is: can we solve this problem more quickly?
There has been a lot of work on Strassen-like algorithms, which say 
that one can use a recursive procedure to decrease the running time to 
$o(n^3)$ time.
For a range of reasons, these algorithms are rarely-used in practice.
They will not be our focus; but, since some of our randomized algorithms will
call traditional algorithms as black boxes, Strassen-like algorithms can 
be used to speed up running times of those black boxes, theoretically at
least.

Here, we will consider a different approach: a randomized algorithm that 
randomly samples columns and rows of the matrices $A$ and $B$.
The key insight is that one should \emph{not} think of matrix multiplication 
as looping over elements in the product matrix and computing, say, the
$(i,j)$th element of the product matrix as the dot product between the 
$i$-th \emph{row} of $A$ and the $j$-th \emph{column} of $B$, as is common.
That is, the usual perspective is that the elements of the product matrix 
should be viewed as 
$$
(AB)_{ik} = \sum_{j=1}^{n} A_{ij}B_{jk} = A_{(i)}B^{(j)}  ,
$$
where each $A_{(i)}B^{(j)} \in \mathbb{R}$ is a number, computed as the
inner product of two vectors in $\mathbb{R}^{n}$.
Instead of this, one should think of matrix multiplication as returning a 
matrix that equals the sum of outer products of \emph{columns} of $A$ and 
the corresponding \emph{rows} of $B$, i.e., as the sum of rank-one matrices.
Recall that
\begin{equation}
\label{eqn:rank-one-sum}
AB = \sum_{i=1}^{n} A^{(i)} B_{(i)} ,
\end{equation}
where each $A^{(i)} B_{(i)} \in \mathbb{R}^{m \times p} $ is an $m \times p$ 
rank-one matrix, computed as the outer product of two vectors in 
$\mathbb{R}^{n}$. 

Viewing matrix multiplication as the sum of outer products \emph{suggests}, 
by analogy with the sum of numbers, that we should sample rank-$1$ 
components, to minimize their size, according to their size.
Recall that, if we were summing numbers, that we could sample (and 
rescale---see below) according to any probability distribution, and in 
particular the uniform distribution, and obtain an unbiased estimator of 
the sum; but that if we want to minimize the variance of the estimator that
we should sample (and rescale) according to the size or magnitude of the 
numbers.
Well, the same is true in the case of matrices.
Since the role of these probabilities will be important in what follows, 
we will leave them unspecified as input to this algorithm, and we will return
below to what probabilities should or could be used in this algorithm.

Given that background, here is our \texttt{BasicMatrixMultiplication} algorithm.

\begin{algorithm}
\caption{The \texttt{BasicMatrixMultiplication} algorithm.}
\label{alg:basic-mm-alg}
\begin{algorithmic}[1]
    \REQUIRE An $m \times n$ matrix $A$, an $n \times p$ matrix $B$, a 
             positive integer $c$, and probabilities $\{p_i\}_{i=1}^{n}$.
    \ENSURE  Matrices $C$ and $R$ such that $CR \approx AB$
    \FOR{$t = 1$ to $c$}
       \STATE Pick $i_t \in \{1,\ldots,n\}$ with probability 
                   $\Probab{i_t=k}=p_k$, in i.i.d. trials, with replacement
       \STATE Set $C^{(t)} = A^{(i_t)}/\sqrt{cp_{i_t}}$ and
              $R_{(t)} = B_{(i_t)}/\sqrt{cp_{i_t}}$.
    \ENDFOR
    \STATE Return $C$ and $R$.
\end{algorithmic}
\end{algorithm}

Basically, what we want to show for this algorithm is that
\begin{eqnarray*}
AB &=& \sum_{i=1}^{n} A^{(i)}B_{(i))} \\
   &\approx& \frac{1}{c} \sum_{t=1}^{c} \frac{1}{p_{i_t}} A^{(i_t)}B_{(i_t)} \\
   &=& CR  .
\end{eqnarray*}
In particular, we will want to show that $\XNorm{AB-CR}$ is small, for
appropriate matrix norms $\xi$.
Not surprisingly, the extent to which we will be able to do this will 
depend strongly on the sampling probabilities $\{p_i\}_{i=1}^{n}$ and the
number of samples $c$, in ways that we will describe in detail.

For much of what follows, it will be convenient to express this and 
related subsequent algorithms in a standardized matrix notation that we will 
call the \emph{sampling matrix formalism}.
To do so, let $S \in \mathbb{R}^{n \times c}$ be a matrix such that 
$$
S_{ij} = \left\{ \begin{array}{l l}
            1 & \quad \text{if the $i$-th column of $A$ is chosen in the 
                            $j$-th independent trial}\\
            0 & \quad \text{otherwise}
         \end{array} \right.
$$
and let $D \in \mathbb{R}^{c \times c}$ be a diagonal matrix such that
$$
D_{tt} = 1/\sqrt{c p_{i_t}}.
$$
With this notation, we can write the output of the 
\texttt{BasicMatrixMultiplication} algorithm and what it is doing as
$C=ASD$, $R=(SD)^TB$, and 
$$
CR = ASD(SD)^TB = A\mathcal{S}\mathcal{S}B \approx AB .
$$
Here, $\mathcal{S}=SD$ is just a way to absorb the diagonal rescaling matrix
into the sampling matrix; we will do this often, and we will often refer to 
it simply as $S$---since one nearly always rescales in a standard way when 
one samples, the meaning should be clear from context.

\subsection{Sampling probabilities and implementation issues}

For approximating the product of two matrices, as well as for all of the
other problems we will consider this semester, one can always sample 
uniformly at random.
Unfortunately, as we will show, by doing so one would do very poorly in 
general, in the sense that it is very easy to construct matrices for which 
uniform sampling will perform extremely poorly.
Thus, a central question in everything we will do will be: how should we 
construct the random sample?

Since $\{p_i\}_{i=1}^{n}$ are essentially importance sampling probabilities, 
informally we would like to choose samples that are more important.
Quantifying this is a little subtle for some of the other problems we will 
consider, but for approximating the product of two matrices, it is quite easy.
Recall that we are basically trying to approximate the product of two 
matrices by sampling randomly rank-one components in the sum
Eqn.~(\ref{eqn:rank-one-sum}).
Thus, by analogy with biasing oneself toward larger terms in the sum of 
numbers, in order to minimize variance, we would like to bias our random 
sample toward rank-one components that are larger.
The notion of size or magnitude of a rank-one matrix that we will use is the 
spectral norm of the rank-one components.
That is, we will choose columns of $A$ and rows of $B$ according to a 
probability distribution that is proportional to $\TNorm{A^{(i)}B_{(i)}}$.
Since this is a rank-$1$ matrix, this spectral norm expression takes a 
particularly simple form:
$$
\TNorm{A^{(i)}B_{(i)}} = \VTTNorm{A^{(i)}} \VTTNorm{B_{(i)}} .
$$
Note that the norm on the left is the matrix spectral norm, while the 
two norms on the right are Euclidean vector norms.
This equality is a consequence of the following simple lemma.
\begin{claim}
Let $x$ and $y$ be column vectors in $\mathbb{R}^{n}$.
Then, $\TNorm{xy^T} = \VTTNorm{x}\VTTNorm{y}$.
\end{claim}
\begin{Proof}
Recall that $\TNorm{\Omega} = \sqrt{\sigma_{max}(\Omega^T\Omega)}$, for a 
matrix $\Omega$.
Thus, 
$ \TNorm{xy^T} = \sqrt{\sigma_{max}(xy^Tyx^T)} 
               = \sqrt{\VTTNormS{x}\VTTNormS{y}} 
               = \VTTNorm{x}\VTTNorm{y} $.
\end{Proof}

Depending on what one is interested in proving, probabilities that depend on 
$A$ in $B$ in other ways might be appropriate, and in a few cases we will 
do so.  
But probabilities that depend on the spectral norm of the rank-one components
have proven to be remarkably useful in the general area of randomized 
numerical linear algebra.

With respect to a few implementation issues, here are some things to note,
when probabilities of different forms are used in the 
\texttt{BasicMatrixMultiplication} algorithm. 
\begin{itemize}
\item
Uniform sampling: one can choose which elements to keep before looking at 
the data, and so one can implement this algorithm in one-pass over the data.
\item
For nonuniform sampling, if one uses 
$p_i = \frac{\|A^{(i)}\|\|B_{(i)}\|}{\sum_{i'=1}^{n}\|A^{(i')}\|\|B_{(i')}\|}$, 
then one pass and $O(n)$ additional space is sufficient to compute the 
sampling probabilities---in that additional space, keep running totals of 
$\|A^{(i)}\|^2$ and $\|B_{(i)}\|^2$, for all $i$, and $O(m+p)$ space in the 
second pass can be used to choose the sample.
\item
For nonuniform sampling, if $B=A^T$ and one uses
$p_i = \frac{\|A^{(i)}\|^2}{\|A\|_F^2}$ as the sampling probabilities, 
then by the select lemma one needs only $O(1)$ additional space (multiplied 
by the number of samples $c$ to be selected) in the first pass to decide
which samples to draw (and still need $O(m+p)$ space in the second pass to 
keep the sample).
\end{itemize}
Actually, the comments in the last bullet are true even if $B \neq A^T$, 
i.e., if we sample based on the norms-squared of $A$ and completely ignore 
information in $B$.
We will see that permitting this flexibility is very helpful in certain 
situations.

\subsection{Initial results for approximating the product of two matrices}
  
Here is our first result for the quality of approximation of the
\texttt{BasicMatrixMultiplication} algorithm.
This lemma holds for any set of sampling probabilities $\{p_i\}_{i=1}^{n}$.
It states that $CR$ is an unbiased estimator for $AB$, element-wise, and it 
provides an expression for the variance of that estimator that depends on 
the probabilities that are used.

\begin{lemma}
\label{lem:multexpvar}
Given matrices $A$ and $B$, construct matrices $C$ and $R$ with the 
\textsc{BasicMatrixMultiplication} algorithm.
Then, 
$$
\Expect{(CR)_{ij}}=(AB)_{ij}
$$ 
and
$$
\Varnce{(CR)_{ij}} = \frac{1}{c}\sum_{k=1}^n \frac{A_{ik}^2 B_{kj}^2}{p_k}
                   - \frac{1}{c}(AB)_{ij}^2   .
$$ 
\end{lemma}
\begin{Proof}
Fix $i,j$. For $t=1,\dots,c$, define
$
X_t = \left( \frac{ A^{(i_t)}B_{(i_t)} }{ cp_{i_t} } \right)_{ij} 
    =        \frac{ A_{ii_t} B_{i_tj}  }{ cp_{i_t} }
$. 
Thus,
$$
\Expect{X_t}   = \sum_{k=1}^n p_k \frac{ A_{ik}B_{kj} }{ cp_k } 
               = \frac{1}{c}(AB)_{ij}   
\mbox{ \ \ and \ \ }
\Expect{X_t^2} = \sum_{k=1}^n \frac{ A_{ik}^2 B_{kj}^2 }{ c^2 p_k }    .
$$
Since by construction
$(CR)_{ij} = \sum_{t=1}^c X_t$, we have
$
\Expect{(CR)_{ij}} = \sum_{t=1}^c \Expect{X_t}
                   = (AB)_{ij}
$.
Since $(CR)_{ij}$ is the sum of $c$ independent random variables,
$
\Varnce{(CR)_{ij}} = \sum_{t=1}^c \Varnce{X_t}
$.
Since $\Varnce{X_t} = \Expect{X_t^2}-\Expect{X_t}^2$ 
we see that 
$$
\Varnce{X_t} = \sum_{k=1}^n \frac{ A_{ik}^2 B_{kj}^2 }{ c^2 p_k }
             - \frac{1}{c^2}(AB)_{ij}^2
$$ 
and the lemma follows.
\end{Proof}

Given Lemma~\ref{lem:multexpvar}, we can provide an upper bound on 
$\Expect{\FNormS{AB-CR}}$ and note how this measure of the error depends 
on the $p_i$'s.

\begin{lemma}
\label{lem:basicmult} 
Given matrices $A$ and $B$, construct matrices $C$ and $R$ with the 
\textsc{BasicMatrixMultiplication} algorithm.
Then,
\begin{equation}
\Expect{\FNormS{AB-CR}} 
   = \sum_{k=1}^n \frac{\VTTNormS{A^{(k)}}\VTTNormS{B_{(k)}}}{c p_k} 
     - \frac{1}{c} \FNormS{AB}   .
\end{equation}
Furthermore, if 
\begin{equation}
p_k = \frac{ \VTTNorm{A^{(k)}} \VTTNorm{B_{(k)}} }{ \sum_{k^\prime=1}^n \VTTNorm{A^{(k^\prime)}} \VTTNorm{B_{(k^\prime)}} }   ,
\label{optimal_probs}
\end{equation}
then 
\begin{equation}
\Expect{\FNormS{AB-CR}} = \frac{1}{c}\left( \sum_{k=1}^n \VTTNorm{A^{(k)}}\VTTNorm{B_{(k)}} \right)^2-\frac{1}{c}\FNormS{AB}   .
\end{equation}
\end{lemma}
\begin{Proof}
First, note that
$$
\Expect{\FNormS{AB-CR}} = \sum_{i=1}^m \sum_{j=1}^p \Expect{\left(AB-CR\right)_{ij}^2} 
                        = \sum_{i=1}^m \sum_{j=1}^p \Varnce{(CR)_{ij}}     .
$$
Thus, from Lemma \ref{lem:multexpvar} it follows that
\begin{eqnarray*}
\Expect{\FNormS{AB-CR}} 
                        &=& \frac{1}{c}\sum_{k=1}^n \frac{1}{p_k} \left(\sum_i A_{ik}^2\right)\left(\sum_j B_{kj}^2\right) -\frac{1}{c}\FNormS{AB} \\ 
                        &=& \frac{1}{c}\sum_{k=1}^n \frac{1}{p_k} \VTTNormS{A^{(k)}} \VTTNormS{B_{(k)}}-\frac{1}{c}\FNormS{AB}   .
\end{eqnarray*}
If the value
$p_k = \frac{ \VTTNorm{A^{(k)}} \VTTNorm{B_{(k)}} }{ \sum_{k^\prime=1}^n \VTTNorm{A^{(k^\prime)}} \VTTNorm{B_{(k^\prime)}} }$
is used in this expression, then 
\begin{eqnarray*}
\Expect{\FNormS{AB-CR}} &=& \frac{1}{c}\left(\sum_{k=1}^n \VTTNorm{A^{(k)}} \VTTNorm{B_{(k)}}\right)^2-\frac{1}{c}\FNormS{AB}   .
\end{eqnarray*}
\end{Proof}

Finally, we can provide the following statement of the manner in which the 
sampling probabilities of the form Eqn.~(\ref{optimal_probs}) are optimal.
Basically, they minimize the expectation of the Frobenius norm of the 
error, and this is equal to the sum of the variances of all of the elements
of the product matrix.

\begin{lemma}
\label{lem:probs-optimize-var}
Sampling probabilities $\{p_i\}_{i=1}^{n}$ of the form 
Eqn.~(\ref{optimal_probs}) minimize $\Expect{\FNormS{AB-CR}}$.
\end{lemma}
\begin{Proof}
To prove that this choice for the $p_k$'s minimizes 
$\Expect{\FNormS{AB-CR}}$ define the function 
$$
f(p_1,\dots p_n) = \sum_{k=1}^n \frac{1}{p_k}\VTTNormS{A^{(k)}}\VTTNormS{B_{(k)}},
$$
which characterizes the dependence of $\Expect{\FNormS{AB-CR}}$ on the $p_k$'s. 
To minimize $f$ subject to $\sum_{k=1}^n p_k =1$, introduce 
the Lagrange multiplier $\lambda$ and define the function 
$$
g(p_1,\dots p_n) = f(p_1,\dots p_n) + \lambda\left(\sum_{k=1}^n p_k-1\right)  .
$$
We then have at the minimum that
$$
0 = \frac{\partial g}{\partial p_i} 
  = \frac{-1}{p_i^2}\VTTNormS{A^{(i)}}\VTTNormS{B_{(i)}} + \lambda   .
$$
Thus, 
$$
p_i = \frac{\VTTNorm{A^{(i)}}\VTTNorm{B_{(i)}}}{\sqrt{\lambda}} 
    = \frac{\VTTNorm{A^{(i)}}\VTTNorm{B_{(i)}}}
           {\sum_{i^\prime=1}^n\VTTNorm{A^{(i^\prime)}}\VTTNorm{B_{(i^\prime)}}} ,
$$
where the second equality comes from solving for $\sqrt{\lambda}$ in 
$\sum_{k=1}^{n-1} p_k = 1$.
That these probabilities are a minimum follows since 
$\frac{\partial^2 g}{{\partial p_i}^2} > 0$ 
$\forall i \mbox{ s.t. } \VTTNormS{A^{(i)}}\VTTNormS{B_{(i)}} > 0$.
\end{Proof}

A few comments about this result.
\begin{itemize}
\item
Many of these results are quite robust to the exact form of the sampling 
probabilities.
For example, if they are a factor of $2$ from optimal, then we get similar 
results if we sample a factor or $2$ or so more.
This flexibility will be important for what we do, and so we will provide 
a precise form of this result next time.
\item
We can use Markov's inequality to ``remove the expectation'' from this bound, 
and in some cases this will be good enough.
In general, however, the sampling complexity will be bad with respect to 
$\delta$, the failure probability parameter.
This will be important for what we do, and so we will spend time to develop
heavier-duty methods to do much better with respect to $\delta$.
\item
This bound is for the Frobenius norm; we will get to the spectral norm later.
Of course, the spectral norm is upper bounded by the Frobenius norm.
Although we loose something in the process, in some cases this won't matter
so much, in the sense that we still get somewhat meaningful results, but 
in other cases it will matter a lot.
For this reason, we will spend time to develop heavier-duty methods to get 
spectral norm bounds.
\item
We can obtain similar but slightly weaker (typically, in terms of the concentration) if we perform the sampling with respect to one or the other matrix, but not both, or if we consider other variants of this basic algorithm.
These will be important for some of our subsequent developments, and so we will revisit them later.
\end{itemize}

\newpage

\section{%
(09/11/2013):  Scalar and Matrix Concentration}

Here, we will give an aside on probabilities, and in particular on various
ways to establish what is know as concentration.
Given some information about a distribution, e.g., its mean or its variance 
or information about higher moments, there are various ways to establish 
bounds on the tails of sums of random variables from that distribution.
That is, there are various ways to establish that estimates are close to 
their expected value, with high probability.  
Today and next time, we will cover several of these methods, both in the 
case where the random variables are scalar or real-valued and when the 
random variables are matrix-valued.
The former can be used to bound that latter, e.g., by bounding every 
element of the random matrix individually, but the latter often provide
tighter bounds in those cases.

Perhaps the simplest method to establish concentration goes by the name 
Markov's inequality.
Although tight, given only information about the mean of a nonnegative 
random variable, it generally gives bounds that are too weak for what we will   
want.
Thus, we will consider more sophisticated method that go by the name Chernoff
(and related) bounds, that provide much stronger bounds, and that amount to 
applying Markov's inequality on the moment generating function of the random 
variable of interest.
Since the Frobenius norm of a matrix is the sum of all the elements of the
matrix, this method---or actually an extension of this basic idea that 
applies to non-independent random variables and that goes by the name 
Hoeffding-Azuma---applied to real-valued random variables will use to bound 
the Frobenius norm of the error in our approximate matrix multiplication 
result.
To get bounds for the spectral norm (tighter than provided by bounding the 
spectral norm by the Frobenius norm), we will have to do something a little 
more sophisticated.
Basically, we will have to consider the analogues of these results for 
matrix-valued random variables.
This will provide tighter bounds, but at the expense of heavier machinery.

There is no particular reading for today, but if the material was too foreign, then take a look at the following reference.
\begin{itemize}
\item
The first few chapters of ``Probability and Computing,'' by Mitzenmacher and Upfal.
\end{itemize}

\subsection{Scalar concentration: Markov's inequality} 

We will start with the following well-known result, known as \emph{Markov's 
inequality}, which provides a bound on the probability that a nonnegative
random variable deviates (on the high side) from its expectation.
Its proof is standard, but we include it for comparison with the 
operator-valued Markov inequality below.

\begin{lemma}[Markov's Inequality]
Let $X$ be a real-valued random variable such that $X \ge 0$.
Then, $\forall a \ge 0$, $\Probab{X \ge a} \le \frac{\Expect{X}}{a}  $.
\end{lemma}
\begin{Proof}
For $a > 0$, let 
$$
\mathcal{X} = \left\{ \begin{array}{ll}
                  1 & \quad \text{if $X\ge a$}\\
                  0 & \quad \text{otherwise}
               \end{array} \right.
$$ 
be the indicator variable for the event that $X\ge a$; 
and note that, since $X\ge a$, it follows that $\mathcal{X} \le \frac{X}{a}$.
Since $\mathcal{X}$ is a $0$-$1$-valued random variable, it follows that
$ \Expect{\mathcal{X}} = \Probab{\mathcal{X}=1} = \Probab{X \ge a} $, 
from which it follows that
$\Probab{X \ge a} =   \Expect{\mathcal{X}} 
                  \le \Expect{\frac{X}{a}} 
                  =   \frac{\Expect{X}}{a}$.
\end{Proof}

\subsection{Applying this to approximate matrix multiplication}

As an aside, consider how this result can be applied back to our bound
from last class on approximating the product of two matrices.
We can use it to provide a result that holds with fixed probability, given
out result in expectation.
To see this, recall that last time we showed that
$$
\Expect{\FNormS{AB-CR}} \le \frac{1}{\beta c}\FNormS{A}\FNormS{B} .
$$
We could remove the expectation from this quantity, but for a cleaner 
comparison with results we will derive below, let's use Jensen's Inequality 
(if $\phi(\cdot)$ is a convex function, then 
$\phi(\Expect{X}) \le \Expect{\phi(X)}$) to get
\begin{equation}
\label{eqn:error-premarkov}
\Expect{\FNorm{AB-CR}} \le \frac{1}{\sqrt{\beta c}}\FNorm{A}\FNorm{B} 
\end{equation}
From this it follows that if the number of samples
$c \ge \beta/\epsilon^2$, then
$ \Expect{\FNorm{AB-CR}} \le \epsilon \FNorm{A}\FNorm{B} $.
To ``remove the expectation'' from Eqn.~(\ref{eqn:error-premarkov}) with 
Markov's inequality, let's set the failure probability to be $\delta$ as 
follows:
$$
\delta 
   = \Probab{ \FNorm{AB-CR} > \frac{\alpha}{\sqrt{\beta c}}\FNorm{A}\FNorm{B} },
$$
where $\alpha$ is a parameter that will determine the how $c$ needs to be
chosen to get a fixed error probability.
Thus, 
\begin{eqnarray*}
\delta \le \frac{\Expect{\FNorm{AB-CR}}}
                  {\frac{\alpha}{\sqrt{\beta c}}\FNorm{A}\FNorm{B}} 
       \le \frac{1}{\alpha}  ,
\end{eqnarray*}
where the first inequality follows by Markov's inequality, and where the
second inequality follows from our result from 
Eqn.~(\ref{eqn:error-premarkov}).
Thus, $\alpha = \frac{1}{\delta}$; and so it follows that 
with probability $\ge 1-\delta$
\begin{eqnarray*}
\FNorm{AB-CR} &\le& \frac{1}{\sqrt{\delta^2\beta c}}\FNorm{A}\FNorm{B} \\
              &\le& \epsilon \FNorm{A}\FNorm{B} ,
\end{eqnarray*}
where the second inequality holds if $c \ge \frac{\beta}{\delta^2\epsilon^2}$.

That dependency on $\delta$ might not be a problem if one wants to choose $\delta = 0.1$.
(Actually it might still be, since it implies that one needs to choose $c$ to be a factor of $100$ larger, but in cases where we just need some event to hold with constant probability, it is typically fine.)
But, it certainly is a problem if one wants, say, $\delta = 10^{-6}$.
In theses cases, we would like the dependence of $c$ on $\delta$ to be $O(\log(1/\delta))$, rather than $\mbox{poly}(1/\delta)$.
For this we need heavier machinery.

(Note that a popular and standard way to boost the success probability is to oversample by a factor of $O(\log(1/\delta))$, if concentration is supported, or to repeat the trials $O(\log(1/\delta))$ times and keep the best result, if there is an easy way to check which of the trials is the best.
These methods won't work if concentration isn't supported and/or if there is not an easy way to check which of the trials is the best.)

\subsection{More scalar concentration}

First, recall the definition of the variance of a random variable.
\begin{definition}
$\Varnce{X} = \Expect{(X-\Expect{X})^{2}}
            = \Expect{X^2}-(\Expect{X})^2$
and $\mbox{StdDev}(X) = \sigma(X) = \sqrt{\Varnce{X}}$.
\end{definition}

Let's say we have information on the variance of $X$, e.g., we might have a 
bound on $\Varnce{X}$.
In this case, we can get stronger result using Chebyshev's inequality.

\begin{lemma}[Chebyshev's Inequality]
$\forall A>0$, 
$\Probab{ | X-\Expect{X} | \ge a } \le \frac{\Varnce{X}}{a^2}$.
\end{lemma}
\begin{Proof}
First, observe that 
$$
\Probab{ | X-\Expect{X} | \ge a } = \Probab{ ( X-\Expect{X} )^2 \ge a^2 } .
$$
Then, since $( X-\Expect{X} )^2$ is a nonnegative random variable, we can 
apply Markov's inequality to get that
$$
\Probab{ | X-\Expect{X} | \ge a } \le \frac{\Expect{( X-\Expect{X} )^2}}{a^2}
                                  = \frac{\Varnce{X}}{a^2} .
$$
\end{Proof}

Of course, there are other variants of Chebyshev's Inequality that are 
parameterized slightly differently.
For example, that $\forall t > 0$, 
$$
\Probab{ | X-\Expect{X} | \ge t \sigma(X) } 
   \le \frac{1}{t^2} ,
$$
or that $\forall t > 0$, 
$$
\Probab{ | X-\Expect{X} | \ge t \Expect{X}} 
   \le \frac{\Varnce{X}}{t^2(\Expect{X})^2} .
$$

Unfortunately, however, this is usually still not good enough for what we 
want.
But, if we have bounds on higher moments of the random variable, then we can 
use ``moment generating function methods'' to get much stronger results that
are qualitatively more like what one would get for the tail behavior of 
Gaussian random variables.
In particular, Chernoff-style bounds are very powerful, providing 
exponentially-decreasing bounds on the tails of the distribution.
Chernoff-style bounds do this basically by applying Markov's inequality on 
the moment generating function of a random variable.

\begin{definition}
The \emph{moment generating function} of a random variable $X$ is
$M_{X}(t) = \Expect{e^{tX}}$.
\end{definition}
We will mostly be interested in the existence and properties of this function
in the neighborhood of $t=0$.
The basic idea of Chernoff-style  bounds is to apply Markov's inequality to 
$e^{tX}$ for a well-chosen value of $t$.
\begin{lemma}[Vanilla Chernoff]
$\forall t > 0$:
$ \Probab{X \ge a} 
     = \Probab{e^{tX} \ge e^{ta} } 
     \le \frac{\Expect{e^{tX}}}{e^{ta}}$.
In particular,
$ \Probab{X \ge a} \le \min_{t>0} \frac{\Expect{e^{tX}}}{e^{ta}}$.
\end{lemma}

The same holds for the other direction.
\begin{lemma}[Vanilla Chernoff, other direction]
$\forall t < 0$:
$ \Probab{X \le a} 
     = \Probab{e^{tX} \le e^{ta} } 
     \le \frac{\Expect{e^{tX}}}{e^{ta}}$.
In particular,
$ \Probab{X \le a} \le \min_{t>0} \frac{\Expect{e^{tX}}}{e^{ta}}$.
\end{lemma}
There are a lot of variants of this basic result, depending on what is known
about the given distribution, how tight a bound one can provide for 
$\Expect{e^{tX}}$, etc.
Here are two versions.
\begin{theorem}[Hoeffding]
Let $\{X_i\}_{i=1}^{n}$ be r.v. such that $X_i \in [a_i,b_i]$, for all $i$, 
and let $X=\sum_{i=1}^{n} X_i$.
Then
$$
\Probab{ |X-\Expect{x}| \ge t }
   \le 2 \exp\left( \frac{-2t^2}{\sum_{i=1}^{n} (b_i-a_i)^2} \right)
$$
\end{theorem}
\begin{theorem}[Bernstein]
Let $\{X_i\}_{i=1}^{n}$ be r.v. such that $\Expect{X}$, 
$\Expect{X^2}=\sigma^2$, $|X|\le M$, $X_i$ are i.i.d. copies.
Then, for all $t > 0$, 
$$
\Probab{ |X-\Expect{x}| \ge t }
   \le \exp\left( \frac{-t^2}{2n\sigma^2 + \frac{4}{3}tM} \right)
$$
\end{theorem}

So, Chernoff-type results provide bounds for large deviations using higher 
moments for sums of independent random variables.

When applied to approximate matrix multiplication, we will be interested 
in providing both spectral and Frobenius norm bounds for the error.
In order to provide spectral norm bounds for randomized matrix 
multiplication, we will need to use matrix versions of these Chernoff-style 
results; but to prove the Frobenius norm bounds we will need an extension 
that deals with random variables that are not quite independent.
These latter extensions are known as Hoeffding-Azuma bounds, and they 
provide Chernoff-like bounds, with no independence assumptions, but assuming 
some sort of bounded difference form holds.
We will do this in a standard way, by construction, and construct a 
martingale difference sequence with differences that are bounded in absolute
value.
To describe how we will do this, we start with the definition of a 
martingale.

\begin{definition}
A sequence of random variables $Z_0,Z_1,\ldots$ is a \emph{martingale with
respect to a sequence} $X_0, X_1, \ldots$ if, $\forall n \ge 0$,
\begin{itemize}
\item
$Z_n = Z_n(X_0,X_1,\ldots,X_n)$, i.e., it is a function of the $X_i$
\item
$\Expect{|Z_n|} < \infty$
\item
$\Expect{Z_{n+1} | X_0 \cdots X_n } = Z_n$
\end{itemize}
A sequence is a \emph{martingale} if it is a martingale with respect to itself.
\end{definition}

The canonical example of a martingale is a gambler who plays a sequence of 
fair games.
In this case, 
let $X_i$ be the amount that the gambler wins in the $i^{th}$ game; and 
let $Z_i$ be the total winnings of the gambler after the $i^{th}$ game.
Since the game is fair, $\Expect{X_i}=0$; and also
$$
\Expect{Z_{i+1} | X_1 \cdots X_i } = Z_i + \Expect{X_{i+1}} = Z_i .
$$
So, in this case, $Z_i$ is a martingale with respect to $X_i$.

Martingales are powerful and ubiquitous in applications of probability, and 
in particular in the area of randomized algorithms, largely since they can 
be formed by nearly ``any'' random variable.
We will use the usual approach, that in particular is common in the theory
of algorithms.
It involves constructing something called the Doob martingale.

\begin{definition}
Let $X_0,X_1,\ldots,X_n$ be a sequence of random variables, and let $Y$ 
be a random variable with $\Expect{|Y|} < \infty$.
(Generally, $Y$ is such that $Y=Y(X_1,\ldots,X_n)$.
Then, consider 
$$
Z_i = \Expect{Y|X_0 \cdots X_i} \quad i=0,1,\ldots,n  .
$$
This is a \emph{Doob Martingale}.
\end{definition}

The first thing to note is that this is a martingale.
\begin{lemma}
$Z_i$ constructed in this way is a martingale w.r.t. $X_i$.
\end{lemma}
\begin{Proof}
\begin{eqnarray*}
\Expect{Z_{i+1}|X_0 \cdots X_i }
   &=& \Expect{ \Expect{ Y | X_0 \cdots X_{i+1}} | X_0 \cdots X_i }  \\
   &=& \Expect{ Y | X_0 \cdots X_i } \\
   &=& Z_i  ,
\end{eqnarray*}
where the second of those inequalities used that 
$\Expect{ Y | X_0 \cdots X_{i+1}}$ is a random variable, and that in 
general $\Expect{v|w} = \Expect{ \Expect{v|u,w} | w }$.
\end{Proof}

In most applications, one starts the Doob martingale with $Z_0=\Expect{Y}$, 
which corresponds to $X_0$ being a trivial random variable, independent of 
$Y$; and then assume that we want to predict the value of $Y$ and that
$Y=Y(X_1,\ldots,X_n)$, then the sequence $Z_0,\ldots,Z_n$ is a sequence of 
more and more refined estimates of the value of $Y$, with $Z_0=\Expect{Y}$, 
going all the way to $Z_n=Y$, in which case we fully know the value of the
random variable.

There are a lot of applications of this idea.  
One big application of Doob martingales in the theory of algorithms is for 
the analysis of algorithms via Chernoff-like tail inequalities, that can 
apply, even when the underlying random variable is not independent.
The basic form goes by the name Azuma-Heoffding.

\begin{theorem}[Azuma-Hoeffding]
Let $X_0, \ldots, X_n$ be a martingale such that $|X_k-X_{k-1}|\le c_k$.
Then, $\forall t \ge 0, \lambda > 0$,
$$
\Probab{ |X_t-X_0| \ge \lambda }
   \le 2 \exp \left( \frac{-\lambda^2}{\sum_{k=1}^{t}c_k^2} \right) .
$$
\end{theorem}

As a corollary, if $|X_k-X_{k-1}| \le c$, then 
$$
\Probab{ |X_t-X_0| \ge \lambda c \sqrt{t} }
   \le 2 \exp \left( -\lambda^2/2 \right) .
$$

\subsection{An aside on SPSD matrices}

Recall that, although Azuma-Hoeffding is what we will use to get 
concentration for the Frobenius norm error of approximate matrix 
multiplication, we will use matrix analogues of Chernoff-style bounds to get 
concentration for the spectral norm error of approximate matrix 
multiplication.
Although these bounds will apply to general matrices---by a relatively 
straightforward trick---they are most easily formulated in terms of 
symmetric positive semi-definite (SPSD) matrices, and so we will review the 
properties of that class of matrices, starting with the definition.

\begin{definition}
A matrix $A\in\mathbb{R}^{n \times n}$ is SPSD if
\begin{itemize}
\item
$x^T A x \ge 0 \quad \forall x\in\mathbb{R}^{n}$
\item
$A = U \Sigma U^T$ is the eigendecomposition, with all $\sigma_i \ge 0$.
\end{itemize}
And it is SPD (symmetric positive definite) if those inequalities are tight, 
i.e., they are strict with no equalities.
\end{definition}

There are several things to note about this definition.
\begin{itemize}
\item
This is a generalization of $\mathbb{R}$ to matrices; and some, but certainly 
not all, properties of real numbers extend to SPSD matrices.
\item
One can define a partial order over the ``cone'' of SPSD matrices:
$A \succ B$ iff $A-B \succ 0$ iff $A-B \in SPD$ and
$A \succeq B$ iff $A-B \succeq 0$ iff $A-B \in SPSD$.
This is not a total order unless we are dealing with $1 \times 1$ matrices, 
i.e., real numbers.
\item
$A \succeq 0$ iff all the eigenvalues are nonnegative.
This is a set of $d$ nonlinear inequalities; but it can be viewed as 
$\infty$-ly many linear inequalities, since $A \succeq 0$ iff 
$\forall \rho$ that are PSD matrices of trace $1$, also known as 
``density operators,'' $\Trace{\rho A} \ge 0$) iff $\forall \pi$ that are 
one-dimensional 
projectors, $\Trace{\pi A} \ge 0$.
\end{itemize}

Since SPSD matrices are a generalization of $\mathbb{R}$, one can generalize
many real functions to them.
In particular, given a function $f:\mathbb{R}\rightarrow\mathbb{R}$, one can:
\begin{itemize}
\item
Define a map on diagonal matrices by applying $f$ to each diagonal entry
\item
Extend $f$ to self-adjoint/Hermitian/symmetric matrices via the eigenvalue 
decomposition 
$f(A) = Q f(\Lambda) Q^T$, where $A=Q \Lambda Q^T$.
\item
Then, the spectral mapping theorem says:
each eigenvalue of $f(A)$ is equal to $f(\lambda)$, for some eigenvalue 
$\lambda$ of $A$.
\end{itemize}
The point is that symmetric and SPSD matrices are \emph{much} more structured
objects than general matrices, and you get much nicer and cleaner results.
We will see the same things for this in general versus symmetric/SPSD matrix
perturbation results, where in the latter case we get much better results.
This is familiar to NLA people, so others don't think general matrices are
so nice.
Also, a lot of data matrices are symmetric or SPSD, or we are interested in
robustness, and so we consider singular vectors/values, rather than eigen
vectors/values via $A \rightarrow A^TA \quad\mbox{and}\quad AA^T$.

We can define the exponential of s.a. matrix $A$ by the spectral mapping
theorem, with $f(\lambda) = e^{\lambda}$, or we can define it as 
$$
\exp(A) = I + \sum_{i=1}^{\infty} \frac{A^i}{i!} 
        = \sum_{i=0}^{\infty} \frac{A^i}{i!}  .
$$
By the spectral mapping theorem, we know that this matrix is PD.
Note that $I+A \preceq e^A$.
Also, note that this expansion of symmetric matrices is more generally-used 
in machine with kernel methods.

Here is a fact.
\begin{fact}
$\Trace{\exp(A)}$ is a convex function; and it is monotone with respect to 
the semidefinite order, i.e., 
$A \preceq B \quad\rightarrow\quad \Trace{\exp(A)} \le \Trace{\exp(B)}$.
Note that the first inequality is over SPSD matrices, while the second 
inequality is over numbers.
\end{fact}

We can define the logarithm as the functional inverse of the matrix 
exponentials: 
$$
\log(\exp(A)) \equiv A \quad\forall\quad\mbox{s.a./Hermitian/symmetric}\quad A  .
$$

\begin{definition}
A function $f$ is \emph{operator monotone} with respect to the semidefinite
order if $0 \preceq A \preceq B$ implies that $f(A) \preceq f(B)$.
(Note that both inequalities are inequalities over SPSD matrices.)
A function $f$ is \emph{operator concave} with respect to the semidefinite
order if $c f(A) + (1-c)f(B) \preceq f(cA+(1-c)B) $, for all PD $A,B$ and for
all $c\in[0,1]$.
\end{definition}
These generalize properties for the analogous things for real numbers, but 
note that operator monotone and operator convex functions are not so common.
But here are some examples we will encounter.
\begin{itemize}
\item
$A \rightarrow \log(A)$ 
and also 
$A \rightarrow A^s$, for $s\in[0,1]$,
are operator monotone and operator con\emph{cave}. 
\item
$A \rightarrow \exp(A)$
and also
$A \rightarrow A^s$, for $s>2$,
are neither operator monotone nor operator con\emph{vex}.
\item
As an aside, 
$A \rightarrow A^s$, for $s\in[0,1]$,
is operator convex, but is not operator monotone.
\end{itemize}
In particular, $A\rightarrow A^{1/2} = \sqrt{A}$ \emph{is} operator monotone, 
which is a fact we will use below.

We should note that, although we are working with symmetric matrices, many
of the results extend easily to general matrices.
In particular, given a rectangular matrix $A\in\mathbb{R}^{m \times n}$, we 
can define a matrix $B\in\mathbb{R}^{(m+n)\times(m+n)}$ as
$$
B = \left[\begin{array}{cc}
       0 & A \\
       A^T & 0 \\
    \end{array}\right]   .
$$
Then, 
$$
B^2 = \left[\begin{array}{cc}
         AA^T & 0 \\
         0 & A^TA \\
      \end{array}\right]   .
$$
And the eigen-vectors/values of $B$ are related to the eigen-vectors/values 
or $A$.
For the eigenvectors, see the first homework.
For the eigenvalues, we will need that 
$\lambda_{max}(B) = \TNorm{B} = \TNorm{A} $.

Two other things to note.
\begin{itemize}
\item
Expectation is a convex combination and the PSD cone is convex, 
so $X \preceq Y \mbox{ a.s. } \rightarrow \Expect{X} \preceq \Expect{Y} $.
\item
Every operator convex function admits a Jensen's inequality, and so since
the matrix square is operator convex, we have that
$\left(\Expect{X}\right)^{2} \preceq \Expect{X^2}$.
\end{itemize}

Finally, for numbers $a,b\in\mathbb{R}$, we have that $e^{a+b}=e^ae^b$.
The matrix exponential does \emph{not} convert sums into products in an
analogous way (unless the matrices commute).
But there is something weaker that will still be good enough for some 
purposes.
\begin{lemma}[Golden-Thompson Inequality]
For $A,B$ that are SPSD matrices, we have that
$$
\Trace{\exp(A+B)} \le \Trace{\exp(A)\exp(B)} .
$$
\end{lemma}

\newpage

\section{%
(09/16/2013):  Concentration and Matrix Multiplication, Cont.}

Today, we will continue with our discussion of scalar and matrix 
concentration, with a discussion of the matrix analogues of Markov's, 
Chebyshev's, and Chernoff's Inequalities.  
Then, we will return to bounding the error for our approximating matrix 
multiplication algorithm.
We will start with using Hoeffding-Azuma bounds from last class to get 
improved Frobenius norm bounds, and then (next time) we will describe how 
to use the matrix concentration results to get spectral norm bounds for 
approximate multiplication.

Here is the reading for today.
\begin{itemize}
\item
Appendix of: Recht, ``A Simpler Approach to Matrix Completion'' 
\item
Oliveira, ``Sums of random Hermitian matrices and an inequality by Rudelson'' 
\item
Drineas, Kannan, and Mahoney, ``Fast Monte Carlo Algorithms for Matrices I: Approximating Matrix Multiplication''
\end{itemize}

\subsection{Matrix Concentration} 

We will now discuss several results having to do with concentration of 
matrix-valued random variables.
We start with a matrix version of the Markov inequality.

\begin{lemma}[Matrix Markov Inequality]
Let $X$ be a random PSD matrix, and let $A$ be a fixed PD matrix.
Then, $\forall A$, 
$\Probab{ X \npreceq A } \le \Trace{ \Expect{X} A^{-1} } $.
\end{lemma}
\begin{Proof}
Consider the random variable $A^{-1/2}XA^{-1/2}$.
Observe that, if $X \npreceq A$, then $A^{-1/2}XA^{-1/2} \npreceq I$.
In this case,
$$
1 < \TNorm{A^{-1/2} X A^{-1/2} } .
$$
Let $\mathcal{X}_{X \npreceq A}$ be the characteristic/indicator function 
of the event $X \npreceq A$.
Then, the claim is that 
$$
\mathcal{X}_{X \npreceq A} \le \Trace{ A^{-1/2}XA^{-1/2} }  .
$$
To prove the claim, observe that the RHS $\ge 0$.
If the LHS $=0$, then we are done.
Otherwise, if the LHS $=1$, then 
$1 < \TNorm{A^{-1/2} X A^{-1/2} } \le \Trace{ A^{-1/2}XA^{-1/2} } $.
So, 
$$
\Probab{X\npreceq A} 
   = \Expect{ \mathcal{X}_{X \npreceq A} }
   \le \Expect{ \Trace{ A^{-1/2}XA^{-1/2} } }
   = \Expect{ \Trace{ XA^{-1} } } 
   = \Trace{ \Expect{ X } A^{-1} }  ,
$$
where the second equality follows from the cyclic properties of the trace, 
and where the last follows since the trace is linear.
\end{Proof}

Although we will not use the matrix version of the Chebyshev inequality in 
what follows, we include it for completeness and for comparison with the 
scalar version.

\begin{lemma}[Matrix Chebyshev Inequality]
Let $X$ be a random PSD matrix, and let $A$ be a fixed PD matrix.
Then, $\forall A$, 
$\Probab{ | X - \Expect{X} | \npreceq A } \le \Trace{ \Varnce{X} A^{-2} } $.
Here, for any matrix symmetric matrix $B$, $|B|$ has eigenvectors the same as $B$ and eigenvalues as the absolute value of those of $B$.
\end{lemma}
\begin{Proof}
First note that $(X-\Expect{X})^{2} \preceq A^2$ implies that 
$ | X-\Expect{X} | \preceq A$.
The reason for this is that $\sqrt{\cdot}$ is operator monotone.
(I.e., while it is obvious for numbers, it is true but non-obvious for matrices.)
So,
\begin{eqnarray*}
\Probab{ | X-\Expect{X} | \npreceq A }
   &\le& \Probab{ (X-\Expect{X})^2 \npreceq A^2 } \\
   &\le& \Trace{ \Expect{ (X-\Expect{X})^2 } A^{-2} } \\
   &=& \Trace{ \Varnce{X} A^{-2} } .
\end{eqnarray*}
\end{Proof}

Next, what we really want to do is get a matrix analogue of the Chernoff 
bound.  
Here is one form of it; we will give more of a history below.

\begin{theorem}[Matrix Chernoff Bound]
Let $X_1,\ldots,X_n$ be independent symmetric random matrices in 
$\mathbb{R}^{d \times d}$, and let $A$ be a fixed PD matrix.
Then, $\forall$ invertible $d \times d$ matrices $T$,
$$
\Probab{ \sum_{k=1}^{n} X_k \npreceq nA }
   \le d \prod_{k=1}^{n} \TNorm{ \Expect{ \exp\left( TX_kT^*-TAT^* \right) } } ,
$$
where $T^*$ denotes the transpose of the (real-valued) matrix $T$.
\end{theorem}
\begin{Proof}
First, by the usual properties of the semi-definite ordering, we have that
\begin{eqnarray*}
\Probab{ \sum_{k=1}^{n} X_k \npreceq nA }
   &=& \Probab{ \sum_{k=1}^{n} (X_k-A) \npreceq 0 } \\
   &=& \Probab{ \sum_{k=1}^{n} T(X_k-A)T^* \npreceq 0 } \\
   &=& \Probab{ \exp\left( \sum_{k=1}^{n} T(X_k-A)T^* \right) \npreceq I_d }  .
\end{eqnarray*}
By combining this with the Matrix Markov Inequality, and since the trace is 
linear, it follows that
\begin{eqnarray*}
\Probab{ \sum_{k=1}^{n} X_k \npreceq nA }
   &\leq& \Trace{ \Expect{ \exp\left( \sum_{k=1}^{n} T(X_k-A)T^* \right) } } \\
   &\leq& \Expect{ \Trace{ \exp\left( \sum_{k=1}^{n} T(X_k-A)T^* \right) } } .
\end{eqnarray*}
Next, observe that we can peel apart the various terms as follows
\begin{eqnarray*}
\Probab{ \sum_{k=1}^{n} X_k \npreceq nA }
   &\leq& \Expect{ \Trace{ \exp\left( \sum_{k=1}^{n-1} T(X_k-A)T^* \right) 
                           \exp\left( T(X_n-A)T^* \right) } } \\
   &=& \Trace{ \Expect{ \exp\left( \sum_{k=1}^{n-1} T(X_k-A)T^* \right) }
                           \Expect{\exp\left( T(X_n-A)T^* \right)}_{n} } \\
   &\leq& \TNorm{ \Expect{\exp\left( T(X_n-A)T^* \right)} } 
          \Expect{ \Trace{ \exp\left( \sum_{k=1}^{n-1} T(X_k-A)T^* \right) }
},
\end{eqnarray*}
where the first line follows from the Golden-Thompson inequality; the 
second line follows from the independence of the $X_k$; and the third 
line follows by strong submultiplicitivity, i.e., since 
$\Trace{AB} \le \Trace{A}\TNorm{B}$, if $A$ and $B$ are SPSD.
By iterating this process it follows that
\begin{eqnarray*}
\Probab{ \sum_{k=1}^{n} X_k \npreceq nA }
   &\leq& \prod_{k=2}^{n} \TNorm{ \Expect{\exp\left( T(X_k-A)T^* \right)} } 
          \Expect{ \Trace{ \exp\left( T(X_1-A)T^* \right) } } \\
   &\leq& d \prod_{k=1}^{n} \TNorm{ \Expect{\exp\left( T(X_k-A)T^* \right)} }  ,
\end{eqnarray*}
where the last line follows since if $A$ is PD, then 
$\Trace{A} =\sum_{i=1}^{d} \lambda_i(A)  \le d \lambda_{max}(A)$,
where $\lambda_i(A)$ is the $i^{th}$ eigenvalue of $A$ and 
where $\lambda_{max}(A)$ is the largest eigenvalue of $A$.
\end{Proof}

We will use this Matrix Chernoff Bound to establish an inequality that we 
will use.
Note that, as in the scalar case, one can get lots of variations, and we will 
use Bernstein version due to Recht.

\begin{theorem}[Noncommutative Bernstein Inequality]
Let $X_1,\ldots,X_L$ be independent zero-mean random matrices of 
dimension $d_1 \times d_2$.
Suppose that 
$ \rho_k = \max\{ \TNorm{\Expect{X_kX_k^*}}, \TNorm{\Expect{X_k^*X_k}} \} $
and that $ \TNorm{ X_k } \le M$ a.s., for all $k$.
Then, $\forall \tau > 0$,
$$
\Probab{ \TNorm{ \sum_{k=1}^{L} X_k} > \tau }
   \le (d_1+d_2) \exp\left( \frac{-\tau^2/2}{\sum_{k=1}^{L} \rho_k^2+M\tau/2} \right)
$$
\end{theorem}

Before the proof, here are a few notes on this result.
\begin{itemize}
\item
If $d_1=d_2=1$, then this is just the $2$-sided version of the standard 
Bernstein Inequality.
\item
If $X_i$ are diagonal, then this is just the standard Bernstein Inequality 
applied and then do a union bound on the diagonal of the matrix sum.
\item
If $\tau \le \frac{1}{M} \sum_{k=1}^{L} \rho_k^2$, then
$RHS \le (d_1+d_2)\exp\left( \frac{-3\tau^2/8}{\sum_{k=1}^{L} \rho_k^2} \right)$.
\end{itemize}

\begin{Proof}
Let 
$ Y_k = \left[\begin{array}{cc}
         0 & X_k \\
         X_k^* & 0 \\
        \end{array}\right]   $.
Then, the $Y_k$ are symmetric random functions, and $\forall k$, we have that
\begin{eqnarray*}
\TNorm{\Expect{Y_k^2}}
   &=& \TNorm{\Expect{ \left[\begin{array}{cc}
                        X_kX_k^* & 0 \\
                        0 & X_k^*X_k \\
                       \end{array}\right] }}  \\
   &=& \max\{ \TNorm{\Expect{X_kX_k^*}}, \TNorm{\Expect{X_k^*X_k}} \} \\ 
   &=& \rho_k^2  .
\end{eqnarray*}
In addition, 
$\sigma_{max}(\sum_{i=1}^{L} X_k ) = \lambda_{max} ( \sum_{k=1}^{L} Y_k )$.
By the Operator Chernoff Theorem, it follows that
\begin{eqnarray*}
\Probab{ \TNorm{ \sum_{k=1}^{L} X_k} > Lt }
   &=& \Probab{ \sum_{k=1}^{L} Y_k \npreceq Lt I }  \\
   &\le& (d_1+d_2) \exp\left(-Lt\lambda\right) \Pi_{k=1}^{L} \TNorm{\Expect{\exp\left( \lambda Y_k \right)}} ,
\end{eqnarray*}
$\forall \lambda > 0$.
Then, $\forall k$, let
$Y_k = U_k \Lambda_k U_k^*$ be the eigenvalue decomposition.
Then, $\forall s> 0$, we have that 
$$
-M^s Y_k^s =   -U_k M^s \Lambda_k^2 U_k^* 
           \le  U_k \Lambda_k^{s+2} U_k^2
           =    Y_k^{s+2}
           \le  U_k M^s \Lambda_k^2 U_k^* 
           = M^s Y_k^2  ,
$$
where $M$ is such that $\TNorm{X}\le M$, forall $k$,
which implies that 
\begin{equation}
\TNorm{\Expect{ Y_k^{s+2} }} \le M^s \TNorm{\Expect{ Y_k^2 }}  .
\label{eqn:int-setp-1}
\end{equation}
For a fixed $k$, we have that 
\begin{eqnarray*}
\TNorm{\Expect{ e^{\lambda Y_k }}}
   &\le& \TNorm{I} + \sum_{j=2}^{\infty} \frac{\lambda^j}{j!} \TNorm{\Expect{ Y_k^j }} \\
   &\le& 1 + \sum_{j=2}^{\infty} \frac{\lambda^j}{j!} \TNorm{\Expect{ Y_k^2 }} M^{j-2}  \\
   &=&   1+\frac{\rho_k^2}{M^2} \sum_{j=2}^{\infty} \frac{\lambda^j}{j!} M^j \\
   &=&   1+\frac{\rho_k^2}{M^2} \left( \exp(\lambda M) -1 -\lambda M \right) \\
   &\le& \exp\left( \frac{\rho_k^2}{M^2}\left( \exp(\lambda M) -1 -\lambda M \right) \right)
\end{eqnarray*}
where the first inequality follows from the triangle inequality and since 
$\Expect{Y_k}=0$; the second inequality follows from 
Eqn.~(\ref{eqn:int-setp-1}); and the last inequality follows since 
$1+x \le e^x$.
Thus, 
$$
\Probab{ \TNorm{ \sum_{k=1}^{L} X_k} > Lt} 
   \le (d_1+d_2) \exp\left( -\lambda L t + \frac{\sum_{k=1}^{L}\rho_k^2}{M^2} \left(  \exp(\lambda M) - 1 - \lambda M \right)  \right)  .
$$
We can minimize this as a function of $\lambda$ by choosing 
$\lambda=\frac{1}{M} \log \left( 1+\frac{tLM}{\sum_{k=1}^{L}\rho_k^2} \right)$, 
from which the result follows by tedious manipulations.
\end{Proof}

\subsection{Back to Frobenius norm matrix multiplication bounds}

We will say that the sampling probabilities of the form
$$
p_k = \frac{ \VTTNorm{A^{(k)}} \VTTNorm{B_{(k)}} }{ \sum_{k^\prime=1}^n \VTTNorm{A^{(k^\prime)}} \VTTNorm{B_{(k^\prime)}} }  
$$
are the \emph{optimal probabilities} since, as we saw before, they minimize 
$\Expect{\FNormS{AB-CR}}$, which is one natural measure of the error caused
by the random sampling process.  
In addition, we will also say that a set of sampling probabilities 
$\left\{ p_i \right\}_{i=1}^{n}$ are \emph{nearly optimal probabilities} if
$$
p_k \ge \frac{ \beta \VTTNorm{A^{(k)}} \VTTNorm{B_{(k)}} }{ \sum_{k^\prime=1}^{n} \VTTNorm{A^{(k^\prime)}} \VTTNorm{B_{(k^\prime)}} }  ,
$$
for some positive constant $\beta \le 1$.
Essentially, if we work with nearly optimal probabilities rather than 
the optimal probabilities, what this says is that we are working with 
probabilities that do not underestimate the optimal probability of choosing 
any column-row pair too much.
The challenge with random sampling algorithms is ensuring that we find 
important samples, and so this is reasonable.
In addition, as we will see below, if $\beta \ne 1$ then we suffer a small 
$\beta$-dependent loss in accuracy.  
That is, we will have to sample a little more, but if we do so then all of
our bounds will work out.
All of the results in which we will be interested will be robust if we work 
with nearly optimal probabilities, as opposed to exactly optimal 
probabilities, and we will gain a great deal of power and flexibility in 
doing so, so we will formulate the remainder of our results this semester in 
terms of nearly optimal probabilities (to such an extent that we will do so 
even when we don't make it explicit).

We now prove, for nearly optimal sampling probabilities, results analogous to 
those of Lemma~2 in Lecture~2.
In addition, we also prove that the corresponding results with the 
expectations removed hold with high probability. 
The proof of the latter will depend on the Hoeffding-Azuma inequality.

\begin{theorem}
\label{thm:BasicMatrixMultiplicationTheorem}
Suppose $A \in \mathbb{R}^{m \times n}$, 
$B \in \mathbb{R}^{n \times p}$, 
$c \in \mathbb{Z}^{+}$ such that $1 \le c \le n$,
and $\left\{ p_i \right\}_{i=1}^{n}$ are such that $\sum_{i=1}^{n} p_i = 1$ and 
such that for some positive constant $\beta \le 1$
\begin{equation}
p_k \ge \frac{\beta \VTTNorm{A^{(k)}}\VTTNorm{B_{(k)}} }{\sum_{k^\prime=1}^{n} \VTTNorm{A^{(k^\prime)}}\VTTNorm{B_{(k^\prime)}} }   .
\label{near_optimal_probs}
\end{equation}
Construct $C$ and $R$ with the \textsc{BasicMatrixMultiplication} algorithm,
and let $C R$ be an approximation to $A B$.
Then,
\begin{equation}
\Expect{ \FNormS{AB-CR} } \le \frac{1}{      \beta c } \FNormS{A}\FNormS{B}  .
\label{matmul_expect_abcr_f2}
\end{equation}
Furthermore, let $\delta \in (0,1)$ and 
$\eta = 1 + \sqrt{(8/\beta)\log(1/\delta)}$.  
Then, with probability at least $1-\delta$,
\begin{equation}
\FNormS{AB-CR} \le \frac{\eta^{2}}{      \beta c } \FNormS{A}\FNormS{B}  .
\label{whp_abcr_f2}
\end{equation}
\end{theorem}
\begin{Proof}
Following reasoning similar to that of Lemma~2 in Lecture~2, and 
using the nearly-optimal sampling probabilities 
in Eqn.~(\ref{near_optimal_probs}), we see that 
\begin{eqnarray*}
\Expect{\FNormS{AB-CR}} &\leq& \frac{1}{c} \sum_{k=1}^n \frac{1}{p_k} \VTTNormS{A^{(k)}} \VTTNormS{B_{(k)}}          \\
                        &\leq& \frac{1}{\beta c} \left( \sum_{k=1}^n  \VTTNorm{A^{(k)}}  \VTTNorm{B_{(k)}} \right)^2 \\
                        &\leq& \frac{1}{\beta c} \FNormS{A} \FNormS{B}   ,
\end{eqnarray*}
where the last inequality follows due to the Cauchy-Schwartz inequality.
Next, we consider removing the expectation.
To do so, define the event ${\mathcal{E}}_2$ to be 
\begin{equation}
\FNorm{AB-CR}  \le \frac{\eta    }{\sqrt{\beta c}} \FNorm{A}\FNorm{B}   
\end{equation}
and note that to prove the remainder of the theorem it suffices to prove that 
$\Probab{{\mathcal{E}}_2} \ge 1-\delta$. 
To that end, note that $C$ and $R$ and thus
$
CR = \sum_{t=1}^{c}\frac{1}{cp_{i_t}}A^{i_t}B_{i_t}
$
are formed by randomly selecting $c$ elements from $\{1,\ldots,n\}$, 
independently and with replacement.  Let the sequence of elements 
chosen be $\left\{i_t\right\}_{t=1}^{c}$.
Consider the function
\begin{equation}
F\left(i_1, \ldots, i_c\right) = \FNorm{AB-CR}   .
\end{equation}
We will show that changing one $i_t$ at a time does not change $F$ too much; 
this will enable us to apply a martingale inequality.
To this end, consider changing one of the $i_t$ to $i_t^\prime$ while keeping 
the other $i_t$'s the same.  Then, construct the corresponding 
$C^\prime$ and $R^\prime$.   
Note that $C^\prime$ differs from $C$ in only a single column and that
          $R^\prime$ differs from $R$ in only a single row.
Thus, 
\begin{eqnarray}
\label{eqn_eea} \FNorm{C R - C^\prime R^\prime} 
                         &=&    \FNorm{ \frac{ A^{(i_t)       }B_{(i_t)       } }{ cp_{i_t       } } -  
                                        \frac{ A^{(i_t^\prime)}B_{(i_t^\prime)} }{ cp_{i_t^\prime} } }           \\
\label{eqn_eeb}          &\leq& \frac{1}{cp_{i_t       }} \FNorm{ A^{(i_t)       }B_{(i_t)       } } +
                                \frac{1}{cp_{i_t^\prime}} \FNorm{ A^{(i_t^\prime)}B_{(i_t^\prime)} }             \\
\label{eqn_eec}          &=&    \frac{1}{cp_{i_t       }} \VTTNorm{ A^{(i_t)       }}\VTTNorm{B_{(i_t)       } } +
                                \frac{1}{cp_{i_t^\prime}} \VTTNorm{ A^{(i_t^\prime)}}\VTTNorm{B_{(i_t^\prime)} }   \\
\label{eqn_eec2}         &\leq& \frac{2}{c} \max_{\alpha}\frac{\VTTNorm{A^{(\alpha)}}\VTTNorm{B_{(\alpha)}}}{p_{\alpha}} .
\end{eqnarray}
(\ref{eqn_eea}) follows by construction
and (\ref{eqn_eec}) follows since $\FNorm{xy^T}=\VTTNorm{x}\VTTNorm{y}$ for 
$x \in \mathbb{R}^{n}$ and $y \in \mathbb{R}^{n}$.  
Thus, using the probabilities (\ref{near_optimal_probs}) and employing 
the Cauchy-Schwartz inequality we see that
\begin{eqnarray}
\label{eqn_eed} \FNorm{C R - C^\prime R^\prime} 
                         &\leq& \frac{2}{\beta c} \sum_{k=1}^n \VTTNorm{A^{(k)}} \VTTNorm{B_{(k)}}                \\
\label{eqn_eee}          &\leq& \frac{2}{\beta c} \FNorm{A}\FNorm{B}   .
\end{eqnarray}
Therefore, using the triangle inequality we see that
\begin{eqnarray}
\nonumber             \FNorm{ A        B        - C        R        } 
                &\le& \FNorm{ A        B        - C^\prime R^\prime } 
                    + \FNorm{ C^\prime R^\prime - C        R        }                                \\
\label{eqn_ffa} &\le& \FNorm{ A        B        - C^\prime R^\prime } + \frac{2}{\beta c}\FNorm{A}\FNorm{B}   .
\end{eqnarray}
By similar reasoning, we can derive
\begin{equation}
\FNorm{ A B - C^\prime R^\prime     } \le 
\FNorm{ A B - C        R        } + \frac{2}{\beta c}\FNorm{A}\FNorm{B}   .
\label{eqn_gga}
\end{equation}
Define $\Delta = \frac{2}{\beta c}\FNorm{A}\FNorm{B}$; thus, 
\begin{equation}
\left| F\left(i_1, \ldots, i_k, \ldots, i_c\right) - F\left(i_1, \ldots, i_k^\prime, \ldots, i_c\right) \right| \le \Delta   .
\end{equation}
Let $\gamma = \sqrt{2c\log(1/\delta)}\Delta$ and consider the associated Doob 
martingale.  
By the Hoeffding-Azuma inequality  
\begin{equation}
\Probab{ \FNorm{AB-CR} \ge \frac{1}{\sqrt{\beta c}}\FNorm{A}\FNorm{B}+\gamma } 
   \le \exp{\left(-\gamma^2/2c\Delta^2\right)} = \delta   
\label{eqn_iii}
\end{equation}
and theorem follows.
\end{Proof} 

An immediate consequence of Theorem \ref{thm:BasicMatrixMultiplicationTheorem} 
is that by choosing enough column-row pairs, the error in the approximation 
of the matrix product can be made arbitrarily small.  
In particular, if $c \ge 1/\beta\epsilon^2$ then by using Jensen's inequality
it follows that
\begin{equation}
\Expect{ \FNorm{AB-CR} } \le \epsilon \FNorm{A}\FNorm{B}
\end{equation}
and if, in addition, $c \ge \eta^2/\beta\epsilon^2$ then with probability at 
least $1-\delta$
\begin{equation}
\FNorm{AB-CR}  \le \epsilon \FNorm{A}\FNorm{B}   .
\end{equation}

In certain applications, we will be interested in an application of 
Theorem~\ref{thm:BasicMatrixMultiplicationTheorem} to the case that
$B=A^T$, i.e., one is interested in approximating $\FNormS{AA^T-CC^T}$. 
In this case, sampling column-row pairs corresponds to sampling columns of 
$A$, and nearly optimal probabilities will be those such that
 $ p_k \ge \frac{\beta\VTTNorm{A^{(k)}}}{\FNorm{A}} $ for some positive 
$\beta \le 1$.
By taking $B=A^T$ and applying Jensen's inequality, we have the following 
theorem as a corollary of Theorem \ref{thm:BasicMatrixMultiplicationTheorem}.

\begin{theorem}
\label{thm:BasicMatrixMultiplicationTheorem_symm}
Suppose $A \in \mathbb{R}^{m \times n}$, 
$c \in \mathbb{Z}^{+}$, $1 \le c \le n$,
and $\left\{ p_i \right\}_{i=1}^{n}$ are such that $\sum_{i=1}^{n} p_i = 1$ and 
such that $p_k \ge \frac{ \beta \VTTNormS{A^{(k)}} }{ \FNormS{A} }$ for some 
positive constant $\beta \le 1$.
Furthermore, let $\delta \in (0,1)$ and 
$\eta = 1 + \sqrt{(8/\beta)\log(1/\delta)}$.  
Construct $C$ (and $R=C^T$) with the \textsc{BasicMatrixMultiplication} 
algorithm, and let $C C^T$ be an approximation to $A A^T$.
Then,
\begin{equation}
\Expect{ \FNorm{AA^T-CC^T} } \le \frac{1}{\sqrt{\beta c}} \FNormS{A}
\label{expect_aacc_f}
\end{equation}
and with probability at least $1-\delta$,
\begin{equation}
\FNorm{AA^T-CC^T}  \le \frac{\eta    }{\sqrt{\beta c}} \FNormS{A}   .
\label{whp_aacc_f}
\end{equation}
\end{theorem}

\newpage

\section{%
(09/18/2013):  Matrix Multiplication, Cont.; and Random Projections} 

Today, we will use the concentration results of the last few classes to go back and make statements about approximating the product of two matrices; and we will also describe an important topic we will spend a great deal more time on, i.e., random projections and Johnson-Lindenstrauss lemmas.
Here is the reading for today.
\begin{itemize}
\item
Dasgupta and Gupta, ``An elementary proof of a theorem of Johnson and Lindenstrauss'' 
\item
Appendix of: Drineas, Mahoney, Muthukrishnan, and Sarlos, ``Faster Least Squares Approximation'' 
\item
Achlioptas, ``Database-friendly random projections: Johnson-Lindenstrauss with binary coins''
\end{itemize}

\subsection{Spectral norm bounds for matrix multiplication} 

Here, we will consider the \texttt{BasicMatrixMultiplication} algorithm, and we will provide a spectral norm bound for the error of the approximation constructed by it.
Recall that, given as input a $m \times n$ matrix $A$ and an $n \times p$ 
matrix $B$, this algorithm randomly samples $c$ columns of $A$ and the 
corresponding rows of $B$ to construct a $m \times c$ matrix $C$ and a 
$c \times p$ matrix $R$ such that $CR \approx AB$, in the sense that some 
matrix norm $||AB-CR||$ is small.
The Frobenius norm bound we established before immediately implies a bound
for the spectral norm, but in some cases we will need a better bound than 
can be obtained in this manner.
Since, in this semester, we will only need a spectral norm bound for the 
special case that $B=A^T$, that is all that we will consider here.

\begin{theorem}
\label{thm:theorem7correct}
Let $A\in\mathbb{R}^{m \times n}$, and consider approximating $AA^T$.
Construct a matrix $C\in\mathbb{R}^{m \times c}$, consisting of $c$ sampled
and rescaled columns of $A$, with the \texttt{BasicMatrixMultiplication} 
algorithm, where the sampling probabilities $\{p_i\}_{i=1}^{n}$ satisfy
\begin{equation}
\label{eqn:defPj}
p_i \geq \beta \frac{\TNormS{A^{(i)}}}{\FNormS{A}}
\end{equation}
for all $i \in [n]$, and for some constant $\beta \in (0,1]$. 
Assume, for simplicity, that $\TNorm{A} \le 1$ and $\FNormS{A} \ge 1/24$, 
let $\epsilon \in (0,1)$ be an accuracy parameter, 
and let 
\begin{equation}
\label{eqn:bound-on-c-spectral}
c \ge \frac{96\FNormS{A}}{\beta\epsilon^2} 
      \log\left(\frac{96\FNormS{A}}{\beta\epsilon^2\sqrt{\delta}} \right) .
\end{equation}
Then, with probability $\ge1-\delta$, we have that 
$$
\TNorm{AA^T-CC^T}\le\epsilon  .
$$
In addition, if we set $\delta=1$ in Eqn.~(\ref{eqn:bound-on-c-spectral}), then
$$
\Expect{\TNorm{AA^T-CC^T}}\leq \epsilon.
$$
\end{theorem}

Before proving this theorem, here are a few things to note about it.
\begin{itemize}
\item
The assumptions on the spectral and Frobenius norms of $A$ are not 
necessary, but instead are only to simplify the expressions.
\item
The assumption on $c$ is important.
That is, whereas the Frobenius norm bound we discussed previously holds for 
any value of $c$ (perhaps yielding weak results if $c$ is too small), here 
we will need to set $c$ to be at least the value of 
Eqn.~(\ref{eqn:bound-on-c-spectral}) for the theorem to hold. 
\item
We can have $\epsilon\FNorm{A}$ on RHS by modifying the sampling complexity.
In particular, this can give the sampling complexity in terms of the stable
rank.
If we define the \emph{stable rank} of a matrix $A$ as 
$\mbox{sr}(S) = \frac{\FNormS{A}}{\TNormS{A}}$, then 
$\mbox{sr}(A) \le \mbox{rank}(A)$, and the stable rank is a more robust 
notion than the usual rank, and bounds parameterized in this way are 
sometimes of interest.
\item
We can generalize this to the product of two different matrices, and we get 
slightly weaker results for $\TNorm{AB-CR}$.
\item
We formulate it this way since we will only need spectral norm bounds for 
approximating matrix products of the form $AA^T$ and since we will use this
theorem by setting, e.g., $\epsilon=1/2$.
\end{itemize}

\begin{Proof}
For the proof of this spectral norm bound, we will need a matrix concentration
result.
For convenience here within the proof, we will use a slightly different 
version of matrix concentration than we proved last time---in particular, 
one due to Oliviera, rather than Recht, which we established last time.
Here, we will simply state that version---it's proof is similar to the 
version we proved last time and so will be omitted.
For completeness, though, here is a brief history of matrix concentration 
bounds and how they are used in randomized numerical linear algebra.
\begin{itemize}
\item
Alshwede-Winter: the original result related to bounding the matrix m.g.f. 
that started this recent flurry of work in this area.
\item
Christofides-Markstron: introduced a matrix version of Heoffding-Azuma
\item
Rudelson and Vershynin: had the original bounds for operator-valued random 
variables that were originally used to bound the spectral norm error.
They bounds had a similar form, but they depended on heavier-duty arguments 
from convex analysis, and they sometimes didn't provide constants, which 
made it awkward for numerical implementation.
\item
Gross, Recht, and Oliviera: several different versions of matrix Chernoff 
bounds.
\item
Tropp: provides a nice review of this line of work.
\end{itemize}
In this proof, we will use the version due to Oliviera (in ``Sums of random 
Hermitian matrices and an inequality by Rudelson''), which we will state but 
not prove.
\begin{lemma}
Let $X_1,\ldots,X_n$ be i.i.d. random column vectors in $\mathbb{R}^{d}$ 
such that $\VTTNorm{X_i}\le M$ a.s. and 
$\TNorm{\Expect{X_iX_i^*}}\le1$.
Then, $\forall t \ge 0$, 
$$
\Probab{\TNorm{\frac{1}{n}\sum_{i=1}^{n}X_iX_i^*-\Expect{X_1X_1^*}} \ge t}
   \le (2n)^2 \exp \left( \frac{-nt^2}{16M^2+8M^2t} \right).
$$
\end{lemma}
To use this in the proof of our spectral norm matrix multiplication bound, 
consider the random sampling algorithm, and note that 
\begin{equation*}
AA^T = \sum_{i=1}^n A^{(i)} A^{{(i)}^T}.
\end{equation*}
We shall view the matrix $AA^T$ as the true mean of a bounded operator 
valued random variable, whereas $CC^T = AS (AS)^T = ASS^TA^T$ will be its 
empirical mean. 
Then, we will apply Lemma 1 of Oliveira. 
To this end, define a random vector $y \in \mathbb{R}^m$ as
\begin{equation*}
\Probab{y = \frac{1}{\sqrt{p_i}}A^{(i)}} = p_i
\end{equation*}
for $i \in [n]$. 
The matrix $C = AS$ has columns 
$\frac{1}{\sqrt{c}}y^1,\frac{1}{\sqrt{c}}y^2,\ldots,\frac{1}{\sqrt{c}}y^c$, 
where $y^1,y^2,\ldots,y^c$ are $c$ independent copies of $y$. 
Using this notation, it follows that
\begin{eqnarray}
\Expect{yy^T} 
\nonumber
   &=& \sum_{i=1}^{n} p_i \frac{1}{\sqrt{p_i}} A^{(i)} \frac{1}{\sqrt{p_i}} A^{{(i)}^T} \\
\nonumber
   &=& \sum_{i=1}^{n} A^{(i)} A^{{(i)}^T} \\
\label{eqn:expectyyt}
   &=& AA^T
\end{eqnarray}
and
\begin{equation*}
CC^T = ASS^TA^T = \frac{1}{c}\sum_{t=1}^c y^t{y^t}^T.
\end{equation*}
Finally, let
\begin{equation}
\label{eqn:defM}
M = \TNorm{y} = \frac{1}{\sqrt{p_i}}\TNorm{A^{(i)}} 
              \le \frac{1}{\sqrt{\beta}} \FNorm{A}  ,
\end{equation}
where the inequality follows by the form of Eqn.~(\ref{eqn:defPj}).
We can now apply Lemma 1, p. 3 of Oliveira. 
Notice that from Eqn.~(\ref{eqn:expectyyt}) and our assumption on the 
spectral norm of $A$, we immediately get that 
\begin{equation*}
\TNorm{\Expect{yy^T}} = \TNorm{AA^T} \leq \TNorm{A}\TNorm{A^T} \leq 1  .
\end{equation*}
Then, Lemma 1 of Oliveira 
implies that
\begin{equation}
\label{eqn:ExpectBound}
\TNorm{CC^T-AA^T} < \epsilon,
\end{equation}
with probability at least 
$1-\left(2c\right)^2\exp\left(-\frac{c\epsilon^2}{16M^2+8M^2\epsilon}\right)$. 
Let $\delta$ be the failure probability of Theorem~\ref{thm:theorem7correct}.
We seek an appropriate value of $c$ in order to guarantee that 
$\left(2c\right)^2 \exp\left(-\frac{c\epsilon^2}{16M^2 + 8M^2 \epsilon}\right) \leq \delta$. 
Equivalently, we need to satisfy
$$
\frac{c}{\ln \left(2c/\sqrt{\delta}\right)} 
   \geq \frac{2}{\epsilon^2}\left(16M^2 + 8M^2\epsilon\right)  .
$$
Recall that $\epsilon < 1$, and by combining Eqn.~(\ref{eqn:defPj}) with the 
above equation, it suffices to choose a value of $c$ such that
$$
\frac{c}{\ln \left(2c/\sqrt{\delta}\right)} 
   \geq \frac{48}{\beta\epsilon^2}\FNormS{A}  ,
$$
or, equivalently,
$$
\frac{2c/\sqrt{\delta}}{\ln \left(2c/\sqrt{\delta}\right)} 
   \geq \frac{96}{\beta\epsilon^2\sqrt{\delta}}\FNormS{A}  .
$$
We now use the fact that for any $\eta \geq 4$, if $x \geq 2\eta \ln \eta$ 
then $\frac{x}{\ln x} \geq \eta$. 
Let $x = 2c/\sqrt{\delta}$, let 
$\eta = 96 \FNormS{A}/\left(\beta \epsilon^2\sqrt{\delta}\right)$, and note 
that $\eta \geq 4$ if $\FNormS{A} \geq 1/24$, since $\beta$, $\epsilon$, and 
$\delta$ are at most $1$. 
Thus, it suffices to set
$$
\frac{2c}{\sqrt{\delta}} 
   \geq 2 \frac{96 \FNormS{A}}{\beta \epsilon^2\sqrt{\delta}}\ln \left(\frac{96 \FNormS{A}}{\beta \epsilon^2\sqrt{\delta}}\right)  ,
$$
which concludes the proof of the theorem.
\end{Proof}

\subsection{Random projections and Johnson-Lindenstrauss lemmas}

Here, we will discuss a related way to perform dimensionality reduction on 
matrices known as random projections, which has strong connections with an
important result known as the Johnson-Lindenstrauss lemma.
Random projections, and in particular the results provided by the JL lemma 
are very powerful, in the sense that the points can be in more general 
metric spaces, etc.
Thus, they have strong connections with random sampling and matrix 
multiplication, which we will make explicit.
The reason we have started with the latter is that since they make more 
explicit the Euclidean vector space structure, which in turn allows us to 
get finer bounds that are more useful for numerical implementations, machine 
learning and data analysis applications, etc.

The general question is one of so-called \emph{dimensionality reduction}, 
i.e., mapping a high-dimensional data set (i.e., a set of data points modeled
as vectors in some high-dimensional, typically but not always, Euclidean 
vector space) to a much lower-dimensional space (again, typically, a 
low-dimensional Euclidean vector space) in such a way that important 
structural properties of the data are preserved.
The most common way to do this, e.g., in statistics and machine learning and
many other related areas, involves choosing a small number of directions
in the original vector space in which the data have high variance.
That is, if one looks at the data and one asks ``What are the directions in 
the high-dimensional space that capture the most amount of variance?'' then 
one is interested in finding those directions.

The most common way to do that is with the SVD, or relatedly PCA.
The SVD, and as a practical matter partial SVDs, i.e., computing a small 
number of singular values and singular vectors, as opposed to the full SVD, 
is moderately but not extremely expensive to compute, and it is useful in 
many situations where it is not obviously-appropriate, e.g., where 
Gaussian-like assumptions underlying it or truncated PCAs are violated.
A lot of what we will be interested in later in the course will be 
computing partial SVDs more quickly than off-the-shelf methods, and we will 
use random sampling and random projections to do this.
Although there are strong connections between random projections and the 
properties of the SVD that involve capturing maximum variance directions, it 
is actually useful to take a step back and consider other types of 
dimensionality reduction methods.
Random projections are most-easily viewed this way, and so we will start with 
that, and we will make the connections with SVD later.
To that end, consider a different type of dimensionality reduction where, 
rather than finding the directions that capture the most variance, the goal 
is to construct some sort of mapping that preserves all ${n \choose 2}$ 
pairwise distances between pairs of data points.

(As we said, there will be connections between preserving pair-wise distances
and capturing variance, but at this point just note that they are two 
different metrics of interest.
For example, one can easily imagine that maximizing variance preserves all 
the pairwise distances ``on average,'' but that a few pairwise distances are 
violated a lot; and conversely that if we force ourselves to preserve every 
pairwise distance, then we might ``overfit'' the data at hand and fail to 
preserve a large-scale measure like overall variance.
We will return to this later, but you should think of the two different
perspectives that we mentioned in the first class: 
preserving every pairwise distance is more natural from the algorithmic 
perspective, where we view the data in front of us as all there is, in which
case we want to preserve the properties on it, and we get worst-case quality
of approximation bounds that depend on the worst-case distortion;
while preserving overall variance might involve sacrificing a few pairwise 
distances and might be more robust in the presence of a bit of noise, and 
so this might be associated with better inferential properties.)

Today, we will consider a very simple but remarkably powerful method to 
do dimensionality reduction that is of the latter flavor, and we will 
describe the results one can prove and the analysis, which go by the name the
Johnson-Lindenstrauss lemma.
We won't use Chernoff bounds directly, and so we won't call our previous 
results as a black box, but we will draw connections later, and we will see
that the proof will use ideas that are very similar to the proofs of 
Chernoff bounds.

Say that we have $n$ points $\{u_i\}_{i=1}^{n}$, each of which is in 
$\mathbb{R}^{d}$, e.g., the rows of an $n \times d$ matrix $A$, and we 
want to find $n$ points $\{v_i\}_{i=1}^{n}$, each of which is in 
$\mathbb{R}^{k}$ such that 
\begin{itemize}
\item
$k \ll d$
\item
\begin{itemize}
\item
$||v_i|| \approx ||u_i||$, for all $i$
\item
$||v_i-v_{i'}|| \approx ||u_i-u_{i'}||$, for all $i,i'$
\end{itemize}
\end{itemize}
Here, $||\cdot||$ refers to the Euclidean norm, i.e., $\VTTNorm{\cdot}$ in 
the above notation.
The first thing to note is that it isn't immediately obvious that such a 
mapping even exists. Never mind that it can be computed efficiently and 
exploited algorithmically.
We will show that such a mapping does exist for the Euclidean norm, but it 
is known that such a mapping does not exist for other norms, e.g., the 
$\ell_1$ norm.
Since the proof of this result is now sufficiently simple to be presented 
in a class, it is worth paying attention to what steps are standard and 
which steps are peculiar to the Euclidean norm and fail to hold for other 
norms.

We will construct a ``random projection,'' and prove a version of the 
Johnson-Lindenstrauss lemma.
Here is a brief history of these methods.
\begin{itemize}
\item
Johnson-Lindenstrauss: proved the result for a random subspace as part of 
a more general result they were interested in proving.
\item
Frankl and Meahara: project onto $k$ random orthogonal vectors.
\item
IM, DG: project onto a matrix whose entries consist of i.i.d. Gaussian random
variables.
\item
Ach: project onto a matrix consisting of $\{\pm1\}$ random variables.
\item
Ailon and Chazelle: construct a ``fast'' Hadamard-based version of JL, in 
which case the projection matrix can be multiplied more quickly than 
vanilla matrix multiplication using fast-Fourier methods.
\item
Sarlos: made explicit the ``subspace JL'' result, basically by putting an 
$\epsilon$-net on a unit ball, to show that JL-like bounds hold for infinitely
many vectors drawn from a low-dimensional subspace, thereby yielding a 
distortion bound of the form we saw with approximate matrix multiplication.
\item
Clarkson and Woodruff: the random projection matrix can be \emph{extremely}
sparse and one can get JL-like results, assuming that the input vectors are
from a low-dimensional subspace
\end{itemize}
Here, we will prove the version for Gaussian random variables:
although it is easier than some of the other versions, similar ideas hold for
the other versions, and we will revisit some of the other versions later.

Before doing that, here is a word about terminology.
In linear algebra and functional analysis, a projection is a linear 
transformation $P$ from a vector space to itself such that $P^2=P$.
In particular, it is idempotent and its eigenvalues are in $\{0,1\}$, i.e., 
equal to either $0$ or $1$.
For example, given a matrix $A$, let $A=U \Sigma V^T$ be its truncated SVD, 
and let $A=QR$ be its QR decomposition.
Then, the projection onto the column space of $A$ is 
$P_A=UU^T=QQ^T$, and it is easy to verify that $P_A = Q (Q^T Q) Q^T$, since 
$Q^TQ$ is a low-dimensional identity.

The JL ``projection'' is more general and is typically not a projection in 
that linear algebraic sense of the word.
(Although it is $\epsilon$-close to a projection in that traditional 
linear algebraic sense of the word, and quantifying this observation is at
the heart of the analysis of many of the random sampling and random 
projection algorithms in RandNLA.)
Why is it not a projection in that sense of the word?
\begin{itemize}
\item
First, it can be applied to arbitrary points in arbitrary metric spaces.
This makes it applicable more generally than Euclidean vector spaces, but it 
makes its use slightly overkill form many data analysis and machine learning 
problems that involve matrices in $\mathbb{R}^{n}$.
\item
Second, in spite of that, it ``looks like'' an orthogonal matrix in many 
ways.
For example, if $P$ is a matrix of i.i.d. Gaussians, then $\mbox{range}(P^TP)$
is a uniformly distributed subspace, but the eigenvalues are \emph{not} in 
$\{0,1\}$.
\item
Third, if the random vectors were exactly orthogonal (as they actually were 
in the original JL constructions), then we would have that the JL projection
was an orthogonal projection and the eigenvalues would be in $\{0,1\}$; but
although this is false for Gaussians, $\{\pm1\}$ random variables, and most 
other constructions, one can prove that the resulting vectors are 
approximately unit length and approximately orthogonal, and for most 
applications of the JL lemma in RandNLA (as well as more generally), this 
is ``good enough.''
\item
Fourth, for $\{\pm1\}$, Hadamard, etc. constructions, they are \emph{not}
even spherically symmetric, so the analysis is messier, but we will be able
to show that they lead to JL projections that are almost orthogonal matrices
and thus projections in the linear algebraic sense of the word.
But the analysis is messier.
\end{itemize}

With those comments in place, here is the version of the JL lemma that we 
will prove in detail.

\begin{lemma}[JL lemma]
Given $n$ points $\{u_i\}_{i=1}^{n}$, each of which is in $\mathbb{R}^{d}$, 
$P\in\mathbb{R}^{d \times k}$ be such that $P_{ij}= \frac{1}{\sqrt{k}}N(0,1)$,
and let $\{v_i\}_{i=1}^{n}$ be points in $\mathbb{R}^{k}$ defined as
$v_i = u_i P$.
Then, if $k \ge \frac{9 \log(n)}{\epsilon^2-\epsilon^3}$, for some 
$\epsilon \in (0,1/2)$, then with probability at least $1/2$, all pairwise 
distances are preserved, i.e., for all $i,i'$, we have
\begin{equation*}
(1-\epsilon)\VTTNormS{u_i-u_i'}
   \le \VTTNormS{v_i-v_i'}
   \le (1+\epsilon) \VTTNormS{u_i-u_i'}   .
\end{equation*}
\end{lemma}

\begin{Proof}
Let $u$ be any fixed vector in $\mathbb{R}^{n}$, and consider $v=uP$.
We will establish results for the expectation of the norm of $v$ as well as
results that state with high probability the norm does not deviate much above
or below the expectation.
Then, we will set parameters to get that a union bound argument means that 
the result holds for ${n \choose 2}$ pairs of points with probability at 
least $1/2$.
First, let's get the expectation.
\begin{eqnarray*}
\Expect{\VTTNormS{v}} 
   &=& \Expect{ \sum_{j=1}^{k} \left( \sum_{i=1}^{d} \frac{1}{\sqrt{k}} u_i P_{ij} \right)^2 }  \\
   &=& \sum_{j=1}^{k} \frac{1}{k} \Expect{ \left( \sum_{i=1}^{d} u_i P_{ij} \right)^2 }  \\
   &=& \sum_{j=1}^{k} \frac{1}{k} \sum_{i=1}^{d} u_i^2 \Expect{P_{ij}^{2}} \\
   &=& \sum_{j=1}^{k} \frac{1}{k} \sum_{i=1}^{d} u_i^2  \\
   &=& \VTTNormS{u}.
\end{eqnarray*}
Most of the equalities are fairly straightforward, but note that the third 
equality follows since $P_{ij} \sim N(0,1)$;
for other JL constructions, establishing this is more complicated, but it 
can be done.

(Note that this derivation basically says that if you take an arbitrary 
unit-length vector in $\mathbb{R}^{d}$ and ``project'' it down to 
$\mathbb{R}^{k}$ by taking random linear combinations, weighted by 
$N(0,1)$ random variables, then the squared length of the resulting vector
is $\frac{k}{d}$, and thus to preserve the length, we have to rescale by 
$\frac{d}{k}$.)

Next, let's bound the probability that the projected vector stretched by 
more than a small amount from the expectation.
To do so, let's define 
$x_j = \frac{1}{\VTTNorm{u}} u^T P^{(j)}$ 
and
$x = k \frac{\VTTNormS{v}}{\VTTNormS{u}} 
   = \frac{1}{\VTTNormS{u}} \sum_{j=1}^{k} ( u^T P^{(j)} )^2 
   = \sum_{j=1}^{k} x_j^2 $.
(Note that this notation is inconsistent with the usual way we use subscripts, 
but we will use it only here in a self-contained way.)
With these definitions, we have that
\begin{eqnarray*}
\Probab{ \VTTNormS{v} \ge (1+\epsilon) \VTTNormS{u} }
   &=& \Probab{ x \ge (1+\epsilon) k }  \\
   &=& \Probab{ e^{\lambda x} \ge e^{ \lambda (1+\epsilon) k } }  \\
   &\le& \frac{ \Expect{ e^{\lambda x} } }{ e^{ \lambda (1+\epsilon) k } }  \\
   &=&   \frac{ \Expect{ e^{\lambda \sum_{j=1}^{k} x_j^2} } }{ e^{ \lambda (1+\epsilon) k } }  \\
   &=&   \frac{ \Pi_{j=1}^{k} \Expect{ e^{\lambda x_j^2 } } }{ e^{ \lambda (1+\epsilon) k } }  \\
   &=&   \left( \frac{ \Expect{ e^{\lambda x_1^2 } } }{ e^{ \lambda (1+\epsilon) } } \right)^{k},
\end{eqnarray*}
where the second equality follows for all $\lambda > 0$ to be chosen later, 
and the rest of the steps should be clear, based on how we derived the 
Chernoff bounds.
To calculate $\Expect{ e^{\lambda x_1^2 } }$, recall that $x_1 \sim N(0,1)$, 
and thus
\begin{eqnarray*}
\Expect{ e^{\lambda x_i^2} } 
   &=& \int_{-\infty}^{\infty} \frac{1}{\sqrt{2\pi}} e^{-t^2/2}e^{-\lambda t^2} dt  \\
   &=& \frac{1}{\sqrt{1-2\lambda}} \int_{-\infty}^{\infty} \frac{\sqrt{1-2\lambda}}{\sqrt{2 \pi}} e^{\frac{t^2}{2}(1-2\lambda)} dt  \\
   &=& \frac{1}{\sqrt{1-2\lambda}} ,
\end{eqnarray*}
for $\lambda < \frac{1}{2}$.
Plugging this into the above, we have that
\begin{eqnarray*}
\Probab{ \VTTNormS{v} \ge (1+\epsilon) \VTTNormS{u} }
   &=&   \left( \frac{ e^{-2 \lambda (1+\epsilon)} }{1-2\lambda} \right)^{k/2}  \\
   &=&   \left( \left(1+\epsilon\right)e^{-\epsilon} \right)^{k/2}  \\
   &\le& \exp\left( -\left(\epsilon^2-\epsilon^3 \right)k/4 \right),
\end{eqnarray*}
where the second equality follows if we choose 
$\lambda = \frac{\epsilon}{2(1+\epsilon)}$, and 
since $1+\epsilon < \exp\left( \epsilon - \left(\epsilon^2-\epsilon^3\right)/2 \right)$ the inequality follows.

Third, let's bound the probability that the projected vector is shrunk by
more than a small amount from the expectation.
The derivation is similar to above derivation, and so we just state the main
steps of it.
\begin{eqnarray*}
\Probab{ \VTTNormS{v} \le (1-\epsilon) \VTTNormS{u} }
   &=& \Probab{ x \le (1-\epsilon) k }  \\
   &=& \Probab{ e^{-\lambda x} \ge e^{ -\lambda (1-\epsilon) k } }  \\
   &\le& \left( \frac{ \Expect{ e^{-\lambda x_1^2 } } }{ e^{ -\lambda (1-\epsilon) } } \right)^{k}  \\
   &=&   \left( \frac{ e^{2 \lambda (1-\epsilon)} }{1+2\lambda} \right)^{k/2}  \quad \mbox{for }\lambda=\frac{\epsilon}{2(1-\epsilon)}  \\
   &=&   \left( \left(1-\epsilon\right)e^{\epsilon} \right)^{k/2}  \\
   &\le& \exp\left( -\left(\epsilon^2-\epsilon^3 \right)k/4 \right).
\end{eqnarray*}

Finally, let's put it all together. 
By combining the above two results, we have that for any one fixed point $u$
that is mapped to $v$, we have that
$$
\Probab{ (1-\epsilon)\VTTNormS{u} \le \VTTNormS{v} \le (1+\epsilon)\VTTNormS{u} }
   \ge 1-2 \exp\left( -\left( \epsilon^2-\epsilon^3 \right) k / 4 \right).
$$
Since we have $n$ points, we want this type of bound to hold for 
${n \choose 2} = \Theta(n^2)$ pairs of points, i.e., for the vectors 
defining the distances between each of the pairs of points.
If we want the failure probability to be $\le \frac{1}{2}$, then by the
union bound, we need that 
$$
2 n^2 \exp\left( -\left(\epsilon^2-\epsilon^3 \right) k/4 \right) < \frac{1}{2}.
$$
That is, we need that 
$$
\left( \epsilon^2-\epsilon^3 \right) k/4 > 2 \log (2n)  ,
$$
and for this it suffices that $k > \frac{9 \log(n)}{\epsilon^2-\epsilon^3}$, 
under the simplifying assumption that $n > 16$.
\end{Proof}

\subsection{Additional comments and thoughts on random projections}

Before proceeding, here are a few remarks on the proof of the JL lemma.
\begin{itemize}
\item
One can view the $n$ points $\{u_i\}_{i=1}^{n}$ as the rows of an 
$n \times d$ matrix $A$. 
In this case, if $u_i = A_{(i)}$ is the $i^{th}$ row of $A$, then
$v_i = (AP)_{(i)}$ is the $i^{th}$ row of the product matrix $AP$.
\item
The failure probability of $1/2$ is for convenience; it can be made to be 
less that $\delta$, for any fixed $\delta > 0$ by adjusting the dimension $k$
to be slightly larger.
\item
All proof of JL lemmas have the same basic structure that we followed above. 
Basically, one proves that the JL mapping doesn't distort any one fixed 
vector too much away from its expectation, and then one performs a union 
bound to show that the result holds for ${n \choose 2}$ pairs of 
vectors---that can be chosen to be the differences between all 
${n \choose 2}$ difference vectors $u_i-u_{i'}$.
(This is basically just the ``random projection theorem'' mentioned below.)
\item
The dependence on $k$ in this lemma is simplified for convenience, e.g., 
the $9$ is suboptimal. 
But the dependence on $n$ is optimal, and the dependence on $\epsilon$ is
optimal up to $\log(1/\epsilon)$ factors.
\item
One of the main points is that the dependence on $k$ is logarithmic in $n$, 
the number of data points, but it is independent of $d$, the formal 
dimensionality of those data points.
In some of what follows, we will be projecting on the left, and sometimes we 
will be projecting on the right.
In addition, if we have a ``rectangular'' matrix, by which we mean that one
dimension is much larger than the other, we will usually be projecting on
the high dimension to make it smaller, but sometimes we will project on the
low dimension to make it even lower.
Plus, in some cases, we will project a subspace, and so we will actually 
preserve distances among an uncountably infinite number of pairs of points, 
basically by using an $\epsilon$ net argument.
For these reasons, as well as the fact that we will use different letters to 
denote the dimensionality of different matrices, one might to lose sight of 
the simple and important point that random projections will allow us to 
project some number of points to a dimension that depends logarithmically on 
the number of original points and is independent of their formal 
dimensionality.
\end{itemize}

Next we will spend a bit of time describing, more informally, what is going
on with the JL lemma and what makes it work, with an eye toward topics that
we will return to throughout the semester.

If we consider $n$ points in $\mathbb{R}^{d}$, as in the JL setup, and we 
want to put them in $\mathbb{R}^{k}$, for some $k \ll d$, the obvious 
naive way to do that is to choose $k$ coordinates u.a.r. and evaluate
pairwise distances based on those $k$ coordinates and hope for the best.
Although naive, that procedure actually works for certain classes of input
vectors that are ``spread out'' in ways that we will quantify later.
The question is: which are properties of input vectors that will cause 
problems for this naive procedure?
Basically, the answer is that two points can be very far apart in Euclidean 
distance, while differing on only a small number of coordinates, or even a 
single coordinate.
In this case, if we sample uniformly, then with high probability we will not
select those coodinates that are important for maintaining pairwise 
distances.
On the other hand, if for all pairs of points, it is the case that all 
coordinates contribute roughly the same amount then this simple procedure 
works.

(Note that essentially this same issue arose in the motivation of the 
random sampling algorithm---there might be a small number of rank-one 
components that were particularly important, and if we don't find them, then
we will get a very poor approximation to the matrix product.
Before the solution was to sample nonuniformly.
Here we will find a different solution.
And later we will discuss connections between these two approaches,
illustrating how the complement one another.)

Motivated by this, the basic idea underlying random projection algorithms
is that, given $n$ points in $\mathbb{R}^{d}$, rather than sampling in the
canonical basis, if we instead apply some sort of ``random rotation'' to the
original point set, then we will get a new ``random basis'' in 
$\mathbb{R}^{d}$, and in that basis things will be ``spread out'' in nice 
ways, so we can sample uniformly.
To see this, note that choosing the first $k$ coordinates after applying a 
random rotation is exactly the same as projecting onto a spherically-symmetric 
$k$-dimensional hyperplane.
So, essentially, randomization in the JL projection protects against problems
with axis-alignment in the canonical basis---which is exactly what the 
nonuniform sampling probabilities ensured against.
If, on the other hand, the input are well-spread-out, then:
either we can have very sparse random projection matrices; 
or we can perform uniform sampling.

As we mentioned above, 
JL actually projected onto a random $k$-dimensional hyperplane.
FM did something similar, but viewed it as projecting onto $k$ orthonormal 
vectors.
If $P$ or $\Pi$ is a proejection matrix that is a $d \times k$
real-valued matrix, then we can construct such a matrix as follows:
for the first row, choose a random uniform vector w.r.t. the Haar measure 
on the sphere $S^{d-1}$, the $(d-1)$-dimensional unit sphere in 
$\mathbb{R}^{d}$ (to choose a random unit vector in $\mathbb{R}^{d}$, choose
$d$ i.i.d. $N(0,1)$ random variables, and normalize the resulting vector 
to have Euclidean norm $1$);
then the second row is a random unit vector from the space orthogonal to the
first vector; 
and so on.
The resulting matrix is a projection onto a random $k$-dimensional 
subspace of $\mathbb{R}^{d}$ with the following properties:
spherical symmetry (i.e., for all orthogonal matrices $U$, $P$ and $PU$ 
have the same distribution);
orthogonality (i.e., the columns of $P$ are orthogonal to each other); 
normality (i.e., the columns of $P$ are unit length).

And what are random subspaces?
A random subspace of dimension equal to $1$ is just a random line through 
the origin.
A random subspace of dimension equal to $k$ is specified by a random line
through the origin, a second random line through the origin orthogonal to 
the first, etc., and their span is the random subspace of interest.

How long are vectors when you project them onto random subspaces?
Let's say we project a fixed unit-length vector $u$ in $\mathbb{R}^{d}$ 
onto a random $k$-dimensional subspace, constructed as above.
Then, by the Pythagorean Theorem, the length squared of the new vector is 
the sum of the lengths squared of the components.
Intuitively, in a random direction the squared length of an arbitrary unit
length vector should be $\sim\frac{1}{d}$ (meaning exactly or approximately
that, e.g., that in expectation and with high probability very near that).
In this case, the squared length of a projection onto a random $k$-dimensional
subspace should be $\sim\frac{k}{d}$.
So, the length or norm of the projected vector should be 
$\sim\sqrt{\frac{k}{d}}$.
The following theorem says that this is the case---in particular, it says 
that the length of the projected vector is very close to this expectation, 
with a failure probability that is exponentially small in $k$.

\begin{theorem}[Random Projection Theorem]
Let $u$ be a fixed unit-length vector in $\mathbb{R}^{d}$, let $V$ be a 
random $k$-dimensional subspace, and let $v$ be the projection of $u$ onto 
$V$.
Then, for all $\epsilon\in(0,1)$, we have that
$$
\Probab{ \left| \VTTNorm{v}-\frac{k}{d} \right| \ge \epsilon\sqrt{\frac{k}{d}} } 
   \le 3 \exp\left( -k\epsilon^2/64 \right)  .
$$
\end{theorem}

Clearly, given this random projection theorem, the simplest form of the JL
lemma follows immediately by a union bound argument.
As for other forms of the JL lemma---and in particular those that are 
algorithmically more appealing like when $P$ consists of i.i.d. Gaussians, 
$\{\pm1\}$ random variables, or structured Hadamard-based transforms---similar
ideas also hold.
Basically, things are not as smooth or nice as with random subspaces but 
similar concentration results can be shown to hold.

IM and DG, in particular, dropped the explicit requirement of normality and
orthogonality.
But, if we choose every element of $P$ i.i.d. $\sim N(0,1)$, as we did in the
above construction, then we get normality and orthogonality in expectation, 
i.e., the squared length of all columns is $1$ in expectation, and the inner
product between all pairs of columns is $0$ in expectation.
This $p$ is still symmetric, but the independence of entries makes it easier
to generate.

(That last statement is actually true in theory, but not necessarily in 
practice.  
That is, if you try to implement a random projection by multiplying an 
input matrix with i.i.d. Gaussians, then the most expensive step can be 
generating the random variables, motivating either more sophisticated 
lower-level methods to generate random variables from Gaussian distributions 
or random projections with structurally simpler entries.
Both have been used in RandNLA, and here we will focus on the latter.)

Ach considered entries that were basically $\{\pm1\}$ random variables, 
thus dropping the spherical symmetry constraint.
In particular, choosing $\{\pm1\}$ random variables, up to rescaling, 
means working with one of the following two distributions.
First, let
$$
P_{ij} = \left\{ \begin{array}{l l}
            1/\sqrt{k}  & \quad \text{w.p.} = 1/2 \\
            -1/\sqrt{k} & \quad \text{w.p.} = 1/2
         \end{array} \right.
$$
and second, let
$$
P_{ij} = \left\{ \begin{array}{l l}
            3/\sqrt{k}  & \quad \text{w.p.} = 1/6 \\
            0           & \quad \text{w.p.} = 2/3 \\
            -3/\sqrt{k} & \quad \text{w.p.} = 1/6
         \end{array} \right.
$$
Note the scaling in these two means that things work out in expectation, and
the analysis shows that they work out with high probability, in a manner 
similar to what we did above.
Working with $\{\pm1\}$ random variables means that you can construct the
entries of $P$ with one random bit per entry, which is easier to generate
since flipping coins is easier than generating Gaussians, and it has other 
advantages that we won't get into.
In addition, the second distribution above (and the analysis, which we won't 
get into, which says that it satisfies JL-like properties) shows that we 
can work with projection matrices that are somewhat sparse.
In this case, $2/3$ of the entries can be zero-ed out, and in worst case
not much more can be zero-ed out, at least in this very simple way, but it 
raises the question of whether and how one can sparsity even more. 

Two other extensions that we will get to in a few classes are the following.
\begin{itemize}
\item
Fast JL: proposed by Ailon and Chazelle, this makes these random 
projection ideas useful for even relatively-quick computations like 
least-squares regression and many related matrix problems.
\item
Subspace JL: originally made explicit by Sarlos, this allows us to make 
statements about an entire subspace of vectors, which will allow us to 
make statements analogous to that the approximate matrix multiplication 
bounds say.
\end{itemize}

\subsection{A random projection algorithm for approximating matrix multiplication}

Recall the \texttt{BasicMatrixMultiplication} algorithm and that, using the sampling matrix formalism, we can write the output of it and what it is doing as $C=ASD$, $R=(SD)^TB$, and
$$
CR = ASD(SD)^TB = A\mathcal{S}\mathcal{S}B \approx AB ,
$$
where $\mathcal{S}=SD$ is just a way to absorb the diagonal rescaling matrix into the sampling matrix.

With this suggestive notation, here is a random projection algorithm for approximating matrix multiplication.
Given as input an $m \times n$ matrix $A$, an $n \times p$ matrix $B$, a positive integer $c$, do the following.
\begin{enumerate}
\item
Let $\Pi$ be an $n \times c$ random projection matrix, as defined above.
\item
Let $C=A\Pi$ and $R=\Pi^TB$ be sketches of the columns of $A$ and rows of $B$.
\item
Compute and return $CR = A\Pi\Pi^TB $.
\end{enumerate}
Depending on how we choose parameters, we can establish similar quality-of-approximation results with this procedure as we saw before.
That is, the matrix multiplication primitive can be done in one to two ways.
\begin{itemize}
\item
In a data-aware manner, in which we perform random sampling with sampling probabilities that depend on the input matrices.
\item
In a data-agnostic manner, in which we perform random projections without looking at the input data.
\end{itemize}
Both of these approaches will be useful later.

\newpage

\section{%
(09/23/2013):  Sampling/Projections for Least-squares Approximation}

In the next several classes, we will be discussing RandNLA algorithms for the least-squares problem.
This is a fundamental problem in linear algebra, and many of the methods in RandNLA are most easily introduced and understood in this relatively-simple setting.
Here is the reading for today.
\begin{itemize}
\item
Chapter 4 of: Mahoney, ``Randomized Algorithms for Matrices and Data'' 
\item
Drineas, Mahoney, Muthukrishnan, and Sarlos, ``Faster Least Squares Approximation'' 
\item
Sarlos, ``Improved Approximation Algorithms for Large Matrices via Random Projections'' 
\end{itemize}

Today, we will start this by covering two topics.
\begin{itemize}
\item
A brief overview of LS problems.
\item
A brief overview of sketching methods for LS problems.
\end{itemize}

\subsection{Some general thoughts on LS problems}

In many applications, we want to find an approximate solution to a problem or set of equations that, for noise reasons or whatever other reasons, does \emph{not} have a solution, or not unrelatedly does not have a unique solution.
A canonical example of this is given by the very overconstrained (i.e., overdetermined) least-squares (LS) problem, and this will be our focus for the next several classes.
Some (but not all) of what we we discuss will generalize to very underconstrained (i.e., undetermined) LS problems, roughly-square LS problems, etc., but here we focus on the simple setup of very overconstrained LS problems.

Let $A \in \mathbb{R}^{n \times d}$ and $b \in \mathbb{R}^{n}$ be given.
If $n \gg d$, in which case there are many more rows/constraints than columns/variables, then in general there does not exist a vector $x$ such that $Ax=b$.
Basically, this is since $b$ may have a part that sits outside the column space of $A$.
That is, $b\in\mathbb{R}^{n}$, but $\mbox{span}(A)$ is a $d$-dimensional subspace of $\mathbb{R}^{n}$, and so with even a little noise, numerical instability, etc., there will be a part of $b$ that is not captured as a linear combination of the columns of $A$.

In this case, a popular way to find the ``best'' vector $x$ such that $Ax \approx b$ is to minimize the norm of the residuals, i.e., to solve $\min_{x\in\mathbb{R}^{d}}||Ax-b||$, where $||\cdot||$ is some norm.
The most popular choice is the Euclidean or $\ell_2$ norm, in which case the LS problem is to minimize the sum of squares of the residual, i.e., to solve
\begin{equation}
\mathcal{Z} = \min_{x\in\mathbb{R}^{d}}\VTTNorm{Ax-b}  .
\label{eqn:ls-vanilla}
\end{equation}
If we let $A^{+}$ denote the Moore-Penrose generalized inverse of $A$, then 
\begin{equation}
x_{opt} = A^{+}b
\label{eqn:ls-vanilla-soln}
\end{equation}
is the solution to the LS problem.
Actually, we should note that $x = A^{+}b + \xi$, where $\xi \perp \mbox{span}(A)$, i.e., where $\xi\in\mathbb{R}^{n}$ is any vector perpendicular to the column span of $A$, solves the LS problem given in 
Eqn.~(\ref{eqn:ls-vanilla}), and the solution given in Eqn.~(\ref{eqn:ls-vanilla-soln}) actually is the minimal-$\ell_2$-norm solution to the LS problem.
Since we will be interested in working with this shortest or minimal-norm solution, we will call it \emph{the} solution to the LS problem.
For most of what we will talk about in the very overconstrained regression problem, worrying about having any components in this perpendicular space will not be a problem, basically since there is not a ``rest of the space'' to deal with.
We will see, however, when we consider the extension of these ideas to low-rank approximation that we will need to be a little more careful in dealing with how the top part of the spectrum of a matrix and its sketch interact with the bottom part of the spectrum of the matrix and its sketch.

This LS problem is ubiquitous and has many well-known interpretations.
A statistical interpretation is that it provides the best linear unbiased estimator to the original problem.
A geometric interpretation is that the solution is simply the orthogonal projection of $b$ onto the $\mbox{span}(A)$.
And so on.
Note that the latter interpretation is basically a statement about the data at hand, while the former interpretation is basically a statement about models and unseen data.
This parallels the algorithmic-statistical approaches we mentioned earlier.
Along these lines, here are two basic questions that people are interested in when considering LS problems.
\begin{itemize}
\item
Algorithmic question:
How long does it take to solve the LS problem ``exactly''?
(By this, we mean, say, to machine precision.)
The answer (roughly) is that the running time in the RAM model to solve the LS problem is $O(nd^2)$ time, and---as we will describe below---this can be accomplished with one of a variety of direct or indirect methods.
\item
Statistical question:
When is solving the LS problem the ``right'' thing to do?
(By this, we mean that it is the optimum for some underlying statistical model.)
The answer (roughly) is that it is when the data are ``nice'' in ways that mean that large-sample theory can be applied, e.g., that there are a large number of small components such that measure concentrates and that there are no small number of components that are particularly important or influential. 
As we will describe below, this can be checked with empirical statistics such as the leverage scores and with other regression diagnostics.
\end{itemize}
We will return to both of these points in detail below.
In particular, in terms of running time, we should be thinking about algorithms that run in $o(nd^2)$ time, and the role of the statistical leverage scores which have been used in regression diagnostics in our worst case algorithms will be particularly important.

To see how to solve the LS problem , we can define a function $f(x) = \VTTNormS{Ax-b} = (Ax-b)^T(Ax-b)$, and to find the minimizer of this function, we can set the derivative equal to zero, $\frac{\partial f}{\partial x}=0$, noting that the second derivative is positive (or SPD, if the matrix $A$ has full column rank).
Then we get $ A^TAx - A^Tb = 0 $, which is just the normal equations,
\begin{equation}
\label{eqn:normal_eqn}
A^TAx = A^Tb  .
\end{equation}
If $A$ has full column rank, then $A^TA$ is square and has full rank, and this is a $d \times d$ system of linear equations with solution 
\begin{equation}
x_{opt} = (A^TA)^{-1}A^Tb  .
\end{equation}
Of course, forming and solving the normal equations in this way is typically not recommended, but it at least provides a form for the solution.
But, in particular, note that this means that $b^{\perp} \equiv b - A x_{opt}$ is orthogonal to $\mbox{span}(A)$, i.e., ${b^{\perp}}^{T}A = 0$, or equivalently since $\mbox{span}(U) = \mbox{span}(A)$, where $U$ is an orthogonal basis for $\mbox{span}(A)$ computed from the SVD or a QR decomposition, we have that ${b^{\perp}}^{T}U=0$.

With respect to the question of how long it takes to solve LS problems, one can use so-called direct methods or so-called iterative methods. 
Here is a rough outline of \emph{direct methods} for solving LS problems.
\begin{itemize}
\item
Cholesky decomposition:
If $A$ is full rank and well-conditioned, then one can use the Cholesky decomposition to compute an upper triangular matrix $R$ such that $A^TA = R^TR$, and then one can solve the normal equations
$R^TRx = A^Tb$.
\item
QR decomposition: 
Somewhat slower but more numerically stable, especially if $A$ is rank-deficient or ill-conditioned, involves computing a QR decomposition $A=QR$ and then solving $Rx=Q^Tb$.
\item
SVD: 
Somewhat more expensive but better still if $A$ is very ill-conditioned, involves computing the SVD, $A= U \Sigma V^T$, where this is the thin or economical SVD (i.e., things that are zeroed-out by singular values are not included), in which case $x_{opt} = V \Sigma^{-1} U^T b$.
\end{itemize}
The complexity of all of theses methods is $O(nd^2)$.
That is, although the numerical properties differ and the constant factors differ, all three classes of algorithms asymptotically take a constant times $nd^2$ time.  
In most cases, using QR is a good tradeoff---but note that in certain large-scale applications, the usual rules, e.g., don't form the normal equations or don't compute the SVD, don't always hold.
In addition, for most of what we will do we are not interested in these details, since we will be computing a randomized sketch and then calling a traditional algorithm as a black box, and so we will treat all of these as similar in that sense.

Another broad class of algorithm for solving LS and other problems are 
\emph{iterative methods}.
We will return to these later.
But here we simply note that many of them boil down to conjugate gradient ideas, and (e.g., with CGNR) they typically have a running time that is something like $O(\kappa(A)\mbox{nnz}(A)\log(1/\epsilon))$ time.
In particular, the running time depends on the error parameter as $O(\log(1/\epsilon))$, and not $\mbox{poly}(1/\epsilon)$, time.

A third broad class of algorithms for solving LS problems use the recursive structure of well-known \emph{Strassen-like methods} for matrix multiplication---basically, those algorithms can also be applied to rectangular LS problems, with similar improvements in worst-case running time.
These are of theoretical interest, and thus they are worth mentioning, but are never used in practice, and so we won't focus on them.

\subsection{Deterministic and randomized sketches for LS Problems}

At a high level, RandNLA algorithms---in general but in particular when applied to LS problems---do one of two thing.
\begin{itemize}
\item
Construct a sketch (with a random sampling or random projection procedure) and solve the subproblem on that sketch with a traditional black box NLA algorithm.
\item
Construct a sketch (with a random sampling or random projection procedure) and use that sketch as a preconditioner for a traditional black box NLA algorithm on the original problem.
\end{itemize}
In both cases, the randomized algorithms interface with traditional NLA algorithms (in two different ways), and so let's start by asking ``What are properties of sketches that lead to good solutions?''

Thus, for the rest of today, we won't specify whether our sketches are deterministic or randomized.
We will be interested in properties of some sketch and how they relate to necessary/sufficient conditions to get good approximate solutions to the original LS problem.
Roughly, we will show certain conditions, and later we will show that one can construct sketches via random sampling/projection that satisfy those conditions quickly.
More generally, it is worth keeping in mind what are properties of the sketches and linear algebra versus what are properties of the randomization.

In this course, we will deal with so-called linear sketches.
Operationally, this means that we can write the operation/action of the sketch as a linear function.
Note, in particular, that random projection matrices and random sampling matrices satisfy this.
The advantages of working with linear sketches include: 
we can take advantage of linear theory stuff (this may be obvious for NLA, but it is actually useful much more generally); and it is easy to update sketches (this matters in streaming, memory constrained environments, etc.).
The disadvantages of working with linear sketches mean that we might loose something, compared to using a broader class of sketches.
Note, however, that a lot of work in ML, e.g., with kernels, say basically that tamely nonlinear stuff can be done linearly.
So, this is an idea that is used more generally, and in general we will work with linear sketches.
But a question worth keeping in the back of your mind is: what are metric spaces that don't embed well, either in general or with linear sketches, since those might have problems.

Now, onto the properties that we want a good sketch to have to help solve linear regression.
Let's let $X$ be an arbitrary sketching matrix, i.e., any matrix.  
By this, we mean an arbitrary matrix (randomized or not, tractable to compute or not, etc.) that we are going to apply to $A$ and $b$ to construct a sketch.
For example, $X$ could be a sampling matrix like we saw before with matrix multiplication sampling algorithms, $X$ could be a dense projection matrix like we saw before, or $X$ could be $S^THD$, $THD$, or other structured random projections.
When forming a sketch, we will be replacing the original LS problem
$$
\mathcal{Z} = \min_{x\in\mathbb{R}^{d}} \VTTNorm{Ax-b}  ,
$$
the solution of which is $x_{opt} = A^+b$, with a sketched LS problem
$$
\tilde{\mathcal{Z}} = \min_{x\in\mathbb{R}^{d}} \VTTNorm{X(Ax-b)}  ,
$$
the solution of which is $\tilde{x}_{opt} =(XA)^{+}Xb$.
And we want to ask what are properties that $X$ needs to satisfy s.t.
\begin{eqnarray*}
\tilde{x}_{opt} &\approx& x_{opt}  \\
\VTTNorm{A\tilde{x}_{opt}} &\approx& \VTTNorm{Ax_{opt}-b}  .
\end{eqnarray*}
Several comments are in order.
\begin{itemize}
\item
The second requirement is a statement and is more common in TCS where one might not even be able to obtain a ``certificate'' for the solution.
The first requirement is usually harder to get in worst-case approximation algorithm theory but usually comes for free in matrix problems with $\ell_2$ objectives.
\item
Moreover, the bound on the vector achieving the optimal solution is typically of greater interest in NLA and ML/data applications, where the vector is used form something downstream such as classification.
\item
For matrix extensions of these ideas, we typically want results for the objective, since we will measure quality by norm reconstruction, rather than capturing an actual set of vectors or an actual subspace.
\item
For $\ell_1$ and other objectives, the connections between objectives and certificates and what one can compute from the other are more tenuous.
\end{itemize}

With this in place, here are two important structural conditions.
Let $A= U \Sigma V^T = U_A \Sigma_A V_A^T$ be the SVD of $A$, and let $b^{\perp} = U_AU_A^TA$ be the part of $b$ sitting outside $\mbox{span}(A)$, and note that 
$\mathcal{Z} = \VTTNorm{Ax_{opt}-b}= \VTTNorm{b^{\perp}}$.
Then, here are two conditions.
\begin{itemize}
\item
\textbf{Condition I:}
\begin{equation}
\label{eqn:lemma1_ass1}
 \sigma_{min}(XU_A) \ge 1/\sqrt{2} .
\end{equation}
\item
\textbf{Condition II:}
\begin{equation}
\label{eqn:lemma1_ass2}
 \VTTNormS{U_AX^TXb^{\perp}} \le \frac{\epsilon}{2}\mathcal{Z}^2 .
\end{equation}
\end{itemize}
Before proceeding, let's consider these conditions; here are several comments.
\begin{itemize}
\item
We have defined this in terms of $U_A$ from the SVD, but we get exactly the same structural conditions for any matrix $Q$, e.g., from the QR decomposition.
Although this issue doesn't matter here so much, it will matter more when we consider the extension of these ideas to low-rank matrix approximation problems.
\item
Although Condition I is that $\sigma_{min}(XU_A) \ge 1/\sqrt{2}$, i.e., we only need a lower bound on the singular values of $\sigma(XU_A)$, all of our constructions will be such that $\| 1-\sigma_i(XU_A) \| \le 1-2^{-1/2}$.
Thus, one should think of $XU_A$ as an approximate isometry (or, more imprecisely, an approximate rotation).
We want that $X$, viewed as a function $f:\mathbb{R}^{n}\rightarrow\mathbb{R}^{r}$, with $r \approx d$, is roughly a rotation/isometry.
In particular, although we zero-out most of the coordinates, the mapping to the remaining are an acute perturbation with respect to the original data.
\item
Condition II states that $Xb^{\perp}=XU_A^{\perp}{U_A^{\perp}}^Tb$ is still roughly orthogonal to $XU_A$.
It is not surprising that we need a condition on the right hand side vector $b$, but it is surprising that we can satisfy this (with random sampling and random projection algorithms) without looking at the right hand side at all, i.e., either data-agnostic random projections or random sampling methods that only depend on information in $A$.
Of course, in certain practical cases, one can sometimes do better by looking at the right hand side, and extensions to $\ell_1$ regression, etc., typically need to look at the right hand side, although sometimes implicitly by defining an augmented matrix 
$ \left[\begin{array}{cc}
         A & -b 
      \end{array}\right] $
and working on that.
\end{itemize}

Here are a few extreme cases to consider.
\begin{itemize}
\item
Let $X=I_n$, viewed as a function $\mathbb{R}^{n}\rightarrow\mathbb{R}^{n}$.
Then, $\sigma_{min}(XU_A) = \sigma_{max}(XU_A) = 1$ (since $XU_A=U_A$), and $U_A^TX^TXb^{\perp}=0$.
In this case, constructing $X$ is ``easy,'' and solving the ``subproblem'' is the same as the original problem and so it ``hard.''
\item
Let $X=U_A^T$, viewed as a function $\mathbb{R}^{n}\rightarrow\mathbb{R}^{d}$.
Then, $\sigma_{min}(XU_A) = \sigma_{max}(XU_A) = 1$ (since $XU_A=I_d$), and $U_A^TX^TXb^{\perp}=0$.
In this case, constructing $X$ is ``hard,'' since it involves computing $U_A^T$ which almost amounts to solving the original problem, but solving the subproblem is easy, I think.
(Similar statements could be made if $X$ was the ``Q'' matrix from a QR decomposition.)
\end{itemize}
So, the goal for what we will be doing will be to construct a sketch that is relatively-easy to construct and such that solving the subproblem is also relatively easy, in the sense that both take $o(nd^2)$ time.

Given all of that, here are our main lemmas for those structural conditions.
Basically, these lemmas say that if we have a sketching matrix $X$ that satisfies those two conditions, then we have a relative-error approximation to the solution to the LS problem, on both the objective and the certificate.
We will do the first (a lemma about being close with respect to the objective function value) now, and we will do the next two (lemmas about being close with respect to the certificate or solution vector) next time.
Note that, for these lemmas, we are interested in quality-of-approximation guarantees for a given sketching matrix $X$, i.e., we don't worry about the time it takes to construct $X$.
We will get to running time considerations soon enough.

\begin{lemma} 
\label{lem:suff_cond1}
Consider the overconstrained least squares approximation problem 
and let the matrix $U_A \in \mathbb{R}^{n \times d}$ contain the top $d$ left singular vectors of $A$. 
Assume that the matrix $X$ satisfies 
conditions~(\ref{eqn:lemma1_ass1}) and~(\ref{eqn:lemma1_ass2}) above, 
i.e., Condition I and Condition II,
for some $\epsilon \in (0,1)$. 
Then, the solution vector $\tilde{x}_{opt}$ to the least squares approximation problem
satisfies:
\begin{equation}
\label{eqn:lemma1_eq3}
\VTTNorm{A\tilde{x}_{opt}-b} \le (1+\epsilon) \mathcal{Z}  .
\end{equation}
\end{lemma}

Before proceeding with the proof of this lemma, here are a few comments.
\begin{itemize}
\item
This lemma is a deterministic statement, i.e., there is no randomness, and it holds for any matrix $X$ that satisfies those two conditions.
Failure probabilities in randomized matrix algorithms will enter into the construction of $X$ and whether $X$ satisfies those two conditions.
\item
We will be mostly interested in worst-case \emph{a priori} bounds, and we will show that $X$ satisfies these two conditions for worst-case input; but one could easily ask for \emph{a posteriori} bounds, by, e.g., sampling/projecting less aggressively and checking if these conditions are satisfied.  
We won't do this for the LS regression problem, but we will consider this approach for low-rank matrix approximation problems, and this is probably the way to implement randomized matrix algorithms more generally.
\end{itemize}

\begin{Proof}
Let us first rewrite the down-scaled regression problem induced by $X$ as
\begin{eqnarray}
\min_{x \in \mathbb{R}^d} \VTTNormS{ Xb - XAx}
\label{eqn:ds1} &=& \min_{y \in \mathbb{R}^d} \VTTNormS{
X(Ax_{opt}+b^{\perp}) -
XA(x_{opt}+y) }                 \\
\nonumber
  &=& \min_{y \in \mathbb{R}^d} \VTTNormS{ Xb^{\perp} - XAy}    \\
\label{eqn:ds2}
  &=& \min_{z \in \mathbb{R}^d} \VTTNormS{ Xb^{\perp} - XU_Az }.
\end{eqnarray}
(\ref{eqn:ds1}) follows since $b=Ax_{opt}+b^{\perp}$ and (\ref{eqn:ds2}) follows since the columns of the matrix $A$ span the same subspace as the columns of $U_A$. Now, let $\tilde x_{opt} \in \mathbb{R}^d$  be such that $U_A z_{opt} = A(\tilde{x}_{opt}-x_{opt})$, and note that $z_{opt}$ minimizes Eqn.~(\ref{eqn:ds2}). The latter fact follows since
$$\VTTNormS{Xb^{\perp} - XA(\tilde x_{opt}-x_{opt})} =
\VTTNormS{Xb^{\perp} + X(b-b^{\perp}) -
XA\tilde{x}_{opt}} = \VTTNormS{Xb - XA\tilde{x}_{opt}}.$$
Thus, by the normal equations~(\ref{eqn:normal_eqn}), we have that
\begin{equation*}
\label{eqn:ds-normal} (XU_A)^TXU_A z_{opt} =
(XU_A)^T Xb^{\perp}.
\end{equation*}
Taking the norm of both sides and observing that under condition~(\ref{eqn:lemma1_ass1}) we have $\sigma_i((XU_A)^TXU_A) = \sigma_i^2(XU_A) \ge 1/\sqrt{2}$, for all $i$, it follows that
\begin{equation}
\label{eqn:z-norm1}
  \VTTNormS{z_{opt}} / 2  \le \VTTNormS{(XU_A)^TXU_Az_{opt}} = \VTTNormS{
(XU_A)^T Xb^{\perp} }.
\end{equation}
Using condition~(\ref{eqn:lemma1_ass2}) we observe that
\begin{equation}
\label{eqn:z-norm2}
   \VTTNormS {z_{opt}} \le \epsilon\mathcal{Z}^2.
\end{equation}

\noindent 
Let us rewrite the norm of the residual vector as
\begin{eqnarray}
\VTTNormS{ b - A\tilde{x}_{opt} } \nonumber
   &=& \VTTNormS{ b - Ax_{opt} + Ax_{opt} - A\tilde{x}_{opt} }  \\
\label{eqn:pfCeq1}
   &=& \VTTNormS{ b - Ax_{opt} } + \VTTNormS{ Ax_{opt} - A\tilde{x}_{opt} } \\
\label{eqn:pfCeq2}
   &=& \mathcal{Z}^{2} + \VTTNormS{ U_Az_{opt}} \\
\label{eqn:pfCeq3}
   &\leq& \mathcal{Z}^{2} + \epsilon \mathcal{Z}^{2} ,
\end{eqnarray}
where (\ref{eqn:pfCeq1}) follows by Pythagoras, since $b - Ax_{opt} = b^\perp$, which is orthogonal to $A$, and consequently to $A(x_{opt} - \tilde{x}_{opt})$; (\ref{eqn:pfCeq2}) follows by the definition of $z_{opt}$ and $\mathcal{Z}$; and (\ref{eqn:pfCeq3}) follows by (\ref{eqn:z-norm2}) and the orthogonality of $U_A$. 
The first claim of the lemma follows since $\sqrt{1+\epsilon} \le 1+\epsilon$.
\end{Proof}

Next time, we will start by proving that the vector achieving the optimum in the subsampled problem is a very good approximation to the vector achieving the optimum in the original problem.

\newpage

\section{%
(09/25/2013):  Sampling/Projections for Least-squares Approximation, Cont.}

We continue with the discussion from last time.
There is no new reading, just the same as last~class.

Recall that last time we provided a brief overview of LS problems and a brief overview of sketching methods for LS problems.
For the latter, we provided a lemma that showed that under certain conditions the solution of a sketched LS problem was a good approximation to the solution of the original LS problem, where good is with respect to the objective function value.
Today, we will focus on three things.
\begin{itemize}
\item
Establishing goodness results for the sketched LS problem, where goodness is with respect to the certificate or solution vector.
\item
Relating these two structural conditions and the satisfaction of the two conditions by random sampling/projection to exact and approximate matrix multiplication algorithms.
\item
Putting everything together into two basic (but still slow---we'll speed them up soon enough) RandNLA algorithms for the LS problem.
\end{itemize}

\subsection{Deterministic and randomized sketches and LS problems, cont.}

Last time we identified two structural conditions, and we proved that if those structural conditions are satisfied by a sketching matrix, then the solution to the subproblem defined by that sketching matrix has a solution that is a relative-error approximation to the original problem, i.e., that the objective function value of the original problem is approximate well.
Now, we will prove that the vector itself solving the subproblem is a good approximation of the vector solving the original problem.
After that, we will show that random sampling and random projection matrices satisfy those two structural conditions, for appropriate values of the parameter settings.

\begin{lemma}
\label{lem:suff_cond2}
Same setup as the previous lemma.
Then
\begin{equation}
\label{eqn:lemma1_eq4}
\VTTNorm{x_{opt}-\tilde{x}_{opt}}
  \leq \frac{1}{\sigma_{min}(A)}\sqrt{\epsilon}\mathcal{Z}  .
\end{equation}
\end{lemma}

\begin{Proof}
If we use the same notation as in the proof of the previous lemma, then $A(x_{opt}-\tilde{x}_{opt})=U_Az_{opt}$. 
If we take the norm of both sides of this expression, we have that
\begin{eqnarray}
\VTTNormS{ x_{opt}-\tilde{x}_{opt} } \label{eqn:pfDeq1}
  &\leq& \frac {\VTTNormS{U_Az_{opt}}} {\sigma_{min}^2(A)} \\
\label{eqn:pfDeq2}
  &\leq& \frac {\epsilon\mathcal{Z}^2} {\sigma_{min}^2(A)},
\end{eqnarray}
where (\ref{eqn:pfDeq1}) follows since $\sigma_{min}(A)$ is the smallest singular value of $A$ and since the rank of $A$ is $d$; and (\ref{eqn:pfDeq2}) follows by
a result in the proof of the previous lemma
and the orthogonality of $U_A$. 
Taking the square root, the second claim of the lemma follows.
\end{Proof}

If we make no assumption on $b$, then~(\ref{eqn:lemma1_eq4}) from Lemma~\ref{lem:suff_cond2} may provide a weak bound in terms of $\VTTNorm{x_{opt}}$. If, on the other hand, we make the additional assumption that a constant fraction of the norm of $b$ lies in the subspace spanned by the columns of $A$, then~(\ref{eqn:lemma1_eq4}) can be strengthened.
Such an assumption is reasonable, since most least-squares problems are practically interesting if at least some part of $b$ lies in the subspace spanned by the columns of $A$.

\begin{lemma}
\label{lem:suff_cond3}
Same setup as the previous lemma, and assume that $\VTTNorm{U_AU_A^Tb} \geq \gamma\VTTNorm{b}$, for some fixed $\gamma \in (0,1]$.
Then, it follows that
\begin{equation}
\VTTNorm{x_{opt}-\tilde{x}_{opt}}
  \leq \sqrt{\epsilon}\left(\kappa(A)\sqrt{\gamma^{-2}-1}\right)\VTTNorm{x_{opt}} .
\end{equation}
\end{lemma}
\begin{Proof}
Since $\VTTNorm{U_AU_A^Tb} \geq \gamma\VTTNorm{b}$, it follows that
\begin{eqnarray}
         \mathcal{Z}^2
\nonumber    &=&    \VTTNormS{b} - \VTTNormS{U_A U_A^T b}          \\
\nonumber    &\leq& (\gamma^{-2}-1) \VTTNormS{U_A U_A^T b}         \\
\nonumber    &\leq&
{\sigma_{\max}^{2}(A)}(\gamma^{-2}-1)\VTTNormS{x_{opt}}  .
\end{eqnarray}
This last inequality follows from $U_AU_A^Tb = Ax_{opt}$, which implies $$ \VTTNorm{U_A U_A^T b} = \VTTNorm{Ax_{opt}} \leq \TNorm{A} \VTTNorm{x_{opt}} = \sigma_{\max}\left(A\right)\VTTNorm{x_{opt}}. $$
By combining this with eqn. (\ref{eqn:lemma1_eq4}) of Lemma~\ref{lem:suff_cond2}, the lemma follows.
\end{Proof}

\subsection{Connections with exact and approximate matrix multiplication}

\subsubsection{An aside: approximating matrix multiplication for vector inputs}

Before continuing with our discussion of LS regression, here is a simple example of applying the matrix multiplication ideas that might help shed some light on the form of the bounds as well as when the bounds are tight and when they are not.

Let's say that we have two vectors $x,y\in\mathbb{R}^{n}$ and we want to approximate their product by random sampling.
In this case, we are approximating $x^Ty$ as $x^TSS^Ty$, where $S$ is a random sampling matrix that, let's assume, is constructed with nearly optimal sampling probabilities.
Then, our main bound says that, under appropriate assumptions on the number $c \ll n$ of random samples we draw, then we get a bound of the form
$$
\FNorm{x^Ty - x^TSS^Ty} \le \epsilon \FNorm{x}\FNorm{y} ,
$$
which, since we are dealing with the product of two vectors simplifies to
$$
\left| x^Ty - x^TSS^Ty \right| \le \epsilon \VTTNorm{x}\VTTNorm{y}  .
$$
The question is: when is this bound tight, and when is this bound loose, as a function of the input data?
Clearly, if $x \perp y$, i.e., if $x^Ty=0$, then this bound will be weak.
In the other hand if $y = x$, then $x^Ty=x^Tx=\VTTNormS{x}$, in which case this bound says that
$$
\left| x^Tx - x^TSS^Tx \right| \le \epsilon \VTTNormS{x}  ,
$$
meaning that the algorithm provides a relative error guarantee on $x^Tx = \VTTNormS{x}$.
(We can make similar statements more generally if we are multiplying two rectangular orthogonal matrices to form a low-dimensional identity, and this is important for providing subspace-preserving sketches.)

The lesson here is that when there is cancellation the bound is weak, and that the scales set by the norms of the component matrices are in some sense real.
For general matrices, the situation is more complex, since subspace can interact in more complicated ways, but the similar ideas goes through.

\subsubsection{Understanding and exploiting these structural conditions via approximate matrix multiplication}

These lemmas say that if our sketching matrix $X$ satisfies Condition I and Condition II, then we have relative-error approximation on both the solution vector/certificate and on the value of the objective at the optimum.
There are a number of things we can do with this, and here we will focus on establishing a prioi running time guarantees for any, i.e., worst-case, input.
But, before we get into the algorithmic details, however, we will outline how these structural conditions relate to our previous approximate matrix  multiplication results, and how we will use the latter to prove our results.

The main point to note is that both Condition I and Condition II can be expressed as approximate matrix multiplications and thus bounded by our approximate matrix multiplication results from a few classes ago.
To see this, observe that a slightly stronger condition than Condition I is that 
$$
\abs{ 1-\sigma_1(XU_A) } \le 2^{-1/2} \quad \forall i  ,
$$
and that $U_A$ is an $n \times d$ orthogonal matrix, with $n \gg d$, we have that $U_A^TU_A=I_d$, and so this latter condition says that $U_A^TU_A \approx U_A^TX^TXU_A$ in the spectral norm, i.e., that
$$
\TNorm{I-(XU_A)^TXU_A} \le 2^{-1/2}  .
$$
Similarly, since $U_A^T b^{\perp}=0$, Condition II says that $0 \approx U_A^TX^TXb^{\perp}$ with respect to the Frobenius norm, i.e., that
$$
\FNormS{U_A^TX^TXb^{\perp}} \le \frac{\epsilon}{2} \mathcal{Z}^{2}.
$$
Of course, this is simply the Euclidean norm, since $b^{\perp}$ is simply a vector.
For general matrices, for the Frobenius norm, the scale of the right hand side error, i.e., the quantity that is multiplied by the $\epsilon$, depends on the norm of the matrices entering into the product. 
But, the norm of $b^{\perp}$ is simply the residual value $\mathcal{Z}$, which sets the scale of the error and of the solution.
And, for general matrices, for the spectral norm, there were quantities that depended on the spectral and Frobenius norm of the input matrices, but for orthogonal matrices like $U_A$, those are $1$ or the low dimension $d$, and so they can be absorbed into the sampling complexity.

\subsubsection{Bounds on approximate matrix multiplication when information about both matrices is unavailable}

The situation about bounding the error incurred in the two structural conditions is actually somewhat more subtle than the previous discussion would imply.
The reason is that, although we might have access of information such as the leverage scores (row norms of one matrix) that depend on $U_A$, we in general don't have access to any information in $b^{\perp}$ (and thus the row norms of it).
Nevertheless, an extension of our previous discussion still holds, and we will describe it now.

Observe that the nearly optimal probabilities 
$$
p_k \ge \frac{ \beta \VTTNorm{A^{(k)}} \VTTNorm{B_{(k)}} }{ \sum_{k^\prime=1}^{n} \VTTNorm{A^{(k^\prime)}} \VTTNorm{B_{(k^\prime)}} }  ,
$$
for approximating the product of two general matrices $A$ and $B$ use information from both matrices $A$ and $B$ in a very particular form.  
In some cases, such detailed information about both matrices may not be available. 
In particular, in some cases, we will be interested in approximating the product of two different matrices, $A$ and $B$, when only information about $A$ (or, equivalently, only $B$) is available.
Somewhat surprisingly, in this case, we can still obtain partial bounds (i.e., similar form, but with slightly weaker concentration) of the form we saw above.
Here, we present results for the \textsc{BasicMatrixMultiplication} algorithm for two other sets of probabilities.

In the first case, to estimate the product $AB$ one could use the probabilities (\ref{non_near_optimal_probs1}) which use information from the matrix $A$ only.
In this case $\FNorm{AB-CR}$ can still be shown to be small in expectation; the proof of this lemma is similar to that of 
our theorem for the nearly-optimal probabilities from a few classes ago, 
except that the indicated probabilities are used.

\begin{lemma}
\label{lem:basicmatmult_nonopt1}
Suppose $A \in \mathbb{R}^{m \times n}$, 
$B \in \mathbb{R}^{n \times p}$, 
$c \in \mathbb{Z}^{+}$ such that $1 \le c \le n$,
and $\left\{ p_i \right\}_{i=1}^{n}$ are such that $\sum_{i=1}^{n} p_i = 1$ and 
such that 
\begin{equation}
p_k \ge \frac{\beta \VTTNormS{A^{(k)}} }{ \FNormS{A} }  
\label{non_near_optimal_probs1}
\end{equation}
for some positive constant $\beta \le 1$.  
Construct $C$ and $R$ with the \textsc{BasicMatrixMultiplication} algorithm, and let $C R$ be an approximation to $A B$.
Then:
\begin{equation}
\Expect{ \FNormS{AB-CR} } \le \frac{1}{      \beta c } \FNormS{A}\FNormS{B}   .
\label{nonopt1_matmul_expect_abcr_f2}
\end{equation}
\end{lemma}

Following the analysis of 
our theorem for the nearly-optimal probabilities from a few classes ago,
we can let $\mathcal{M}=\max_{\alpha}\frac{\VTTNorm{B_{(\alpha)}}}{\VTTNorm{A^{(\alpha)}}}$, let $\delta \in (0,1)$ and let $\eta=1+\frac{\FNorm{A}}{\FNorm{B}}\mathcal{M}\sqrt{(8/\beta)\log(1/\delta)}$, in which case it can be shown that, with probability at least $1-\delta$:
$$
\FNormS{AB-CR} \le \frac{\eta^{2}}{      \beta c } \FNormS{A}\FNormS{B}   .
$$
Unfortunately, the assumption on $\mathcal{M}$, which depends on the maximum ratio of two vector norms, is sufficiently awkward that this result is not  useful.
Nevertheless, we can still remove the expectation from Eqn.~(\ref{nonopt1_matmul_expect_abcr_f2}) with Markov's inequality, paying the factor of $1/\delta$, but without any awkward assumptions, assuming that we are willing to live with a result that holds with constant probability.
This will be fine for several applications we will encounter, and when we use Lemma~\ref{lem:basicmatmult_nonopt1}, this is how we will use it.

We should emphasize that for most probabilities, e.g., even simple probabilities that are proportional to (say) the Euclidean norm of the columns of $A$ (as opposed to the norm-squared of the columns of $A$ or the product of the norms of the columns of $A$ and the corresponding rows of $B$), we obtain much uglier and unusable expressions, e.g., we get awkward factors such as $\mathcal{M}$ above.
Lest the reader think that any sampling probabilities will yield interesting results, even for the expectation, here are the analogous results if sampling is performed u.a.r.
Note that the scaling factor of $\sqrt{\frac{n}{c}}$ is \emph{much} worse than anything we have seen so far---and it means that we would have to choose $c$ to be \emph{larger} than $n$ to obtain nontrivial results, clearly defeating the point of random sampling in the first place.

\begin{lemma}
\label{lem:basicmatmult_nonopt4}
Suppose $A \in \mathbb{R}^{m \times n}$, 
$B \in \mathbb{R}^{n \times p}$, 
$c \in \mathbb{Z}^{+}$ such that $1 \le c \le n$,
and $\left\{ p_i \right\}_{i=1}^{n}$ are such that 
\begin{equation}
p_k =  \frac{1}{n}   .
\label{non_near_optimal_probs4}
\end{equation}
Construct $C$ and $R$ with the \textsc{BasicMatrixMultiplication} algorithm, and let $C R$ be an approximation to $A B$.
Then:
\begin{equation}
\Expect{ \FNorm{AB-CR} } 
   \le \sqrt{\frac{n}{c}}\Sqrt{\sum_{k=1}^{n}\VTTNormS{A^{(k)}}\VTTNormS{B_{(k)}}}  .
\label{nonopt4_matmul_expect_abcr_f}
\end{equation}
Furthermore, let $\delta \in (0,1)$ and 
$\gamma = \frac{n}{\sqrt{c}}\sqrt{8\log\left(1/\delta\right)}
          \max_{\alpha}\VTTNorm{A^{(\alpha)}}\VTTNorm{B_{(\alpha)}}$;
then with probability at least $1-\delta$:
\begin{equation}
\FNorm{AB-CR} 
   \leq \sqrt{\frac{n}{c}}
        \Sqrt{\sum_{k=1}^{n}\VTTNormS{A^{(k)}}\VTTNormS{B_{(k)}}}
     + \gamma      .
\label{nonopt4_whp_abcr_f}
\end{equation}
\end{lemma}

\subsection{Random sampling and random projection for LS approximation}

So, to take advantage of the above two structural results and bound them with our matrix multiplication bounds, we need to perform the random sampling with respect to the so-called \emph{statistical leverage scores}, which are defined as $\VTTNorm{U_{(i)}}$, where $U_{(i)}$ is the $i^{th}$ row of any orthogonal matrix for $\mbox{span}(A)$.
If we normalize them, then we get the \emph{leverage score probabilities}:
\begin{equation}
p_i = \frac{1}{d} \VTTNormS{ U_{(i)} } .
\end{equation}
These will be important for our subsequent discussion, and so there are several things we should note about them.
\begin{itemize}
\item
Since $U$ is an $n \times d$ orthogonal matrix, the normalization is just the lower dimension $d$, i.e., $d=\FNormS{U}$.
\item
Although we have defined these scores i.t.o. a particular basis $U$, they don't depend on that particular basis, but instead they depend on $A$, or actually on $\mbox{span}(A)$.
To see this, let $P_A = AA^+$ be a projection onto the span of $A$, and note that $P_A= QRR^{-1}Q=QQ^T$, where $R$ is any square non-singular orthogonal transformation between orthogonal matrices for $\mbox{span}(A)$.
So, in particular, up to the scaling factor of $\frac{1}{d}$, the leverage scores equal the diagonal elements of the projection matrix $P_A$:
\begin{eqnarray*}
\left(P_A\right)_{ii} &=& \left(U_AU_A^T\right)_{ii} = \VTTNormS{{U_A}_{(i)}}  \\
                      &=& \left(Q_AQ_A^T\right)_{ii} = \VTTNormS{{Q_A}_{(i)}}.  
\end{eqnarray*}
Thus, they are equal to the diagonal elements of the so-called hat matrix.
\item
These are scores that quantify \emph{where} in the high-dimensional space $\mathbb{R}^{n}$ the (singular value) information in $A$ is being sent (independent of what that information is).
\item
They capture a notion of leverage or influence that the $i^{th}$ constraint has on the LS fit.
\item
They can be very uniform or very nonuniform.
E.g., if 
$ U_A = \left[\begin{array}{c}
           I  \\
           0 \\
        \end{array}\right]   $, then they are clearly very nonuniform, but if $U_A$ consists of a small number of columns from a truncated Hadamard matrix or a dense Gaussian matrix, then they are uniform or nearly uniform.
\end{itemize}

With that in place, here we will present two algorithms that compute relative-error approximations to the LS problem.

First, we start with a random sampling algorithm, given as Algorithm~\ref{alg:slow-sample-ls-alg}.

\begin{algorithm}
\caption{A ``slow'' random sampling algorithm for the LS problem.}
\label{alg:slow-sample-ls-alg}
\begin{algorithmic}[1]
    \REQUIRE An $n \times d$ matrix $A$, with $n \gg d$, an $n$-vector $b$
    \ENSURE  A $d$-vector $\tilde{x}_{opt}$
    \STATE Compute $p_i = \frac{1}{d}\VTTNormS{U_{(i)}}$, for all $i\in[n]$, 
           from the QR or the SVD.
    \STATE Randomly sample $r \gtrsim O(\frac{d \log d}{\epsilon})$ rows of 
           $A$ and elements of $b$, rescaling each by $\frac{1}{rp_{i_t}}$, 
           i.e., form $SA$ and $Sb$.
    \STATE Solve $\min_{x\in\mathbb{R}^{d}}\VTTNorm{SAx-Sb}$ with a black 
           box to get $\tilde{x}_{opt}$.
    \STATE Return $\tilde{x}_{opt}$.
\end{algorithmic}
\end{algorithm}

For this algorithm, one can prove the following theorem.
The idea of the proof is to combine the structural lemma with matrix multiplication bounds that show that under appropriate assumptions on the size of the sample, etc., that the two structural conditions are satisfied.
\begin{theorem}
Algorithm~\ref{alg:slow-sample-ls-alg} returns a 
$(1\pm\epsilon)$-approximation to the LS objective and an 
$\epsilon$-approximation to the solution vector.
\end{theorem}

Next, we start with a random projection algorithm, given as Algorithm~\ref{alg:slow-project-ls-alg}.

\begin{algorithm}
\caption{A ``slow'' random projection algorithm for the LS problem.}
\label{alg:slow-project-ls-alg}
\begin{algorithmic}[1]
    \REQUIRE An $n \times d$ matrix $A$, with $n \gg d$, an $n$-vector $b$
    \ENSURE  A $d$-vector $\tilde{x}_{opt}$
    \STATE Let $S$ be a random projection matrix consisting of scaled i.i.d. 
           Gaussians, $\{\pm1\}$, etc., random variables.
    \STATE Randomly project onto $r \gtrsim O(\frac{d \log d}{\epsilon})$ rows,
           i.e., linear combination of rows of $A$ and elements of $b$.
    \STATE Solve $\min_{x\in\mathbb{R}^{d}}\VTTNorm{SAx-Sb}$ with a black 
           box to get $\tilde{x}_{opt}$.
    \STATE Return $\tilde{x}_{opt}$.
\end{algorithmic}
\end{algorithm}

For this algorithm, one can prove the following theorem.
As before, the idea of the proof is to combine the structural lemma with the random projection version of matrix multiplication bounds that are in the first homework to show that under appropriate assumptions on the size of the sample, etc., that the two structural conditions are satisfied.
\begin{theorem}
Algorithm~\ref{alg:slow-project-ls-alg} returns a 
$(1\pm\epsilon)$-approximation to the LS objective and an 
$\epsilon$-approximation to the solution vector.
\end{theorem}

We are not going to go into the details of the proofs of these two theorems, basically since they will parallel proofs of ``fast'' versions of these two results that we will discuss in the next few classes.
But, it is worth pointing out that you do get good quality-of-approximation bounds for the LS problem with these algorithms.
The problem is the running time.
Both of these algorithms take at least as long to run (at least in terms of worst-case FLOPS in the RAM model) as the time to solve the problem exactly with traditional deterministic algorithms, i.e., $\Theta(nd^2)$ time.
For Algorithm~\ref{alg:slow-sample-ls-alg}, the bottleneck in running time is the time to compute the leverage score importance sampling distribution exactly.
For Algorithm~\ref{alg:slow-project-ls-alg}, the bottleneck in running time is the time to implement the random projection, i.e., to do the matrix-matrix multiplication associated with the random projection, and since we are projecting onto roughly $d \log d$ dimensions the running time is actually $\Omega(nd^2)$.
Thus, they are ``slow'' since they are slower than a traditional algorithms---at least in terms of FLOPs in an idealized RAM model, but note that they may, and in some cases are, faster on real machines, basically for communication reasons, and similarly they might be faster in parallel-distributed environments.
In particular, the random projection is just matrix-matrix multiplication, and this can be faster than doing things like QR or the SVD, even if the FLOP count is the same.
But, we will focus on FLOPS and so we want algorithms to runs in $o(nd^2)$ time.
We will use structured or Hadamard-based random projections, which can be implemented with Fast Fourier methods, so that the overall running time will be $o(nd^2)$.
There will be two ways to do this:
first, call a black box (the running time bottleneck of which is a Hadamard-based random projection) to approximate the leverage scores, and use them as the importance sampling distribution; and
second, do a Hadamard-based random projection to uniformize the leverage scores and sample uniformly.

In the next few classes, we will get into these issues.
Why random projections satisfy matrix multiplication bounds might be a bit of a mystery, partly since we have focused less on it, so we get into the details of two related forms of the random projection.
Also, the black box to approximate the leverage scores might be surprising, since it isn't obvious that they can be computed quickly, so we will get into that.
All of the results we will describe will also hold for general random sampling with exact leverage scores and general random projections, but we will get into the details for the fast versions, so we can make running time claims for analogues of the fast sampling and projection versions of above two algorithms.

\newpage

\section{%
(09/30/2013):  Fast Random Projections and FJLT}

Today, we will discuss a particular form of random projections known as structured random projections or the FJLT that are ``fast'' in that one can use fast Fourier methods to apply them quickly to arbitrary or worst case input.
We will be able to use this to speed up both random projection as well as random sampling RandNLA algorithms for a wide variety of problems.
Here is the reading for today.
\begin{itemize}
\item
Ailon and Chazelle, ``The fast Johnson–Lindenstrauss transform and approximate nearest neighbors'' 
\item
Matousek, ``On variants of the Johnson-Lindenstrauss lemma'' 
\item
Drineas, Magdon-Ismail, Mahoney, and Woodruff, ``Fast approximation of matrix coherence and statistical leverage'' 
\end{itemize}

Today we will focus on two things.
\begin{itemize}
\item
An introduction to fast Fourier/Hadamard-base random projection methods.
\item
An introduction to how to use these methods for LS approximation.
\end{itemize}

\subsection{Background on ``fast'' random projections methods}

Let's start with the basic AC Hadamard-based FJLT.
To set the context, recall that applying a random projection matrix consisting of i.i.d Gaussians or i.i.d. $\{\pm1\}$ random variables to an arbitrary input vector take ``matrix multiplication time,'' since we have to actually implement the random projection.
(By this, we mean the time to perform an in general dense matrix-vector multiplication---which is \emph{not} via Strassen-like algorithms, except for purely theoretical considerations.)
For an $n \times d$ matrix, since this involves projecting with an $r \times n$ matrix, where $r \gtrsim d$, this is $\Theta(ndr)$ time, with the usual matrix-vector product methods.
Unfortunately, this is at least as expensive as solving the original LS problem exactly, since this takes $\Theta(nd^2)$ time.

But while working with vanilla Gaussian-based random projections it might not be necessary.
In particular, we saw that the point of preprocessing the input with a random projection is to make the input data ``nice,'' in that the eigenvector or singular vector mass is spread out among all the coordinates, in which case uniform sampling, sparse projections, etc. perform well.
This leads to the question: 
\begin{itemize}
\item
Can we preprocess (or ``precondition'') the input data, so that the data are ``nice'' in the same or in a similar sense, but do it faster?
\end{itemize}
The answer is ``Yes.''
There are a range of tradeoffs and details here, depending on what exactly is one's goal, e.g., best theory, best implementations, assumptions on the input, etc., but this opens the door to improved randomized matrix algorithms for a wide variety of problems.
We will now turn to this topic.

To provide an overview of fast random projection methods, let's start with the following definition of a Johnson Lindenstrauss Transform, which is of much more general interest.
\begin{definition}
Given an $\epsilon > 0$ and $n$ points $\{x_i\}_{i=1}^{n}$, where $x_i\in\mathbb{R}^{d}$, an $\epsilon$-JLT (an $\epsilon$-Johnson Lindenstrauss Transform), denoted $\Pi \in \mathbb{R}^{r \times d}$ is a projection of the points into $\mathbb{R}^{r}$ such that 
$$
\left(1-\epsilon\right ) \VTTNormS{x_i}
   \le \VTTNormS{ \Pi x_i }  
   \le \left(1+\epsilon\right ) \VTTNormS{ x_i }   .
$$
\end{definition}
JLTs are of interest in a wide range of algorithmic applications, basically since they provide a way to embed input data a discrete set of $n$ data points into lower dimension without sacrificing too much in terms of distance information between pairs of those points.

In RandNLA, the notion of a JL Transform is usually not \emph{directly} useful, since we are typically more interested in subspaces than in discrete sets of points.
Fortunately, this definition can be generalized to all points in the subspace, and this is sometimes called SubspaceJL.
The basic idea is to show that the usual JL result holds with exponentially high probability for each pair of points and then put an $\epsilon$ net on the unit ball.


Here is an important point.
One can view this (and related) SubspaceJL result in one of two complementary ways. 
\begin{itemize}
\item
As an approximate matrix multiplication result applied to special input, where both matrices are an orthogonal matrix spanning the same space.
\item
As a generalization of the usual JL lemma from a finite set of vectors to a specially-structured infinite set of vectors.
\end{itemize}
One perspective of the other is more useful, depending on the situation.

Here, i.e., for the ``fast'' random projection methods we will discuss today and in the next few classes, we will be interested in the stronger requirement that we get similar JLT bounds, but in addition that we can compute the JLT quickly. 

\begin{definition}[Fast Subspace JL]
Given an $\epsilon > 0 $ and an orthogonal matrix $U\in\mathbb{R}^{n \times d}$, viewed ad $d$ vectors in $\mathbb{R}^{n}$.
A FJLT projects vectors from $\mathbb{R}^{n} \rightarrow \mathbb{R}^{r}$ s.t.the orthogonality of $U$ is preserved, and it does it quickly.
I.e., $\Pi \in \mathbb{R}^{r \times n}$ is an $\epsilon$-FJLT if 
\begin{itemize}
\item
$\TNorm{ I_d - U^T \Pi^T \Pi U } \le \epsilon$
\item
$\forall x \in \mathbb{R}^{n \times d}$, we can compute $ \Pi x $ in $O(n d \log(r))$ time.
\end{itemize}
\end{definition}

The original construction in this area is due to AC (although fast Fourier ideas have certainly existed much longer), but there are many others.
Theoretically, they are all to a first approximation the same; but practically, there can be a big difference between them.
Here is the original AC construction.
(BTW, the notation is inconsistent with what we use above and below, as it is taken from the AC journal paper.)
Let $\Pi = PHD$, where
\begin{itemize}
\item
$P$ is a sparse JL matrix or a uniform sampling matrix with a few extra dimensions.
To make things specific, let $P\in\mathbb{R}^{n \times n}$ have elements
\[
\Pi_{ij} = \left\{ \begin{array}{l l}
                      0 & \quad \text{with probability $1-q$}\\
                      N(0,q^{-1})      & \quad \text{with probability $q$}
                   \end{array} 
           \right.  ,
\]
where $q = \min\{ 1, \Theta \left( \frac{\log^2(n)}{d} \right) \}$.
\item
$H$ is structured, so that we can apply Fast Fourier methods to compute it quickly.
Basically, it spreads out ``spiky'' vectors.
\item
$D$ is a random $\{\pm 1 \}$ matrix that basically is used to put bad cases, i.e., localizing delocalized vectors, into the failure probability.
\end{itemize}
Here are some notes on this AC construction.
\begin{itemize}
\item
$P$ is a very sparse matrix---in expectation, only a fraction of the elements are nonzero.
\item
One might hope to use $P$ directly on $x$, but $\VTTNorm{ Px }$ is too large (to be a usual JL) for certain ``bad'' inputs, basically when $x$ is a very sparse matrix, since then we don't get sufficient concentration.
\item
If $x$ is ``smooth,'' i.e., if the mass is roughly uniformly spread out, then $\VTTNorm{ Px }$ does get good concentration.
\item
The mapping $HD$ ensures that $HDx$ is smooth and not too ``spiky.''
\item
$HD$ is an orthogonal matrix (exactly), meaning in particular that the Euclidean norm of vectors to which it is applied doesn't change.
\item
So, basically, $HD$ preconditions $x$ before we apply $P$.
\end{itemize}

Here, we will make precise the sense in which $HD$ ``spreads out'' input vectors.
(We basically take this particular form from the AC journal paper.)

\begin{lemma}
Fix a set $X$ of $n$ vectors in $\mathbb{R}^{d}$.
Then, with probability $ \ge 1 - \frac{1}{20}$, we have that
\[
\max_{x \in X} \VINorm{ HDX } = O\left( \sqrt{ \frac{\log(n)}{d} } \right).
\]
\label{lem:ACflatten}
\end{lemma}
\begin{Proof}
Assume w.l.o.g. that $\|x\|_2=1$.
Fix some $x \in X$, and define the random variable
\[
u = HDx = \left( u_1, \ldots, u_d \right)^T  .
\]
Note that $u_i$ is of the form $\sum_{i=1}^{d} a_i x_i$, where each $a_i = \frac{\pm1}{\sqrt{d}}$ is chosen uniformly and independently.

Then, we can apply a Chernoff argument in the usual way.
\begin{eqnarray}
\nonumber
\Expect{ e^{tdu_i} } 
   &=& \Pi_i \Expect{ e^{tda_ix_i} } \\
\nonumber
   &=& \Pi_i \Expect{cosh(t\sqrt{d_i}x_i)} \\
\label{eqn:tmp1}
   &\le& \exp\left( t^2 d \VTTNormS{x} / 2 \right).
\end{eqnarray}
Hence, $\forall s > 0$, by applying Markov's Inequality and plugging $t=sd$ into (\ref{eqn:tmp1}), we have that
\begin{eqnarray}
\nonumber
\Probab{ |u_1| } 
   &=& 2 \Probab{ e^{2du_1} \ge e^{s^2d} } \\
\nonumber
   &\le& 2 \Expect{ e^{sdu_1} }/e^{s^2d} \\
\nonumber
   &\le& 2 e^{s^2d\VTTNormS{x}/2 - s^2d} \\
\nonumber
   &=& 2 e^{-s^2d/2} \\
\nonumber
   &\le&\frac{1}{20nd}  , 
\end{eqnarray}
for $s = \Theta\left( \sqrt{ \frac{\log(n)}{d} } \right)$.
By performing a union bound over all $nd < n^2$ coordinates of the vectors 
\[
\{ HDx : x \in X \}   ,
\]
it follows that 
\[
\max_{x \in X} \VINorm{ HDx } = O\left( \sqrt{ \frac{\log(n)}{d} } \right)  ,
\]
which establishes the lemma.
\end{Proof}

Since the pre-processed or preconditioned vector is flat, in the sense made precise by that lemma, we can now do one of two things.
\begin{itemize}
\item
Sample uniformly, and oversample a little bit since the uniform sampling probabilities are not exactly optimal.
\item
Apply very sparse random projections, which is sufficient to get concentration since the input vectors are flat.
\end{itemize}
The first of these is a sampling procedure, i.e., involves choosing only a single row, and the second of these involves taking a linear combination of a small number of rows.
Nevertheless, when coupled with the $HD$ preprocessing, both of these procedures achieve JL-type results and thus can be meaningfully interpret as performing random projections.

\subsection{Applying these ideas to LS}

Before describing these algorithms, let's start with a lemma that quantifies the manner in which $HD$ uniformizes the information in the left singular subspace of $A$.
Note that this is very similar to Lemma~\ref{lem:ACflatten}---basically, it applies Lemma~\ref{lem:ACflatten} to orthogonal vectors that define the singular subspace of a given tall input matrix $A$.

\begin{lemma}
Let $U \in\mathbb{R}^{n \times d}$ be an orthogonal matrix (spanning the column space of an $n \times d$ matrix $A$), and let $HD$ be the $n \times n$ Randomized Hadamard Transform.
Then, with probability $\ge 0.95$, we have that 
\[
\max_{i \in [n]} \VTTNormS{\left( HDU \right)_{(i)}} \le \frac{2d\log(40nd)}{n}  .
\]
\end{lemma}
\begin{Proof}
Following the above lemma, which states that, for a fixed $j \in [d]$ and a fixed $i \in [n]$, 
\[
\Probab{ \left| \left(HDU^{(j)}\right)_{i} \right| > s} \le 2 e^{-s^2n/2}.
\]
(Note that we have $n$ and $d$ reversed; ugh.)
Let $s = \sqrt{2n^{-1}\log(40nd)}$.
So, then we have that
\[
\Probab{ \left| \left( HDU^{(j)} \right)_{i} \right| \ge \sqrt{ 2n^{-1}\log(40nd) }  } \le \frac{1}{20nd}.
\]
From the standard union bound, this implies that with probability $\ge 1-\frac{1}{20}$ we have that
\[
\left| \left( HDU^{(j)} \right)_{i} \right| \ge \sqrt{ 2n^{-1}\log(40nd) },
\]
$\forall i \in [n] , j \in [d]$.
Since 
\[
\VTTNormS{ \left( HDU \right)_{(i)} } = \sum_{j=1}^{d} \left( HDU^{(j)} \right)_{i}^{2} \le \frac{ 2d\log(40nd) }{n},
\]
the lemma then follows.
\end{Proof}

Now, let's now apply these ideas in the context of LS.
Here are three three related fast LS algorithms.
\begin{itemize}
\item
Random projection.
\begin{itemize}
\item
Multiply by $HD$ to approximately uniformize the leverage scores, and then sample uniformly.
\item
Multiply by $HD$ to approximately uniformize things, and then use sparse projection matrix like above.
\end{itemize}
\item
Random sampling.
\begin{itemize}
\item
Use FJLT (either of those two projection-based procedures) to compute approximations to the leverage scores and sample w.r.t. those approximations.
\end{itemize}
\end{itemize}
We will describe these results in more detail in the next two classes.

\newpage

\section{%
(10/02/2013):  Fast Random Projections and FJLT, Cont.}

We continue with the discussion from last time.
There is no new reading, just the same as last~class.

Today, we will do the following.
\begin{itemize}
\item
Show that the two structural conditions required for good LS approximation are satisfied by FJLT projections.
\item
State the algorithms for fast LS approximation via random projection as well as via random sampling.
\item
Describe in some more detail the connection between SubspaceJL methods and randomized matrix multiplication.
\end{itemize}

As a reminder of where we are, to see how we will use these fast random projections and FJLT for RandNLA algorithms, recall from a few classes ago that there were two conditions that were sufficient to obtain relative-error approximation.
As a reminder, here are those two conditions.
\begin{itemize}
\item
\textbf{Condition I:}
\begin{equation}
\label{eqn:lemma1_ass1_RESTATED}
 \sigma_{min}(XU_A) \ge 1/\sqrt{2} .
\end{equation}
\item
\textbf{Condition II:}
\begin{equation}
\label{eqn:lemma1_ass2_RESTATED}
 \VTTNormS{U_AX^TXb^{\perp}} \le \frac{\epsilon}{2}\mathcal{Z}^2 .
\end{equation}
\end{itemize}

Recall also from the last few classes that we can solve LS problems ``slowly'' with RandNLA methods in one of two ways. 
\begin{itemize}
\item
\textbf{Algorithmic Approach I:} with random sampling (which needs some sort of algorithm to compute importance sampling probabilities), or 
\item
\textbf{Algorithmic Approach II:} with random projections (which is like uniform sampling if we first preprocess to ``flatten out'' information in singular value spaces).
\end{itemize}

In this class and the next, we will show how we can combine these results: in particular, we can establish both conditions with both random sampling and random projection algorithms in $o(nd^2)$ time.
Let's start with random projections.

\subsection{Establishing the two conditions for fast random projections}

Let's start with Condition I, i.e., let's start by establishing a lemma that says that if we uniformly sample in the ``randomly rotated'' basis then the singular values are all close to $1$.
(This lemma will be an important ingredient more generally in what follows---basically it means that we are dealing with what is known as a ``subspace embedding.'')

\begin{lemma}
Let $\mathcal{S}$ be a uniform sampling and rescaling matrix.
And recall that $\VTTNormS{ \left( HDU \right)_{(i)} } \le \frac{ 2d\log(40d) }{n}$, for all $i$.
Then if we sample $r \gtrsim O\left( d \log(nd) \log(d \log(nd))  \right)$, then with probability $\ge 0.95$ we have that
\[
| 1-\sigma_i^2\left(\mathcal{S}HDU_A \right) | \le 1-\frac{1}{\sqrt{\epsilon}}.
\]
\end{lemma}
\begin{Proof}
The idea of the proof is that $HD$ approximately uniformness the leverage scores, so that uniform sampling is approximately optimal, if we are willing to oversample by a factor $1/\beta$, where $\beta \in (0,1]$ quantifies how far from uniform is the leverage score distribution.

In more detail, since $U_A^TDH^THDU_A = I_d$, we have that 
\begin{eqnarray*}
| 1-\sigma_i^2\left(\mathcal{S}HDU_A \right) | 
   &=& | \sigma_i\left( U_A^TDH^THDU_A \right) - \sigma_i\left( U_A^TDH^T\mathcal{S}^T\mathcal{S}HDU_A \right) |  \\
      &\le& \TNorm{ U_A^TDH^THDU_A - U_A^TDH^T\mathcal{S}^T\mathcal{S}HDU_A }  .
\end{eqnarray*}
Consider the matrix $\left(HDU_A\right)^T$.
Since $H$, $D$, and $U_A$ are orthogonal matrices, it follows that 
\begin{eqnarray*}
\TNorm{ HDU_A } &=& 1 \\
\FNorm{ HDU_A } &=& \FNorm{U_A} = \sqrt{d}  .
\end{eqnarray*}
Let $\beta = \left( 2 \log(40nd) \right)^{-1}$, in which case 
\[
\frac{1}{n} \ge \beta \frac{ \VTTNormS{ \left( HDU_A \right)_{(i)} } }{ \FNormS{ HDU_A } }   ,
\]
$\forall i \in [n]$.

We can now apply the following spectral norm bound theorem:
\begin{itemize}
\item 
If $\TNorm{A}=1$, $\FNorm{A} \ge \frac{1}{24}$, $p_i \ge \beta \frac{ \VTTNormS{A^{(i)}} }{ \FNormS{A} }$, and $c \ge \frac{ 96\FNormS{A} }{ \beta\epsilon^2 } \log\left( \frac{ 96\FNormS{A} }{ \beta\epsilon^2\sqrt{\delta} } \right)$, then with probability $\ge 1-\delta$, we have that $\TNorm{AA^T-CC^T}\le\epsilon$.
\end{itemize}
By applying this theorem with $\epsilon=1-\frac{1}{\sqrt{2}}$ and $\delta=\frac{1}{20}$, then with probability $\ge 0.95$, we have that 
\[
\TNorm{ U_ADH^THDU_A - U_ADH^T\mathcal{S}\mathcal{S}^THDU_A } \le 1-\frac{1}{\sqrt{2}}  ,
\]
thus establishing the lemma.
\end{Proof}

The above lemma basically says that the first Condition I is satisfied in the randomly rotated space.
(Alternatively, it establishes a Subspace JL result.)

Next, let's give a lemma that says that Condition II is satisfied.
The following condition will establish the second condition.

\begin{lemma}
Assume that $\VTTNormS{ \left( HDU \right)_{(i)} } \le \frac{ 2d\log\left( 40nd \right) }{n}$, for all $i$.
Let $r \gtrsim 40d\log(40nd)/\epsilon$.
Then, with probability $\ge 0.9$, we have that 
\[
\VTTNorm{ \left(\mathcal{S}^THDU_A\right)^TS^THDb^{\perp} } \le \frac{\epsilon\mathcal{Z}^2}{2}.
\]
\end{lemma}
\begin{Proof}
Recall that $b^{\perp} = U_A^{\perp}U_A^{\perp T}b$ and that $\mathcal{Z} = \VTTNorm{b^{\perp}}$.
Since $\VTTNormS{ U_A^TDH^THDb^{\perp} } = \VTTNormS{ U_A^Tb^{\perp} }=0$, it follows that 
\[
\VTTNormS{ \left(\mathcal{S}^THDU_A \right)^T \mathcal{S}HDb^{\perp} } 
   = \VTTNormS{ U_A^TDH^T\mathcal{S}\mathcal{S}^THDb^{\perp} - U_A^T D H^THDb^{\perp} }.
\]
We will apply the following approximate matrix multiplication result with probabilities depending on only one matrix:
\begin{itemize}
\item
If $p_k \ge \beta \frac{ \VTTNormS{A^{(k)}} }{ \FNormS{A} }$, then $ \Expect{\FNormS{AB-CR}} \le \frac{1}{\beta c} \FNormS{A}\FNormS{B} $.
\end{itemize}
Let's let $\beta = \left( 2 \log(40nd) \right)^{-1}$, and so for all $i$ we have that
\[
\frac{1}{n} \ge \beta \frac{ \VTTNormS{ \left( HDU_A \right)_{(i)} } }{ \FNormS{HDU_A} }.
\]
So, we have that 
\[
\Expect{ \VTTNormS{ \left(\mathcal{S}HDU_A\right)^{T}\mathcal{S}^THDb^{\perp} } } \le \frac{1}{\beta r} \FNormS{HDU_A}\VTTNormS{HDb^{\perp}} \le \frac{d\mathcal{Z}^2}{\beta r},
\]
where we have used that $\FNormS{HDU_A}=d$.
By applying Markov's Inequality, we have that with probability $\ge 0.9$ we have that
\[
\VTTNormS{ \left(\mathcal{S}HDU_A\right)^{T}\mathcal{S}^THDb^{\perp} } \le \frac{ 10d\mathcal{Z}^{2} }{\beta r}.
\]
So, if $r \ge 20 \beta^{-1} d/\epsilon$, then with that value of $\beta$, the lemma follows.
\end{Proof}

So, since these two lemmas were established with the fast Hadamard-based rotations, we now we have all the ingredients for our first ``fast'' LS approximation algorithm.

\subsection{Fast LS approximation}

Here, we will describe a random projection algorithm and a random sampling algorithm for approximating the solution to LS that are fast in the sense that they run in $o(nd^2)$ time.

\subsubsection{Fast LS approximation via random projections}

Here, we will present our first ``fast'' LS approximation algorithm.

Given as input a matrix $A\in\mathbb{R}^{n \times d}$ a vector $b\in\mathbb{R}^{n}$, and a number $\epsilon\in(0,1)$, do the following.
\begin{enumerate}
\item
Let $r = O\left( d \left(\log(d)\right)\left(\log(n)\right) + \frac{d \log(n)}{\epsilon} \right)$.
(This holds if $d \le n \le e^d$, and we can get messier expressions more generally.)
\item
Let $S$ be an $r \times n$ uniform sampling matrix, i.e., it has one nonzero per row, selected u.a.r, with value equal to $\sqrt{n/r}$ and zero otherwise; and choose each of the $r$ rows in i.i.d. trials with replacement.
\item
Let $H\in\mathbb{R}^{n \times n}$ be a normalized Hadamard matrix.
\item
Let $D\in\mathbb{R}^{n \times n}$ be a diagonal matrix with $\{\pm 1\}$ entries u.a.r.
\item
Compute and return
\[
\tilde{x}_{opt} = \left(\mathcal{S}^{T}HDA\right)^{+}\mathcal{S}^{T}HDb \in \mathbb{R}^{d}  .
\]
\end{enumerate}

Here is a theorem that we can establish about this algorithm.

\begin{theorem}
The above algorithm gives a vector $x_{opt}$ such that with probability $\ge 0.8$ we have that 
\begin{itemize}
\item
$\VTTNorm{ A\tilde{x}_{opt} - b } \le \left( 1+\epsilon \right) \mathcal{Z}$
\item
$\VTTNorm{ \tilde{x}_{opt} - x_{opt} } \le \sqrt{\epsilon}\kappa(A) \sqrt{\gamma^{-2}-1} \VTTNorm{x_{opt}} $.
\end{itemize}
The running time of this algorithm is 
\[
O\left( nd \log\left(d/\epsilon\right) + d^3\left(\log(d)\right)\left(\log(n)\right) + \frac{d^3\log(n)}{\epsilon} \right)
\]
(if $d \le n \le e^d$, with a similar but messier expression otherwise).
\end{theorem}
\begin{Proof}
Define the following three events:
\begin{eqnarray*}
\mathcal{E}_1 &=& \mbox{event that leverage scores are uniformized}  \\
\mathcal{E}_2 &=& \mbox{event that singular values approximately equal one}  \\
\mathcal{E}_3 &=& \mbox{event that the second matrix multiplication result holds}  
\end{eqnarray*}
Each of these holds with constant probability, so we can then apply the union bound, which establishes the quality-of-approximation claims.

To do $HA$ takes $O(nd \log(r))$ time, and solving the subproblem takes $O(rd^2)$ time; working though the exact details of those expressions establishes the running time claim.
\end{Proof}

More generally, there are several types of ``fast'' random projection algorithms that take the following form.
Given a matrix $A\in\mathbb{R}^{n \times d}$ and a vector $b\in\mathbb{R}^{d}$,
\begin{enumerate}
\item
Let $\Pi$ be any other FJLT.
\item
Project $A$ and $b$ onto roughly $r \sim \frac{d \log(d)}{\epsilon}$ rows.
\item
Solve $\min_x\VTTNorm{\Pi A x - \Pi b}$.
\end{enumerate}
In these cases, it can generally be established theorems of the above form that state that you get a $(1\pm\epsilon)$ approximation in roughly $O(nd\log(r))$ time.

\subsubsection{Fast LS approximation via random sampling}

Next, let's mention (we'll get into more detail on this next time) a ``fast'' random sampling algorithm.
The basic idea is the following.
Given a matrix $A\in\mathbb{R}^{n \times d}$ and a vector $b\in\mathbb{R}^{d}$,
\begin{enumerate}
\item
Let $\{p_i\}_{i=1}^{n}$ be $1\pm\epsilon$ approximations to the leverage, computed in $o(nd^2)$ time with a black box that we will describe next class.
\item
Sample $r \gtrsim \frac{d \log(d)}{\epsilon}$ constraints with probability depending on $p_i$ to construct a nonuniform probability sampling matrix $S$
\item
Solve $\min_x\VTTNorm{ S A x - S b}$.
\end{enumerate}
Again, one can show that you obtain a $(1\pm\epsilon)$ approximation in roughly $O(nd\log(r))$ time.
\begin{itemize}
\item
The reason for the running time is that the running time bottleneck for the approximate leverage score computation boils down to a random projection.
\item
The quality of approximation comes since Condition I and Condition II can be established with this sampling procedure; and the running time is what it is since that is how long it takes to approximate the leverage scores with the black box that we will describe below.
\end{itemize}
We just mention this now---we'll go into more detail on this next time, as well as discuss why one might prefer sampling versus projection methods.

\subsection{More on SubspaceJL and randomized matrix multiplication}

Here we will go into 
more detail about the idea of SubspaceJL (Subspace Johnson-Lindenstrauss).
This was implicit in what we were doing before, and here we will make it explicit.

First, recall the definition of a JL transform.
\begin{definition}
Given $\epsilon > 0$, $n$ points $\{x_i\}_{i=1}^{n}\in\mathbb{R}^{d}$, an $\epsilon$-JLT is a $\Pi\in\mathbb{R}^{r \times d}$ such that
\[
\left(1-\epsilon\right)\VTTNormS{x_i}
   \le \VTTNormS{\Pi x_i } 
   \le \left(1+\epsilon\right)\VTTNormS{x_i}.
\]
\end{definition}
As we mentioned before, there are several different constructions for this.

In RandNLA applications, we typically don't want to preserve approximately the distances between a point set of $n$ points, but instead we want to preserve approximately the geometry of an entire subspace.
(In addition, we will want it to be ``fast,'' in the sense we use the term before.)
That motivates the following definition.

\begin{definition}
Given $\epsilon>0$, an orthogonal matrix $U\in\mathbb{R}^{n \times d}$, where $n \gg d$, $\Pi\in\mathbb{R}^{r \times d}$ is an $\epsilon$-FJLT or a fast subspace JL, if
\begin{itemize}
\item
$\TNorm{I_d - U^T\Pi^T\Pi U } \le \epsilon$
\item
$\forall X\in\mathbb{R}^{n \times d}$, we can compute $\Pi X$ in $O(nd\log(r))$ time.
\end{itemize}
\end{definition}

From this definition, it is clear that the subspace embedding (as opposed to running time) property of an $\epsilon$-FJLT is a special case (applied to the situation $B=A^T=U$, an orthogonal matrix) of matrix multiplication.  
(Note that, in this special case, the additive error bounds for matrix multiplication become relative error.)
So, randomized matrix multiplication with exact or approximate leverage scores leads to quality-of-approximation bounds of an $\epsilon$-FJLT.
(With the approximate leverage score computation algorithm we will discuss next time, we can also satisfy the running time requirements of an $\epsilon$-FJLT.)

Note that, in this case, i.e., when doing sampling, we are looking at the input matrix to construct the importance sampling probabilities. 
While that satisfies the definition of an $\epsilon$-FJLT, as stated, that is not how TCS people typically think about the problem, since TCS typically thinks about and hopes for data-oblivious embeddings/projections.
(In other settings, there are other algorithmic advantages in doing that.)
Importantly, since random projections uniformize leverage scores to permit uniform sampling in the randomly rotated basis, we can achieve this.
This holds for both slow and fast random projections, with appropriate parameter setting; and this may be viewed as providing data-oblivious approximate matrix multiplication (via projections rather than sampling).
Here is a lemma we can show about this with FJLTs.

\begin{lemma}
Let $\hat{H}_1 = 1$, $\hat{H}_{2n} = \left( \begin{array}{cc} \hat{H}_n & \hat{H}_n \\ \hat{H}_n &  -\hat{H}_n \end{array} \right)  $, and $H_n = \hat{H}/\sqrt{n}$.
Let $\Pi = S^THD$, where $S^T$ is a uniform sampling and rescaling operator that chooses $r$ rows from $HD$, and let $U\in\mathbb{R}^{n \times d}$ be a (fixed but arbitrary) orthogonal matrix.
Then, if $r \ge O\left( \frac{d\log(nd)}{\epsilon^2}\log\left( \frac{d\log(nd)}{\epsilon^2} \right) \right)$, then with probability $\ge 0.9$ it follows that
$\Pi$ is an $\epsilon$-FJLT for $U$.
\end{lemma}
\begin{Proof}
Combine the two results:
\begin{itemize}
\item
The previous lemma that says: $\max_{i\in[n]} \VTTNormS{\left(HDU\right)_{(i)}} \le \frac{2d\log(40nd)}{n}$.
\item
The previous lemma that says: $ | 1 - \sigma_i^2\left( S^THDU \right) | \le 1-\frac{1}{\sqrt{2}} $.
\end{itemize}
The lemma follows.
\end{Proof}

There are many different constructions that, for appropriate parameter settings, satisfy the Subspace JL property, including the following.
\begin{itemize}
\item
``fast'' Hadamard-based constructions.
\item
``slow'' Gaussians, $\{\pm1\}$ r.v.s, etc.
\item
Random sampling matrices with, e.g., $p_i \sim \frac{1}{d} \VTTNormS{U_{(i)}}$.
This can be computed exactly which is ``slow,'' or this can be computed approximately with an algorithm we will discuss next time which is ``fast.''
\end{itemize}

\subsection{An aside for next class}

Next time, we will go into more detail on the random sampling algorithm.
In particular, we are going to prove that we can compute the leverage scores quickly, if we are willing to settle for approximations.
To do so, we will use the following lemma (although we could prove it in other related ways).
This lemma states several related results that are related to having a good subspace embedding and that are needed to get good bounds for LS-related problems, and it highlights a key property of the Moore-Penrose generalized inverse that is false in general but that is true when the subspace is preserved.

\begin{lemma}
Let $A\in\mathbb{R}^{n \times d}$, with $n \gg d$, and let $\mbox{rank}(A)=d$, nd let $A=U \Sigma V^T$ be the SVD of $A$, and let $S$ satisfy the Subspace JL property.
(E.g., it could be a sampling matrix constructed with leverage-based sampling probabilities, or it could be a data-agnostic random projection matrix, that is either fast or slow.)
Then,
\begin{itemize}
\item
$\mbox{rank}(SA) = \mbox{rank(SU)} = \mbox{rank}(U) = \mbox{rank}(A) = d$
\item
$\TNorm{ \Sigma_{SU} - \Sigma_{SU}^{-1}} \le \epsilon$
\item
$\left(SA\right)^{+} = V\Sigma\left(SU\right)^{+}$
\item
$\TNorm {\left(SU\right)^{+} - \left(SU\right)^{T}}  = \TNorm{ \Sigma_{SU} - \Sigma_{SU}^{-1}}$.
\end{itemize}
\end{lemma}
\begin{Proof}
For the first claim, note that $\forall i \in [\rho]$ we have that
\begin{eqnarray*}
\abs{ 1-\sigma_i^2(SU) } 
   &=&  \abs{ \sigma_i(U^TU) - \sigma_i( U^TS^TSU ) }  \\
   &\le& \TNorm{ U^TU - U^TS^TSU } \\
   &\le& \epsilon  ,
\end{eqnarray*}
with appropriate parameters if $r \gtrsim \frac{d \log(d)}{\epsilon}$.
For the second claim, note that
\begin{eqnarray*}
\TNorm{ \Sigma_{SU}^{-1} - \Sigma_{SU} } 
   &=& \max_{ij \in [\rho]} \abs{ \sigma_i(SU) - \frac{1}{\sigma_j(SU)} } \\
   &=& \max_{ij \in [\rho]} \frac{\abs{ \sigma_i(SU)\sigma_j(SU)-1 }}{\abs{\sigma_j(SU)}} \\
   &\le& \max_{j \in [\rho]} \frac{\abs{\sigma_j^2(SU)-1}}{\sigma_j(SU)} \\
   &\le& \frac{\TNorm{U^TU-U^TS^TSU}}{\sqrt{1-\TNorm{U^TU-U^TS^TSU}}} ,
\end{eqnarray*}
where the last inequality follows since 
$$
\frac{1}{\sigma_i(SU)} \le \frac{1}{\sqrt{1-\TNorm{U^TU-U^TS^TSU}}} ,
$$
which follows since $\abs{1-\sigma_i(SU)} \le \TNorm{U^TU-U^TS^TSU}$.
For the third claim, note that
\begin{eqnarray}
\nonumber
\left(SA\right)^{\dagger}
   &=& \left(SU_A\Sigma_AV_A^T\right)^{\dagger} \\
   \nonumber
   &=& \left( U_{SA}\Sigma_{SA}V_{SA}^T\Sigma_A V_A^T \right)^{\dagger} \\
   \label{eqn:any-orthogonal-matrix}
   &=& V_A \left( \Sigma_{SU} V_{SU}^T \Sigma_A \right) U_{SA}^T \\
   \label{eqn:since-rank-preserved}
   &=& V_A \Sigma_A^{-1} V_{SU} \Sigma_{SU}^{-1} U_{SU}^T \\
   \nonumber
   &=& V_A \Sigma_A^{-1} \left(SU\right)^{\dagger}  .
\end{eqnarray}
(Note that $\left(SA\right)^{\dagger}=V_A \Sigma_A^{-1} \left(SU\right)^{\dagger}$ might seem intuitive, given the behavior of inverses, but it is false in general, and it it a common mistake to assume that it it true of generalized inverses.
In particular, note that Eqn~(\ref{eqn:since-rank-preserved}) holds since rank is preserved; otherwise it is false, and the claims of the lemma fail to hold.
On the other hand, Eqn~(\ref{eqn:any-orthogonal-matrix}) holds for any orthogonal matrices.)
For the fourth claim, note that
\begin{eqnarray*}
\TNorm{ \left(SU\right)^{\dagger} - \left(SU\right)^T } 
   &=& \TNorm{ \left(U_{SU}\Sigma_{SU}V_{SU}^T\right)^{\dagger} - \left(U_{SU}\Sigma_{SU}V_{SU}^T\right)^T } \\
   &=& \TNorm{ V_{SU} \left( \Sigma_{SU}^{-1} - \Sigma_{SU} \right) U_{SU}^T } \\
   &=& \TNorm{ \Sigma_{SU}^{-1} - \Sigma_{SU} } ,
\end{eqnarray*}
where the last claim follows since $V_{SU}$ and $U_{SU}$ have orthogonal columns.
The lemma then follows.
\end{Proof}

So, this lemma is just another way to prove the same LS result.
(In fact, this was the way that we first established the relative-error LS result.)
Basically, it establishes several senses in which $\left(SU\right)^{\dagger} \approx \left(SU\right)^T$ if rank is preserved.

This lemma was stated for completeness, since we could prove the leverage score result directly.
Alternatively, we could prove the previous theorem with this result.
Part of the reason for doing it this way is that I didn't have a chance to get a uniform set of notes before class, but part of the reason for this is also since using this following lemma is perhaps more natural for NLA people as opposed to TCS people, since it highlights the role of the singular structure of the input matrix.

\newpage

\section{%
(10/07/2013):  Fast Random Projections and FJLT, Cont.}

We continue with the discussion from last time.
There is no new reading, just the same as last~class.

Today, we will cover the following.
\begin{itemize}
\item
We will describe a fast algorithm to compute very fine approximations to the leverage scores of an arbitrary tall matrix.
\item
We will describe a few subtleties to extend this basic algorithm to non-tall matrices, which is of interest in extending these LS ideas to low-rank matrix approximation.
\item
We will describe how to use this algorithm in a fast random sampling algorithm for the LS problem (that has a running time that is essentially the same as the fast random projection algorithm that we discussed last time).
\end{itemize}
Today will basically wrap up our worst-case theory discussion for the LS problem.
Next time, we will start to discuss how these ideas can be used in practice in high-quality implementations for the overdetermined LS problem.

\subsection{Computing leverage scores quickly for a tall matrix}

The leverage scores for a tall $n \times d$ matrix $A$, with $n \gg d$, are the diagonal elements of the projection matrix onto the column span of $A$.
Equivalently, they are the Euclidean norms of the rows of \emph{any} $n \times d$ orthogonal matrix $U$ spanning the column span of $A$.
Thus, a straw-man algorithm is to perform a QR decomposition of the SVD on $A$ and read off the leverage scores.
This takes $O(nd^2)$ time, and so this represents what our fast algorithm should beat.

The basic idea behind the fast algorithm to approximate leverage quickly is to use the ``R'' matrix, not from a QR decomposition of the original matrix $A$, but from a random sketch of $A$ of the form $\Pi_1 A$, where $\Pi_1$ is a FJLT.
Equivalently, the idea is to use a ``randomized sketch'' of the form $ A \left (\Pi_1 A\right)^{\dagger} \Pi_2 $, where $\Pi_1$ is an FJLT and $\Pi_2$ is a JLT.
To see this, recall that if $\ell_i$ is the $i^{th}$ leverage score, then
\begin{eqnarray}
\ell_i 
   = \VTTNormS{U_{(i)} }  
   = \VTTNormS{ e_i^T U }  
   \label{eqn:hard1}
   &=& \VTTNormS{ e_i UU^T } \\
   &=& \VTTNormS{ e_i AA^{\dagger} }  
   \label{eqn:hard2}
   = \VTTNormS{ \left(AA^{\dagger} \right)_{(i)} }  .
\end{eqnarray}
Viewed this way, the hard part of computing $\ell_i$ via Eqn.~(\ref{eqn:hard1}) is to compute the $U$ matrix, which takes $O(nd^2)$ time; and that hard part of computing $\ell_i$ via Eqn.~(\ref{eqn:hard2}) is first to compute $A^{\dagger}$ and second to do the matrix multiplication between $A$ and $A^{\dagger}$, each of which take $O(nd^2)$ time.

While Eqn.~(\ref{eqn:hard2}) might seem more difficult to work with, we can insert random projections at appropriate places to speed up both steps.
In particular, we have the following.
\begin{eqnarray*}
\ell_i 
   &=& \VTTNormS{ e_i AA^{\dagger} }  \\
   &\approx& \VTTNormS{ e_i A \left(\Pi_1A\right)^{\dagger} } = \hat{\ell}_i \\
   &\approx& \VTTNormS{ e_i A \left(\Pi_1A\right)^{\dagger} \Pi_2 } = \tilde{\ell}_i .
\end{eqnarray*}
In these expressions, $A \in \mathbb{R}^{n \times d}$, $\Pi_1 \in \mathbb{R}^{r_1 \times n}$, where $r_1 = O\left(\frac{d\log(d)}{\epsilon}\right)$, in which case $\Pi_1A \in \mathbb{R}^{r_1 \times d}$ and $\left(\Pi_1A\right)^{\dagger} \in \mathbb{R}^{d \times r_1}$.

That is, we compute the pseudo inverse of the smaller matrix $\Pi_1A$, rather than $A$.

But, computing the product of $A$ and $\left(\Pi_1A\right)^{\dagger}$ takes $O(ndr)$ time, which is $\Omega(nd^2)$ time, since $ r \approx \frac{d\log(d)}{\epsilon}$.
On the other hand, we only need estimates of the Euclidean norms of the rows, and so we can do a second random projection by $\Pi_2 \in \mathbb{R}^{r_1 \times r_2}$, where $r_2= O\left(\log(n)\right)$, which if the matrices are multiplied in the proper order is faster.

(As an aside, we note that the sketch $A\left(\Pi_1 A \right)^{\dagger} \Pi_2$ can be used in other ways, e.g., to estimate the dot products between different rows of $U$, which is of interest since the so-called cross leverage scores are defined to be $c_{ij} = U_{(i)}^TU_{(j)}$, i.e., the off-diagonal elements of the projection matrix.
Also, the so-called coherence is $\gamma = \max_i \ell_i$, although it is sometimes defined to be $\max_{ij} c_{ij}$.)

With this motivation, here is the main algorithm for computing approximations to the leverage scores quickly.

\begin{algorithm}
\caption{The \texttt{FastApproximateLeverageScores} algorithm.}
\label{alg:fast-approx-lev-scores-alg}
\begin{algorithmic}[1]
    \REQUIRE $A\in\mathbb{R}^{n \times d}$, with SVD $A=U \Sigma V^T$, and $\epsilon\in(0,1/2]$.
    \ENSURE  $\tilde{\ell}_i$, for $i\in[n]$
    \STATE
    Let $\Pi_1 \in \mathbb{R}^{r_1 \times n}$ be an $\epsilon$-FJLT for $U$, with $r_1 = \Omega\left( \frac{d\log(n)}{\epsilon^2} \log\left( \frac{d\log(n)}{\epsilon^2} \right) \right)$.
    \STATE
    Compute $\Pi_1 A \in \mathbb{R}^{r_1 \times d}$ and its QR decomposition or SVD.  Let $R \in \mathbb{R}^{d \times d}$ be the ``R'' matrix from QR or $\Sigma_{\Pi_1A}V_{\Pi_1A}^T$ from SVD.
    \STATE
    View the rows of $AR^{-1} \in \mathbb{R}^{n \times d}$ as $n$ vectors in $\mathbb{R}^{d}$, and let $\Pi_2\in\mathbb{R}^{d \times r_2}$ be an $\epsilon$-JLT for $n^2$ vectors (the $n$ vectors and the ${n \choose 2}$pairwise sums), with $r_2 = O\left( \frac{\log(n)}{\epsilon^2} \right)$.
    \STATE 
    Construct $\Omega = AR^{-1}\Pi_2$.
    \STATE
    For all $i\in[n]$, compute and return $\tilde{\ell}_i = \VTTNormS{ \Omega_{(i)} }$.
\end{algorithmic}
\end{algorithm}

Here is the main theorem that we can establish for this algorithm.

\begin{theorem}
The \texttt{FastApproximateLeverageScores} algorithm returns $\tilde{\ell}_{(i)}$ such that 
$$
\abs{ \ell_{i} - \tilde{\ell}_{i} }  \le \epsilon \ell_{i}  ,
$$
for all $i\in[n]$, with constant probability.
The running time is
$$
O\left( nd\log\left( d/\epsilon \right) + nd\epsilon^{-2}\log(n) + d^3\epsilon^{-2}\log(n) \log\left(d\epsilon^{-1} \right) \right)
$$
(assuming that $d \le n \le e^d$, with a more complicated expression otherwise).
\label{thm:fast-lev-approx}
\end{theorem}

Before presenting the main proof, let's start with an outline of the proof, which will follow the discussion above.
The algorithm compute $\tilde{\ell}_i = \VTTNormS{ \tilde{U}_{(i)} }$, where $\tilde{U}_{(i)} = e_i^T A \left( \Pi_1 A \right)^{\dagger} \Pi_2$.
The first thing to note is that we need to be not naive about the order of operations, e.g., don't multiply $AR^{-1}\Pi_2$ in the wrong order; and also that $\Pi_2$ is included only to improve the (worst case FLOPS) running time.
That is, the quality-of-approximation result holds if $\Pi_2$ is not there, and in certain implementations the algorithm might be faster/better without it.
So, if we can establish what we want without the $\Pi_2$, then including it is a JL transformation and doesn't change the distances (in particular the norms of the rows) too much.

So, the result will follow if 
$$
e_iA \left( \Pi_1A\right)^{\dagger} \left( \left( \Pi_1 A \right)^{\dagger} \right)^T A^T e_j \approx e_i UU^T e_j
$$
and if $\Pi_1A$ is efficient to compute.
But since $\Pi_1$ is an FJLT, to compute the quantity $\left(\Pi_1A\right)^{\dagger}$ takes $O\left( nd\log\left( r_1 \right) + r_1 d^2 \right)$ time.
Then, by the structural lemma from the end of last class, we have that
$$
\left( \Pi_1 A \right)^{\dagger} = V \Sigma^{-1} \left( \Pi_1 U \right)^{\dagger} ,
$$
which recall is false in general but holds when rank is preserved.
So,
$$
e_i A \left (\Pi_1 A \right)^{\dagger} \left( \left( \Pi_1A \right)^{\dagger} \right)^T A^T e_j
   = e_i U \left(\Pi_1 U \right)^{\dagger} \left( \left( \Pi_1 U \right)^{\dagger} \right)^{T} U^T e_j  .
$$
Since $\Pi_1$ is an FJLT, we have that 
$$
\left(\Pi_1 U \right)^{\dagger} \left( \left( \Pi_1 U \right)^{\dagger} \right)^{T} \approx I_d  ,
$$
from which the theorem will follow.

Note also that this analysis is more general, e.g., it shows that we can compute the large cross-leverage scores. 
To extract them, however, takes additional work, which basically amounts to using a ``heavy hitter'' algorithm to find the large ones without looking at all ${n \choose 2}$ possibilities.
This can be done---see DMMW---but we won't describe it here.

Now onto the proof of Theorem~\ref{thm:fast-lev-approx}.

\begin{Proof}
We will condition our analysis on the following two events, each of which holds with constant probability if we choose parameters correctly.
\begin{eqnarray*}
\text{Event 1}: & & \Pi_1\in \mathbb{R}^{r_1\times n} \mbox{ is an } \epsilon\mbox{-FJLT for } U \\
\text{Event 2}: & & \Pi_2 \in \mathbb{R}^{r_1 \times r_2} \mbox{ is an } \epsilon\mbox{-JLT for the }{n \choose 2}\mbox{ points we described}  .
\end{eqnarray*}
Then, let's define the following ``hatted'' and ``tilded'' quantities
\begin{eqnarray*}
\hat{U}_{(i)} = e_iA\left(\Pi_1A\right)^{\dagger} & & \quad \hat{\ell}_i = \VTTNormS{\hat{U}_{(i)}}   \\
\tilde{U}_{(i)} = e_iA\left(\Pi_1A\right)^{\dagger}\Pi_2 & & \quad \tilde{\ell_i} = \VTTNormS{\tilde{U}_{(i)}}  ,
\end{eqnarray*}
where the hatted quantities correspond to one level of approximation (with $\Pi_1$) and the tilded quantities correspond to a second level of approximation (with $\Pi_2$).
The theorem will follow if we show that 
$$
U_{(i)}^T U_{(j)} \approx \hat{U}_{(i)}^T \hat{U}_{(j)}
$$
and also that 
$$
\hat{U}_{(i)}^T \hat{U}_{(j)} \approx \tilde{U}_{(i)}^T \tilde{U}_{(j)}
$$
with appropriate parameters.

More precisely, we need to establish the following two results.

\begin{lemma} 
For all $i,j\in[n]$, we have that
$$
\abs{ U_{(i)}^TU_{(j)} - \hat{U}_{(i)}^T\hat{U}_{(j)} } \le \frac{\epsilon}{1-\epsilon} \VTTNorm{ U_{(i)} } \VTTNorm{ U_{(j)} }.
$$
\label{lem:first-approx}
\end{lemma}

\begin{lemma} 
For all $i,j\in[n]$, we have that
$$
\abs{ \hat{U}_{(i)}^T\hat{U}_{(j)} - \tilde{U}_{(i)}^T\tilde{U}_{(j)} } \le 2\epsilon \VTTNorm{ \hat{U}_{(i)} } \VTTNorm{ \hat{U}_{(j)} }.
$$
\label{lem:second-approx}
\end{lemma}

Before proving these two lemmas, let's say what they mean and how they imply the results of the theorem.

First, observe that Lemma~\ref{lem:first-approx} says that the leverage scores (for $i=j$; respectively, the cross-leverage scores for $i \ne j$) are preserved to within relative (respectively, additive) error, i.e., the action of $\Pi_1$ doesn't distort them too much.
(Viewing this in terms of approximate matrix multiplication, if $i=j$ then the two matrices/vectors are aligned with no cancellation, while if $i \ne j$ they will in general not be, and so there might be cancellation, in which case obtaining relative error bounds isn't possible.)

Second, Lemma~\ref{lem:second-approx} says the same thing for the action of applying $\Pi_2$.
In particular, by Lemma~\ref{lem:first-approx} it follows that 
$$
\abs{ \ell_i - \hat{\ell}_i } \le \frac{\epsilon}{1-\epsilon}\ell_i  ,
$$
and by Lemma~\ref{lem:second-approx} it follows that 
$$
\abs{ \hat{\ell}_i - \tilde{\ell}_i } \le 2\epsilon \hat{\ell}_i  .
$$

Finally, by combining these results, it follows that 
\begin{eqnarray*}
\abs{ \ell_i - \tilde{\ell}_i } 
   &\le& \abs{ \ell_i - \hat{\ell}_i } + \abs{ \hat{\ell}_i - \tilde{\ell}_i }   \\
   &\le& \left( \frac{\epsilon}{1-\epsilon} + 2\epsilon \right) \ell_i   \\
   &\le& 4 \epsilon \ell_i ,
\end{eqnarray*}
where the first inequality follows from the triangle inequality, the second inequality follows by applying the two lemmas, and the last inequality follows since $\epsilon \le 1/2$.
From this, the quality-of-approximation claims of the theorem theorem follow.
So, let's prove those two lemmas.

\begin{Proof}[of Lemma~\ref{lem:first-approx}]
Let $A=U \Sigma V^T$, and recall that $\left(\Pi_1 A \right)^{\dagger} = V \Sigma^{-1} \left( \Pi_1 U \right)^{\dagger}$.
Then, 
\begin{eqnarray*}
\hat{U}_{(i)}^T\hat{U}_{(j)}
   &=& e_i A \left( \Pi_1 A \right)^{\dagger} \left(\left( \Pi_1A \right)^{\dagger}\right)^{T} A^T e_j   \\
   &=& e_i U \Sigma V^T V \Sigma^{-1} \left( \Pi_1 U \right)^{\dagger} \left(\left( \Pi_1U \right)^{\dagger}\right)^{T} \Sigma^{-1} V^TV \Sigma U^T e_j   \\
   &=& e_i U \left( \Pi_1 U \right)^{\dagger} \left(\left( \Pi_1U \right)^{\dagger}\right)^{T} U^T e_j.
\end{eqnarray*}
Thus, it follows that
\begin{eqnarray*}
\abs{ U_{(i)}^T U_{(j)} - \hat{U}_{(i)}^T\hat{U}_{(j)} } 
   &=& \abs{ e_i UU^T e_j - e_i U  \left( \Pi_1 U \right)^{\dagger} \left(\left( \Pi_1U \right)^{\dagger}\right)^{T} U^T e_j  } \\
   &=& \abs{ e_i U \left(  I -  \left( \Pi_1 U \right)^{\dagger} \left(\left( \Pi_1U \right)^{\dagger}\right)^{T} \right) U^T e_j } \\
   &\le& \TNorm{ I -  \left( \Pi_1 U \right)^{\dagger} \left(\left( \Pi_1U \right)^{\dagger}\right)^{T} } \VTTNorm{ U_{(i)} } \VTTNorm{ U_{(j)} }   .
\end{eqnarray*}
Since $\Pi_1U = U_{\Pi_1U}\Sigma_{\Pi_1U}V_{\Pi_1U}^T$ and $ \left( \Pi_1 U \right)^{\dagger} \left(\left( \Pi_1U \right)^{\dagger}\right)^{T} = V_{\Pi_1U}\Sigma_{\Pi_1U}^{-2} V_{\Pi_1U}^T$, it then follows that
\begin{eqnarray*}
\abs{ U_{(i)}^T U_{(j)} - \hat{U}_{(i)}^T\hat{U}_{(j)} } 
   &=& \TNorm{ I - V_{\Pi_1U}\Sigma_{\Pi_1U}^{-2} V_{\Pi_1U}^T }  \VTTNorm{ U_{(i)} } \VTTNorm{ U_{(j)} }  \\
   &=& \TNorm{ I - \Sigma_{\Pi_1U}^{-2} }  \VTTNorm{ U_{(i)} } \VTTNorm{ U_{(j)} }  \\
   &=& \frac{\epsilon}{1-\epsilon} \VTTNorm{ U_{(i)} } \VTTNorm{ U_{(j)} } ,
\end{eqnarray*}
which establishes the lemma.
\end{Proof}

\begin{Proof}[of Lemma~\ref{lem:second-approx}]
Since $\Pi_2$ is and $\epsilon$-JLT, it preserves the norm of $n^2$ vectors.
Let $x_i = \hat{U}_{(i)} / \VTTNorm{ \hat{U}_{(i)} }$, and consider the following $n^2$ vectors:
\begin{eqnarray*}
x_i && i\in[n] \\
x_i+x_j && i,j\in[n], i\ne j  .
\end{eqnarray*}
By the $\epsilon$-JLT property, and since $\VTTNorm{x_i}=1$, it follows that:
\begin{eqnarray*}
1-\epsilon \le \VTTNorm{ x_i \Pi_2 } \le 1+\epsilon  & & \forall i  \\
\left(1-\epsilon\right)\VTTNormS{x_i+x_j} \le  \VTTNormS{ \left(x_i + x_j\right) \Pi_2 } \le \left(1+\epsilon\right)\VTTNormS{x_i+x_j} & & \forall i.j \mbox{ s.t. } i \ne j  .
\end{eqnarray*}
If we combine these, expand the squares, and use that $\VTTNormS{\alpha+\beta}=\VTTNormS{\alpha}+\VTTNormS{\beta}+2\alpha^T\beta$, for vectors $\alpha$ and $\beta$, and use that $\VTTNorm{x}=1$, we have that
$$
x_i^Tx_j -2\epsilon \le \left( x_i\Pi_2 \right)^T \left( x_j \Pi_2 \right) \le x_i^Tx_j + 2\epsilon   .
$$
If we multiply through by $\VTTNorm{ \hat{U}_{(i)} }\VTTNorm{ \hat{U}_{(j)} }$ and then use the homogeneity of the inner product, we get that
$$
\hat{U}_{(i)}^T\hat{U}_{(j)} - 2\epsilon \VTTNorm{ \hat{U}_{(i)} }\VTTNorm{ \hat{U}_{(j)} }
   \le \left( \hat{U}_{(i)}\Pi_2\right)^{T} \left( U_{(j)}\Pi_2\right)
   \le \hat{U}_{(i)}^T\hat{U}_{(j)} + 2\epsilon \VTTNorm{ \hat{U}_{(i)} }\VTTNorm{ \hat{U}_{(j)} }  ,
$$
which establishes the lemma.
\end{Proof}

For the running time, here is a summary.
\begin{itemize}
\item
Computing $\Pi_1A$ takes time $O\left( nd\log(r_1) \right)$.
\item
Computing the SVD of $\Pi_1A$ takes time $O\left( r_1d^2 \right)$.
\item
Computing $R^{-1}\Pi_2$ takes time $O\left( r_2 d^2 \right)$.
\item
Premultiplying by $A$ takes time $O\left( ndr \right)$.
\end{itemize}
Thus, the overall running time is $O\left( nd\log(r_1) + ndr_2 + r_1d^2 + r_2d^2 \right)$, which by the choice of the various parameters is $O\left( nd\log(n)+d^3\log(n)\log(d) \right)$ time.

Combining these results establishes the theorem.
\end{Proof}

\subsection{Using the fast leverage score approximation algorithm for a fast random sampling algorithm for the LS problem}

We have already seen a fast random projection-based algorithm for the LS problem that runs in $o(nd^2)$ time as well as a slow random sampling algorithm that used the exact leverage scores as an importance sampling distribution.
Not surprisingly, we can use the fast approximations to the leverage scores to get a fast random sampling-based algorithm for the LS problem that runs in $o(nd^2)$ time.
Here is that algorithm.

\begin{algorithm}
\caption{A ``fast'' random sampling algorithm for the LS problem.}
\label{alg:fast-sample-ls-alg}
\begin{algorithmic}[1]
    \REQUIRE An $n \times d$ matrix $A$, with $n \gg d$, an $n$-vector $b$, and an error parameter $\epsilon\in(0,1/2)$.
    \ENSURE  A $d$-vector $\tilde{x}_{opt}$
    \STATE 
    Run the \texttt{FastApproximateLeverageScores} algorithm (with $\epsilon=1/2$ as input to that algorithm) to get $2$ approximations to all of the leverage scores of $A$; rescale them to form an importance sampling distribution $\{p_i\}_{i=1}^{n}$.
    \STATE 
    Randomly sample $r \gtrsim O(\frac{d \log d}{\epsilon})$ rows of $A$ and elements of $b$, using $\{p_i\}_{i=1}^{n}$ as the importance sampling distribution, rescaling each by $\frac{1}{\sqrt{rp_{i_t}}}$, i.e., form $SA$ and $Sb$.
    \STATE 
    Solve $\min_{x\in\mathbb{R}^{d}}\VTTNorm{SAx-Sb}$ with a black box to get $\tilde{x}_{opt}$.
    \STATE 
    Return $\tilde{x}_{opt}$.
\end{algorithmic}
\end{algorithm}

For this algorithm, we can prove the following.
\begin{theorem}
For this algorithm,  the output is a vector $\tilde{x}_{opt}$ such that with probability $\ge 0.8$:
\begin{itemize}
\item
$\VTTNorm{ A\tilde{x}_{opt} - b } \le \left( 1+\epsilon \right) \mathcal{Z}$
\item
$\VTTNorm{ \tilde{x}_{opt} - x_{opt} } \le \sqrt{\epsilon}\kappa(A) \sqrt{\gamma^{-2}-1} \VTTNorm{x_{opt}} $
\end{itemize}
In addition, the running time is $O\left( nd\log(n)+d^3\log(n)\log(d) \right)$.
\end{theorem}
The quality-of-approximation claims follow since using the approximate leverage scores an importance sampling distribution leads to the two structural conditions being satisfied.
The running time claims follow since the running time bottleneck for this algorithm is the running time of the \texttt{FastApproximateLeverageScores} algorithm (and the running time bottleneck of that algorithm is applying the random projection), which is $O\left( nd\log(n)+d^3\log(n)\log(d) \right)$ (it is this since we have set $\epsilon=1/2$ in the input to that algorithm).

\subsection{Computing leverage scores quickly for an arbitrary matrix}

A question that will arise once we consider RandNLA algorithms for low-rank matrix approximation is whether this fast leverage score approximation algorithm extends to compute the leverage scores of general ``fat'' matrices, i.e., matrices where both dimensions are large and we are interested in an approximation with respect to a low dimension defined by a low rank parameter.

The short answer is yes.  
The longer answer is that there are some subtleties.
(There are also subtleties in applying random projections to ``fat'' matrices that we will also consider.  In both cases, the subtleties have to do with the interaction between the space spanning the top part of the spectrum of the matrix and the space spanning the bottom part of the spectrum of the matrix.)
Here, we will briefly describe some of those subtleties, without going into too much detail (if you want more detail, see the DMMW paper).

We have seen that, when applied to a tall matrix, random projections flatten out leverage scores to permit sparse projections or uniform sampling.
It is also the case that, when applied to fat matrices, random projections uniformize things.
But, what things? 
And, in particular, is there a notion of leverage, so we can view a random projection as preprocessing or preconditioning so uniform sampling is appropriate?

Consider a general $n \times d$ matrix $A$, i.e., for which $n \approx d$, and write $A=U \Sigma V^T$.
In this case, $U$ and $V$ are in general square, and so have both orthonormal rows and columns.
Thus, the definition of leverage we have presented before is uninteresting, since it is always uniform.
On the other hand, if we project that matrix $A$ onto $\ell \gtrsim k$ dimensions, where $k$ is some explicit or implicit rank parameter, then we are really interested in the non uniformity structure on the top part of the spectrum of $A$.

This motivates the following definition.
\begin{definition}
Given a matrix $A\in\mathbb{R}^{n \times d}$, where $n \approx d$ and $k \ll \min\{n,d\}$.
Then, the \emph{leverage scores, relative to the best rank $k$ approximation to $A$} are 
$$
p_i = \frac{1}{k} \VTTNormS{ \left(U_k\right)_{(i)} }  ,
$$
if $A=U_k\Sigma_kV_k^T+U_k^{\perp}\Sigma_k^{\perp}V_k^{\perp}$ is the decomposition of $A$ into the best rank $k$ approximation and the residual.
\end{definition}
Note that the way we have defined it, $\sum_{i=1}^{n} p_i = 1$.
We could have defined quantities $\ell_i$ without the normalization, in which case we would have $\sum_{i=1}^{n} \ell_i = \FNormS{U_k}=k$, which is where the $k$ in the denominator of the equation in the definition comes from.

The basic idea of the approximate leverage score algorithm extends to this case, with the following caveat:
the problem of computing the leverage scores relative to the best rank $k$ approximation to $A$ is an ill-posed problem, in that a minor change in the problem input can completely change the answer.
(On the other hand, if we used those perturbed leverage scores in a low-rank approximation algorithm, they would still obtain good quality of approximation bounds, but they would identify a somewhat different subspace.) 
To see this, consider 
$$
A = I_n \quad\Rightarrow\quad U_k \mbox{ is not unique since there are }{n \choose k}\mbox{ equivalent choices}
$$
More generally, consider the matrix
\[
A = \left( \begin{array}{cc} I_k & o \\
                             0 & \left(1-\gamma\right) I_{n-k}  
           \end{array}
    \right)  ,
\]
which is parameterized by a number $\gamma >0$.
As $\gamma\rightarrow0$, it isn't possible to distinguish the top $k$ directions.

There are two common solutions to this.
\begin{itemize}
\item
Parameterize the problem in terms of the ``spectral gap,'' i.e., in terms of $\gamma = \sigma_k^2-\sigma_{k+1}^{2}$.
This is theoretically convenient, but it is awkward and typically represents an unrealistic assumption.
\item
Compute the leverage for a space that is ``near'' the best rank $k$ space.
This is more involved, but it uses ideas from subspace iteration methods that make the connections with random projection (and in particular high-precision random projection) algorithms clearer.
\end{itemize}

Here is a definition of nearness that DMMW used in the latter approach.

\begin{definition}
Given a matrix $A\in\mathbb{R}^{n \times d}$, a rank parameter $k \ll \min\{n,d\}$, let $A_k$ be the best rank $k$ approximation to $A$.
Then, 
$$
S_{\epsilon} = \{ X\in\mathbb{R}^{n \times d} : \mbox{rank}(X)=k \mbox{ and } \XNorm{A-X}\le\left(1+\epsilon\right)\XNorm{A-A_k}  \}
$$
is a set of subspaces near (in a sense quantified by $\XNorm{\cdot}$) the best rank $k$ approximation to $A$.
\end{definition}
Here $\XNorm{\cdot}$ represents a unitarily-invariant norm; clearly, for different norms, one can expect different quality-of-approximations.

Given this, we can define a weaker notion of leverage as follows.

\begin{definition}
Call the numbers $\hat{p}_i$, for all $i\in[n]$ \emph{$\beta$-approximations to the normalized leverage scores of $A_k$} if there exists an $X\in S_{\epsilon}$ such that $\hat{p}_i \ge \frac{\beta}{k}\VTTNormS{ \left(U_X\right)_{(i)} }$ and $\sum_{i=1}^{n}\hat{p}_i=1$, where $U_X\in\mathbb{R}^{n \times k}$ is the matrix of left singular vectors of $X$.
\end{definition}

Given this notion, we can compute approximations to these scores in ``random projection time.''
We haven't described fat matrices and low-rank matrix approximation yet, but this time to implement a random projection can often be made (in theory and/or in practice) to be faster than an SVD or QR computation.
In addition, random projections uniformize these quantities---and the extent to which these are uniformized depends on details like the number of iterations in subspace iterative methods that also determine the quality of low-rank random projection methods.
We won't go into detail on these topics, but we will return to some of them below.

\newpage

\section{%
(10/09/2013):  Randomized Least-squares Approximation in Practice}

During this class and the next few classes, we will describe how the RandNLA theory from the last few classes can be implemented to obtain high-quality implementations in practice.
Here is the reading for today and the next few classes.
\begin{itemize}
\item
Avron, Maymounkov, and Toledo, ``Blendenpik: Supercharging LAPACK's Least-Squares Solver'' 
\item
Avron, Ng, and Toledo, ``Using Perturbed QR Factorizations to Solve Linear Least-Squares Problems''
\end{itemize}

Today, in particular, we will do three things.
\begin{itemize}
\item
We will provide an overview of some of the implementation challenges.
\item
We will go into more detail on forward error versus backward error questions.
\item
We will go into more detail on preconditioning and $\epsilon$-dependence issues.
\end{itemize}

\subsection{Overview of implementation challenges}

There are several issues that must be dealt with to implement these RandNLA ideas in practice.
By this, we mean that we will want to implement these algorithms as is done in NLA, e.g., in LAPACK.
Different issues arise when we implement these algorithms in a data center or a database or a distributed environment or a parallel shared memory environment; and different issues arise when we implement and apply these algorithms in machine learning and data analysis applications.
We will touch on the latter two use cases to some extent, but our main focus here will be on issues that arise in providing high-quality implementations on moderately large problems to solve problems to (say) machine precision on a single machine.

Here are the main issues.
\begin{itemize}
\item
\textbf{Forward versus backward error.}
TCS worst-case bounds deal with forward error bounds in one step, but in NLA this is done via a two step process, where one considers the posedness of a problem and then the stability of an algorithm for that problem. 
This two step approach is less general than the usual one step approach in TCS that makes forward error claims on the objective function, but it leads to finer bounds when it works.
\item
\textbf{Condition number issues.}
So far, we have said almost nothing about condition number issues, except to note that a condition number factor entered depending on whether we were interested in relative error on the objective versus certificate, but they are very important in finite precision arithmetic and in practical implementations.
\item
\textbf{Dependence on $\epsilon$ parameter.}
A dependence of $1/\epsilon$ or $1/\epsilon^2$ is natural in TCS and is fine if we view $\epsilon$ as fixed and not too large, e.g., $0.1$ (and the $1/\epsilon^2$ is a natural bottleneck for Monte Carlo methods due to law of large numbers considerations), but this is actually exponential in the number of accuracy bits (since a number of size $n$ can be represented in roughly $\log(n)$ bits) and it is a serious problem if we want to obtain high precision, e.g., $10^{-10}$.
\end{itemize}

Here are some additional issues, that arise especially when dealing with the third issue above.
\begin{itemize}
\item
\textbf{Hybrid and preconditioned iterative methods.}
Here, we go beyond simply doing random sampling and random projection where we call a black box on the subproblem.
\item
\textbf{Failure probability issues.}
In some cases, it is preferable to have the failure probability enter into the running time that it takes to get a good solution and not whether or not the algorithm gets a good solution at all.
This has been studied in TCS (under the name Las Vegas algorithms); and, fortunately, it meshes well with the condition number and iterative considerations mentioned above.
\item
\textbf{Downsampling more aggressively.}
In code, it is difficult to sample $O\left(d\log(d)\right)$ rows, i.e., to loop from $1$ to $O\left(d\log(d)\right)$, where the constant in the big-O is unspecified.
Instead, one typically wants to loop from $1$ to (say) $2d$ or $4d$.
At that point, there are worst-case examples where the algorithm might fail, and thus one needs to deal with the situation where, e.g., there are a small number of rows with large leverage that are lost since we downsample more aggressively.
Again, we will see that---while a serious problem for worst-case TCS-style analysis---this meshes well with iterative algorithms as used in NLA.
\end{itemize}
We will deal with each of these issues in turn.

\subsection{Forward versus backward error.}

Let's start with the following definition.
\begin{definition}
A problem $P$ is well-posed if:
the solution exists; 
the solution is unique; 
and the solution depends continuously on the input data in some reasonable topology.
\end{definition}
We should note that this is sometimes called \emph{well-conditioned} in NLA, but the concept is more general.
(For example, $P$ doesn't need to be a LS or low-rank matrix problem---it could be anything such as the MaxCut Problem or the $k$-Sat Problem or whatever.)

The point here is that, if we work with matrix problems with real-valued continuous variables, even for a well-posed or well-conditioned problem, certain algorithms that solve the problem ``exactly,'' e.g., with infinite precision, perform poorly in the presence of ``noise'' introduced by truncation and roundoff errors.

This leads to the idea of the numerical stability \emph{of an algorithm}.
Let's consider an algorithm as a function $f$ attempting to map input data $X$ to output data $Y$; but, due to roundoff errors, random sampling, or whatever, the algorithm actually maps input data $X$ to output data $Y^{*}$.
That is, the algorithm ``should'' return $Y$, but it actually returns $Y^{*}$.
In this case, we have the following.
\begin{itemize}
\item
\textbf{Forward error.}
This is $\Delta Y = Y- Y^{*}$, and so this is the difference between the exact/true answer and the answer that was output by the algorithm.
(This is typically what we want to bound, although note that one might also be interested in some function of $Y$, e.g., the objective function value in an optimization, rather than that argmin.)
\item
\textbf{Backward error.}
This is the smallest $\Delta X$ such that $f\left( X + \Delta X \right) = Y^{*}$, and so this tells us what input data the algorithm that we ran actually solved exactly.
\end{itemize}
In general, the forward error and backward error are related by a problem-specific complexity measure, often called the condition number as follows:
$$
\abs{\text{forward error}} \le \abs{\text{condition number}} \times \abs{\text{backward error}} .
$$
In particular, backward stable algorithms provide accurate solutions to well-conditioned problems.

TCS typically bounds the forward error directly in one step, while NLA bounds the forward error indirectly in two steps, i.e., by considering only well-posed problems and then bounding the backward error for those problems.
That typically provides finer bounds, but it is less general, e.g., since it says nothing about ill posed problems.

In light of this discussion, observe that the bounds that we proved the a few classes ago bound the forward error in the TCS style.
In particular, recall that the bounds on the objective are of the form
\begin{equation}
\VTTNorm{A\tilde{x}_{opt}-b} \le \left(1+\epsilon\right)\VTTNorm{Ax_{opt}-b}  , 
\label{eqn:tcs-forward-error-certificate}
\end{equation}
and this implies that
\begin{eqnarray}
\VTTNorm{ \tilde{x}_{opt}-x_{opt} } 
   \nonumber
   &\le& \sqrt{1-\gamma^2} \kappa\left(A\right) \sqrt{\epsilon} \VTTNorm{x_{opt}} \\
   \label{eqn:tcs-forward-error}
   &=& \tan\left(\theta\right)\kappa\left(A\right) \sqrt{\epsilon} \VTTNorm{x_{opt}}  ,
\end{eqnarray}
where $\theta = \cos^{-1}\left( \frac{ \VTTNorm{ Ax_{opt} } }{ \VTTNorm{b} } \right)$ is the angle between the vector $b$ and the column space of $A$.

This is very different than the usual stability analysis which is done in NLA which is done in terms of backward error as follows.
Consider the approximate solution $\tilde{x}_{opt}$ (usually in NLA this is different than the exact solution in exact arithmetic, e.g., due to roundoff errors; but in RandNLA it is different, even in exact arithmetic, since we solve a random subproblem of the original problem), and consider the perturbed problem that it is the exact solution to.
That is, 
\begin{equation}
\tilde{x}_{opt} = \text{argmin} \VTTNorm{ \left(A + \delta A \right) x + b }  ,
\label{eqn:perturbed-eqn-error}
\end{equation}
where $\XNorm{ \delta A } \le \tilde{\epsilon} \XNorm{A}$.
(We could of course include a perturbed version of $b$ in Equation~(\ref{eqn:perturbed-eqn-error}).)
By standard NLA methods, Equation~(\ref{eqn:perturbed-eqn-error}) implies a bound on the forward error
\begin{equation}
\VTTNorm{  \tilde{x}_{opt} - x_{opt} } 
   \le \left( \kappa\left(A\right) + \frac{ \kappa^2\left(A\right) \tan\left(\theta\right) }{ \eta } \right) \tilde{\epsilon} \VTTNorm{x_{opt}}  ,
\label{eqn:nla-forward-error}
\end{equation}
where $\eta = \frac{ \TNorm{A}\VTTNorm{x} }{ \VTTNorm{Ax} }$.
Importantly, Equation~(\ref{eqn:tcs-forward-error}) and Equation~(\ref{eqn:nla-forward-error}), i.e., the two different forward error bounds on the vector or certificate achieving the optimal solution, are not obviously comparable.

There are some but very few results on the numerical stability of RandNLA algorithms, and this is a topic of interest, especially when solving the subproblem to achieve a low-precision solution. 
If RandNLA methods are used as preconditions for traditional algorithms on the original problem, then this is less of an issue, since they inherit the properties of the traditional iterative algorithms, plus something about the quality of the preconditioning (which is probably less of an issue).

\subsection{Preconditioning, the dependence on $\epsilon$, and related issues}

There are two broad methodologies for solving sparse and dense linear systems: direct methods and iterative methods.
The two classes of methods are complementary, and each comes with pros and cons, some of which are listed here.
\begin{itemize}
\item
\textbf{Direct methods.}
These methods typically involve factoring the coefficient matrix into the product of simpler matrices whose inverses are easier to apply.
For example, $A=LU$ in general, $A=LL^T$ for SPSD matrices, and $A=QR$ or $A=U \Sigma V^T$ for overdetermined/rectangular problems.
These are generic, robust, predictable, and efficient; but they can have limited scalability, they may run out of memory, they may be too slow for large and especially sparse matrices, etc.
\item
\textbf{Iterative methods.}
These methods typically involve starting with an approximate solution and iteratively refining it, in the simplest case by doing matrix-vector products.
These often scale much better to large and/or sparse problems; but they can be more fragile and can be slower than direct methods for many inputs.
\end{itemize}
The issue about iterative methods being slower depends on several factors, but there is usually some sort of problem-specific complexity measure, e.g., the condition number, that determines the number of iterations, and this can be large in worst-case.

A partial solution that can over this difficulty is to use something called a preconditioner and then work with preconditioned iterative methods.
This comes at the expense of a loss of generality since a given preconditioner in general doesn't apply to every problem instance.
Among other things, preconditioning opens the door to hybridization: we can use direct methods to construct a preconditioner for an iterative methods, e.g., use an incomplete decomposition to minimize fill-in and then apply an iterative method.
This can lead to improved robustness and efficiency, while not sacrificing too much generality.

While perhaps not obvious from the discussion so far, randomization as used in RandNA can be combined with ideas like hybridization to obtain practical iterative LS solvers.
Basically, this is since low-precision solvers, e.g., those obtained when using $\epsilon=1/2$ (or worse), provide a pretty good solution that can be computed pretty quickly.
In particular, this involves using a randomized algorithm to construct a preconditioned for use in a deterministic iterative solver.

Let's say a few words about why this is the case.
Recall that Krylov subspace iterative methods for solving large systems of linear equations treat matrices as black boxes and only perform matrix-vector multiplications.
Using this basic operation, they find an approximate solution inside the Krylov subspace:
$$
K_n\left(A,b\right) = \{ b,Ab,A^2b, \ldots,A^{n-1}b \} .
$$
For example: 
(1) CG, for SPSD matrices;
(2) LSQR, for LS problems, which is like CG on the normal equations; and
(3) GMRES, for general matrices.

While it's not the most common perspective, think of a preconditioned as a way to move between the extremes of direct and iterative solvers, taking advantage of the strengths of each.
For example, if we have an SPSD matrix $A$ and we want to do CG and use a preconditioner $M$, then instead of solving the original problem
$$
Ax = b  ,
$$
we might instead solve the preconditioned problem
$$ 
M^{-1}Ax = M^{-1}b  .
$$

(Alternatively, for the LS problem, instead of solving the original problem
$$
\min_x \VTTNorm{ Ax-b }  ,
$$
we might instead solve the preconditioned problem
$$
\min_y \VTTNorm{ AR^{-1}y - b }  ,
$$
where $Rx=y$.)

For simplicity, let's go back to the CG situation (although the basic ideas extend to overdetermined LS problems, which will be our main focus, as well as to more general problems, which we won't discuss).
There are two straw men to consider in preconditioning.
\begin{itemize}
\item
If $M=A$, then then the solver is basically a direct solver.
In particular, constructing the preconditioner requires solving the original problem exactly, and since this provides the solution there is no need to do any iteration.
\item
If $M=I$, then the solver is basically an unpreconditioned iterative solver.
The ``preconditioning phase'' takes no time, but the iterative phase is no faster than without preconditioning.
\end{itemize}
The goal, then, is to find a ``sweet spot'' where $M$ is not too expensive to compute and where it is not too hard to form its inverse.
In particular, we want $\kappa\left(M^{-1}A\right)$ to be small and also that $M^{-1}$ is easy to compute and apply, since then we have a good preconditioner.
(The extension of this to the overdetermined LS problem is that $AR^{-1}$ is quick to compute and that $\kappa\left(AR^{-1}\right)$ is small, where $R$ is a matrix not-necessarily from QR.)

The way we will use this is that we will use the output of a low-precision random projection or random sampling process to construct a preconditioned.
(In particular, we will compute a QR decomposition of $\Pi A$, where $\Pi$ is a random projection or random sampling matrix.)
That is, rather than solving the subproblem on the projection/sampling sketch, we will use it to construct a preconditioner.
If we obtain a very good sketch, e.g., one that provides a $1\pm\epsilon$ relative error subspace embedding, then it will also provide a very good $1\pm\epsilon$ preconditioner.
That works; but, importantly, it is often overkill, since we can get away with lower quality preconditioners.

To understand how this works, we need to get into a few details about how to analyze the quality of preconditioners.
The simplest story is that if the eigenvalue ratio, i.e., the condition number, of the problem is small, then we have a good preconditioner.
And if we sample/project onto $O\left(d\log(d)\right)$ dimensions, then we will satisfy this.
But we can also get good quality preconditioners with many fewer samples.
To do this, we need to know a little more than just the eigenvalue ratio, since controlling that is sufficient but not quite necessary to have a good preconditioner, so let's get into that.

Again, for simplicity, let's consider the case for SPSD matrices.
(It is simpler, and via LSQR, which is basically CG on the normal equations, most of ideas go through to the LS problem.  We will point out the differences at the appropriate points.)
Analyzing preconditioners for SPSD matrices is usually done in terms of generalized eigenvalues and generalized condition numbers.
Here is the definition.

\begin{definition}
Let $A,B \in \mathbb{R}^{n \times n}$, then $\lambda = \lambda\left(A,B\right)$ is a \emph{finite generalized eigenvalue} of the matrix pencil/pair if there exists a vector $v \ne 0$ such that
$
\left\{ \begin{array}{l}
                    Av = \lambda B v \\
                    Bv \ne 0     
                 \end{array}
         \right.  .
$
\end{definition}
Given this generalized notion of eigenvalue, we can define the following generalized notion of condition number.
\begin{definition}
Let $A,B \in \mathbb{R}^{n \times n}$ be two matrices with the same null space.
Then the \emph{generalized condition number} is 
$$
\kappa\left(A,B\right) = \frac{ \lambda_{max}\left(A,B\right) }{ \lambda_{min}\left(A,B\right) }  .
$$
\end{definition}

These notions are important since the behavior of preconditioned iterative methods is determined by the clustering of the generalized eigenvalues, and the number of iterations is proportional to the condition number.
\begin{itemize}
\item
For CG, the convergence is in $O\left( \sqrt{ \kappa\left(A,M\right) } \right)$ iterations.
\item
For LSQR, if $A$ is preconditioned by a matrix $R$, then convergence is in $O\left( \sqrt{ \kappa\left( A^TA,R^TR \right) } \right)$ iterations.
\end{itemize}
Next time, we'll discuss how these ideas can be coupled with the RandNLA algorithms we have been discussing.

\newpage

\section{%
(10/14/2013):  Randomized Least-squares Approximation in Practice, Cont.}

We continue with the discussion from last time.
There is no new reading, just the same as last~class.

Today, we will focus on three things.
\begin{itemize}
\item
We will describe condition numbers and how RandNLA algorithms can lead to good preconditioning.
\item
We will describe two different ways that randomness can enter into the parameterization of RandNLA problems.
\item
We will describe the Blendenpik RandNLA LS solver.
\end{itemize}

\subsection{Condition numbers and preconditioning in RandNLA algorithms}

Recall that we are interested in the conditioning quality of randomized sketches constructed by RandNLA sampling and projection algorithms.

For simplicity of comparison with the Blendenpik paper, I'll state the results as they are stated in that paper, i.e., with a Hadamard-based projection, and then I'll point out the generalization (e.g., to leverage score sampling, to other types of projections, etc.) to other related RandNLA sketching methods.

Recall that after pre-multiplying by the randomized Hadamard transform $H$, the leverage scores of $HA$ are roughly uniform, and so one can sample uniformly.
Here is a definition we have mentioned before.
\begin{definition}
Let $A\in\mathbb{R}^{n \times d}$ be a full rank matrix, with $n > d$ and let $U\in\mathbb{R}^{n \times d}$ be an orthogonal matrix for $\mbox{span}(A)$.
Then, if $U_{(i)}$ is the $i^{th}$ row of $U$, the \emph{coherence} of $A$ is 
$$
\mu\left(A\right) = \max_{i \in[n]} \VTTNormS{ U_{(i)} }.
$$
\end{definition} 
That is, the coherence is---up to a scaling that isn't standardized in the literature---equal to the largest leverage score.
Equivalently, up to the same scaling, it equals the largest diagonal element of $P_A=A\left(A^TA\right)^{\dagger}A^T$, the projection matrix onto the column span of $A$.
Defined this way, i.e., not normalized to be a probability distribution, possible values for the coherence are
$$
\frac{d}{n} \le \mu\left(A\right) \le 1  .
$$
Thus, with this normalization:
\begin{itemize}
\item
If $\mu\left(A\right) \gtrsim \frac{d}{n}$, then the coherence is small, and all the leverage scores are roughly uniform.
\item
If $\mu\left(A\right) \lesssim 1$, then the coherence is large, and the leverage score distribution is very nonuniform (in that there is at least one very large leverage score).
\end{itemize}

The following result (which is parameterized for a uniform sampling process) describes the relationship between the coherence $\mu\left(A\right)$, the sample size $r$, and the condition number of the preconditioned system.

\begin{lemma}
Let $A\in\mathbb{R}^{n \times d}$ be a full rank matrix, and let $S\in\mathbb{R}^{r \times n}$ be a uniform sampling operator.
Let 
$$
\tau = C \sqrt{ m \mu\left(A\right) \log(r)/r }, 
$$
where $C$ is a constant in the proof.
Assume that $\delta^{-1}\tau < 1$, where $\delta$ is a failure probability.
Then, with probability $\ge 1-\delta$, we have that
$$
\mbox{rank}\left(SA\right) = d  ,
$$
and is the QR decomposition of $SA$ is $SA = \tilde{Q}\tilde{R}$, then
$$
\kappa\left(A\tilde{R}^{-1}\right) = \frac{ 1+\delta^{-1}\tau }{ 1-\delta^{-1}\tau }   . 
$$
\label{lem:blendenpik-coherence}
\end{lemma}

Before proceeding with the proof, here are several things to note about this result.
\begin{itemize}
\item
We can obtain a similar result on the condition number if we sample non-uniformly based on the leverage scores, and in this case the coherence $\mu\left(A\right)$ (which could be very large, rendering the results as stated trivial) does \emph{not} enter into the expression.
This is of interest more generally, but we'll state the result for uniform sampling for now.
The reason is that the Blendenpik solver does a random projection which uniformizes (approximately) the leverage scores, i.e., it preprocesses the input matrix to have a small coherence.
\item
Also, $\kappa\left(A\right)$, i.e., the condition number of the original problem instance, does \emph{not} enter into the bound on $\kappa\left(A\tilde{R}^{-1}\right)$.
\item
If we are willing to be very aggressive in downsampling, then the condition number of the preconditioned system might \emph{not} be small enough.
In this case, all is not lost---a high condition number might lead to a large number of iterations of LSQR, but we might have a distribution of eigenvalues that is not too bad and leads to a number of iterations that is not too bad.
In particular, the convergence of LSQR depends on the full distribution of the singular value of $\kappa\left(AR^{-1}\right)$ and not just the ratio of the largest to smallest (considering just that ratio leads to sufficient but not necessary conditions), and if only a few singular values are bad then this can be dealt with.
We will return to this topic below.
\item
As we will see, the proof of this lemma directly uses ideas, e.g., subspace preserving embeddings, that were introduced in the context of low-precision RandNLA solvers, but it uses them toward a somewhat different aim.
\end{itemize}

\begin{Proof}[of Lemma~\ref{lem:blendenpik-coherence}]
To prove the lemma, we need the following specialization of a result we stated toward the beginning of the semester.
Again, since the lemma is formulated in terms of a uniform sampling process, we state the following lemma as having the coherence factor ($\mu\left(U\right)$), which is necessary when uniform sampling is used.
We saw this approximate matrix multiplication result before when the uniform sampling operator was replaced with a random projection operator or a non-uniform sampling operator, and in both cases the coherence factor did not appear.
\begin{lemma}
Let $U\in\mathbb{R}^{n \times d}$ be an orthogonal matrix, and let $S\in\mathbb{R}^{n \times n}$ be a uniform sampling-and-rescaling operator.
Then
$$
\Expect{ \TNorm{ I - U^TS^TSU } } \le C \sqrt{ \frac{ m \mu\left(U\right) \log(r) }{ r } }   .
$$
\end{lemma}

Since $SU$ is full rank, so too is $SA$ full rank.
Then, we can claim that 
$$
\kappa\left(SU\right) = \kappa\left( A \tilde{R}^{-1} \right)  .
$$ 
To prove the claim, recall that 
\begin{eqnarray*}
SU &=& U_{SU} \Sigma_{SU} V_{SU}^{T} \quad \mbox{by definition}   \\
SA &=& \tilde{Q}\tilde{R} \quad \mbox{by definition}   \\
U_{SU} &=& \tilde{Q}W, \quad \mbox{for a $d \times d$ unitary matrix $W$, since they span the same space.}
\end{eqnarray*}
In this case, it follows that 
\begin{eqnarray*}
\tilde{R} 
   &=& \tilde{Q}^{T} S A   \\
   &=& \tilde{Q}^{T} S U \Sigma V^T   \\
   &=& \tilde{Q}^{T} U_{SU} \Sigma_{SU} V_{SU}^T \Sigma V^T   \\
   &=& W \Sigma_{SU} V_{SU}^{T} \Sigma V^T   .
\end{eqnarray*}
From this (and since $\Sigma_{SU}$ is invertible, by the approximate matrix multiplication bound, since we have sampled sufficiently many columns), it follows that
\begin{eqnarray*}
A\tilde{R}^{-1}
   &=& U \Sigma V^T V \Sigma^{-1} V_{SU} \Sigma_{SU}^{-1} W^T  \\
   &=& U V_{SU} \Sigma_{SU}^{-1} W^T  .
\end{eqnarray*}
From this, it follows that
\begin{eqnarray*}
\TNorm{ \left(A \tilde{R}^{-1}\right)^{T} A \tilde{R}^{-1} }
   &=& \TNorm{ W \Sigma_{SU}^{-1} V_{SU}^T U^TU V_{SU} \Sigma_{SU}^{-1} W^T  }   \\
   &=& \TNorm{ W \Sigma_{SU}^{-2} W^T }   \\
   &=& \TNorm{ \Sigma_{SU}^{-2} }   \\
   &=& \TNormS{ \Sigma_{SU}^{-1} }  ,
\end{eqnarray*}
where the penultimate equality follows since $W$ is orthogonal and the last equality follows since $\Sigma_{SU}$ is diagonal.
Similarly, 
$$
\TNorm{ \left( \left( A\tilde{R}^{-1} \right)^{T} \left( A\tilde{R}^{-1} \right) \right)^{-1} } = \TNormS{ \Sigma_{SU} }.
$$
This, using that $\kappa\left(\alpha\right) = \left( \TNorm{ \alpha^T\alpha } \TNorm{ \left( \alpha^T \alpha \right)^{-1} } \right)^{1/2}$ for a matrix $\alpha$, if follows that
\begin{eqnarray*}
\kappa\left(A\tilde{R}^{-1}\right) 
   &=& \left( \TNorm{ \left(A\tilde{R}^{-1}\right)^T A \tilde{R}^{-1} } \TNorm{ \left( \left(A\tilde{R}^{-1}\right)^T A \tilde{R}^{-1} \right)^{-1} } \right)^{1/2}   \\
   &=& \left( \TNormS{ \Sigma_{SU}^{-1} } \TNormS{ \Sigma_{SU} }  \right)^{1/2}   \\
   &=& \TNorm{ \Sigma_{SU}^{-1} } \TNorm{ \Sigma_{SU} }  \\
   &=& \kappa\left(SU\right)  .
\end{eqnarray*}
This establishes the claim.

So, given that claim, back to the main proof.
Recall that 
$$
\Expect{ \TNorm{ I - U^TS^TSU } } \le \tau   .
$$
By Markov's Inequality, it follows that
$$
\Probab{ \TNorm{ I - U^TS^TSU } \ge \delta^{-1}\tau } \le \delta  .
$$
This, with probability $\ge 1-\delta$, we have that
$$
\TNorm{ I - U^TS^TSU } < \delta^{-1} \tau < 1   ,
$$
and thus $SU$ is full rank.
\end{Proof}

Next, recall that every eigenvalue $\lambda$ of $U^TS^TSU$ is the Rayleigh quotient of some vector $x \ne 0$, i.e.,
\begin{eqnarray*}
\lambda
   &=& \frac{ x^TU^TS^TSUx }{ x^Tx } \\
   &=& \frac{ x^Tx - x^T\left( U^TS^TSU - I \right) x }{ x^Tx }  \\
   &=& 1+\eta  ,
\end{eqnarray*}
where $\eta$ is the Rayleigh quotient of $I-U^TS^TSU$.

Since this is a symmetric matrix, it's singular values are the absolute eigenvalues.
Thus, $\abs{\eta} < \delta^{-1}\tau$.
Thus, all the eigenvalues of $U^TS^TSU$ are in the interval $\left(1-\delta^{-1}\tau,1+\delta^{-1}\tau\right)$.
Thus, 
$$
\kappa\left(SU\right) \le \sqrt{ \frac{ 1+\delta^{-1}\tau }{ 1-\delta^{-1}\tau } }   .
$$

\subsection{Randomness in error guarantees versus in running time}

So far, we have been describing algorithms in the ``deterministic running time and probabilistic error guarantees'' framework.
That is, we parameterize/formulate the problem such that we guarantee that we take not more than 
$$
O\left( f\left( n,\epsilon,\delta \right) \right)    
$$
time, where $f\left(n,\epsilon,\delta)\right)$ is some function of the size $n$ of the problem, the error parameter $\epsilon$, and a parameter $\delta$ specifying the probability with which the algorithm may completely fail.

That is most common in TCS, where simpler analysis for worst-case input is of interest, but in certain areas, e.g., NLA and other areas concerned with providing implementations, it is more convenient to parameterize/formulate the problem in a ``probabilistic running time and deterministic error'' manner.
This involves making a statement of the form
$$
\Probab{ \mbox{Number of FLOPS for $\le \epsilon$ relative error } \ge f \left(n,\epsilon,\delta\right) } \le \delta  .
$$
That is, in this case, we can show that we are guaranteed to get the correct answer, but the running time is a random quantity.

A few things to note.
\begin{itemize}
\item
These algorithms are sometimes known in TCS as Las Vegas algorithms, to distinguish them from Monte Carlo algorithm that have a deterministic running time but a probabilistic error guarantee.
Fortunately, for many RandNLA algorithms, it is relatively straightforward to convert a Monte Carlo type algorithm to a Las Vegas type algorithm.
\item
This parameterization/formulation is a particularly convenient when we downsample more aggressively than worst-cast theory permits, since we can still get a good preconditioner (since a low-rank perturbation of a good preconditioner is still a good preconditioner) and we can often still get good iterative properties due to the way the eigenvalues cluster.
In particular, we might need to iterate more, but we won't fail completely.
\end{itemize}

\subsection{Putting it all together into the Blendenpik algorithm}

With all this in place, here is the basic Blendenpik algorithm.

\begin{algorithm}
\caption{The \texttt{Blendenpik} algorithm.}
\label{alg:blendenpik-alg}
\begin{algorithmic}[1]
    \REQUIRE 
    $A\in\mathbb{R}^{n \times d}$, and $b\in\mathbb{R}^{n}$.
    \ENSURE  
    $\tilde{x}_{opt}\in\mathbb{R}^{d}$
    \WHILE{Not returned}
    \STATE
    Compute $HDA$ and $HDb$, where $HD$ is one of several Randomized Hadamard Transforms.
    \STATE
    Randomly sample $\gamma d / n$ rows of $A$ and corresponding elements of $b$, where $\gamma \approx 2d$, and let $S$ be the associated sampling-and-rescaling matrix.
    \STATE
    Compute $SHDA = QR$.
    \STATE
    $\tilde{\kappa} = \kappa_{estimate}\left(R\right)$, with LAPACK's DTRCON routine.
    \IF{$\tilde{\kappa}^{-1} > 5 \epsilon_{mach}$} 
    \STATE{$\tilde{x}_{opt} = \text{LSQR}\left( A,b,R,10^{-14} \right)$ and return}
    \ELSIF{Number of iterations $>3$} 
    \STATE{Call LAPACK and return}
    \ENDIF
    \ENDWHILE
\end{algorithmic}
\end{algorithm}

We will go into more detail on how/why this algorithm works next time (in terms of condition number bounds, potentially loosing rank, since $r=o\left(d\log(d)\right)$, etc.), but here are a few final notes for today.
\begin{itemize}
\item
Depending on how aggressively we downsample, the condition number $\kappa\left( A R^{-1}\right)$ might be higher than $1+\epsilon$.
If it is too much larger, then LSQR converges slowly, but the algorithm does not fail completely, and it eventually converges.
In this sense, we get Las Vegas guarantees.
\item
Since the downsampling is very aggressive, the preconditioner can actually be rank-deficient or ill-conditioned.
The solution that Blendenpik uses is to estimate the condition number with a condition number estimator from LAPACK, and if it is $\ge \epsilon_{mach}^{-1}/5$ then randomly project and sample again.
\item
In Blendenpik, they use LSQR, but one could use other iterative procedures, and one gets similar results.
\end{itemize}
We will go into more detail on these topics next time.

\newpage

\section{%
(10/16/2013):  Randomized Least-squares Approximation in Practice, Cont.}

We continue with the disucssion from last time.
There is no new reading, just the same as last~class.

Today, we will focus on three things.
\begin{itemize}
\item
We will discuss issues with good implementations that arise when downsampling more aggressively than worst-case theory needs.
\item
We will talk more about generalized eigenvalues and their relationship with perturbations of good preconditioners.
\item
We will talk about how those ideas are used in RandNLA solvers.
\end{itemize}
(This will wrap up our discussion of least squares.  Next time, we will move onto RandNLA algorithms for low-rank matrix approximation.)

\subsection{Issues with good implementations}

Recall that last time we were discussing Blendenpik, which provides an implementation of a RandNLA algorithm to solve the very overdetermined LS problem.
There is a bit of a theory-practice gap, and looking at how Blendenpik addresses some of those issues is illustrative more generally for other problems such as low-rank matrix approximation.

The basic idea is that rather than doing a QR decomposition of $A$, do a QR decomposition on $\Pi A$, where $\Pi$ is a FJLT (or some other, e.g., data-aware subspace embedding), and then use the $\tilde{R}^{-1}$ from QR on the subproblem as a preconditioner for an iterative algorithm on the original problem.
We saw that if $\Pi A = \tilde{Q}\tilde{R}$ then $\kappa\left(A\tilde{R}^{-1}\right) = \kappa\left(SU\right)$.
If we sample ``enough,'' i.e., $\Omega\left(d\log(d)/\epsilon\right)$, then this condition number is $\le 1+\epsilon$ and the very good subspace embedding provides a very good preconditioner. 

While this sampling complexity is basically necessary if we want to use $\Pi A$ and $\Pi b$ to solve the subproblem and obtain $1\pm\epsilon$ approximation guarantees, this might be overkill, if these methods are coupled with an iterative algorithm.
This is fortunate, since in practice one typically down-samples more aggressively s.t. $\kappa\left(A\tilde{R}^{-1}\right)$ is still somewhat large (and/or that the sample looses rank, which essentially means that it is infinite).
In this situation, the worst-case TCS theory fails, but we we might still be able to use these RandNLA methods to construct a good preconditioner for a traditional NLA iterative algorithm.
Even if we are not in such an extreme situation, constructing preconditioners, whether with deterministic or randomized methods, is expensive, and there is a tradeoff between very good preconditioners that need very few iterations and moderately good preconditioners that are less expensive but need more iterations.
Having control over this tradeoff is very important in practice.

Said another way, here is the main issue.
\begin{itemize}
\item
What if there are just a very few rows with very bad leverage scores (and we sample uniformly), or (if we sample non-uniformly) we get estimates of the leverage scores (via the fast leverage score algorithm, except that we down-sample more aggressively in the $\Pi_1$ projection inside that algorithm) that are good for most of the leverage scores but severely underestimate the leverage of a small number of rows?
In this case, the following is true.
\begin{itemize}
\item
We don't have a good subspace embedding, and so solving the subproblem does not lead to a good solution, in worst case analysis.
\item
Since we don't have a good subspace embedding, when using the sketch as a preconditioner, the condition number of the preconditioner is large (or infinite, if rank is lost), and so a naive bound that uses the condition to bound the number of iterations leads to poor results.
\item
We can often still use this as a preconditioner for iterative methods such as LSQR and get good convergence in a small number of iterations in practice.
Thus, it is still a reasonably good preconditioner, and there is theory to explain this---basically, the reason is since a small perturbation of it is a good preconditioner, and we will discuss that now.
\end{itemize}
\item
The basic idea here is that if there are a few rows that are missed (either with uniform sampling since they have high leverage, or with nonuniform sampling if we down-sample too aggressively), then the number of large singular values is bounded by the condition number of the large rows.
Having just a few large rows means that $\kappa\left(A\tilde{R}^{-1}\right)$ is large, but is doesn't much affect the convergence properties of LSQR.
\item
Somewhat more precisely, the ``R'' factor from a QR decomposition of a perturbation $\tilde{A}$ of $A$ is effective as a preconditioner for $A$; if is $A$ is poorly conditioned, then it is ok if $\tilde{A}$ is well-conditioned.
Ditto for $A^TA$ and $\tilde{A}^T\tilde{A}$.
Results of this form hold in general; and, for RandNLA, $A$ is the matrix from the sample, and $\tilde{A}$ is some other matrix that isn't explicitly constructed.
\end{itemize}
Now, we'll go into more detail on this.
In particular, we will outline some of the ideas from the ANT paper, but we will skip some of the details that are less relevant.

\subsection{Using perturbed QR factorizations to solve linear LS problems}

Before we go into some of the details, let's describe the notion of generalized eigenvalues and generalized condition numbers that we mentioned a few classes ago.

Recall that if we use CG with a preconditioner $M$, then we solve $Ax=b$ by solving $M^{-1}Ax=M^{-1}b$.  
If $M=A$, the we are basically solving the linear system with a direct method, while of $M=I$, then we are basically solving it with unpreconditioned iterative method.
So, the goal is to find an $M$ such that $M$ is easy to compute and $\kappa\left(M^{-1}A\right)$ is small.
Recall also the definitions of generalized eigenvalues and generalized condition numbers.

\begin{definition}
Let $S,T \in \mathbb{R}^{n \times n}$, then $\lambda = \lambda\left(S,T\right)\in\mathbb{R}^{n}$ is a \emph{finite generalized eigenvalue} of the matrix pencil $\left(S,T\right)$ if there exists a vector $v \ne 0$ such that
$
\left\{ \begin{array}{l}
                    Sv = \lambda T v \\
                    Tv \ne 0
                 \end{array}
         \right.  .
$
In addition, $\infty$ is an \emph{infinite generalized eigenvalue} of $\left(S,T\right)$ if there exists a $v \ne 0$ such that
$
\left\{ \begin{array}{l}
                    Tv = 0 \\
                    Sv \ne 0
                 \end{array}
         \right.  .
$
Note that $\infty$ is a eigenvalue of $\left(S,T\right)$ iff $0$ is an eigenvalue of $\left(T,S\right)$.
\end{definition}

\begin{definition}
The finite and infinite eigenvalues of a pencil are \emph{determined eigenvalues}, i.e., the eigenvector uniquely determines the eigenvalue.
If $Sv = Tv = 0$ for some $v \ne 0$, then $v$ is an \emph{indeterminate eigenvector}, since $Sv = \lambda T v$, for all $\lambda\in\mathbb{R}$.
We can denote the set of determined eigenvalues of $\left(S,T\right)$ by $\Lambda\left(S,T\right)$.
\end{definition}

\begin{definition}
Let $S,T\in\mathbb{R}^{n \times n}$ have the same null space.
The {generalized condition number} is
$$
\kappa\left(S,T\right) = \frac{ \lambda_{max}\left(S,T\right) }{ \lambda_{min}\left(S,T\right) }  ,
$$
where the $\max$ and $\min$ are over the determined eigenvalues of $\left(S,T\right)$.
\end{definition}

Here is a fact.
The behavior of preconditioned iterative methods is determined by the clustering of the generalized eigenvectors, and the number of iterations is bounded by a quantity that is proportional to the generalized condition number.
\begin{itemize}
\item
CG converges in $O\left( \sqrt{ \kappa\left(A,M\right) } \right)$ iterations.
\item
LSQR on $A$ preconditioned by $R$ converges in $O\left( \sqrt{ \kappa\left(A^TA,R^TR\right) } \right)$ iterations.
\end{itemize}
Note that these bounds provide sufficient conditions; but since these bounds  come from a more refined analysis that depends on the entire spectrum, they are not necessary.
We now turn to a theory that provides a more refined analysis.

In particular, here we describe a theory (from the ANT paper) that is more general than RandNLA preconditioning but which can be applied directly to RandNLA preconditioning.

Let $A$ be a matrix and let 
$
\hat{A} = \left( \begin{array}{c} A  \\
                             B 
           \end{array}
    \right)  .
$
Then, 
\begin{eqnarray*}
\left(\hat{A}^{T}\hat{A}\right)^{-1}A^TA 
   &=& \left(A^TA+B^TB\right)^{-1}A^TA   \\
   &=& \left(A^TA+B^TB\right)^{-1} \left( A^TA + B^TB - B^TB \right)   \\
   &=& I -  \underbrace{ \left(A^TA+B^TB\right)^{-1} B^TB }_{= \Omega}   .
\end{eqnarray*}

Here is a fact:
$$
\mbox{rank}\left(\Omega\right) \le \mbox{rank}\left(B\right)  .
$$
So, in particular, if the matrix $B$ is low-rank, then the matrix $\left(\hat{A}^{T}\hat{A}\right)^{-1}A^TA$ is a low-rank perturbation of the identity $I$.

The important consequence of this is that a symmetric rank-$k$ perturbation of the identity $I$ has $\le k$ non-unit eigenvalues.
In exact arithmetic, this is sufficient to guarantee the convergence in $k$ iterations of several Krylov methods.
So, in particular, for the LS problem, the Cholesky factor of $\hat{A}^{T}\hat{A}$, which is the $R$  matrix from the QR decomposition of $\hat{A}$, is a good LS preconditioner for $A$.

The same analysis extends to other types of perturbations, e.g., to the case when the perturbation is such that rows of $A$ are dropped.
The Avron, Ng, and Toledo paper generalized this to other matrix perturbations.
\begin{itemize}
\item
To when $\hat{A}$ is singular.
\item
To when rows are removed instead of added.
\item
To when columns are exchanged.
\item
To preconditioners for $\hat{A}$ other than the $R$ factor.
\end{itemize}
They also bound the size of the non-unit eigenvalues, which is important when $A$ is rank deficient.

Observe that the generalized spectrum of $\left(A^TA,A^TA\right)$ is very simple: the pencil has $\mbox{rank}\left(A\right)$ eigenvalues that are $1$ and the rest are indeterminate.
In light of this, let's describe the spectra of the following perturbed pencils.
\begin{itemize}
\item
$\left(A^TA,A^TA+B^TB-C^TC\right)$
\item
$\left(A^TA,\hat{A}^{T}\hat{A}\right)$, when
$
A = \left( \begin{array}{cc} D & E  
           \end{array}
    \right)  
$
and
$
\hat{A} = \left( \begin{array}{cc} D & F  
           \end{array}
    \right)  
$.
\end{itemize}
The perturbations of $A^TA$ shift some of the eigenvalues of $\left(A^TA,A^TA\right)$.
Let's call the eigenvalues that move away from $1$ \emph{runaway} eigenvalues.
We will analyze the runaway eigenvalues, which govern the convergence of LSQR when a factorization or approximation of a perturbed matrix is used as a preconditioner.

To start simple, let's give a result that bounds the number of runaway eigenvalues (and other aspects of the spectrum) when we add/subtract a symmetric product from a matrix.

\begin{theorem}
Let $A\in\mathbb{R}^{m \times n}$, $B\in\mathbb{R}^{k \times n}$, $C\in\mathbb{R}^{r \times n}$, for some $1 \le k+r \le n$, and define 
$
\chi = \left( \begin{array}{c} B  \\
                             C 
           \end{array}
    \right)  .
$

Then, 
\begin{itemize}
\item
In the pencil $\left(A^TA,A^TA+B^TB-C^TC\right)$, at most $\mbox{rank}\left(\chi\right) \le k+r$ generalized determined eigenvalues may be different than $1$.
\item
If $1$ is not a generalized eigenvalue in the pencil $\left(B^TB,C^TC\right)$, and if $A^TA+B^TB-C^TC$ is full rank, then:
(1) the pencil $\left(A^TA,A^TA+B^TB+C^TC\right)$ does not have indeterminate eigenvectors, and
(2) the multiplicity of eigenvalue $1$ is exactly $\mbox{dim-null}\left(\chi\right) \ge n-k-r$, and
(3) the multiplicity of the zero eigenvalue is exactly $\mbox{dim-null}\left(A\right)$.
\item
The sum pencil $\left(A^TA,A^TA+B^TB\right)$ cannot have an infinite eigenvalue, and all of its eigenvalues are in the interval $[0,1]$.
\end{itemize}
\end{theorem}

Similar results can be obtained for modifying $A$ in other ways, e.g., a set of columns of $A$.
Here is one such result.
To state it, denote the columns of $A$ that are not modified by $D$, and denote the modified columns before and after by $E$ and $F$, respectively.

\begin{theorem}
Let $D\in\mathbb{R}^{m \times n}$, $E\in\mathbb{R}^{m \times k}$, and $F\in\mathbb{R}^{m \times k}$, for some $1 \le k < n$.
In addition, let
\begin{eqnarray*}
A &=& \left( \begin{array}{cc} D & E  
           \end{array}
    \right)  \in\mathbb{R}^{m\times(n+k)}    \\
\hat{A} &=& \left( \begin{array}{cc} D & F  
           \end{array}
    \right)  \in\mathbb{R}^{m\times(n+k)}   .
\end{eqnarray*}
Then in the pencil $\left(A^TA,\hat{A}^{T}\hat{A}\right)$, at least $n-k$ of the generalized finite eigenvalues are equal to $1$.
\end{theorem}

Similarly, 
\begin{itemize}
\item
If a preconditioner $M$ is effective for a matrix $A^TA$, then it is also effective for the perturbed matrices $A^TA+B^TB-C^TC$ and also $\hat{A}^{T}\hat{A}$.
\item
If the rank of the matrices $B$, $C$, $E$, and $F$ is low, then most of the generalized eigenvalues of the perturbed preconditioned system will be bounded by the extreme generalized eigenvalues of the unperturbed preconditioned system.
\item
I.e., the number of runaway eigenvalues is small, but the non-runaway eigenvalues are not necessarily at $1$, and they can move in an interval whose size determines the condition number of the original preconditioned system.
\end{itemize}

\begin{theorem}
Let $A\in\mathbb{R}^{m \times n}$, $B\in\mathbb{R}^{k \times n}$, and $C\in\mathbb{R}^{r \times n}$, for some $1 \le k +r < n$.
Let $M\in\mathbb{R}^{n \times n}$ be SPSD, and assume
\begin{eqnarray*}
\mbox{null}\left(M\right) & \subseteq & \mbox{null}(A^TA)  \\
\mbox{null}\left(M\right) & \subseteq & \mbox{null}(B^TB)  \\
\mbox{null}\left(M\right) & \subseteq & \mbox{null}(C^TC) 
\end{eqnarray*}
Then, if we assume that 
$$
\alpha \le \lambda_1\left(A^TA,M\right) \le \lambda_{\mbox{rank}(M)}\left(A^TA,M\right) \le \beta  ,
$$
then it follows that
\begin{eqnarray*}
\alpha 
   &\le& \lambda_{r+1}\left( A^TA+B^TB-C^TC,M \right) \\
   &\le& \lambda_{\mbox{rank}(M)-k}\left( A^TA+B^TB-C^TC,M \right)  \\
   &\le& \beta.
\end{eqnarray*}
\end{theorem}

\begin{corollary}
Let $A\in\mathbb{R}^{d \times n}$, $B\in\mathbb{R}^{k \times n}$, for $1 \le k \le n$, be full rank matrices, and let $M\in\mathbb{R}^{n \times n}$ be SPSD.
If the eigenvalues of $\left(A^TA,M\right)$ are in the interval $\left(\alpha,\beta\right)$, where $\alpha$ and $\beta$ are numbers, then the $rank(M)-k$ smallest eigenvalues of $\left(A^TA + B^TB,M\right)$ are in the same interval $\left(\alpha,\beta\right)$.
\end{corollary}

Similar results can be stated when columns are modified and for other types of perturbations.

\subsection{Back to our preconditioned RandNLA LS solver}

There are several applications of these ideas for preconditioned LS solvers: drop rows for sparsity; updating (adding rows) and down-dating (drop rows); adding rows to help solve rank deficient problems.
We will describe how they are used in the Blendenpik solver; other RandNLA solvers like LSRN do similar things.

A key aspect of implementations is that they project onto (say) $2d$ rows, rather than (say) $10 d\log(d)/\epsilon$ rows.
In that case, we might loose rank or have other problems; but the theory we just outlined means that we can still obtain a good preconditioner.
Here is an outline of how that happens.

What if we project onto (say) $2d$ rows, so that we don't uniforms the leverage scores, i.e., so that there are still a few bad coherence rows? 
Relatedly, what if the input matrix has only a very few high leverage rows and that we miss them in the random sample?
Then, the sample can (sometimes) still lead to a good preconditioner.
The basic reason for this is that a few rows with large norm may allow a few singular values of the preconditioned system $A\hat{R}^{-1}$ to be very large, but the number of large singular values is bounded by the number of large rows (and those can be dealt with with a few extra iterations of the iterative method).
That is, a few large singular values can cause the condition number of $AR^{-1}$ to be large, and they can even lead to subspace-nonpreservation, leading the worst-case bounds to fail, but they don't much affect the convergence of LSQR.

Here is a basic lemma.

\begin{lemma}
Let $A\in\mathbb{R}^{n \times d}$, with $n \ge d$, and suppose that 
$
A = \left( \begin{array}{c} A_1  \\
                             A_2 
           \end{array}
    \right)  
$, 
where $A_2$ has $\ell \le \min\{n-d,d\}$.
In addition, let $S\in\mathbb{R}^{k\times(n-\ell)}$ be a matrix such that $SA_1$ is full rank, and let its QR decomposition be $SA_1 = \hat{Q}\hat{R}$.
Then, at least $n-\ell$ of the singular values of $A\hat{R}^{-1}$ are in the interval $\left( \sigma_{min}\left( A_1\hat{R}^{-1} \right), \sigma_{max}\left( A\hat{R}^{-1} \right) \right)$.
\end{lemma}
\begin{Proof}
The singular values $\sigma_i\left( A_1\hat{R}^{-1} \right)$ are the square roots of the generalized eigenvalues of 
$$
\left( A_1^TA_1,\left(SA_1\right)^{T}SA_1 \right) ;
$$
and the singular values $\sigma_i\left( A\hat{R}^{-1} \right)$ are the square roots of the generalized eigenvalues of 
$$
\left( A_1^TA_1+A_2^TA_2,\left(SA_1\right)^{T}SA_1 \right) .
$$
The matrix $A^TA = A_1^TA_1+A_2^TA_2$ is a rank $\ell$ perturbation of $A_1^TA_1$.
So, by the corollary above, we know that at least $d-\ell$ eigenvalues of $\left( A^TA,\left(SA_1\right)^{T}SA_1 \right)$ are in the interval between the smallest and largest of the generalized eigenvalues of $\left( A_1^TA_1, \left(SA_1\right)^{T}SA_1 \right)$.
\end{Proof}

So, what this lemma says is that as long as the number of runaway eigenvalues is small, then we can still use it as a good preconditioner.
If we precondition more aggressively, etc., such that we loose rank, then similar ideas apply if we perturb the matrix to make it full rank.
We won't go into the details now, except to say that there are many other variants of this possible, e.g., see the related solver LSRN.

\newpage

\section{%
(10/21/2013):  Additive-error Low-rank Matrix Approximation}

Today, we will shift gears and begin to discuss RandNLA algorithms for low-rank matrix approximation.  
We will start with additive-error low-rank matrix approximation with sampling and projections.
These are of interest historically and since they illustrate several techniques (norm-squared sampling, simple linear algebraic manipulations, the use of matrix perturbation theory, etc.), but they are much coarser than more recent finer bounds that can be obtained.
Importantly, these additive-error bounds can be improved (and we will get to this soon).
Depending on whether one is interesting in random sampling or random projection algorithms, the improvement comes either in the algorithm or in the analysis.
Understanding this improvement will lead to a structural condition that extends the structural conditions for rectangular least squares problems to one for general ``fat'' matrices.
As we will see, this condition underlies many of the theoretically and/or practically most interesting RandNLA algorithms for low-rank approximation.

Here is the reading for today and the next class.
\begin{itemize}
\item
Drineas, Kannan, and Mahoney, ``Fast Monte Carlo Algorithms for Matrices II: Computing Low-Rank Approximations to a Matrix'' 
\item
Deshpande and Vempala, ``Adaptive Sampling and Fast Low-rank Matrix Approximation''
\end{itemize}

Today, in particular, we will cover the following topics.
\begin{itemize}
\item
Basics of low-rank matrix approximation.
\item
Two simple matrix perturbation theory results.
\item
An overview of RandNLA methods for low-rank approximation.
\item
A basic random sampling algorithm and a quality-of-approximation result for it.
\end{itemize}

\subsection{Basics of Low-rank Matrix Approximation}

Since we are going to shift gears now and talk about how to use randomized algorithms to compute low-rank approximation of matrices, we will start with a brief overview of low-rank matrix approximation.
Hopefully, this should just be a review to set notation, and in fact we have already covered some of these topics in our discussion of regression for very rectangular matrices, but we describe it here in detail since some of the details are different for matrices where both the number of rows and the number of columns are very large and of comparable size.

(BTW, this has mattered less in TCS and ML, where one is typically interested in quality-of-approximation metrics that depend only on how well one reproduces the top part of the spectrum, but it has mattered more in areas such as NLA and scientific computing, where one is also interested in controlling how the top and bottom parts of the singular subspace of the matrix versus approximated matrix interact.)

If $A \in \mathbb{R}^{m \times n}$, then there exist orthogonal matrices $ U=[u^{1} u^{2} \ldots u^{m}]\in \mathbb{R}^{m \times m} $ and $ V=[v^{1} v^{2} \ldots v^{n}]\in \mathbb{R}^{n \times n} $ where $\left\{u^{t}\right \}_{t=1}^m \in \mathbb{R}^m$ and $\left\{v^{t}\right \}_{t=1}^n \in \mathbb{R}^n$ are such that
\begin{equation}
U^TAV=\Sigma=\mbox{Diag}(\sigma_1,\ldots,\sigma_\rho)   ,
\end{equation}
where $\Sigma \in \mathbb{R}^{m \times n}$, $\rho=\min\{m,n\}$ and $\sigma_1 \geq \sigma_2 \geq \ldots \geq \sigma_\rho \geq 0$.  
Equivalently, 
$$
A =U{\Sigma}V^T   .
$$  
The three matrices $U$, $V$, and $\Sigma$ constitute the Singular Value Decomposition (SVD) of $A$.
The $\sigma_i$ are the singular values of $A$ and the vectors $u^{i}$, $v^{i}$ are the $i$-th left and the $i$-th right singular vectors, respectively.  
The columns of $U$ and $V$ satisfy the relations $ Av^{i} = \sigma_iu^{i} $ and $ A^Tu^{i} = \sigma_i v^{i} $.  
For symmetric positive definite (or semi-definite) matrices the left and right singular vectors are the same. 
The singular values of $A$ are the non-negative square roots of the eigenvalues of $A^TA$ and of $AA^T$.  
Furthermore, the columns of $U$, i.e., the left singular vectors, are eigenvectors of $AA^T$ and the columns of $V$, i.e., the right singular vectors, are eigenvectors of $A^TA$.

The SVD can reveal important information about the structure of a matrix.
If we define $r$ by
$ \sigma_1\geq \sigma_2\geq \ldots\geq \sigma_r>\sigma_{r+1}=\ldots=\sigma_{\rho}=0 $,
then $\mbox{rank}(A)=r$,
$\mbox{null}(A)=\mbox{span}\{v^{r+1},\ldots,v^{\rho}\}$, and
$\mbox{range}(A)=\mbox{span}\{u^{1},\ldots,u^{r}\}$.
In particular, if we let $U_r \in \mathbb{R}^{m \times r}$ denote the matrix consisting of the first $r$ columns of $U$, $V_r \in \mathbb{R}^{r \times n}$ denote the matrix consisting of the first $r$ columns of $V$, and $\Sigma_r \in \mathbb{R}^{r \times r}$ denote the principal $r\times r$ sub-matrix of $\Sigma$, then
\begin{equation}
A = U_r\Sigma_r V_r^T=\sum_{t=1}^r \sigma_t u^{t}{v^{t}}^T   . 
\label{svdeq} 
\end{equation}

(BTW, this is the usual linear algebraic notion of rank.  Note, however, that one can also define other notions of ``soft rank,'' sometimes called the ``effective rank'' and typically defined as the ratio of the Frobenius to spectral norm, both defined below, and this is sometimes of greater interest in machine learning and data analysis applications.  We won't explicitly cover that much, but we note that many of the analysis tools we do discuss are also useful more or less directly for dealing with this softer notion of rank.)

Note that this dyadic decomposition property given in Eqn.~(\ref{svdeq}) provides a canonical description of a matrix as a sum of $r$ rank one matrices of decreasing importance.
If $k \leq r$ and we define
\begin{equation}
\label{rank_k_approx}
A_k = U_k\Sigma_k V_k^T=\sum_{t=1}^k \sigma_t u^{t}{v^{t}}^T,
\end{equation}
then $A_k = U_kU_k^TA =\left(\sum_{t=1}^k u^{t}{u^{t}}^T\right)A$ and $A_k = AV_kV_k^T = A\left(\sum_{t=1}^k v^{t}{v^{t}}^T\right)$, i.e., $A_k$ is the projection of $A$ onto the space spanned by the top $k$ singular vectors of $A$.  
Furthermore, the distance (as measured by both $\TNorm{\cdot}$ and $\FNorm{\cdot}$) between $A$ and any rank $k$ approximation to $A$ is minimized by $A_k$, i.e.,
\begin{equation}
\label{Best_2_Norm_Approx} 
\min_{D \in \mathbb{R}^{m \times n}:\mbox{rank}(D)\leq k}\TNorm{A-D}  =  \TNorm{A-A_k} = \sigma_{k+1}(A)               
\end{equation}
and 
\begin{equation}
\label{Best_F_Norm_Approx} 
\min_{D \in \mathbb{R}^{m \times n}:\mbox{rank}(D)\leq k}\FNormS{A-D} = \FNormS{A-A_k} = \sum_{t=k+1}^r\sigma_{t}^2(A)   .
\end{equation}
Thus, $A_k$ constructed from the $k$ largest singular triplets of A is the optimal rank $k$ approximation to $A$ with respect to both $\| \cdot \|_F$ and $\| \cdot \|_2$.
(It is actually optimal with respect to the more general class of unitarily-invariant norms.)
More generally, one can also show that $ \TNorm{A} = \sigma_1 $ and that$ \FNormS{A} = \sum_{i=1}^r \sigma_i^2 $.

We reviewed the above since our main m.o. will be that we will project onto a set of random columns (or random linear combinations of columns) that is not the optimal set of columns, and we will show that the error incurred is not much worse than the optimal set of columns.

Finally, let's conclude this linear algebra review of general fat matrices with a brief word about the running time of computing the SVD and of low-rank approximations to the SVD.
It might seem like this should be a simple question with a simple answer, but it is actually a rather complicated topic, and the answer depends on your background and your perspective as to what's important to count and what is acceptable to be ignored.
Here are the key points for us to keep in mind.
\begin{itemize}
\item
One can compute the full SVD of a general matrix $A$, in infinite precision arithmetic, with various direct methods, in $\Theta\left(n^3\right)$ time (or in $\min \{ mn^2,m^2n \} $ time, if $A$ is an $m \times n$ matrix) operations in the RAM model (which is actually \emph{not} such a good model here).
One can compute a rank-$k$ low-rank approximation to $A$ in that same $\Theta\left(n^3\right)$ time by computing the full SVD and keeping only those top $k$ components of interest.
\item
Of course, that naive approach throws away a lot, but as an inequality the running time is $O\left(n^3\right)$ time.
Alternatively, one must read the full input, and so the running time is $\Omega\left(n^2\right)$ for general matrices and $\Omega\left( \mbox{nnz}(A) \right)$, where $\mbox{nnz}(A)$ is the number of non-zero entries in the matrix $A$, for sparse matrices.
\item
Even aside from roundoff issues, this $O\left(n^3\right)$ running time provides a straw-man comparison, in the sense that one almost never needs the full SVD.
Indeed, one almost never needs the full rank-$k$ approximation to the SVD.
Instead one typically only computes what one needs, e.g., a basis for the left or right singular subspace, the top $k$ singular values, all or some of the diagonal elements for a projection matrix onto the span of $A$, an orthogonal matrix spanning a subspace close to the top $k$ left or right singular subspace, etc.
\item
In these cases, some sort of iterative algorithm is typically used, and the running time of these iterative algorithms depends on lots of details about the matrix, the implementations, whether the matrix is represented explicitly or implicitly, whether communication is expensive, etc.
(You got a feeling for some of these things when we discussed Blendenpik, but the situation is much more complex for general matrices.)
\item
We won't dwell on these issues too much, since we aren't focused primarily on implementations and since different disciplines say very different things about this.
For the development of RandNLA algorithms (and matrix algorithms in machine learning and data analysis more generally) it is at least as important to understand some of the details/subtleties of what different research areas say is important for running time as it is to know any simple statement of running time.
\item
All that being said and as a rule of thumb (and ignoring things like condition number and other related issues), think of the running time as being roughly $\Theta\left(mnk\right)$, where the matrix $A$ is of size $m \times n$ and where one is interested in a rank $k$ approximation.
Informally, you have to touch each entry once, and you have to touch each dimension once.
Also, roughly, if the matrix is sparse, then the $mn$ can be replaced with $\mbox{nnz}(A)$.
(We will see that, roughly, randomized algorithms can be used to improve the $k$ to $\log(k)$, but they also have many other benefits---robustness, parallelizability, simplicity, etc.---that are at least as important as the running time improvements.)
\end{itemize}

\subsection{Two Results from Matrix Perturbation Theory}

Matrix perturbation theory has to do with how properties of a matrix such as its spectrum change (or are perturbed) as the elements of the matrix are varied.
It is a large area, with many varied applications, and we will need only a very small part of it.
In particular, from the perturbation theory of matrices it is known that the size of the difference between two matrices can be used to bound the difference between the singular value spectrum of the two matrices.  
More precisely, if $A,E \in \mathbb{R}^{m \times n}, m \ge n$, then: 
\begin{equation}
\max_{t: 1 \le t \le n}|\sigma_{t}(A+E)-\sigma_t(A)|\leq \TNorm{E}
\label{eqn_GVL8.6.2}
\end{equation}
and
\begin{equation}
\sum_{k=1}^{n}( \sigma_k(A+E) - \sigma_k(A) )^{2} \le \FNormS{E}   .
\label{eqn_hoffman_wielandt}
\end{equation}
The latter inequality is known as the Hoffman-Wielandt inequality.

Note that both of these results are of the form of the $\ell_p$ norm of the difference of the singular values of a matrix and a perturbed version of the matrix is bounded above by a matrix norm that equals the $\ell_p$ norm of the singular values of the perturbation.
(These two results are for $p=\infty$ and $p=2$, respectively.)

Note also that neither of these bounds depends on the structure of the perturbation.
Given the large fraction of RandNLA algorithms that boil down to matrix perturbation results such as these two results, and given that the perturbations from random sampling or random projection that RandNLA algorithms perform are quite structured, it is of interest to see if one can get finer bounds by taking advantage of the structured form of the perturbation.

\subsection{Randomization for Low-rank Matrix Approximation}

In the randomized algorithms for low-rank approximation that we will discuss, there will be several ``knobs,'' and the details of how these knobs are handled are important.
Not only will those details make a big difference for how successful various algorithms are in theory and/or in practice, but---if not given appropriate attention---those same details can be the source of a great deal of confusion about how different algorithms related to each other.

The reason for this latter comment is that different research communities find it more or less natural/convenient to fiddle with different knobs in different ways, and (relatedly) different research communities find it more or less natural/convenient to ask for different types of quality-of-approximation guarantees.
This has led to a confusing array of algorithms, which are often superficially quite different, but which in reality have very strong (algorithmic or statistical or structural) similarities and connections.
In the next few weeks, we will try to focus on these commonalities, trying to highlight structural properties responsible for seemingly-different results.

As with our description of algorithms for least-square approximation, we will discuss algorithms for low-rank matrix approximation first in terms of basic random sampling algorithms, and then in terms of extensions to random projection algorithms.

Here are examples of different perspectives that people adopt on these algorithms.
\begin{itemize}
\item
\textbf{Additive-error bounds versus relative-error bounds.}
Let's say that we would like to quantify how well a matrix $C$, or perhaps the best rank $k$ approximation to $C$ if $C$ has more than $k$ columns, captures the top part of the spectrum of a matrix $A$, and (for now) let's say that we are interested in the error with respect to the Frobenius norm.
One type of bound one could hope for is to show that
\begin{equation}
\FNorm{A-P_{C_k}A} \le \FNorm{A-P_{U_k}A } + \epsilon \FNorm{A}  .
\label{eqn:sample-add-err-F}
\end{equation}
In the theory of algorithms, bounds of the form (\ref{eqn:sample-add-err-F}) are known as \emph{additive-error bounds}, the reason being that the ``additional'' error (above and beyond that incurred by the SVD) is bounded above by an additive factor of the form $\epsilon$ times the scale $\FNorm{A}$.

Bounds of this form are very different and in general weaker than when the additional error enters as a multiplicative factor, such as when the error bounds are of the form 
$$
 ||A-P_{C_k}A|| \le f(m,n,k,\eta) ||A-P_{U_k}A ||   , 
$$
where $f(\cdot)$ is some function and $\eta$ represents other parameters of the problem.
Bounds of this type are of greatest interest when $f(\cdot)$ does not depend on $m$ or $n$, in which case they are known as a \emph{constant-factor bounds}, or when they depend on $m$ and $n$ only weakly.
The strongest bounds are when $f = 1+\epsilon$, for an error parameter $\epsilon$, \emph{i.e.}, when the bounds are of the form
\begin{equation}
\FNorm{A-P_{C_k}A} \le (1+\epsilon) \FNorm{A-P_{U_k}A}  .
\label{eqn:sample-rel-err-F}
\end{equation}
These \emph{relative-error bounds} are the gold standard (in TCS, but not necessarily in other areas), since the scale of the additive error becomes the residual error itself, and they provide a \emph{much} stronger notion of approximation than additive-error or weaker multiplicative-error bounds.
\item
\textbf{Other notions of reconstruction quality.}
Eqn.~(\ref{eqn:sample-add-err-F}) and Eqn.~(\ref{eqn:sample-rel-err-F}) measure in two different ways how much of $A$ is captured by the sample $C$, and they do so by measuring a particular norm (the Frobenius norm) of the difference between two matrices.
Of course, one might be interested in other norms, e.g., the spectral norm, which is the largest singular value and thus the $\ell_{\infty}$ norm of the vector of singular values, or the trace/nuclear norm, which is the sum of the singular values and thus the $\ell_1$ norm on the vector of singular values.
(The Frobenius norm is the $\ell_2$ or Euclidean norm on the singular value vector.)
There are still other norms of interest.
Alternatively, one might want to measure the quality in some other way, e.g., with respect to a divergence or whatever.
\item
\textbf{Reconstructing the matrix versus other notions of approximation quality.}
Eqn.~(\ref{eqn:sample-add-err-F}) and Eqn.~(\ref{eqn:sample-rel-err-F}) make statements about how well $C$ captures the information in the top part of this spectrum of $A$.
This is reasonable, but sometimes one is also interested in other things.
(For example, we saw this before in $\ell_2$ regression, where we wanted not just relative error on the objective function value, but we also wanted to say that the actual solution vectors were close.)
In the case of low-rank approximation via $U_k$, the matrix consisting of the top $k$ left singular vectors of $A$, in addition to capturing the maximum amount of $A$ with respect to any unitarily-invariant matrix norm, $U_k$ is also ``good'' for other reasons: the columns of $U_k$ are orthogonal to each other and thus maximally ``spread out,'' the matrix $U_k$ is exactly orthogonal to the matrix $U_{k,\perp}$, where the latter is the matrix consisting of the bottom $m-k$ singular vectors, and thus the approximation is maximally far from the optimal residual subspace, etc.

In more general low-rank matrix approximation methods, these considerations manifest themselves by, e.g., asking for an interpolative decomposition, where the condition number of the sample matrix $C$ is relatively good, asking for a rank-revealing decomposition, where one shows that the singular values of the part of the matrix that is not captured are not too large, that there is not much overlap between the sample and the bottom $m-k$ singular directions of $A$, etc.
Importantly, while these notions are all vaguely related and in many cases coincide when considering the exactly optimal SVD-based low-rank approximation, approximately optimizing one of them often says very little about exactly or approximately optimizing another one of them.
Depending on the downstream application, one or the other of these objectives might be of greatest interest.

These observations hold in general, e.g., with deterministic algorithms like rank-revealing QR decompositions, but we will be mostly interested in how they hold for randomized algorithms via using random sampling or random projections.
We will discuss how to deal with some of these issues with RandNLA algorithms.
As we will see, in some cases this difference manifests itself in the algorithm, while in other cases it manifests itself in the analysis.
\item
\textbf{Sampling versus projection.}
As we saw before, roughly, random projections correspond to uniform sampling in randomly rotated spaces.  
The same holds true, again at one level of granularity, in the case of randomized algorithms for low-rank approximation.
Indeed, that perspective is often a helpful way to think about similarities between seemingly-different problems and algorithms, and so we will emphasize that perspective.
But, if we get greedy in optimizing various factors (e.g., as needs to be done with the oversampling parameter when providing high-quality implementations), then sometimes it is better to do it directly and not view it as this two step process.
Alternatively, it is sometimes convenient (e.g., in scientific computing) to view a random projection as providing an estimator for the range space of a matrix; and it is sometimes convenient (e.g., in TCS) to view a random projection as providing a data-oblivious or data-agnostic subspace embedding.
Whether projections directly ``boil down'' to sampling uniformly in a randomly-rotated space or do so only indirectly and approximately, it is helpful to think of sampling and projections on a similar footing and providing two different types of randomized ``sketching'' matrices.
\item
\textbf{More aggressive downsampling.}
Sometimes, we are interested in sampling fewer than roughly $O(k \log k /\epsilon^2)$ or even $O(k/\epsilon)$ columns.
For example, we might want to sample \emph{exactly} $k$ columns, or we might want to project onto \emph{exactly} $k+p$ columns, where $p$ is a small integer like $5$ or $10$.
In these cases, the simplest analysis typically fails for worst-case matrices, and this is basically for coupon collector reasons, but this might be ok if there is a quick way to check whether some property has been satisfied.
This is sometimes of interest by themselves and sometimes for their numerical properties (as we saw with LS), and we need to control different structures---consider CSSP and ``slow'' extensions to projections.
\item
\textbf{Reproducing the data versus reproducing hypothesized data.}
Eqn.~(\ref{eqn:sample-add-err-F}) and Eqn.~(\ref{eqn:sample-rel-err-F}) are statements about how well an algorithm does with respect to the data sitting in front of us.
This is a very natural thing to ask for in NLA and TCS.
Alternatively, one might be interested in how well the algorithm does with respect to hypothesized but unseen data.
This latter approach is more natural in statistics and machine learning.
The MSE, the usefulness of the low-rank approximation in a prediction task such as kernel ridge regression, etc., are all examples of such a metrics, and there are many others.
\end{itemize}

\subsection{The \textsc{LinearTimeSVD} Algorithm}

We will start with a very simple random sampling algorithm.
Given an $m \times n$ matrix $A$, we wish to choose columns of $A$ such that the projection of the matrix onto those columns ``captures'' as much of the matrix as possible, i.e., that is a basis for a space close to the space spanned by the top singular vectors of the matrix.
Thus, in particular, if $A$ is well approximated by a low-rank matrix, then we would like $A \approx P_{span(C)}A$, where $C$ is a matrix consisting of the chosen columns and where $P_{span(C)}$ is a projection onto the column space of $C$.
To this end, the \textsc{LinearTimeSVD} algorithm randomly samples a small number of columns from an input matrix, and it returns an approximation to the singular values and left singular vectors of that matrix.
This algorithm is somewhat too simple to have found widespread use in practice (for reasons we will discuss), but it is important historically, and it is pedagogically convenient since its analysis will illustrate several important concepts.

\begin{algorithm}
\caption{The \textsc{LinearTimeSVD} Algorithm. }
\label{alg:linear-time-svd-alg}
\begin{algorithmic}[1]
    \REQUIRE An $m \times n$ matrix $A$, integers $c, k$ s.t.  
             $1 \le k \le c \le n$, and a probability distribution
             $\left\{ p_i \right\}_{i=1}^{n}$.
    \ENSURE  An $m \times k$ orthogonal matrix $H_k$ and numbers 
             $\sigma_t(C), t = 1,\ldots,k$.
    \FOR{$t = 1$ to $c$}
       \STATE Pick $i_t \in {1,\ldots,n}$ with 
              $\Probab{i_t = \alpha} = p_\alpha$, $\alpha=1,\ldots,n$.
       \STATE Set $C^{(t)} = A^{(i_t)}/\sqrt{c p_{i_t}} $.
    \ENDFOR
    \STATE Compute $C^T C$ and its singular value decomposition; say 
           $C^T C = \sum_{t=1}^{c} \sigma_t^2(C) y^t {y^t}^T$.
    \STATE Compute $h^t = C y^t / \sigma_t(C) $ for $t = 1,\ldots,k$.
    \STATE Return $H_k$, where $H_k^{(t)} = h^t$, and 
           $\sigma_t(C), t = 1,\ldots,k$.
\end{algorithmic}
\end{algorithm}

We have formulated this algorithm to say that it returns the top $k$ left singular vectors of $C$, but we could have just returned the matrix $C$.
(By that, we mean that the quality-of-approximation and running time claims that we discuss today and next time work for both $C$ and $C_k$.
If we were interested in different objectives, like we just discussed above, then the difference between $C$ and $C_k$ can become important.
We will revisit this issue later with other low-rank approximation algorithms.)

The point here is that we want to say that the matrix $C$ is in some sense close to the matrix $A$. 
It is not immediately obvious how to make that comparison, however, given that the two matrices have different dimensions.
Note, though, that the ambient dimensionality of the range space of both matrices is the same, i.e., $\mathbb{R}^{m}$, and so we will say that they are similar if their left singular subspaces are similar, or relatedly if $CC^T \approx AA^T$.
(Other notions of similarity are certainly possible, but this notion says that the two matrices have a similar correlational structure on the non-sampled dimension, and one important aspect of this notion is that we can relate it back to approximation algorithms for the matrix multiplication primitive.)

Before presenting our quality-of-approximation results, here is a summary of the running time of this algorithm.
\begin{itemize}
\item
If we work with probabilities that are approximately proportional to the squared Euclidean norms of the columns of $A$, as in Eqn.~(\ref{eqn:norm-squared}) below, then one pass and $O(c)$, where $c$ is the number of random samples to be drawn, i.e., the number of independent counters that are run in parallel in the pass efficient model, additional space and time are needed to choose the indices of the columns to choose.
(Of course, if we work with uniform sampling probabilities, then we can choose the columns to be sampled without even looking at the data and store the indices of those columns in $O(c)$ additional space.)
\item
Given the indices of the columns to be sampled, then one additional pass and $O(mc)$ additional space and time is needed to select the columns from $A$ and construct the matrix $C$.
\item
Given the matrix $C$, then computing $C^TC$ requires $O(mc^2)$ additional space and time, and computing the SVD of $C^TC$ requires $O(c^3)$ additional space and time.
\item
Given the SVD of $C^TC$, then computing $H_k$ requires $k$ matrix-vector multiplications, for a total of $O(mck)$ additional space and time.
\item
So, on the whole, if $c,k = O(1)$, then $O(m)$ additional space and time are needed.
That is, the \textsc{LinearTimeSVD} algorithm has additional running space and time that is linear in the dimensionality of the data/features and not in the size or number of non-zeros of the matrix.
\end{itemize}

Next, we will be interested in establishing quality-of-approximation results for this algorithm.
To separate clearly the effect of linear algebraic structure from the effect of randomization on the quality-of-approximation claims, we will first do this for general sampling probabilities, and we will then specialize the result to the case that the probabilities depend on the Euclidean norms squared of the columns of $A$.
In the former case, the additional error, above and beyond that incurred by the best rank-$k$ approximation, will depend on $||AA^T-CC^T||_{\xi}$, for $\xi=\{2,F\}$.
Then, we will call our previous matrix multiplication results to bound that additional error.
So, the choice of columns will enter only in the form of an approximate matrix multiplication bound.

Let's start with what we can establish for the Frobenius norm error.

\begin{theorem}
\label{thm:LinearSVDFNorm}
Suppose $A \in \mathbb{R}^{m \times n}$ and let $H_k$ be constructed from the \textsc{LinearTimeSVD} algorithm.
Then,
$$
\FNormS{A-H_kH_k^TA} \le \FNormS{A-A_k} + 2 \sqrt{k} \FNorm{AA^T-CC^T}   .
$$
\end{theorem}

\begin{Proof}
Recall that for matrices $X$ and $Y$, $\FNormS{X}=\Trace{X^TX}$, 
$\Trace{X+Y}=\Trace{X}+\Trace{Y}$, and also that $H_k^TH_k=I_k$.  
Thus, we may express $\FNormS{A-H_kH_k^TA}$ as:
\begin{eqnarray}
\nonumber      \FNormS{A-H_kH_k^TA} &=& \Trace{(A-H_kH_k^TA)^T(A-H_kH_k^TA)}            \\
\nonumber                           &=& \Trace{A^TA-2A^TH_kH_k^TA+A^TH_kH_k^TH_kH_k^TA} \\
\nonumber                           &=& \Trace{A^TA} - \Trace{A^TH_kH_k^TA}             \\
\label{eqn_aaa}                     &=& \FNormS{A} - \FNormS{A^TH_k}    .
\end{eqnarray}
We may relate $ \FNormS{A^TH_k} $ and $ \sum_{t=1}^{k} \sigma_t^2(C) $ by the 
following:
\begin{eqnarray}
\nonumber     \left| \FNormS{A^TH_k} - \sum_{t=1}^{k} \sigma_t^2(C) \right|
                 &\le& \sqrt{k} \Sqrt{\sum_{t=1}^{k} \left( \VTTNormS{A^Th^t} - \sigma_t^2(C)    \right)^2 } \\
\nonumber        &=&   \sqrt{k} \Sqrt{\sum_{t=1}^{k} \left( \VTTNormS{A^Th^t} - \VTTNormS{C^Th^t} \right)^2 } \\
\nonumber        &=&   \sqrt{k} \Sqrt{\sum_{t=1}^{k} \left( {h^t}^{T}(AA^T-CC^T)h^t               \right)^2 } \\
\label{eqn_ccc}  &\le& \sqrt{k} \FNorm{AA^T-CC^T}   .
\end{eqnarray}
The first inequality follows by applying the Cauchy-Schwartz inequality;
the last inequality follows by writing $AA^T$ and $CC^T$ with respect to a 
basis containing $\{h^t\}_{t=1}^{k}$.
By again applying the Cauchy-Schwartz inequality, noting that 
$\sigma_t^2(X) = \sigma_t(XX^T)$ for a matrix $X$, and applying the 
Hoffman-Wielandt inequality, (\ref{eqn_hoffman_wielandt}), we may also relate 
$\sum_{k=1}^{k}\sigma_t^2(C)$ and $\sum_{k=1}^{k}\sigma_t^2(A)$ by the 
following:
\begin{eqnarray}
\nonumber      \left| \sum_{t=1}^{k} \sigma_t^2(C) - \sum_{t=1}^{k} \sigma_t^2(A) \right|
                  &\leq& \sqrt{k} \Sqrt{\sum_{t=1}^{k} \left( \sigma_t^2(C)  - \sigma_t^2(A)  \right)^2 } \\
\nonumber         &=&    \sqrt{k} \Sqrt{\sum_{t=1}^{k} \left( \sigma_t(CC^T) - \sigma_t(AA^T) \right)^2 } \\  
\nonumber         &\leq& \sqrt{k} \Sqrt{\sum_{t=1}^{m} \left( \sigma_t(CC^T) - \sigma_t(AA^T) \right)^2 } \\
\label{eqn_eeee}  &\le&  \sqrt{k} \FNorm{CC^T-AA^T}   .
\end{eqnarray}
Combining the results of (\ref{eqn_ccc}) and (\ref{eqn_eeee}) allows us to 
relate $ \FNormS{A^TH_k} $ and $ \sum_{t=1}^{k} \sigma_t^2(A) $ 
by the following:
\begin{equation}
\| \FNormS{A^TH_k} - \sum_{t=1}^{k} \sigma_t^2(A) \| 
   \le 2 \sqrt{k} \FNorm{AA^T-CC^T}   .
\label{eqn_fff}
\end{equation}
Combining (\ref{eqn_fff}) with (\ref{eqn_aaa}) yields the theorem.
\end{Proof}

Let's conclude today with several observations about this theorem.
\begin{itemize}
\item
This theorem says that the error in any low-rank approximation, above and beyond that provided by the best rank-$k$ approximation, can be related to an error in approximating the product of two matrices.
Thus, if we can make that matrix multiplication error small, then we get a good low-rank matrix approximation.
\item
In particular, if we use probabilities $\left\{ p_i \right\}_{i=1}^{n}$ that are close to the Euclidean norms squared of $A$, in the sense that 
\begin{equation}
\label{eqn:norm-squared}  
p_i \ge \beta \frac{ \VTTNormS{A^{(i)}} }{ \FNormS{A} }  ,
\end{equation}
for some positive $\beta \le 1$ (e.g., just set $\beta=1$ and use the Euclidean norms squared of the columns of $A$), then one has that worst-case additive-error bounds of the form 
\begin{equation}
\label{eqn:additive-frob}
\FNormS{A-H_kH_k^TA} \le \FNormS{A-A_k} + \epsilon \FNormS{A}
\end{equation}
hold in expectation and with high probability, if one chooses $c \gtrsim k/\epsilon^2$ in the algorithm.
\item
If the Euclidean norms of the columns of $A$ are approximately uniform, then uniform sampling is approximately optimal, in the sense that the probabilities $p_i = 1/n$ are close to the probabilities $p_i =  \VTTNormS{A^{(i)}} / \FNormS{A}$, e.g., in the sense that $\beta$ is not too small (and thus $1/\beta$ is not too large).
Then, with an appropriately small choice of $\beta$ that can be absorbed into the sampling complexity, one can get bounds of the form Eqn.~(\ref{eqn:additive-frob}) with $c \gtrsim k/\beta\epsilon^2$ (for a value of $\beta$ which doesn't make this expression too large). 
\item
On the other hand, if the uniform sampling probabilities are very different than the norm-squared probabilities, then one must choose $\beta$ to be very small (and thus $1/\beta$ to be very large) to get bounds of the form Eqn.~(\ref{eqn:additive-frob}).
For example, $c \gtrsim k/\beta\epsilon^2$---where the value of $\beta$ makes this very large.
In particular, this can be a very large number of uniformly-sampled columns if $\beta$ depends on $n$ (in theory) of if, say, $\beta = 1/1000$ (in practice).
\item
Alternatively, one can sample $c \gtrsim k/\epsilon^2$ columns uniformly, i.e., where $c$ has no $\beta$ dependency, and obtain bounds of the form
\begin{equation}
\label{eqn:additive-frob-bad}
\FNormS{A-H_kH_k^TA} \le \FNormS{A-A_k} + \epsilon n^2  .
\end{equation}
This is also an additive-error bound, but the scale of the additive error is \emph{much} worse that before---so much worse, in fact, that bounds with additive scale factor don't even provide a qualitative guide to the practical performance of the algorithm.
\end{itemize}

\newpage

\section{%
(10/23/2013):  Additive-error Low-rank Matrix Approximation, Cont.}

We continue with the discussion from last time.
There is no new reading, just the same as last~class.

Today, in particular, we will cover the following topics.
\begin{itemize}
\item
A spectral norm bound for reconstruction error for the basic low-rank approximation random sampling algorithm.
\item
A discussion of how similar bounds can be obtained with a variety of random projection algorithms.
\item
A discussion of possible ways to improve the basic additive error bounds.
\item
An iterative algorithm that leads to additive error with much smaller additive scale.
This will involve using the additive error sampling algorithm in an iterative manner in order to drive down the additive error quickly as a function of the number of iterations.
\end{itemize}

\subsection{Reconstruction Error for Low-rank Approximation, Cont.}

Recall what we did last time: we introduced the \textsc{LinearTimeSVD} algorithm, and we proved a result that characterized the Frobenius norm error in terms of an approximation of the product of two matrices.
Let's now provide a similar spectral norm error bound and show how both results can be used with the right sampling probabilities to get additive error bounds.

We start with the following result, which characterizes the reconstruction error with respect to the spectral norm.
Note that the factor $\sqrt{k}$ that we had before is not present.

\begin{theorem}
\label{thm:LinearSVDTNorm}
Suppose $A \in \mathbb{R}^{m \times n}$ and let $H_k$ be constructed from the \textsc{LinearTimeSVD} algorithm.
Then,
$$
\TNormS{A-H_kH_k^TA} \le \TNormS{A-A_k} + 2          \TNorm{AA^T-CC^T}   .
$$
\end{theorem}

\begin{Proof}
Let $ \mathcal{H}_k = \mbox{range}(H_k) = \mbox{span}\{h^1, ..., h^k\}$ and 
$\mathcal{H}_{m-k}$ be the orthogonal complement of $\mathcal{H}_k$.
Let $x \in \mathbb{R}^{m}$ and let $x = \alpha y +\beta z$ where 
$y \in \mathcal{H}_k$, $z \in \mathcal{H}_{m-k}$, and $\alpha^2 + \beta^2 = 1$; 
then, 
\begin{eqnarray}
\nonumber  \TNorm{A-H_kH_k^TA} &=& \max_{x \in \mathbb{R}^{m}, \|x\| = 1}\left\|x^T(A-H_kH_k^TA)\right\|                             \\
\nonumber                      &=&    \max_{y \in \mathcal{H}_k, \|y\|=1, z \in \mathcal{H}_{m-k}, \|z\|=1, \alpha^2 + \beta^2 = 1}
                                                                            \left\|(\alpha y^T + \beta z^T)(A-H_kH_k^TA)\right\|   \\
\label{eqn_hhe}                &\leq& \max_{y \in \mathcal{H}_k,     \|y\|=1} \left\|y^T(A-H_kH_k^T A)\right\|         
                                    + \max_{z \in \mathcal{H}_{m-k}, \|z\|=1} \left\|z^T(A-H_kH_k^T A)\right\|                       \\
\label{eqn_hhf}                &=& \max_{z \in \mathcal{H}_{m-k}, \|z\|=1} \left\|z^T A \right\|   .
\end{eqnarray}
(\ref{eqn_hhe}) follows since $\alpha, \beta \leq 1$ and (\ref{eqn_hhf}) follows 
since $y \in \mathcal{H}_k$ and $z \in \mathcal{H}_{m-k}$.
We next bound (\ref{eqn_hhf}):
\begin{eqnarray}
\nonumber       
   \VTTNormS{z^TA} & = & z^TCC^Tz            + z^T\left(AA^T-CC^T\right)z \\
\label{eqn_ggg}       
                  &\le& \sigma_{k+1}^{2}(C) +   \TNorm{AA^T-CC^T}        \\
\label{eqn_ggh}      
                  &\le& \sigma_{k+1}^{2}(A) + 2 \TNorm{AA^T-CC^T}        \\
\label{eqn_ggi}      
                  &\le& \TNormS{A-A_k}      + 2 \TNorm{AA^T-CC^T}         .
\end{eqnarray}
(\ref{eqn_ggg}) follows since $\max_{z \in \mathcal{H}_{m-k}}\|z^TC\|$ occurs 
when $z$ is the $(k+1)$-st left singular vector, i.e., the maximum possible in 
the $\mathcal{H}_{m-k}$ subspace.
(\ref{eqn_ggh}) follows since $\sigma_{k+1}^{2}(C) = \sigma_{k+1}(CC^T)$ and
since by 
the spectral norm variant of the Hoffman-Wielandt inequality
we have that
$ \sigma_{k+1}^{2}(C) \leq \sigma_{k+1}(AA^T) + \TNorm{AA^T-CC^T} $;
(\ref{eqn_ggi}) follows since $\TNorm{A-A_k}=\sigma_{k+1}(A)$.
The theorem then follows by combining (\ref{eqn_hhf}) and (\ref{eqn_ggi}).
\end{Proof}

This result for the spectral norm error, as well as the result we derived in the last class for the Frobenius norm error, holds for any set of sampling probabilities $\left\{ p_i \right\}_{i=1}^{n}$.  
That is, the choice of sampling probabilities and thus the choice of columns enters the approximation quality bound for $\|A-H_kH_k^TA\|_{\xi}^2$ only via a term $\|AA^T-CC^T\|_\xi$ that is of the form of an approximate matrix multiplication product. 

For completeness, we state the following theorem, in which we specialize the sampling probabilities to be those that are nearly optimal.
By choosing enough columns, we obtain an additive-error low rank approximation to the matrix $A$, and the additional error in the approximation of the SVD can be made arbitrarily small.  

\begin{theorem}
\label{thm:LinearSVD_corr}
Suppose $A \in \mathbb{R}^{m \times n}$, let $H_k$ be constructed from the 
\textsc{LinearTimeSVD} algorithm by sampling $c$ columns of $A$ with 
probabilities $\left\{ p_i \right\}_{i=1}^{n}$ such that
$p_i \ge \beta \VTTNormS{A^{(i)}} / \FNormS{A} $ for some positive 
$\beta \le 1$, and let $\eta = 1 + \sqrt{(8/\beta) \log(1/\delta)}$.
Let $\epsilon > 0$.
If $c \ge 4k/\beta\epsilon^2$, then 
\begin{equation}
\Expect{ \FNormS{A-H_kH_k^TA} } \le \FNormS{A-A_k} + \epsilon \FNormS{A}   ,
\label{linear_expect_fnorm_eps}
\end{equation}
and if $c \ge 4k\eta^2/\beta\epsilon^2$ then with probability at least $1 - \delta$ 
\begin{equation}
\FNormS{A-H_kH_k^TA} \le \FNormS{A-A_k} + \epsilon \FNormS{A}   .
\label{linear_whp_fnorm_eps}
\end{equation}
In addition, if $c \ge 4/\beta\epsilon^2 $, then 
\begin{equation}
\Expect{ \TNormS{A-H_kH_k^TA} } \le \TNormS{A-A_k} + \epsilon \FNormS{A}   ,
\label{linear_expect_2norm_eps}
\end{equation}
and if $ c \ge 4\eta^2/\beta\epsilon^2 $ then with probability at least $1 - \delta$ 
\begin{equation}
\TNormS{A-H_kH_k^TA} \le \TNormS{A-A_k} + \epsilon \FNormS{A}   .
\label{linear_whp_2norm_eps}
\end{equation}
\end{theorem}

Note that the norm on the approximate matrix multiplication error in Theorem~\ref{thm:LinearSVDTNorm} and in our theorem from last class is the same as the norm on the low-rank approximation we are interested in.
That is, spectral/Frobenius norm on the matrix multiplication term if we are interested in spectral/Frobenius on the low-rank error.
When specializing to nearly-optimal probabilities, for $\FNorm{\cdot}$, we use our previous bound on $\FNorm{AA^T-CC^T}$.
For $\TNorm{\cdot}$, we use that $\TNorm{\cdot} \le \FNorm{\cdot}$ to get an additive error spectral norm, the scale of which depends on the Frobenius norm of the matrix.
(This is weak in some sense, but we can't expect a relative-error bound for the spectral norm when choosing a small number of columns.)
Alternatively, we note that one could use our previous bound on $\TNorm{AA^T-CC^T}$, which would provide spectral norm bounds, under assumptions on the number of samples and/or parameterized in terms of the stable rank.
We do not do that here, but others have considered it, and it could be of interest, under assumptions on the input matrices.

\subsection{Low-rank Approximation via Random Projections}

So far, we have been mostly discussing \emph{random sampling algorithms} for low-rank matrix approximation. 
To obtain interesting results, these algorithms need to use a data-dependent importance sampling distribution, and then they need to choose parameters such that they achieve the needed measure concentration.
Very similar ideas extend to \emph{random projection algorithms} for low-rank matrix approximation, and we can derive similar bounds by using a wide range of random projection constructions.
Essentially, these use a data-independent transformation that ``rotates'' the input to a random basis where the norm-squared importance sampling distribution is uniformized.

To see this, note that the error for our random sampling algorithm for low-rank matrix approximation depends on $\left\| AA^T-ASS^TA^T \right\|$, where $S$ is a sketching matrix that has a special form that is the ``sampling matrix'' constructed by our LinearTimeSVD algorithm.
But, nowhere in the analysis of our theorems did we use the fact that this sketching matrix had any particular form.
Indeed, we have seen before that we can get similar matrix multiplication bounds by using random projection matrices such as matrices consisting of i.i.d. Gaussian entries, $\{\pm1\}$ entries, structured Hadamard matrices, input-sparsity-time projections, etc.
So, if $S$ is a random projection matrix, with parameters chosen appropriately, then we can get the same additive-error bounds, if we have bounds on $\left\| AA^T-ASS^TA^T \right\|$ (which we do).
I won't go through the details on this here, since you will do this in detail in the second homework.

I should note that a variant of this random projection algorithm has actually been implemented and used in several high performance scientific computing applications.
We will discuss this below, along with modifications to it that are necessary to bridge the theory-practice gap.
But it is important to note that it has been used, not due to the additive-error bounds, which are actually rather weak, but instead since much stronger $1\pm\epsilon$ bounds are possible.
Let's take a step back and ask what exactly is this random projection doing.
Essentially, what it is doing is applying JL ideas to the columns of $A$, which is why we get additive-error guarantees.
The improvement we will get to later in the semester applies JL ideas to a different set of vectors associated with the columns of $A$---essentially, to the truncated subspace vectors that are gotten by an orthogonal matrix spanning the top part of the spectrum.

Said another way, by applying JL ideas on the columns of $A$, the analysis of the algorithm is weaker than possible.
Random projections uniformize a lot of things, only one of which is the norms of input matrices.
To see this, we will introduce a more sophisticated random sampling algorithm, which will also achieve $1\pm\epsilon$ bounds for the Frobenius norm reconstruction error.
This will involve sampling with respect to the empirical statistical leverage scores of the input matrix.
Thus, for that improved random sampling algorithm, we will be putting the nonuniformity into the algorithm, while for the improved random projection algorithm, we will obtain improved results by performing a more refined analysis.

\subsection{Toward Better Bounds for Low-rank Approximation}

Before we do that, let's ask what are possible extensions of these ideas of choosing columns according to their size/norm.
\begin{itemize}
\item
Find a more sophisticated ``univariate statistic,'' meaning a score assigned to each column/row, to sample with respect to (and one that is hopefully still tractable to compute exactly or approximately).
This will involve using the statistical leverage scores.
This approach has gained a lot of traction, both in theory and in numerical implementation practice and in machine learning and data analysis applications.
In addition, these ideas can be used directly as the basis for other random projection ideas that are also used in theory and in practice, essentially since random projections preprocess or precondition to uniformize these scores.
We will cover these methods, starting next class.
\item
Iteratively choose sets of columns according to their ``size'' relative to what is not captured yet.
Since this approach is iterative, the columns chosen in successive trials are dependent on previous trials, and thus there is no simple ``univariate statistic'' associated with the columns that says that they are all the ``same'' in some sense, e.g., sampled from the same distribution.
In spite of that, this is a randomized or softer version of popular greedy heuristics, and not surprisingly this can do quite well in practice.
We will cover this next, and we will show that the additional error drops off very quickly, in the sense that with the right parameters it drops off exponentially in the number of rounds.
\item
Choose sets of $k$ columns according to the ``size'' of that set, e.g., proportional to the volume of the parallelpiped or simplex that they define.
This is not a univariate statistic, but it is a $k$-variate statistic, in that it depends on sets of columns/rows of cardinality $k$.
This method is intractable for most notions of best.
That being said, note that RVW, DRVW, DV show that the previous iterative approach can approximate this method, and thus this method can get a $1\pm\epsilon$ approximation that is ``fast'' in at least a theoretical sense.
These ideas have not gained widespread traction, in theory and certainly not in applications, and so we will not focus on them.
\end{itemize}

\textbf{Remark.}
It is an open question, and one with likely practical significance, whether one can use the iterative method to approximate the leverage scores, and, relatedly, what exactly are the connections between the leverage scores and the notions of volume that are used in the third bullet.

\subsection{An Iterative Additive-error Low-rank Approximation Algorithm}

Here, we will describe a variant of the iterative algorithm of RVW, DRVW, DV.
For simplicity, we will describe a variant that does \emph{not} filter the data through a rank-$k$ space.
(Note that the previous additive-error algorithm didn't need to, but it did filter through a low-rank space, and that is a stronger result.)

The \textsc{SelectColumnsSinglePass} algorithm takes as input a matrix $A$ and a number $c$ of columns to choose.
It returns as output a matrix $C$ such that the columns of $C$ are chosen from the columns of $A$ in $c$ i.i.d. trials by sampling randomly according to the probability distribution (\ref{eqn:probs_singlepass}).
More formally, for an $m \times n$ matrix $A$ and a multiset $S \subseteq \{1,\ldots,n\}$, let $C = A_S$ denote the $m \times |S|$ matrix whose columns are the columns of $A$ with indices in $S$.
The \textsc{SelectColumnsSinglePass} constructs the multiset $S$ by randomly sampling according to (\ref{eqn:probs_singlepass}) and returns the matrix $C=A_S$.
Note that this is basically just the same algorithm we had before, just parameterized a little differently, e.g., probabilities are inside the algorithm, the algorithm returns the matrix $C$ rather than just the top $k$ singular vectors, and consequently the quality-of-approximation theorem won't filter the matrix through a rank $k$ space.

\begin{algorithm}
\caption{The \textsc{SelectColumnsSinglePass} Algorithm. }
\label{alg:three-loop-singlepass}
\begin{algorithmic}[1]
    \REQUIRE An $m \times n$ matrix $A$, and an integer $c$ s.t. $1\le c\le n$. 
    \ENSURE  An $m \times c$ matrix $C$, s.t. $CC^+ A \approx A$. 
    \STATE Compute (for some positive $\beta \leq 1$) probabilities 
           $\{p_i\}_{i=1}^{n}$ s.t.
           \begin{equation}
           p_i \ge \beta \VTTNormS{A^{(i)}}/\FNormS{A}     ,
           \label{eqn:probs_singlepass} 
           \end{equation}
           where $A^{(i)}$ is the $i$-th column of $A$ as a column vector.
    \STATE $S = \{\}$.
    \FOR{$t = 1$ to $c$}
       \STATE Pick $i_t \in \{1,\ldots,n\}$ with $\Probab{i_t=\alpha}=p_\alpha$.
       \STATE $S = S \cup \{i_t\}$.
    \ENDFOR
    \STATE Return $C = A_S$.
\end{algorithmic}
\end{algorithm}

The \textsc{SelectColumnsSinglePass} algorithm is so-named since, given probabilities of the form (\ref{eqn:probs_singlepass}), the matrix $C$ can be constructed in one pass over the (externally-stored) data matrix $A$.
The following theorem is our main quality-of-approximation result for the\textsc{SelectColumnsSinglePass} algorithm.

\begin{theorem}
\label{thm:ConstructC_onepass} 
Suppose $A \in \mathbb{R}^{m \times n}$, and let $C$ be the $m \times c$ matrix 
constructed by sampling $c$ columns of $A$ with the 
\textsc{SelectColumnsSinglePass} algorithm.
If $\eta = 1 + \sqrt{(8/\beta) \log(1/\delta)}$ for any $0 < \delta < 1$, 
then, with probability at least $1 - \delta$,
\begin{equation}
\FNormS{A-CC^+A} \leq \FNormS{A-A_k} + \epsilon \FNormS{A}   ,
\label{eqn:linear_whp_fnorm_eps}
\end{equation}
if $c \geq 4 \eta^2 k /(\beta \epsilon^2)$.
\end{theorem}
\begin{Proof}
Let the $m \times c$ matrix $\hat{C}$ be that matrix whose columns consist of 
appropriately rescaled copies of the columns of $C$, as discussed in 
conjunction with the \textsc{LinearTimeSVD} algorithm of
the last class.
First, note that since $CC^{+} = P_C = P_{\hat{C}} = \hat{C}\hat{C}^{+}$ is a 
projection onto the full column space of $C$, it follows that
\begin{equation}
\FNormS{A-CC^{+}A} 
   \le \FNormS{A-P_{\hat{C},k}A}     .
\label{eqn1_pf1}
\end{equation}
The theorem follows by combining this with the results of 
the last class.
\end{Proof}

One final comment on this algorithm.
The relationship of this algorithm with the \textsc{LinearTimeSVD} algorithm should also be emphasized.
In the \textsc{LinearTimeSVD} algorithm, the columns of $A$ that are sampled by the algorithm are {\em scaled} prior to being included in $C$, by dividing each sampled column by a quantity proportional to the square root of the probability of picking it.
This scaling allows one to prove that the top $k$ singular values of the matrix $\hat{C}$, i.e., the scaled version of $C$, and the top $k$ singular values of $A$ are close. 
Additionally, it allows one to prove that under appropriate assumptions
\begin{equation}
\XNormS{A-P_{\hat{C},k}A} \le \XNormS{A-A_k} + \epsilon \FNormS{A}    ,
\label{eqn:lineartimeSVD}
\end{equation}
in both expectation and with high probability, for both the spectral and Frobenius norms, $\xi=2,F$.
Here, in the projection matrix to the full space spanned by the columns of $C$, namely $P_C = CC^+ = \hat{C}\hat{C}^+ = P_{\hat{C}}$ rather than $P_{\hat{C},k}$.
Clearly, any scaling of the columns of $C$ does not affect this full projection matrix.

Next, we will choose columns in multiple rounds, where in each round we choose $c$ columns.
So, this is a randomized version of a greedy algorithm that chooses the next column based on who has the largest residual.
This algorithm was first presented by RVW, and it was extended by DRVW, DV.
In particular, Rademacher, Vempala and Wang provided the first proof of a theorem in which the additional error drops exponentially with the number of passes.  
In more detail, they proved that there exists a rank $k$ matrix in the subspace spanned by $C$ that satisfies (in expectation) a bound of the form (\ref{eqn:linear_multipass_fnorm_eps}). 
Thus, by Markov's inequality, they obtain a bound of the form (\ref{eqn:linear_multipass_fnorm_eps}) that holds with probability at least $1-\bar{\delta}$ if $c = O(t^2/\bar{\delta})$.
The proof below is simpler.
In addition, observe that it obtains (\ref{eqn:linear_multipass_fnorm_eps}) with probability at least $1-\bar{\delta}$ if $c = O(t\log(t/\bar{\delta}))$.

The \textsc{SelectColumnsMultiPass} algorithm takes as input a matrix $A$, a number $t$ of rounds to perform, and a number $c$ of columns to choose per round.
It returns as output a matrix $C$ such that the columns of $C$ are chosen from the columns of $A$ in the following manner.
There are $t$ rounds, and each round consists of $2$ passes over the data.
In the first round, let $\ell=1$.
Sampling probabilities of the form (\ref{eqn:probs_singlepass}) are computed in the first pass of the first round, and in the second pass a multiset $S_1$ of columns of $A$ is picked in $c$ i.i.d. trials by sampling according to the probabilities (\ref{eqn:probs_singlepass}).
For each subsequent round $\ell = 2,\ldots,t$, sampling probabilities of the form (\ref{eqn:probs_multipass}) are constructed that depend on the lengths of the columns of the the $m \times n$ matrix $E_{\ell}$ that is the residual of the matrix $A$ after subtracting the projection of $A$ on the subspace spanned by the columns sampled in the first $\ell-1$ rounds.

More formally, let the indices of the columns that have been chosen in the first $\ell-1$ rounds form the multiset $\{S_1, S_2, \ldots, S_{\ell-1}\}$ (where the multiset of columns $S_i$ were chosen in the $i$-th round) and let $C_{\ell-1} = A_{\{S_1, S_2, \ldots, S_{\ell-1}\}}$ denote the $m \times |S_1||S_2|\cdots|S_{\ell-1}|$ matrix whose columns are the columns of $A$ with indices in $\{S_1, S_2, \ldots, S_{\ell-1}\}$.
Then,
\begin{equation}
E_{\ell} 
   = A - A_{\{S_1,\ldots,S_{\ell-1}\}} A_{\{S_1,\ldots,S_{\ell-1}\}}^+ A      
   = A - C_{\ell-1} C_{\ell-1}^+ A                                    .
\end{equation}
Sampling probabilities of the form (\ref{eqn:probs_multipass}) are then constructed in the first pass of each round $\ell = 2,\ldots,t$, and $c$ columns are chosen from $A$ by sampling in $c$ i.i.d. trials according to the probabilities (\ref{eqn:probs_multipass}) in the second pass of each round $\ell = 2,\ldots,t$.
(Note that if, by definition, $E_1 = A$, then for $\ell=1$ the sampling probabilities (\ref{eqn:probs_multipass}) are the same as those of (\ref{eqn:probs_singlepass}).)

\begin{algorithm}
\caption{The \textsc{SelectColumnsMultiPass} Algorithm. }
\label{alg:three-loop-multipass}
\begin{algorithmic}[1]
    \REQUIRE An $m \times n$ matrix $A$, and an integer $c$ s.t. $1\le c\le n$, 
             and a positive integer $t$. 
    \ENSURE  An $m \times c$ matrix $C$, s.t. $CC^+ A \approx A$. 
    \STATE $S = \{\}$.
    \FOR{$\ell = 1$ to $t$}
       \IF{$\ell == 1$} 
          \STATE $E_1 = A$.
       \ELSE 
          \STATE $E_{\ell} = A - A_S A_S^+ A$.
       \ENDIF 
       \STATE Compute (for some positive $\beta \leq 1$) probabilities 
              $\{p_i\}_{i=1}^{n}$ s.t.  
              \begin{equation}
              p_i \ge \beta \VTTNormS{E_{\ell}^{(i)}}/\FNormS{E_{\ell}}    ,
              \label{eqn:probs_multipass} 
              \end{equation}
              where $E_{\ell}^{(i)}$ is the $i$-th column of $E_{\ell}$ as a 
              column vector.
       \FOR{$t = 1$ to $c$}
          \STATE Pick $i_t \in \{1,\ldots,n\}$ with 
                 $\Probab{i_t = \alpha} = p_\alpha$.
          \STATE $S = S \cup \{i_t\}$.
       \ENDFOR
    \ENDFOR
    \STATE Return $C = A_S$.
\end{algorithmic}
\end{algorithm}

The \textsc{SelectColumnsMultiPass} algorithm is so-named since, given probabilities of the form (\ref{eqn:probs_multipass}), $c$ columns can be extracted in one pass over the (externally-stored) data matrix $A$.
Then, of course, in each round the probabilities $\{p_i\}_{i=1}^{n}$ that are used by the algorithm may be computed with one pass over the data and $O(1)$ additional space.
The algorithm is thus efficient in the Pass Efficient Model.

Here are several things to note.
This algorithm takes $t$ rounds, and each round is $2$ passes over the data.
In the first round, it computes the simple sampling probabilities $p_i = \frac{\VTTNormS{A^{(j)}}}{\FNormS{A}}$ in the first pass and then pulls out the actual columns in the second pass.
In the second and subsequent rounds, ditto, except that the sampling probabilities depend on the length of the columns of the matrix $E$, that is the residual after you subtract the projection of $A$ onto the subspace spanned by the columns in the first $\ell-1$ rounds.
That is, if $\{S_1,S_2,\ldots,S_{\ell-1}\}$ is a multiset chosen in the first $\ell-1$ rounds, then $C_{\ell-1} = A_{S_1 S_2 \cdots A_{\ell-1}}$ is the $m \times |S_1||S_2| \cdots |S_{\ell-1}|$ matrix with columns of $A$ that has indices in $\{ S_1, S_2, \cdots, S_{\ell} \}$.
So, 
$E_{\ell} 
   = A - A_{\{S_1,\ldots,S_{\ell-1}\}} A_{\{S_1,\ldots,S_{\ell-1}\}}^+ A      
   = A - C_{\ell-1} C_{\ell-1}^+ A                                    $.

\begin{theorem}
\label{thm:ConstructC_multipass} 
Suppose $A \in \mathbb{R}^{m \times n}$ and let $C$ be the $m \times tc$ matrix constructed by sampling $c$ columns of $A$ in each of $t$ rounds with the \textsc{SelectColumnsMultiPass} algorithm. 
If $\eta = 1 + \sqrt{(8/\beta) \log(1/\delta)}$ for any $0 < \delta < 1$, then, with probability at least $1 - t\delta$,
\begin{equation}
\label{eqn:linear_multipass_fnorm_eps} 
\FNormS{A-CC^+ A} 
   \leq \frac{1}{1-\epsilon}\FNormS{A-A_k} + \epsilon^t \FNormS{A},
\end{equation}
if $c \geq 4\eta^2 k/(\beta \epsilon^2)$ columns are picked in each of the $t$ rounds. 
\end{theorem}

Recall that we will go with the following, which is a simpler and improved proof, compared with that of RVW.

\begin{Proof}
The proof will be by induction on the number of rounds $t$. 
Let $S_1$ denote the set of columns picked at the first round, and let
$C^1 = A_{S_1}$. 
Thus, $C^1$ is an $m \times c$ matrix, where 
$c \ge 4\eta^2 k/(\beta\epsilon^2)$.
By Theorem \ref{thm:ConstructC_onepass} and since $1 < 1/(1-\epsilon)$ for 
$\epsilon > 0$, we have that
\begin{equation}
\FNormS{A - C^1\left(C^1\right)^+ A} 
   \leq \frac{1}{1-\epsilon} \FNormS{A-A_k} + \epsilon \FNormS{A}
\label{eqn1_pf2}
\end{equation}
holds with probability at least $1-\delta$, thus establishing the base case of 
the induction.

Next, let $(S_1,\ldots,S_{t-1})$ denote the set of columns picked in the
first $t-1$ rounds and let $C^{t-1} = A_{(S_1,\ldots,S_{t-1})}$. 
Assume that the proposition holds after $t-1$ rounds, i.e., assume that by 
choosing $c \ge 4\eta^2 k/(\beta\epsilon^2)$ columns in each of the first $t-1$
rounds, we have that
\begin{equation}
\FNormS{A-C^{t-1}\left(C^{t-1}\right)^+ A} 
   \leq \frac{1}{1-\epsilon}\FNormS{A-A_k} + \epsilon^{t-1} \FNormS{A}
\label{eqn2_pf2_ind_hypoth} 
\end{equation}
holds with probability at least $1-(t-1)\delta$.

We will prove that it also holds after $t$ rounds. 
Let $ E_t = A - C^{t-1} \left(C^{t-1}\right)^+ A $ be the residual of the 
matrix $A$ after subtracting the projection of $A$ on the subspace spanned by 
the columns sampled in the first $t-1$ rounds.
(Note that it is $\FNormS{E_t}$ that is bounded by (\ref{eqn2_pf2_ind_hypoth})).
Consider sampling columns of $E_t$ at round $t$ with probabilities
proportional to the square of their Euclidean lengths, i.e., according to 
(\ref{eqn:probs_multipass}), and let $Z$ be the matrix of the columns of $E_t$ 
that are included in the sample. 
(Note that these columns of $E_t$ have the same span and thus projection as the 
corresponding columns of $A$ when the latter are restricted to the residual 
space.)
Then, by choosing at least $c \ge 4\eta^2 k/(\beta \epsilon^2)$ columns of 
$E_t$ in the $t$-th round we can apply Theorem \ref{thm:ConstructC_onepass} to 
$E_{t}$ and get that
\begin{equation}
\FNormS{E_t - Z Z^+ E_t} 
   \leq \FNormS{E_t - \left(E_t\right)_k} + \epsilon \FNormS{E_t}
\label{eqn3_pf2}
\end{equation}
holds with probability at least $1 - \delta$.
By combining (\ref{eqn2_pf2_ind_hypoth}) and (\ref{eqn3_pf2}) we see that
if at least $4\eta^2 k/(\beta \epsilon^2)$ columns are picked in each of the 
$t$ rounds then
\begin{equation}
\FNormS{E_t - Z Z^+ E_t} 
   \leq \FNormS{E_t - \left(E_t\right)_k} 
      + \frac{\epsilon}{1-\epsilon}\FNormS{A-A_k} 
                     + \epsilon^{t} \FNormS{A}
\label{eqn3_pf22}
\end{equation}
holds with probability at least $1-t\delta$.
The theorem thus follows from (\ref{eqn3_pf22}) if we can establish that
\begin{eqnarray}
\label{eqn4_pf2}
E_t - ZZ^+ E_t 
   &=& A - C^{t  } \left(C^{t  }\right)^+ A             \\
\label{eqn6_pf2} 
\FNormS{E_t - \left(E_t\right)_k} 
   &\leq& \FNormS{A - A_k}                               .
\end{eqnarray}
But (\ref{eqn4_pf2}) follows from the definition of $E_t$, 
since
$C^{t} \left(C^{t}\right)^+ 
   = C^{t-1} \left(C^{t-1}\right)^+ + ZZ^+ $
by the construction of $Z$, and
since $ZZ^+ C^{t-1} \left(C^{t-1}\right)^+ = {\bf 0}$.
To establish (\ref{eqn6_pf2}), and thus the theorem, notice that 
\begin{eqnarray}
\label{eqn9a_pf2}
\FNormS{E_t - \left(E_t\right)_k}
   &=&   \FNormS{        \left( I - C^{t-1} \left(C^{t-1}\right)^+ \right) A 
                 - \left(\left( I - C^{t-1} \left(C^{t-1}\right)^+ \right) A\right)_k}              \\
\label{eqn9b_pf2}
   &\le& \FNormS{        \left( I - C^{t-1} \left(C^{t-1}\right)^+ \right) A 
                 -       \left( I - C^{t-1} \left(C^{t-1}\right)^+ \right) A_k}                     \\
\label{eqn9c_pf2}
   &=& \FNormS{\left( I - C^{t-1} \left(C^{t-1}\right)^+ \right) \left(A-A_k\right)}              \\
\label{eqn9d_pf2}
   &\le& \FNormS{A-A_k}         .
\end{eqnarray}
(\ref{eqn9a_pf2}) follows by definition of $E_t$, 
(\ref{eqn9b_pf2}) follows since 
$\left( I - C^{t-1} \left(C^{t-1}\right)^+ \right) A_k$ is a rank $k$ matrix, 
but not necessarily the optimal one, 
(\ref{eqn9c_pf2}) follows immediately, and
(\ref{eqn9d_pf2}) follows since $I - C^{t-1} \left(C^{t-1}\right)^+$ is a 
projection.
\end{Proof}

This algorithm and theorem demonstrate that by sampling in $t$ rounds and by judiciously computing sampling probabilities for picking columns of $A$ in each of the $t$ rounds, the overall error drops {\em exponentially} with $t$. 
This is a substantial improvement over the results of Theorem \ref{thm:ConstructC_onepass}. 
In that case, if $c \ge 4 \eta^2 k t / (\beta \epsilon^2)$ then the additional additive error is $(\epsilon/\sqrt{t})\FNormS{A}$. 
Note also that although we have described this as an iterative additive-error low-rank matrix approximation algorithms, it becomes a relative-error approximation if the number of iterations depends on the stable rank (which is not known a priori, but which can in some senses be estimated).

\newpage

\section{%
(10/28/2013):  Relative-error Low-rank Matrix Approximation}

Today, we will start to discuss how to improve the rather coarse additive-error low-rank matrix approximation algorithms from the last two classes to obtain much better results for low-rank matrix approximation.
Importantly, ``better'' means very different things to different research communities, and thus we will discuss several different notions of better.
We will start by describing how to improve the additive-error bounds we have been discussing to relative-error low-rank matrix approximation.
Here is the reading for today.
\begin{itemize}
\item
Drineas, Mahoney, and Muthukrishnan, ``Relative-Error CUR Matrix Decompositions''
\end{itemize}

Today, in particular, we will cover the following topics.
\begin{itemize}
\item
Various types of column/row-based low-rank approximations.
\item
Different classes of approaches to choosing good columns/rows.
\item
A generalization of the LS regression algorithm to fat matrices that are low rank.
\item
Basic results for CX/CUR low-rank decompositions that achieve relative-error reconstruction guarantees on the top part of the spectrum with respect to the Frobenius norm error.
\end{itemize}

\subsection{Low-rank Approximations via Columns/Rows}

We are now going to consider much better (in the sense of TCS) low-rank matrix approximations---these approximations will come with so-called relative-error, or $1\pm\epsilon$, approximation guarantees.
This will be an improvement over the additive-error algorithm presented in the last class since the scale of the additional error for these improved algorithms will be the base residual, i.e., the norm of the part of the matrix that is not captured by the best rank $k$ approximation.
As with our discussion of $\ell_2$ regression and additive-error low-rank matrix approximation algorithms, our discussion here will start with algorithms that randomly sample actual columns and/or rows, and then it will consider the extension of these ideas to algorithms that perform random projections.

\paragraph{CX matrix decompositions.}
We will start with the following definition.

\begin{definition}
Given a matrix $A\in\mathbb{R}^{m \times n}$, for a matrix $C\in\mathbb{R}^{m \times c}$, consisting of $c$ actual columns of $A$, the matrix $A^{\prime} = CX$ is a \emph{column-based matrix approximation} or a \emph{CX matrix decomposition}, for any matrix $X \in\mathbb{R}^{c \times n}$.
\end{definition}

Things to note about this definition.
\begin{itemize}
\item
First, although we have defined it in general, we are most interested in the case that $c \ll n$.
\item
Second, one may think of this as a decomposition that decomposes $A$ into a small number of ``dictionary elements'' or ``basis columns,'' each of which is an actual column of $A$.
Although this is of less interest from an abstract perspective, it can be very important in certain applications.
\item
Third, if we choose $X=C^{+}A$, where $C^{+}$ is the Moore-Penrose generalized inverse of $C$, then this is the ``best'' CX approximation, for that $C$, with respect to reconstruction error for any unitarily invariant matrix norm.
\item
Finally, many algorithms, e.g., traditional QR decompositions, can be thought of a providing CX decompositions---just keep the columns that are used in the Gram-Schmidt procedure of the algorithm.
The question of interest will be what quality-of-approximation results can be proven for the procedure.
\end{itemize}

\paragraph{CUR matrix decompositions.}
We will next consider the following definition.

\begin{definition}
Given a matrix $A\in\mathbb{R}^{m \times n}$, for a matrix $C\in\mathbb{R}^{m \times c}$, consisting of $c$ actual columns of $A$ and a matrix $R\in\mathbb{R}^{r \times n}$, consisting of $r$ actual rows of $A$, the matrix $A^{\prime} = CUR$ is a \emph{column-row-based matrix approximation} or a \emph{CUR matrix decomposition}, for any matrix $U\in \mathbb{R}^{c \times r}$.
\end{definition}

Things to note about this definition.
\begin{itemize}
\item
First, it is sometimes called a pseudoskeletal decomposition, and it has also been called a generalized Nystr\"{o}m approximation.
\item
Second, it is also a CX decomposition, on the original matrix as well as on its transpose, where the matrix $X$ has a particular structure of $X=UR$.
\item
Third, in terms of its singular value structure, since both $C$ and $R$ contain singular value information from the original matrix $A$, the matrix $U$ should contain ``inverse of $A$'' information, in order to provide a good low-rank approximation.
For example, it could consist of the generalized inverse of the intersection of $C$ and $R$, or a related quantity that is quicker to~compute.
\item
Finally, while we have defined CX and CUR in terms of actual columns, the analysis will try to quantify how well $C$ and $R$ represent the column space and row space of $A$.
If $C$ and $R$ were permitted to be matrices consisting of linear combinations of columns, then one choice (and the best choice with respect to unitarily invariant matrix norms) would be to set them to be the left and right singular vectors, in which case the middle matrix would be the matrix of singular values.
\end{itemize}

\paragraph{Nystr\"{o}m approximations.}
We will next consider the following definition, which generalizes the notion of CUR decompositions to SPSD matrices.
(SPSD matrices are of interest in many applications.  In machine learning, in particular, they are known as kernels.)

\begin{definition}
Given a matrix $A\in\mathbb{R}^{n \times n}$ that is SPSD, for a matrix $C\in\mathbb{R}^{m \times c}$, consisting of $c$ actual columns of $A$, let $U = W^{+}$, where the $c \times c$ matrix $W$ is the intersection of $C$ with itself, then the matrix $A^{\prime} = CUC^T = CW^{+}C^T$ is a low-rank \emph{Nystr\"{o}m approximation} to $A$.
\end{definition}

Things to note about this definition.
\begin{itemize}
\item
First, this is just a CUR approximation for a SPSD matrix, where one chooses the same columns and rows and a symmetric middle matrix, in order to preserve the SPSD property.
\item
Second, the name Nystr\"{o}m comes because decompositions of this form have been used historically in scientific computing, but there has been interest in decompositions of this form in recent years in ML.
\item
Third, it is not immediate, given the usual constructions of CUR, that one can choose some columns and rows and preserve the SPSD property; alternatively, if you choose the same columns and rows, then you preserve the SPSD property, but the analysis that leads to good quality-of-approximation bounds typically fails.
\end{itemize}
The reason for these points is that the analysis boils down to a LS approximation, and thus one gets an asymmetry in the choice of $C$ and then $R$.
Relatedly, it is difficult to certify that a matrix is SPSD, unless one has the square root, the computation of which we want to avoid, or unless the matrix has some strong property like being diagonally dominant.
One can do this with uniform sampling and preserve the SPSD property, and this empirically works well in certain cases, but it can perform very poorly in general.
Alternatively, one can get around some of these problems by working within the SPSD sketching model and appealing to the ``implicit square roor'' result due to Gittens.

\paragraph{More general sketching models.}
Although the three previous definitions were in terms of actual columns, if we view those columns as $C=AS$, where $A$ is the original matrix and $S$ is a sampling matrix, it is fair to ask how sensitive the results are to the particular form of $S$.
For example, could it be a projection matrix, or could it be some other deterministic matrix?
In general, the answer is Yes.
We discussed this last week, when we described how the analysis for the additive-error low-rank sampling algorithms extends more-or-less immediately to additive-error bounds for random projection algorithms, and it holds more generally. 
(You also got a chance to work through it in detail on the homework.)
We will describe this more generally below, but for now we just state this result for SPSD matrices. 

\begin{definition}
Let $A\in\mathbb{R}^{n \times n}$ be SPSD.
Let $S\in\mathbb{R}^{n \times \ell}$, with $\ell \ll n$, be a sketching matrix (e.g., a sampling matrix or a projection matrix or anything else).
Let $C=AS$, and let $W=S^TAS$.
Then, a matrix $A^{\prime}$ of the form $A^{\prime}=CW^{+}C^T$ is a low-rank approximation with $\mbox{rank}(A^{\prime})\le k$ that we call the \emph{SPSD sketching model}.
\end{definition}

Things to note about this definition.
\begin{itemize}
\item
First, as stated, this is not guaranteed to be numerically stable, so we can truncate the small singular values and/or regularize/smooth them out in different ways.
This has been done in both ML and scientific computing, although the details of how the do it and the exact results that are established differ somewhat.
\item
Second, Nystr\"{o}m is an example of this, if $S$ is a sampling matrix, but this can include random projections also, and other things also, if $S$ is more general.
\item
Third, Gittens has shown that if you work with a low-rank approximation satisfying this, then in a certain precise sense, you are implicitly working on the square root of $A$; this means that you can get much stronger bounds than naive methods like uniform sampling.
We will get back to this later.
\end{itemize}

\paragraph{Pictorial illustration.}
CX and CUR decompositions take a particularly nice form, which is illustrated as follows.
\begin{eqnarray}
\label{eqn:cx-cur-1}
\underbrace{\left(
   \begin{array}{ccccc}
   &&&& \\
   &&&& \\
   &&A&& \\
   &&&& \\
   &&&&
   \end{array}
\right)}_{m \times n}
&\approx&
\underbrace{\left(
   \begin{array}{ccc} && \\  &&\\ &C& \\ && \\ && \end{array}
\right)}_{m \times c}
\underbrace{\left(
   \begin{array}{ccccc}  &&&& \\ &&X&& \\ &&&&  \end{array}
\right)}_{c \times n}   \\
\label{eqn:cx-cur-2}
&\approx&
\underbrace{\left(
   \begin{array}{ccc} && \\  &&\\ &C& \\ && \\ && \end{array}
\right)}_{m \times c}
\underbrace{ \left(
   \begin{array}{ccc}  && \\ &U& \\ &&  \end{array}
\right) }_{c \times r}
\underbrace{\left(
   \begin{array}{ccccc}  &&&& \\ &&R&& \\ &&&&  \end{array}
\right)}_{r \times n} .
\end{eqnarray}
For Nystr\"{o}m approximations, the decomposition is of course more symmetric than for general matrices.

\subsection{Approaches to Choosing Good Columns and Rows}

Let's consider how to construct a decomposition of the form $A \approx CX$ or $A \approx CUR$.
In particular, there are several questions.
First, how do you choose ``good'' columns $C$?
(For now, we will focus on good in the sense that we want to reproduce well the top part of the spectrum of the original matrix, but we will return later when we discuss the Column Subset Selection Problem to other notions of a good set of columns.)
Among the answers are:
\begin{itemize}
\item
Perform a QR decomposition, and keep the associated columns.
Often this actually perform fairly well empirically.
Most of the bounds are for the spectral norm of the reconstruction error (although there are a bunch of variants, e.g., vanilla QR, rank-revealing QR, etc. that can guarantee other properties).
But note that this method can get ``stuck'' in ``corners,'' which operationally basically means that they need to make smarter pivot rule decisions in order to get better results.
\item
Perform a greedy iterative algorithm.
This also often does reasonably well in practice, either being very greedy, like with methods like GreedyPursuit, etc., or greedy in a softer randomized way, e.g., don't keep the column at each step that is best according to some metric, but instead keep columns that are biased toward those that are best by that metric.
These methods can be more difficult to analyze, leading to weaker~bounds.
\item
Perform random sampling, e.g., by choosing a small number of columns with probabilities proportional to $\VTTNormS{A^{(i)}}$.
This is pass efficient, and you get additive error bounds, which as we have said is good but certainly not great.
\item
Perform random iterative sampling with probabilities that depend on the norm of the residual.
The iterative algorithm we discussed is a version of this.
It is weakly pass efficient, in that it takes a number of rounds that is ``small,'' in that it depends on $k$, and you get additive error bounds with a scale that improves exponentially with the number of rounds.
\item
Construct some sort of (potentially more expensive) score over the columns of $A$ and use those an an importance sampling distribution.
In particular, given the $1\pm\epsilon$ relative-error bounds for $\ell_2$ regression, perhaps that can be used to get $1\pm\epsilon$ relative-error bounds for low-rank approximation problems.
\end{itemize}
Second, given $C$, how do you choose ``good'' columns $R$?
Among the answers are:
\begin{itemize}
\item
Ignore $C$ and choose rows $R$ using one of the methods described above by looking at $A^T$.
Often, in this case, you can still construct $U$ such that if you have additive-error bounds for $C$ then you get additive-error bounds for $CUR$.
Alternatively, often you can construct $U$ such that if you have $1\pm\epsilon$ bounds originally, then you get $2\pm\epsilon$ bounds by combining them in this relatively naive way.
\item
Perhaps we can use information in $C$ and take advantage of $\ell_2$ regression result to get rows that are good with respect to $C$ (which are good with respect to $A$) and combine them in order to get $1\pm\epsilon$ bounds for $CUR$.
\end{itemize}
In fact, we will be able to do this last suggestion.
The basic idea is that we will choose columns $C$ that are relative-error good approximations to the best rank $k$ approximation to $A$; and then choose rows $R$ that are relative-error good approximations for those columns $C$.
(BTW, this strategy that is asymmetric with respect to rows and columns explains why it is difficult---which as we have said is overcomeable---to do very good Nystr\"{o}m approximation of general SPSD matrices.)
In particular, the choice for each of these two steps is a special case of a generalization of the over-determined $\ell_2$ regression problem.
Thus, let's consider that problem.

\subsection{Generalized $\ell_2$ Regression}

The approximation symbol in Eqn.~(\ref{eqn:cx-cur-1}) and Eqn.~(\ref{eqn:cx-cur-2}) is meant to be more than suggestive.
In particular, in addition to providing an approximation to $A$ in some vague sense, the analysis for relative-error CX and CUR decompositions (which we will get to now) proceeds by showing that (1) one can find a small number of columns $C$ that are relative-error-good with respect to the columns that define the space that is the best rank $k$ approximation to $A$, and (2) one can then find a small number of rows $R$ that are relative-error-good with respect to those columns $C$.
Thus, in both cases, the approximation is a generalization of the relative-error random sampling algorithm for the least-squares problem.

Thus, since our bounds for CX and CUR will in fact boil down to a generalization of the over-determined $\ell_2$ regression problem we discussed earlier, we will start with that.
Recall that before we considered solving 
$$
\min_x \VTTNormS{Ax-b}  ,
$$
where $A$ is a ``tall'' matrix and $b$ is a vector.
Here, we will consider the generalization where $A$ is an arbitrarily-sized matrix with rank no greater than $k$, and we will also consider the generalization of the case where the right and side is a general matrix $B$ rather than a vector.
Here is our algorithm.

\begin{algorithm}
\caption{The \textsc{Generalized $\ell_2$ Regression} Algorithm. }
\label{alg:generalized-L2}
\begin{algorithmic}[1]
    \REQUIRE An $m \times n$ matrix $A$, with rank no greater than $k$,
             an $m \times p$ matrix $B$, an integers $c$ s.t.  
             $1 \le k \le c \le m$, and a probability distribution
             $\left\{ p_i \right\}_{i=1}^{m}$.
    \ENSURE  An $n \times p$ matrix $\tilde{X}_{opt}$ and a 
             number $\tilde{\mathcal{Z}}\in\mathbb{R}$.
    \STATE Use the probabilities $\{p_i\}_{i=1}^{m}$ to form a random sampling matrix 
           $S\in\mathbb{R}^{m \times c}$ and a diagonal rescaling matrix $D$.
    \STATE Construct $DS^TA$ and $DS^TB$.
    \STATE Solve the subsampled problem with a black box to get
           \begin{eqnarray}
           \tilde{X}_{opt} &=& (DS^TA)^{+}DS^TB \\
           \tilde{\mathcal{Z}} &=& \min_{X\in\mathbb{R}^{n \times p}} \FNorm{DS^TB-DS^TA\tilde{X}_{opt}}.
           \end{eqnarray}
\end{algorithmic}
\end{algorithm}

Here is the main theorem that we can prove about this algorithm.
Note that is is the obvious generalization of the over-determined $\ell_2$ regression result.

\begin{theorem}
\label{thm:ls_bound}
Suppose $A \in \mathbb{R}^{m \times n}$ has rank no greater than $k$, 
$B \in \mathbb{R}^{m \times p}$,
$\epsilon \in (0,1]$, and let
$
{\cal Z} 
   = \min_{X \in \mathbb{R}^{n \times p}} \FNorm{B - AX} 
   = \FNorm{B-AX_{opt}}     
$,
where $X_{opt}=A^+B=A_k^+B$.
Run 
Algorithm~\ref{alg:generalized-L2}
with any sampling probabilities
of the form
\begin{equation}
\label{eqn:ass1}
p_i \geq \beta
         \frac{             \VTTNormS{\left(U_{A,k}\right)_{(i)}}}
              {\sum_{j=1}^m \VTTNormS{\left(U_{A,k}\right)_{(j)}}}     
       = \frac{\beta}{k} \VTTNormS{\left(U_{A,k}\right)_{(i)}}          ,
   \hspace{0.25in}
   \mbox{   }
   \forall i \in [m]   ,
\end{equation}
for some $\beta \in (0,1]$, and assume that the output of the algorithm is a 
number $\tilde{\cal Z}$ and an $n \times p$ matrix $\tilde{X}_{opt}$.
If exactly $ r = 3200k^2/\beta\epsilon^2$ rows are chosen with the 
\textsc{Exactly($c$)} algorithm, then with probability at least $0.7$:
\begin{eqnarray}
\label{eqn:result2}
\FNorm{B - A\tilde{X}_{opt}}
   &\leq& \left( 1+\epsilon \right){\cal Z},           \\
\label{eqn:result3}
\FNorm{X_{opt} - \tilde{X}_{opt}}
   &\leq& \frac{\epsilon}{\sigma_{\min}(A_k)} {\cal Z}     .
\end{eqnarray}
If, in addition, we assume that 
$\FNorm{U_{A,k}U_{A,k}^TB} \geq \gamma \FNorm{B} $,
for some fixed $\gamma \in (0,1]$, then with probability at least $0.7$:
\begin{equation}
\label{eqn:result4}
\FNorm{X_{opt} - \tilde{X}_{opt}}
   \leq \epsilon \left( \kappa(A_k)\sqrt{\gamma^{-2}-1} \right) \FNorm{X_{opt}}     .
\end{equation}
Similarly, under the same assumptions, if $r=O(k \log k/\beta\epsilon^2)$ rows 
are chosen in expectation with the \textsc{Expected($c$)} algorithm, then with 
probability at least $0.7$,  
(\ref{eqn:result2}), (\ref{eqn:result3}), and (\ref{eqn:result4}) hold.
\end{theorem}

Things to note about this result.
\begin{itemize}
\item
Clearly, the factors of $3200$ and so on are artifacts of the analysis and the particular (now out of date) bounds that were used to establish this result.
We include them here in the statement of this result for ease of comparison with the DMM paper.
\item
Also, the proof of this theorem is a pretty immediate generalization of our previous analysis of very over-determined $\ell_2 $regression, and in particular it boils down to two approximate matrix 
multiplication bounds that generalize the previous structural results.
(The role of the low dimension in the previous result is replaced here with the exactly low-rank space that captures all of the matrix $A$.)
Thus, we won't provide it here.
\item
We will apply this result to do CX and CUR on general matrices $A$ with general rank (and by extension relative-error low-rank random projection algorithms), and there will be no assumption of being exactly low-rank on those matrices.
We will prove that those CX/CUR results work by appealing to this generalization of $\ell_2$ regression that works for matrices of exactly rank $k$.
(Actually, the rank can be less, and there are some numerical issues there, but we won't go into them.)
So, think of this rank restriction as being inside the analysis of sampling and projection algorithms on arbitrary matrices (for the analysis of CX/CUR/projections/etc.), and it is \emph{not} an assumption about the input.
\item
As with the rectangular regression problem, if the other sketching operators satisfy those two conditions, then the same results go through.
In particular, we can use random projections of appropriate sized, as well as  CX, CUR, and Nystr\"{o}m approximation results below.
One can view this as a modification of Algorithm~\ref{alg:generalized-L2} to hold for general sketching matrices or as preprocessing the input with a random projection based preconditioning.
\end{itemize}

\subsection{CX and CUR Decompositions of General Matrices}

Next, we will use the generalized $\ell_2$ regression result to get very fine relative-error bounds on CX and CUR decompositions.
(By extension, this will also give relative-error bounds random projection algorithms for low-rank matrix approximation, when the dimensions of the projection are chosen appropriately.  We won't go into those here, but see the homework for details.)
We'll first describe a few related algorithms and establish quality-of-approximation bounds, and then we will discuss running time considerations.

Here is a randomized algorithm for constructing CX matrix decompositions.
The algorithm takes as input a matrix $A\in\mathbb{R}^{m \times n}$, a rank parameter $k$, and an error parameter $\epsilon$, and it returns as output a matrix $C\in\mathbb{R}^{m \times c}$.
It does the following steps.
\begin{enumerate}
\item
Compute (exactly or approximately) the distribution $\{p_i\}_{i=1}^{n}$, where
$ p_i = \frac{1}{k} \VTTNormS{\left( U_{A,k} \right)_{(i)}} .$
\item
Using $\{p_i\}_{i=1}^{n}$ as an importance sampling distribution, construct a random sampling matrix $S_C\in\mathbb{R}^{n \times c}$ and a diagonal rescaling matrix $D_C\in\mathbb{R}^{c \times c}$.
\item
Construct $C = AS_C D_C \in \mathbb{R}^{m \times c}$, a matrix consisting of a small number of columns of $A$.
\end{enumerate}

Here is what we can prove regarding this CX algorithm.
\begin{theorem}
Let $A\in\mathbb{R}^{m \times n}$, and let $k\in\mathbb{Z}^{+}$.
If we call the above algorithm with $c = O\left( k \log(k)/\epsilon^2 \right)$, then with constant probability we have that
$$
\FNorm{ A-CC^{\dagger}A } \le \left(1+\epsilon\right) \FNorm{ A-A_k }  .
$$
\end{theorem}
\begin{Proof}
Since $C=AS_CD_C$, we have that $X_{opt} = C^{\dagger}A$ is the matrix that minimized $\FNorm{ A-CX }$.
Then, we have the following chain of equalities and inequalities.
\begin{eqnarray*}
\FNorm{ A-CC^{\dagger}A } 
   &=& \FNorm{ A-\left(AS_CD_C\right)\left(AS_CD_C\right)^{\dagger} A } \\
   &\le&  \FNorm{ A-\left(AS_CD_C\right)\left(P_{A_k}AS_CD_C\right)^{\dagger} P_{A_k}A }  \quad\mbox{(where $P_{A_k}=U_{A,k}U_{A,k}^T$)}   \\
   &=& \FNorm{ A-\left(AS_CD_C\right)\left(A_kS_CD_C\right)^{\dagger} A_k } \\
   &\le& \left(1+\epsilon\right)\FNorm{ A - A A_k^{\dagger} A_k }  \quad\mbox{(by the generalized LS result)} \\
   &=& \left(1+\epsilon\right)\FNorm{ A - A_k }  .
\end{eqnarray*}
\end{Proof}

\textbf{Remark.}
This holds with constant probability, but that probability can be boosted to $1-\delta$ using standard methods.
Also, for simplicity, this is stated such that $A^{\prime} = CC^{\dagger}A$ might have rank $>k$, but actually the following even stronger result holds.
If we consider $A^{\prime\prime} = C\left(P_{A,k}C\right)^{\dagger}P_{A,k}A$, then the analysis of this theorem can also be used to show that $A^{\prime\prime}$ is a CX approximation, such that it has rank no greater than $k$ and also that it is also a $1\pm\epsilon$ relative-error approximation.

Next, let's consider how to extend this CX result to CUR decompositions.
We'll show a trivial way that gives a weaker $2+\epsilon$ constant-factor approximation, and then we'll show a non-trivial way that gives stronger $1+\epsilon$ relative-error approximation.

Here is the weaker randomized algorithm for construction CUR matrix decompositions.  
It basically amounts to calling the previous algorithm on $A$ and $A^T$ separately. 
The algorithm takes as input a matrix $A\in\mathbb{R}^m \times n$, a rank parameter $k$, and an error parameter $\epsilon$, and it returns as output matrices $C$, $U$, and $R$.
It does the following steps.
\begin{enumerate}
\item
With $c = O\left( k \log(k)/\epsilon^2 \right)$, call the previous algorithm on $A$ to get a matrix $C\in\mathbb{R}^{m \times c}$.
\item
With $r = O\left( k \log(k)/\epsilon^2 \right)$, call the previous algorithm on $A^T$ to get a matrix $R\in\mathbb{R}^{r \times n}$.
\item
Let $U=C^{\dagger}AR^{\dagger}$.
\end{enumerate}
Note that $U$ clearly has the singular value structure of the pseudo-inverse of $A$; this is true more generally, but this weaker construction makes it very obvious.
 
 Here is what we can prove regarding this weaker CUR algorithm.
 \begin{theorem}
Let $A\in\mathbb{R}^{m \times n}$, and let $k\in\mathbb{Z}^{+}$.
If we call the above algorithm with $c = O\left( k \log(k)/\epsilon^2 \right)$ and $r = O\left( k \log(k)/\epsilon^2 \right)$, then with constant probability we have that
$$
\FNorm{ A-CUR } \le \left(2+\epsilon\right) \FNorm{ A-A_k }  .
$$
\end{theorem}
\begin{Proof}
\begin{eqnarray*}
\FNorm{ A-CUR }
   &=& \FNorm{ A - CC^{\dagger} A R^{\dagger}R }  \\
   &\le& \FNorm{ A-CC^{\dagger}A } + \FNorm{CC^{\dagger}A - CC^{\dagger}AR^{\dagger}R }  \quad\mbox{(by submultiplicitivity)}  \\
   &\le& \FNorm{ A-CC^{\dagger}A } + \FNorm{A - AR^{\dagger}R }  \quad\mbox{(since $CC^{\dagger}$ only decreases the norm)}  \\
   &=& \FNorm{ A-P_C A} + \FNorm{ A-A P_R } \\
   &\le& \left(2+\epsilon\right) \FNorm{ A-A_k }.
\end{eqnarray*}
\end{Proof}

That factor of $\left(2+\epsilon\right)$ might not matter if these algorithms were applied to matrices that were really \emph{very} well approximated by a low-rank matrix, but they are often applied to matrices that are only \emph{moderately} low-rank, in which case that factor is much larger and can matter a lot.
Also, the increase is ``real'' in that the choice of columns and rows is uncoupled, which in many practical applications introduces a lot of additional error.
To remedy this, we are interested in coupling the choice of $C$ and $R$, as this will permit us to obtain $1+\epsilon$ relative-error approximation for columns and rows together.

Here is the stronger randomized algorithm for construction CUR matrix decompositions.  
It basically amounts to calling the CX algorithm to choose columns from $A$ and then calling the same CX algorithm on $C^T$ to get columns of $C^T$. 
These columns of $C^T$ are rows of $C$, and the algorithm keeps the corresponding rows of $A$.
The following algorithm takes as input a matrix $A\in\mathbb{R}^{m \times n}$, a matrix $C\in\mathbb{R}^{m \times c}$ consisting of $c$ columns of $A$, a rank parameter $k$, and an error parameter $\epsilon$, and it returns as output matrices $R\in\mathbb{R}^{r \times n}$ consisting of $r$ rows of $A$, a matrix $W\in\mathbb{R}^{c \times r}$ consisting of the corresponding $r$ rows of $C$, and a matrix $U\in\mathbb{R}^{r \times c}$.
It does the following steps.
\begin{enumerate}
\item
Compute probabilities $p_i = \frac{1}{c}\VTTNormS{ \left( U_C^T \right)^{(i)}}$, for all $i\in[m]$.
\item
Construct a sampling matrix $S_R$ and a diagonal rescaling matrix $D_R$.
\item
Construct $R=D_RS_R^TA$, consisting of a few rescaled rows of $A$, and return it.
\item
Construct $W=D_RS_R^TC$, consisting of a few rescaled rows of $C$, and return it.
\item
Let $U=W^{\dagger}$, and return it.
\end{enumerate}

Here is what we can prove regarding this algorithm
\begin{theorem}
Given matrices $A$ and $C$.
If we choose $r = O\left( c \log(c)/\epsilon^2 \right)$, then 
$$
\FNorm{ A-CUR } \le \left(1+\epsilon \right) \FNorm{ A-CC^{\dagger}A }  .
$$
\end{theorem}
\begin{Proof}
Consider the problem of approximating the solution to 
$$
\min_{X\in\mathbb{R}^{c \times n}} \FNorm{ CX-A }  
$$
by randomly sampling rows from $C$ and $A$.
Then, we have that
$$
\FNorm{ A - \underbrace{C}_C \underbrace{ \left( D_R S_R^T C \right)^{\dagger} }_U \underbrace{ D_RS_R^TA }_R } \le \left(1+\epsilon\right)\FNorm{A-CC^{\dagger}A}
$$
which establishes the result.
\end{Proof}

\textbf{Remark.}
Clearly, we can combine the two previous results, which gives us the stronger CUR matrix decomposition, as follows:
\begin{eqnarray*}
\FNorm{A-CUR} 
   &\le& \left(1+\epsilon\right) \FNorm{ A-CC^{\dagger} A } \\
   &\le& \left(1+\epsilon\right)^{2} \FNorm{ A-A_k }  \\
   &\le& \left(1+\epsilon^{\prime}\right) \FNorm{ A-A_k}  .
\end{eqnarray*}

\paragraph{Running time.}
Let's say a few words about the running time of these $\left(1+\epsilon\right)$ relative-error CX and CUR matrix decompositions.
The bottleneck to both of these algorithms is the computation of the importance sampling probabilities, which depend on the leverage scores relative to the best rank-$k$ approximation to $A$.
Thus, naively, one could spend $\Theta\left(n^3\right)$ time, computing the full SVD and use that to compute the leverage scores.
One might hope to compute an approximation to the best rank $k$ approximation to $A$ and use the leverage scores from that.
In this case, the running time of both of these algorithms boils down to the time to compute a low-rank approximation to $A$.
As we have seen, this is roughly $O\left(mnk\right)$ for deterministic iterative methods and roughly $\left(mn\log(k)\right)$ for randomized methods.
(Actually, it could be even faster, depending on the values of parameters, if one uses the input-sparsity-time projection algorithms that we are not going to be able to cover).
That basically works, meaning essentially that CX/CUR decompositions can be computed in ``random projection time''. 
This is true in theory as well as in practical implementations.
There are, however, some subtleties (that appear even for traditional deterministic algorithms for approximating subspaces) that we should point out.

The basic issue is that the problem of computing the leverage scores relative to the best rank $k$ approximation to a matrix is \emph{not} a well-posed problem.
If there is a strong eigenvalue gap assumption or if the matrix is rectangular, then it is, but it is not in general.
(This is also true for approximating subspaces, e.g., with traditional deterministic iterative methods.)
To see this, recall that a possible matrix of left singular vectors is an identity matrix, and one could have the top $k$ singular values be $1$ and the bottom $n-k$ singular values be slightly less than $1$.
In this case, if any of the singular values that is less than $1$ ``swaps'' with one of those that is equal to $1$, which could happen with a very small perturbation, then the corresponding leverage scores relative to the best rank $k$ approximation to $A$ would change completely.
To deal with this issue, we instead ask for leverage scores that are good relative to some subspace that is close to the best rank $k$ approximation to $A$.
(BTW, this is similar to the solution employed by traditional deterministic algorithms for approximating subspaces.)

Here is the definition of close subspaces.

\begin{definition}
\label{def:close-rank-k-subspaces}
Given a matrix $A\in\mathbb{R}^{m \times n}$ and a rank parameter $k \ll \min\{ m,n \}$, let $A_k$ be the best rank $k$ approximation to $A$.
Let $S_{\epsilon}$ be the set of rank $k$ matrices that are a good approximation to $A$, in the sense that 
$$
S_{\epsilon} = \left\{ X \in \mathbb{R}^{m \times n} : \mbox{rank}(X)=k \mbox{  and  } \XNorm{ A-X } \le \left(1+\epsilon\right) \XNorm{ A-A_k } \right\}  ,
$$
where $\XNorm{\cdot}$ is a matrix norm.
\end{definition}

\textbf{Remark.}
Note that the notion of closeness here depends on the norm used to measure closeness.
One obtains somewhat different results depending on whether one uses the spectral versus the Frobenius norm.

Given this definition, here is a notion of approximate leverage scores, that is approximate not only in that individual elements can be up to a factor of $\beta$ off, but also that they can be only approximate with respect to some subspace that is close to the best rank $k$ approximation to $A$.

\begin{definition}
We will call the numbers $\hat{p}_i$ (for all $ i \in [m] $) the $\beta$-approximate normalized leverage scores of $A$ relative the best rank $k$ approximation to $A$ if there exists a matrix $X \in S_{\epsilon}$ such that 
$$
\hat{p}_i \ge \frac{\beta}{k} \VTTNormS{ \left( U_X \right)_{(i)} }
\quad\mbox{and}\quad
\sum_{i=1}^{m} \hat{p}_i = 1  ,
$$
where here $U_X\in\mathbb{R}^{n \times k}$ is a matrix of left singular vectors of $X$.
\end{definition}

Here is an algorithm to approximate these scores.
Basically, it does a random projection to construct a tall matrix, and then it calls the previous fast algorithm to approximate the leverage scores of tall matrices.
This algorithm takes as input a matrix $A\in\mathbb{R}^{m \times n}$, with $\mbox{rank}(A)=\rho$, and a rank parameter $k \ll \rho$, and it returns as output a vector of numbers $\{p_i\}_{i=1}^{m}$ that is a probability distribution.
The algorithm does the following steps.
\begin{enumerate}
\item
Construct a random projection matrix $\Pi \in \mathbb{R}^{n \times 2k}$ with i.i.d. Gaussian entries.
\item
Compute the matrix $B=\left(AA^T\right)^{q}A \Pi \in \mathbb{R}^{m \times 2k}$.
\item
Compute approximations to the leverage scores of the ``tall'' matrix $B$ with the previous algorithm.
Let $\hat{\ell}_i$ be the returned approximations.
\item
Return $p_i = \frac{\hat{\ell}_i}{ \sum_{j=1}^{m} \hat{\ell}_j }$.
\end{enumerate}

Here is what we can say about this algorithm.
\begin{itemize}
\item
This algorithm computes normalized scores that are $1\pm\epsilon$ approximations to the leverage scores of the best rank $k$ approximation to $A$, with constant probability.
\item
If $q=0$, then this provides bounds with respect to the Frobenius norm notion of closeness in Definition~\ref{def:close-rank-k-subspaces}; while if $q>0$, then this can be used to provide bounds with respect to the spectral norm notion of closeness in Definition~\ref{def:close-rank-k-subspaces}.
\item
The precise theoretical statement of the running time of this algorithm is rather complex, and it depends on whether one is interested in Frobenius or spectral norm approximations of nearby subspaces from Definition~\ref{def:close-rank-k-subspaces}. 
\item
Empirically, with reasonably-good implementations, if $q=0$, then the algorithm takes roughly ``random projection time'', since that is the computational bottleneck, and while one can get reasonable reconstruction error, the actual leverage scores relative to the best rank $k$ space are poorly approximated.
If $q$ is a small integer, then the algorithm takes somewhat longer, but one gets better spectral norm bounds and one approximates the actual leverage scores relative to the best rank $k$ space quite well.
Clearly, additional iterations beyond that can be slower even than traditional deterministic methods.
See one of the sections in the long version of the Gittens-Mahoney Nystr\"{o}m paper for details on these empirical claims.
\end{itemize}

\newpage

\section{%
(10/30/2013):  Toward Randomized Low-rank Approximation in Practice}

Today, we will continue with the discussion of improved low-rank matrix approximation algorithms by describing a slightly different but much more powerful structural result that will allow us to reparameterize the low-rank approximation problem to obtain improved results both in theory and in practice.
Here is reading for today.
\begin{itemize}
\item
Lemma 2 (of arXiv-v2, or Lemma 4.2 of SODA) of: Boutsidis, Mahoney, and Drineas ``An Improved Approximation Algorithm for the Column Subset Selection Problem'' 
\item
Theorem 9.1 of: Halko, Martinsson, and Tropp, ``Finding structure with randomness: Probabilistic algorithms for constructing approximate matrix decompositions''
\end{itemize}

In particular, today we will cover the following topics.
\begin{itemize}
\item
A discussion of theory-practice gap issues in low-rank matrix approximation algorithms.
\item
A finer structural result that we will use in the next few classes to bridge that gap.
\end{itemize}

\subsection{Some Challenges for Low-rank Matrix Approximation in Practice}

As with the LS problem and algorithms, here we also want to understand how these theoretical ideas for randomized low-rank matrix approximation can be used in practice.
As we will see, just as with the LS problem and algorithms, the basic ideas do go through to practical situations, but some of the theory must be modified in certain ways.
Among the issues that will come up for the randomized low-rank approximation situation are the following.
\begin{itemize}
\item
It might be too expensive to sample $O\left(\frac{k\log(k)}{\epsilon^2}\right)$ rows/columns, and it might be difficult to do so if the constant in the big-O is left unspecified.
Instead, we might want to choose exactly $k$, or we might want to choose $k+p$, where $p$ is a small integer such as $5$ or $10$.
\item
In many applications, and in particular in those that are particularly interested moderate- to high-precision low-rank matrix approximation, e.g., numerical analysis and scientific computing applications, there are other goals of interest.
For example, given a good approximation to an orthogonal basis $Q$ approximating $A$, one might want to find other types of matrix decompositions (e.g., various QR decompositions, thin SVDs, interpolative decompositions,~etc.).
\item
One might want to parameterize problems/algorithms in terms of fixed rank version (where the input is a rank parameter, which is the approach we have taken), or one might want to parameterize problems/algorithms in terms of a fixed precision version (roughly, fix a pre-specified precision level, e.g., near machine precision, and look for an approximation that provides that numerical error).
\item
If the spectrum decays somewhat slowly but not very slowly, then one might be interested in doing some sort of power iteration, which will help the spectrum to decay more quickly, and it might be of interest to incorporate this process directly into the algorithm.
\item
Rather than asking for a priori worst-case error bounds, one might be interested in doing a posteriori error estimation and deciding whether to continue with the algorithm based on the output of that estimation procedure.
\end{itemize}

We will briefly describe all of these issues---many of the issues are similar to those that arose when we discussed how RandNLA algorithms for the LS problem work in practice, but here we are considering the low-rank matrix approximation problem---but before we do that, let's give a more refined structural result.
This result gives improved results in general; and, in particular, it makes it easier to perform these extensions.

\subsection{A More Refined Structural Result for Low-rank Approximation}

Recall that when we discussed the LS problem, we described a deterministic structural result, and then we showed how random sampling and random projections interface to that result.
Moreover, how the randomization interfaced to that structure differed for algorithms that obtained the best results in worst-case theory versus those that obtained the best results in practice.
For the versions of the low-rank approximation problem that we discussed in the last class, i.e., the $1\pm\epsilon$ relative-error sampling and projection algorithms, we just related them to the LS problem.
Thus, we really didn't take into account the low-rank structure, e.g., how the top and bottom subspaces of the input matrix interacted, in a particularly refined way.
The reason was two-fold: (1) we were only interested in how the sample reproduced the top part of the spectrum and the top subspace of the matrix; and (2) we were willing to oversample to a level sufficient to obtain worst-case bounds. 
If we are interested in obtaining more refined results, as is common in practice, then we need a more refined structural result that takes into account how the top and bottom part of the spectrum of a matrix interact.

To do that, observe that there are actually two related ways that we can break up the generalized LS problem.
Given a matrix $A\in \mathbb{R}^{m \times n}$, where $\mbox{rank}(A)=k$ and a matrix $B\in\mathbb{R}^{m \times p}$, consider the generalized LS problem:
$$
\mbox{argmin}_{X\in\mathbb{R}^{n \times p}} \XNorm{ Z^TAX-Z^TB } = \left(Z^TA\right)^{\dagger}Z^TB  ,
$$
where $Z^TU$ is full rank (i.e., the rank $=k$).
Then, we can split up the expression $\XNorm{ A X_{opt}-B }$ in one of two ways.
\begin{itemize}
\item
\begin{eqnarray*}
\XNorm{ A \left(Z^TA\right)^{\dagger}Z^TB - B } 
   &\le& \XNorm{ U^{\perp}{U^{\perp}}^{T} B }  \\
   &+& \XNorm{ U^TZZ^TU^{\perp}{U^{\perp}}^{T}B } \\ 
   &+& \max_i \left|  \sigma_i\left(Z^TU\right) - \sigma_i^{-1}\left(Z^TU\right) \right| \XNorm{ Z^TU^{\perp} {U^{\perp}}^{T} B }.
\end{eqnarray*}
\item
\begin{eqnarray*}
\XNorm{ A \left(Z^TA\right)^{\dagger}Z^TB - B } 
   \le \XNorm{ U^{\perp}{U^{\perp}}^{T} B } + \XNorm{ \left(U^TZ\right)^{\dagger}Z^TU^{\perp}{U^{\perp}}^{T} B }.
\end{eqnarray*}
\end{itemize}
Note that these two correspond to a generalization of the two related ways that we proved the tall LS result.
Here, though, the two different ways to split up this expression will lead to two different structural results.
One is the immediate generalization of the LS result that can be used to get $(1+\epsilon)$ relative-error bounds on the top part of the spectrum that we saw in the last class.
The other can be used to do that, but it is more general; in particular, it can be used to get a more refined structural result that leads to better algorithms for the CSSP as well as for random projection algorithms with very aggressive downsampling.

The main issue is that the generalized LS algorithm we had assumes that the matrix is exactly rank $k$, which essentially means that it is exactly rectangular and just artificially fat.
Then, we applied it to arbitrary matrices by carefully wedging projection matrices at various places, but the consequence of this is that we only got control on the top part of the spectrum.
Now, let's do better by getting a structural result that says how the sampling operator interacts with both the top and bottom part of the spectrum.
This structural result will hold for any sketching/sampling/projection matrix, and the randomness will enter only through it, so in that sense it will decouple the linear algebraic structure from the randomness.

Here is the basic setup.
Let $A\in\mathbb{R}^{m \times n}$, and let its SVD, $A=U \Sigma V^T$, be represented as
\begin{equation*}
A = U
\left( \begin{array}{cc} \Sigma_1 & 0 \\
                         0 & \Sigma_2
       \end{array}
\right)
\left( \begin{array}{c} V_1^T \\ V_2^T \end{array} \right)
\end{equation*}
where $\Sigma_k$ is the $k \times k$ diagonal matrix consisting of the top $k$ singular values, $\Sigma_2$ is the $(\min\{m,n\}-k)\times(\min\{m,n\}-k)$ diagonal matrix consisting of the bottom $\min\{m,n\}-k$ singular values, $V_1^T$ and $V_2^T$ are the matrices of the associated singular vectors, etc.
(Note that we are using subscripts differently/inconsistently with respect to how we used them before, as well as how we will use them later; here, ``$1$'' and ``$2$'' refer to the top and bottom part of the spectrum, respectively.)

In this case, assume that we have the sketching matrix $S \in \mathbb{R}^{\ell\times k}$, which could be a sampling or projection or some other matrix, and where $\ell \ge k$.
For example, $\ell=k$ or $\ell = k+p$ for $p=5$ or $p=10$, or $\ell = O\left( k\log(k)/\epsilon^2\right)$ are three regimes of particular interest to us. 
Then, we can define
\[
\left\{ \begin{array}{l l}
                    \Omega_1 = V_1^TS \\
                    \Omega_2 = V_2^TS
                 \end{array}
         \right.  
\]
to be the perturbed version of the singular subspaces.
To obtain good low-rank matrix approximation, we will want to control the singular subspaces of $\Omega_1$ and $\Omega_2$.
In the absence of sketching, they are orthogonal, i.e., $V_2^TV_1=0$, and thus we will want to show that the sketched versions of the subspaces are approximately orthogonal.
This is different than before, where we just needed to show that 
\[
\TNorm{ \Omega_1\Omega_1^T - I } = 
\TNorm{ V_1^TSS^TV_1 - I } < 1/2 ,
\]
i.e., that the top part of the subspace is well-behaved.
That is, here we want to control both the top and bottom part of the spectrum as well as how they interact with each other via the sketching matrix $S$.

To do this, let $C=AS$, in which case we can write
\begin{equation*}
C 
= U
\left( \begin{array}{c} \Sigma_1V_1^TS \\ \Sigma_2V_2^TS \end{array} \right) 
= U
\left( \begin{array}{c} \Sigma_1\Omega_1^T \\ \Sigma_2\Omega_2^T \end{array} \right)   ,
\end{equation*}
where $\Sigma_1V_1^TS$ is $k \times \ell $ and $\Sigma_2V_2^TS$ is $(n-k) \times \ell$.
(Note that $C$ does not need to be actual columns, unless $S$ is a sampling matrix, but instead it is any sketch of the columns.)

If $Q$ is an orthonormal basis for the range of $C$ (in this discussion, we are \emph{not} filtering through the best rank $k$ approximation to $C$, which corresponds to the ``easier'' situation before), then $QQ^T = P_C$, and we want to bound
\[
\XNorm{ A-QQ^TA } = \XNorm{ \left(I-P_C\right) A }  .
\]

One can then prove the following, which is our main structural result for low-rank matrix approximation via randomized algorithms.

\begin{theorem}
Given the above setup, then assuming that $\Omega_1 = V_1^TS$ has full rank, then 
\[
\XNorm{ \left(I-P_C\right) A } \le \XNorm{ A-A_k } + \XNorm{ \Sigma_2 \Omega_2 \Omega_1^{\dagger} }  ,
\]
where $\Omega_1 = V_1^TS$ and $\Omega_2 = V_2^TS$.
\end{theorem}

\textbf{Remark.}
This structural result was first established and proven by Boutsidis et al. in the context of the Column Subset Selection Problems, and it was reproved with more complicated methods by Halko et al. in the context of parameterizing random projection algorithms for high-quality implementations. 
Gittens, Gu, and several others have used it since then in one form or another.
That and other prior work which used this structural result only established if for the spectral and Frobenius norms, but it actually holds for any unitarily-invariant norm.
This result is due to Drineas and Mahoney, but we haven't published it yet, so I'll include it here.

\textbf{Remark.}
The $\Omega_2 \Omega_1^{\dagger}$ term describes the interaction between the top and bottom part of the spectrum. 
The ``unsketched'' version of this is $V_2^T{V_1^T}^{\dagger} = V_2^TV_1 = 0$, in which case there is no interaction between these orthogonal subspaces.

\textbf{Remark.}
The assumption that $\Omega_1$ is full rank is very nontrivial.
Indeed, the entire point of using leverage-based sampling or random projections for the overdetermined LS problem is to ensure that.
Here, it holds for worst-case input if we use leverage-based sampling or if we use random projections, with parameters set appropriately.
Of course, if one can do an after-the-fact check to confirm that it is true (which is what one often does in practice), then one can use this theorem.

\begin{Proof}[of theorem]
First note that 
\begin{equation}
\label{eqn:lr-struct-eq1}
\XNorm{ A - P_C A } = \XNorm{ A - AS \left(AS\right)^{\dagger} A }
\end{equation}
and also that
\begin{equation}
\label{eqn:lr-struct-eq2}
\left(AS\right)^{\dagger} = \mbox{argmin}_{X\in\mathbb{R}^{k \times n}} \XNorm{ A-ASX }
\end{equation}
and also that these two results hold for any unitarily invariant matrix norm.
So, we can replace $\left(AS\right)^{\dagger}$ in (\ref{eqn:lr-struct-eq1}) with any other $k \times n$ matrix and replace the equality ($=$) with an inequality ($\le$).
In particular, we will replace $\left(AS\right)^{\dagger}A$ with $\left(A_kS\right)^{\dagger}A_k$.
Doing this, we get the following.
\begin{eqnarray*}
\XNorm{ A-P_CA } 
   &=& \XNorm{ A - AS\left(AS\right)^{\dagger} A }   \\
   &\le& \XNorm{ A - AS\left(A_kS\right)^{\dagger} A_k }   \\
   &=& \XNorm{ A -  A_k + A_k - \left(A-A_k+A_k\right)S\left(AS\right)^{\dagger} A }   \\
   &\le& \underbrace{ \XNorm{ A_k - A_kS\left(A_kS\right)^{\dagger}A_k } }_{\gamma_1} + \underbrace{ \XNorm{A-A_k} }_{\gamma_2} + \underbrace{ \XNorm{ \left(A-A_k\right)S\left(A_kS\right)^{\dagger}A_k } }_{\gamma_3}  .
\end{eqnarray*}
Let's bound each of those three terms.
Since $\gamma_2$ is simply $\XNorm{A-A_k}$, we'll bound the other two terms.
First, bound $\gamma_1$ as follows:
\begin{eqnarray*}
\gamma_1
   &=& \XNorm{ A_k - A_k S \left( A_kS \right)^{\dagger} A_k }  \\
   &=& \XNorm{ A_k - A_k S \left( U_k \Sigma_k V_k^T S \right)^{\dagger} A_k }  \\
   &=& \XNorm{ A_k - A_k S \left( V_k^T S \right)^{\dagger} \left( U_k \Sigma_k  \right)^{\dagger} A_k }  \quad\mbox{(since both $V_k^TS$ and $U_k\Sigma_k$ are full rank)}  \\
   &=& \XNorm{ A_k - U_k\Sigma_k \underbrace{V_k^T S \left( V_k^T S \right)^{\dagger}}_{I_k} \underbrace{\left( U_k \Sigma_k  \right)^{\dagger} U_k \Sigma_k}_{I_k} V_k^T }  \\
   &=& \XNorm{ A_k - U_k \Sigma_k V_k^T }  \\
   &=& 0  .
\end{eqnarray*}
Next, bound $\gamma_3$ as follows:
\begin{eqnarray*}
\gamma_3
   &=& \XNorm{ \left(A-A_k\right)S\left(A_kS\right)^{\dagger}A_k } \\
   &=& \XNorm{ \left(A-A_k\right)S\left(U_k\Sigma_kV_k^TS\right)^{\dagger}A_k } \\
   &=& \XNorm{ \left(A-A_k\right)S\left(V_k^TS\right)^{\dagger}\left(U_k\Sigma_k\right)^{\dagger}A_k } \quad\mbox{(since both matrices are full rank)} \\
   &=& \XNorm{ U_{k,\perp}\Sigma_{k,\perp}V_{k,\perp}^{T}S\left(V_k^TS\right)^{\dagger} \underbrace{\left(U_k\Sigma_k\right)^{\dagger}U_k\Sigma_k}_{I_k}V_k^T } \\
   &=& \XNorm{ U_{k,\perp}\Sigma_{k,\perp}V_{k,\perp}^{T}S\left(V_k^TS\right)^{\dagger} V_k^T } \\
   &\le& \XNorm{ \Sigma_{k,\perp}V_{k,\perp}^{T}S\left(V_k^TS\right)^{\dagger} }. \quad\mbox{(since the orthogonal matrices can be dropped)}
\end{eqnarray*}
Here, we use $U_{k,\perp}$, $\Sigma_{k,\perp}$, $V_{k,\perp}$ to refer to the parts of $U$, $\Sigma$, and $V$ that are orthogonal to the best rank $k$ approximation to $A$.

The theorem then follows.
\end{Proof}

Two final remarks.
First, you can prove the generalization of this result to when there is a square on the norm using more sophisticated methods.
Second, you can prove the generalization of this result to when $A_k = AV_kV_k^T$ is replaced with $AYY^T$ is any approximation to $V_k$.
This is of interest, since one can then choose $Y$ to be any approximation to $V_k$, e.g., one constructed with a random projection algorithm.

\newpage

\section{%
(11/04/2013):  Randomized Low-rank Approximation in Practice, Cont.}

Today, we will continue with the discussion from last time.
Again, from last time, here is the reading for today.
\begin{itemize}
\item
Boutsidis, Mahoney, and Drineas ``An Improved Approximation Algorithm for the Column Subset Selection Problem'' 
\item
Halko, Martinsson, and Tropp, ``Finding structure with randomness: Probabilistic algorithms for constructing approximate matrix decompositions''
\end{itemize}
Recall that last time we described a more refined structural result for low-rank matrix approximation.
Today and next class we will describe how to use it for two seemingly very different types of low-rank matrix approximation.

In particular, today we will cover the following topics.
\begin{itemize}
\item
The Column Subset Selection Problem.
\end{itemize}

\subsection{Column Subset Selection Problem}

Last time, we talked about a basic structural result for low-rank matrix approximation via RandNLA algorithms, and we said how it was crucial for obtaining improved results in theory and in practice when one wants to sample $o(k \log(k))$, e.g., exactly $k$ or $k+p$ for a small positive integer $p$, number of columns.
The first example of this---historically as well as in terms of what we will do this semester---has to do with the so-called Column Subset Selection Problem (CSSP).

The CSSP is the problem of choosing the best set of \emph{exactly} $k$ columns from an input matrix $A$.
Importantly, this problem is intractable (in the sense of TCS complexity theory) for nearly every interesting formalization of best, e.g., to get columns that capture the most mass, to get columns that are maximally uncorrelated, to get columns that span a parallelepiped of maximum volume, etc.
The reason is that asking for actual columns is in many ways very different than asking for the set of linear combinations of columns that are best by any of these notions.
The leading eigenvectors or singular vectors are best in terms of maximizing variance, being maximally uncorrelated, etc, and this is fundamental to linear algebra and Euclidean spaces, but asking for the best set of actual columns is a much more combinatorial problem.
Here, we will consider the following version of the~CSSP.

\begin{definition}
Given $A\in\mathbb{R}^{m \times n}$ and $k\in\mathbb{Z}^{+}$, choose the $k$ columns of $A$ to form the matrix $C\in\mathbb{R}^{m \times k}$ such that the residual $\XNorm{A-P_CA}$, where (say) $\xi \in\{2,F,*\}$, is minimized over all ${n \choose k}$ choices for $C$.
\end{definition} 

This is a nice version of the problem for us for the following reasons.
Basically, it is asking to capture mass on the top part of the spectrum, so it is doing what TCS-based versions of the low-rank approximation problem are doing, but it is also asking for exactly $k$ columns, which is doing something very different and more like the usual NLA approach.
Relatedly, we can use it to highlight several of the differences between NLA and TCS perspectives.
Let's do that now.

In NLA:
\begin{itemize}
\item
The algorithms have historically been almost exclusively deterministic, typically greedy (depending on things like pivot rule decisions, etc.) algorithms.
\item
There are strong connections between the CSSP and QR, RRQR, etc. algorithms, and one typically chooses exactly $k$ columns with these procedures.
\item
There is a strong emphasis on conditioning, backward error analysis, and constant factors in the running time.
\item
Most of the effort focuses on getting good spectral norm bounds.
\end{itemize}

In TCS:
\begin{itemize}
\item
There has been a long tradition of randomization used inside the algorithm, in general, with RandNLA projection algorithms, and also for problems like CX/CUR that have some similarities with the CSSP.
\item
One typically keeps more than $k$ column, e.g., $O\left(k \log(k)/\epsilon^2\right)$ columns, where the big-O hides sometimes unknown factors.
\item
Most of the effort focuses on getting good Frobenius norm bounds.
\end{itemize}

Importantly, it isn't obvious how to combine these two very different approaches.
For example, when running typical NLA algorithms, if one looks at the details of the pivot rule decisions, etc., then it isn't clear from the analysis that the bounds will improve if one keeps a few extra columns.
Alternatively, when running typical TCS algorithms, one often can't get worst case bounds by keeping fewer columns, typically for rather basic reasons such as reduction to the coupon collector problem.

Here, we will describe an algorithm for the CSSP that combines these two perspectives in a nontrivial manner.
It is a two-stage algorithm, where in the first stage one chooses a random sample with a TCS-style approach, and where in the second stage one cuts back to exactly $k$ columns by running a QR on the chosen columns.
(Actually, there is an important subtlety, where the QR is done on the sampled version of the matrix of right singular subspace, and where the corresponding columns from the original matrix are kept.)

Here is the algorithm.
The following algorithm takes as input a matrix $A\in\mathbb{R}^{m \times n}$ and a number $k \in\mathbb{Z}^{+}$, and it returns as output a matrix $C\in\mathbb{R}^{m \times k}$ consisting of exactly $k$ columns of $A$.
\begin{enumerate}
\item
Initial stage
\begin{enumerate}
\item
Compute (exactly or approximately) the top $k$ right singular vectors of $A$, call them $V_k$.
\item
From that orthogonal basis, compute (exactly or approximately) the importance sampling probabilities $\{p_i\}_{i=1}^{n}$, where
\[
p_i = \frac{1}{k}\VTTNormS{ \left( V_k \right)_{(i)} } .
\]
\item
Let $c=\Theta\left( k \log(k) \right)$ (where there is no dependence on $\epsilon$, e.g., set $\epsilon=1/2$ in the previous sampling algorithm).
\end{enumerate}
\item
Randomized Stage
\begin{enumerate}
\item
For $t\in[c]$, choose an integer from $[n]$ with probability $p_i$, and if $i$ is chosen then keep the scaling factor $1/\sqrt{cp_i}$.
Form the sampling matrix $S_1$ and the diagonal rescaling matrix~$D_1$.
\end{enumerate}
\item
Deterministic Stage
\begin{enumerate}
\item
Run a deterministic QR algorithm, e.g., an algorithm of Pan or the Gu-Eisenstat algorithm, on $V_k^TS_1D_1$ to get exactly $k$ columns from $V_k^TS_1D_1$, thereby forming the sampling matrix $S_2$.
\item
Return the corresponding $k$ columns of $A$, i.e., return $C=AS_1S_2$.
\end{enumerate}
\end{enumerate}

Here are several comments regarding this algorithm.
\begin{itemize}
\item
For simplicity, we described this algorithm in terms of the exact right singular vectors.
The proof of the main quality-of-approximation theorem for this algorithm uses the main structural result for low-rank approximation, and that result is robust to using an approximate basis for the right/left singular subspace.
Thus, the running time of this algorithm could be improved by approximating that subspace, say, with a random projection algorithm.
\item
The running time bottleneck for this algorithm is the time to compute $V_k^T$, either exactly or approximately.
\item
The sampling probabilities given in the algorithm work for the Frobenius and Trace norms.
For the spectral norm, one should use exact or approximate probabilities of the form
\begin{eqnarray*}
p_i
   &=& \frac{ \frac{1}{2} \VTTNormS{ \left( V_k \right)_{(i)} } }{ \sum_{j=1}^{n} \VTTNormS{ \left( V_k \right)_{(j)} } } + \frac{ \frac{1}{2} \VTTNormS{ \left( A-A_k \right)^{(i)} } }{ \sum_{j=1}^{n} \VTTNormS{ \left( A-A_k \right)^{(j)} }  } \\
   &=& \frac{ \VTTNormS{ \left( V_k \right)_{(i)} } }{ 2k } + \frac{ \VTTNormS{ A^{(i)} } - \left( AV_kV_k^T \right)^{(i)} }{ 2 \left( \FNormS{A}-\FNormS{AV_kV_k^T} \right) }  .
\end{eqnarray*}
While this looks complicated, these probabilities can computed from only the top part of the spectrum of $A$, i.e., a knowledge of $V_k$ (exactly or approximately) suffices to compute them.
It is an open question whether these more complicated probabilities are necessary or simply a weakness of the analysis.
\item
Note that since the algorithm returns exactly $k$ columns, there is no need to worry about rescaling and thus the algorithm doesn't do that.
\end{itemize}

Here is the main theorem that we can prove about this algorithm.
\begin{theorem}
Let $A\in\mathbb{R}^{m \times n}$, and let $k\in\mathbb{Z}^{+}$.
If we run the above CSSP algorithm, then with constant probability, the following hold.
\begin{eqnarray*}
\TNorm{ A-P_CA } &\le& \Theta\left( k \log^{1/2}(k)\right) \TNorm{A-A_k} + \Theta\left(k^{3/4}\log^{1/4(k)}\right)\FNorm{A-A_k} \\
\FNorm{ A-P_CA } &\le& \Theta\left( k\log^{1/2}(k)\right)\FNorm{A-A_k} \\
\|A-P_CA \|_{*} &\le& \|A-A_k\|_{*} + \Theta\left( k^{3/2}\log^{1/2}(k)\right)\FNorm{A-A_k}.
\end{eqnarray*}
\end{theorem}

\begin{Proof}
We will prove the theorem by establishing a sequence of lemmas.
\begin{lemma}
Given $S_1$ and $D_1$ from the algorithm, with constant (say, at least $0.9$) probability, 
\[
\sigma_k\left(V_k^TS_1D_1\right) \ge \frac{1}{2}  .
\]
In particular, $V_k^TS_1D_1$ has full rank.
\end{lemma}
\begin{Proof}[of lemma]
To bound $\sigma_k\left(V_k^TS_1D_1\right)$, we bound
\[
\TNorm{  V_k^TS_1D_1 D_1S_t^T V_k^T - I_k }  ,
\]
showing that it is $\le 1/2$.
Note that our sampling probabilities that are used by the algorithm are approximately optimal for this, i.e., $p_i \ge \frac{1}{2k} \VTTNormS{ \left( V_k^T \right)_{(i)} }$, and so we have that $\beta=1/2$.
We can set $\epsilon=1/2$, and the lemma follows by the approximate matrix multiplication theorem.
\end{Proof}

(Note that the proof of that lemma is essentially the same argument that we used for over-determined regression problems, except that there we were sampling rows, using the leverage scores defined by the column space, while here we are sampling columns, so we use leverage scores defined by the best rank $k$ approximation to the row space.)

\begin{lemma}
\[
\XNorm{A-P_CA} \le \XNorm{A-A_k} + \sigma_k^{-1}\left(V_k^TS_1D_1\right) \XNorm{ \left(A-A_k\right)S_1D_1 }.
\]
\end{lemma}
\begin{Proof}[of lemma]
First, observe that since we are projecting onto exactly $k$ columns, rescaling doesn't matter, and so we can rescale or not as convenient.
Next we have that
\begin{eqnarray*}
A-P_CA 
   &=& A - AS_1S_2 \left(AS_1S_2\right)^{\dagger} A  \\
   &=& A - AS_1D_1S_2 \left(AS_1D_1S_2 \right)^{\dagger} A \\
   &=& A - AS \left(AS \right)^{\dagger} A  ,
\end{eqnarray*}
where $S = S_1D_1S_2 \in \mathbb{R}^{n \times k}$ is a sketching matrix representing the two steps of the algorithm.
By the main structural lemma for low-rank approximation, we have that
\begin{eqnarray*}
\XNorm{ A-P_C A } 
   &\le& \XNorm{A-A_k} + \XNorm{ \Sigma_{k,\perp} V_{k,\perp} S \left( V_k^TS \right)^{\dagger} }  \\
   &=& \XNorm{A-A_k} + \XNorm{ U_{k,\perp}\Sigma_{k,\perp} V_{k,\perp} S \left( V_k^TS \right)^{\dagger} }  \\
   &=& \XNorm{A-A_k} + \XNorm{ \left(A-A_k\right) S \left( V_k^TS \right)^{\dagger} }  \\
   &\le& \XNorm{A-A_k} + \XNorm{ \left(A-A_k\right) S } \TNorm{ \left( V_k^TS \right)^{\dagger} }  \quad\mbox{(by strong submultiplicitivity)}   \\
   &\le& \XNorm{A-A_k} +  \sigma_k^{-1}\left(V_k^TS_1D_1\right) \XNorm{ \left(A-A_k\right)S_1D_1 }  .
\end{eqnarray*}
The last line follows since $S=S_1D_1S_2$, since $S_2$ is orthogonal, and since $V_k^TS_1D_1$ has full rank.
\end{Proof}

\begin{lemma}
With constant probability (say, $\ge 0.9$), we have that $\sigma_k^{-1}\left(V_kS_1D_1S_2\right) \le 2\sqrt{2k\left(c-k\right) + 1}$.
\end{lemma}
\begin{Proof}[of lemma]
This follows since we call the analysis of the Gu-Eisenstat QR routine.
\end{Proof}

(We would get a different result here if we used a different QR algorithm, other than that of Gu-Eisenstat, at that step of the algorithm.)

\begin{lemma}
With constant probability (say, $\ge 0.9$), we have that
\begin{eqnarray*}
\TNorm{  \left(A-A_k\right)S_1D_1 } &\le& \TNorm{A-A_k} + \frac{12}{c^{1/4}}\FNorm{A-A_k}  \\
\FNorm{ \left(A-A_k\right)S_1D_1 } &\le& \sqrt{10}\FNorm{A-A_k}  \\
\| \left(A-A_k\right)S_1D_1 \|_{*} &\le& \sqrt{10c} \FNorm{A-A_k}  .
\end{eqnarray*}
\end{lemma}
\begin{Proof}[of lemma]
This is straightforward, so we will omit.
\end{Proof}

The theorem follows by combining the results from these lemmas.
\end{Proof}

\newpage

\section{%
(11/06/2013):  Randomized Low-rank Approximation in Practice, Cont.}

We continue with the discussion from last time.
There is no new reading, just the same as last~class.

In particular, today we will cover the following topics.
\begin{itemize}
\item
Practical random projection algorithms.
\end{itemize}

\subsection{Practical Random Projections}

Let's revisit random projections in light of the main structural result for low rank approximation.
We will be particularly interested in modifications to what we have been discussing to make random projection algorithms implementable.
As a trivial but important point, we can't write code to perform a loop over a dummy variable from $1$ to $O\left(k\log(k)/\epsilon^2\right)$, where the constant in the big-O is left unspecified, and we don't want to choose the upper index arbitrarily.
Also, in practice, we don't want to be quite so cavalier about constant factors as we typically are in theory.
As it turns out, the same structural result that is used in the CSSP can also be used to parameterize random projection algorithms in a manner that make them more easily implementable, and it also provides finer control over several other issues of interest in practice.
We turn to that now.

Recall that random projections can be used to get additive-error low-rank approximations.
At root, this was accomplished by applying subspace JL or approximate matrix multiplication ideas to the columns/rows of $A$.
Relatedly, observe that random projections rotate the input data to a random basis where the squared-norms of the columns are approximately uniform, and thus where uniform sampling can be used.
We also saw that by using a more sophisticated importance sampling distribution the random sampling algorithms can lead to relative-error guarantees, when parameters are chosen appropriately.
In addition, since it was in the homework, we didn't go through the details in class, but random projection algorithms also lead to relative-error low-rank approximation.
If you recall, at root this was accomplished by applying subspace JL or approximate matrix multiplication ideas to the rows of the truncated singular vectors.
Thus, both sampling and projection algorithms can be improved from additive-error to relative-error---the former are improved by improving the algorithm, and the latter are improved by improving the analysis.
In particular, the latter happened since random projections uniformize a lot of things (both row norms squared as well as leverage scores), and thus the original analysis was weak.

Given this, one might wonder whether random projections uniformize other things related to the basic structural result underlying low-rank matrix approximation.
Not unrelatedly, in NLA and scientific computing, one often thinks of $\epsilon$ as machine precision, and one is more interested in obtaining results with respect to the spectral norm than with respect to the Frobenius norm or trace norm, and one often uses iterative algorithms to accomplish this.
As with iterative RandNLA LS solvers, here too for low-rank matrix approximation, we can couple basic random projections with traditional numerical methods to lead to good implementations.
There are several other related issues that we will discuss today, but the best implementations exploit the basic structural result in important ways.
We will review this here.

Most high-quality implementations for low-rank matrix approximation have been done in the context of scientific computing, where the end goal is to obtain a basis.
The reason for that is that, once one has an exact or approximate basis $Q$ for the range space of the matrix (think: operator), one can do a lot of other things of interest with it.
(It is really still unresolved whether this will be as useful a primitive in building a foundation of linear algebraic code for machine learning and data analysis applications, since in many of those applications the requirements are very different.)
Here are several matrix decompositions of interest that can be computed with such a basis.
\begin{itemize}
\item
Pivoted QR decomposition.
\item
Eigenvalue decomposition.
\item
SVD.
\end{itemize}
All of these deal with the (numerical) rank/range of a matrix.
Often one uses a \emph{truncated} form of these decompositions to get low-rank matrix approximations of the general form
\[
\underbrace{A}_{m \times n} \approx \underbrace{B}_{m \times k} \underbrace{C}_{k \times n}  ,
\]
where $k$ is the numerical rank.
(Here, $B$ and $C$ are some matrices of the given dimension, with no necessary relationship to columns of rows or the way we used these letters before.)
If $k \ll \min\{m,n\}$, then these decompositions allow $A$ (essentially \emph{all} of $A$, with no loss, since $k$ is often chosen to be the the numerical rank) to be stored cheaply and/or to be multiplied quickly with other vectors/matrices.

One way these algorithms and decompositions are used in scientific computing is via a two step procedure as follows.
\begin{enumerate}
\item
Step 1:
Compute an approximate basis for the range of $A$, i.e., compute an orthogonal matrix $A$ such that $A\approx QQ^TA$ (where, again, the notion of approximation captured by $\approx$ is often very strong, e.g., that they are the same up to numerical precision).
\item
Step 2: 
Given this orthogonal matrix $Q$, use $Q$ to compute QR, SVD, etc. for $A$.
\end{enumerate}

As before with the LS problem, where we did one of two general approaches (use the sketch directly by solving a subproblem on the sketch, or use the sketch indirectly to construct a preconditioner to solve the original problem), so too here for the low rank approximation problem we can either work directly with the sketch (what we have been discussing so far) or use the sketch to couple to other traditional numerical algorithms (as we will describe now).
Here are several differences to keep in mind.
\begin{itemize}
\item
We will allow $Q$ to have ``extra'' dimensions, i.e., if we want to have a target numerical rank of $k$, then we will allow $Q$ to have $\ell=k+p$, where $p$ is a small positive integer, columns, and we will not ``filter'' through the rank $k$ space.
This corresponds to the ``easier'' case we saw before of not filtering the low-rank approximation through an exactly rank $k$ space.
Importantly, though, keeping those extra dimensions will provide us with big gains in ways that aren't immediately obvious.
\item
Since we will call traditional iterative algorithms on fat matrices, we need to be more careful about how the top and bottom parts of the spectrum interact, i.e., how the subspaces corresponding to the top and bottom parts of the spectrum interact.
In particular, this means that we need to go beyond just reproducing the error with respect to the top of the spectrum, instead taking into account the other criteria we discussed.
The basic structural result will make this easier.
\end{itemize}

An issue we will discuss now is problem parameterization.
(We saw an analogous issue of problem parameterization with the LS algorithms.)
\begin{itemize}
\item
In TCS, it is more common to do a \emph{fixed rank $k$ approximation}, where one fixes the rank to be $k$ and then asks for the best or a good approximation with respect to that value of $k$.
For example: given a matrix $A\in\mathbb{R}^{m \times n}$, and numbers $k\in\mathbb{R}^{+}$ and $p\in\mathbb{Z}^{+}$, construct a matrix $Q\in\mathbb{R}^{m \times(k+p)}$ such that 
\[
\XNorm{ A-QQ^TA } \approx \min_{\mbox{rank}(B)\le k} \XNorm{A-B}  ,
\] 
where $\XNorm{\cdot}$ is some norm, e.g., the spectral or Frobenius or trace norm.
In TCS, $k$ is assumed to be part of the input.
In machine learning and data analysis applications, it is typically determined by some sort of model selection rule.
In either case, there is no expectation that it captures the full numerical rank of the matrix.
\item
In NLA and scientific computing, it is more common to do a \emph{fixed precision $\epsilon$ approximation}, where one fixed the acceptable value of the error, e.g., to be machine precision, and then chooses $k$ to get below that error level.
For example: given a matrix $A\in\mathbb{R}^{m \times n}$ and an error parameter  $\epsilon > 0$, construct a matrix $A$ consisting of $k=k(\epsilon)$ columns such that 
\[
\XNorm{A-QQ^TA} \le \epsilon  ,
\]
where in these applications (especially if $\epsilon$ is set to machine precision) $\XNorm{\cdot}$ refers typically to the spectral norm.
\end{itemize}
One way to deal with the difference in problem parameterization is to see that algorithms for the fixed rank approximation can often be adapted to solve the fixed precision approximation problem.
Importantly, though, just as we saw with the LS problem, the precise form of the original worst-case bounds typically do \emph{not} go through to the new problem parameterization, but it is typically the case that some variant of those bounds can be established.

That being said, the basic idea to convert a fixed rank approximation problem to a related fixed precision approximation problem is to build the basis $Q$ incrementally (which explains the difficulty of obtaining worst-case bounds).
Here is an example of such an algorithm.
\textsc{AlgorithmFixedRank} takes as input $A$, $k$, and $\ell$ or $p$.
Then, it does the following.
\begin{itemize}
\item
Let $\Pi$ be a random projection matrix consisting of Gaussian random variables.
\item 
Let $C=A\Pi \in\mathbb{R}^{m \times \ell}$.
\item
Form $Q\in\mathbb{R}^{m \times \ell}$, an orthonormal basis for the span of $C$.
\end{itemize}
It is not essential that $\Pi$ consists of Gaussian random variables, but among scientific computing implementations it is most common, so we have stated it that way.
(All the expected consequences follow---it is worse on worst-case FLOPS, it may or may not be faster on particular matrices depending on the size, aspect ratio, pipelining issues, etc.---but it is more convenient to work with when the matrix $A$ is only implicitly represented and/or can be applied quickly to an arbitrary vector.)
One reason it is common to use Gaussian random variables is that in very successful scientific computing applications of RandNLA, e.g., geophysics and certain areas of PDEs, one only has an implicit representation of the matrix $A$, but that representation can be quickly applied to arbitrary vectors, and thus much of the benefit of input-sparsity-time or fast Hadamard-based projections is lost.

To make this work, one uses the fixed precision $\epsilon$ approximation, and thus one constructs the basis incrementally.
HMT describe a probabilistic error estimator (that was introduced by LWMR, WLR, TR) that can be used as part of the incremental iteration.
If the exact approximation error is $\TNorm{\left(I-QQ^T\right)A}$, then the algorithm is the following.
\begin{itemize}
\item
Draw a sequence of $r$ $N(0,1)$ random vectors, call them $g^{(i)}$, where $r$ is chosen empirically to balance the tradeoff between additional computational cost and the reliability of the estimator.
\end{itemize}
Then, one can establish the following lemma.
\begin{lemma}
\label{lem:prob-err-est}
\[
\TNorm{I-QQ^TA} \le 10 \sqrt{\frac{2}{\pi}} \max_{i\in[r]} \TNorm{ \left(I-QQ^T\right)Ag^{(i)} }.
\]
\end{lemma}
\begin{Proof}
The lemma follows easily from the following lemma.
\begin{lemma}
Let $B\in\mathbb{R}^{m \times n}$, and fix $r\in\mathbb{Z}^{+}$ and $\alpha\in(0,1)$.
If we draw $\{ g^{(i)}, i\in[r] \}$ Gaussian vectors, then with probability $\ge 1-\alpha^r$, we have that
\[
\TNorm{B} \le \frac{1}{\alpha}\sqrt{\frac{2}{\pi}}\max_{i\in[r]} \VTTNorm{Bg^{(i)}}.
\]
\end{lemma}
We won't go though the details of the proof of either of these.
\end{Proof}

Here are several remarks.
\begin{itemize}
\item
To do this error estimate requires a small number of additional matrix-vector products (which is often relatively cheap).
\item
In practice, the way this would be implemented is as follows: one would make an underestimate of the rank and then add more samples as necessary.
\item
BTW, I have not worked with this error estimator myself, and I have heard mixed reviews about it, in particular that its variability might practically be too large for most applications of interest, especially outside scientific computing applications.
So, if you want to use it, then look into how it performs for you.
\end{itemize}

The last comment aside, we could implement this probabilistic error estimator as part of the main algorithm.
Here is the combined algorithm.
\textsc{AlgorithmCombined} takes as input a matrix $A$ and an error parameter $\epsilon$, and it does the following steps.
\begin{enumerate}
\item
Let $Q^{(0)}$ be an empty basis matrix.
\item
For $i=1,2,3,\ldots$, do the following.
\begin{enumerate}
\item
Draw $n \times 1$ Gaussian random vector $g^{(i)}$ and set $y^{(i)} = A g^{(i)}$.
\item
Compute $\hat{q}^{(i)} = \left(I-Q^{(i-1)}{Q^{(i-1)}}^T \right) y^{(i)} $.
\item
Normalize $q^{(i)} = \hat{q}^{(i)} / \VTTNorm{ \hat{q}^{(i)} }$ and form the matrix $Q^{(i)} = \left[ Q^{(i-1)} q^{(i)} \right]$.
\end{enumerate}
\end{enumerate}
To determine when to stop, note that the vectors $\hat{q}^{(i)}$ are vectors that appear in the bound of Lemma~\ref{lem:prob-err-est}, and so one stopping rule is to stop the look when the error $\epsilon^{\prime} = \frac{\epsilon}{10\sqrt{2\pi}}$.
(There are some numerical issues that we are ignoring, but this captures the main ideas.)


Recall that, in the motivating scientific computing applications where these implemented algorithms have been most fully developed, one wants to use the basis $Q$ to construct other decompositions. 
Then, given a matrix $Q$ of size $n\times(k+p)$, we want to get various decompositions with it.
So, assume that we are given matrices $B$ and $C$ such that 
\[
\TNorm{A-BC} \le \epsilon ,
\]
where $\mbox{rank}(B) = \mbox{rank}(C) = k$.
For example, this could be gotten with a fixed precision variant of a random projection algorithm, as just described.
Then, how can we use it to compute other factorizations, with comparable additional error?
Here is how to do it.

\begin{itemize}
\item
\textbf{Pivoted QR.}
Given the matrix $A=\mathbb{R}^{m \times n}$, there is the decomposition $A=QR$, where $Q\in\mathbb{R}^{m \times \ell}$ is orthogonal, and $R$ is---up to a permutation---an upper triangular matrix.
In some cases, this process is stopped early, and we keep fewer than all the columns.

To construct a partial QR decomposition, do the following.
\begin{enumerate}
\item
Compute QR factorization of $B$, i.e., $B=Q_1R_1$.
\item
Form the product $D=R_1C$, and compute the QR factorization of $D$, i.e., $D=Q_2R$.
\item
Form the product $Q = Q_1Q_2$.
\end{enumerate}
The result of this is an orthogonal $Q$ and a matrix $R$ that is---up to permutation---upper-triangular such that $\TNorm{A-QR}\le\epsilon$.
\item
\textbf{SVD.}
$A=U \Sigma V^T$.
To construct a partial SVD, do the following.
\begin{enumerate}
\item
Compute QR factorization $B$ such that $B=Q_1R_1$.
\item
Form the product $D=R_1C$, and do an SVD to get $D=U_2\Sigma V^T$.
\item
Form the product $U=Q_1U_2$.
\end{enumerate}
The result of this are matrices $U$, $\Sigma$, and $V^T$ such that $\TNorm{A-U\Sigma V^T} \le \epsilon$.
\item
\textbf{Interpolative Decomposition.}
Given the matrix $A=\mathbb{R}^{m \times n}$, there is an index set $J =[j_1,\cdots,j_k]$ such that $A=A_{(:,J)}X$, with $X\in\mathbb{R}^{k \times n}$, where the $k \times n$ matrix $X$ contains an $k \times k$ identity matrix $I$, i.e., $X_{(:,J)}=I$.
Note that this is NP-hard to compute (I think, still to check), but there exist algorithms to compute a relaxation of it such that $X$ has entries with magnitude bounded by $2$ (and this leads to good conditioning properties).
This interpolative decomposition can also be computed from $A\approx BC$, but we won't describe it here.
\end{itemize}

Another issue is that for some matrices the spectrum might decay rather slowly, and this can slow down Krylov-based iterative methods.
In this case, RST
suggested essentially to do a ``power method'' to enhance the decay of the spectrum.
This reduces the weight of the small singular values, without changing the singular vectors and singular subspaces, and so it gets better convergence with iterative methods.
So, we can apply random sampling/projection to $B=\left(AA^T\right)^{q}A$, observing~that 
\[
B = \left(AA^T\right)^{q}A = U \Sigma^{2q+1}V^T   .
\]
Note that the matrix $B$ has the same singular vectors as $A$, but its singular values decay faster as $\sigma_i(B) = \sigma_i(A)^{2q+1}$.
So, this method requires doing more matrix-vector multiplications, but it is much more accurate.
Thus, if the original basis is within a factor, call it $\gamma$, of optimum, then this gives a basis that is within a factor of $\gamma^{1/(2q+1)}$ of optimum, i.e, it is exponentially fast with the number of iterations.

We will analyze this below, and this will make use of our basic structural result in an important way.
In particular, the main structural result suggests that the performance of algorithms depends on how the top and bottom part of the spectrum of $A$ interact with the sketching matrix, in the way just described.
Here is a lemma.

\begin{lemma}
Let $A\in\mathbb{R}^{m \times n}$, and let $S\in\mathbb{R}^{n \times \ell}$ be a sampling matrix, and let $Z=BS$.
Then, 
\[
\TNorm{ \left(I-P_Z\right)A } \le \TNorm{ \left(I-P_Z\right)B }^{1/(2q+1)}  .
\]
\end{lemma}
\begin{Proof}
We will skip the proof, but it is based on a variant of the spectral radius framework.
\end{Proof}

To see how this lemma \emph{applies} to the previous result, recall that 
\[
\XNorm{ I-P_{AS}A } \le \XNorm{A-A_k} + \XNorm{ \Sigma_2 \Omega_2 \Omega_1^{+} }  ,
\]
where recall that $\Omega_2 = V_2^TS$ and $\Omega_1 = V_1^TS$.
Thus, 
\[
\XNorm{I-P_{AS}A} \le \left( 1+\XNorm{ \Omega_1\Omega_2^{+} } \right) \sigma_{k+1}  ,
\]
where this is true for the spectral norm (still to check: is it true for other norms).
But from the above lemma, we have that (and still to check: is it true for other norms):
\begin{eqnarray*}
\XNorm{I-P_{BS}A } 
   &\le& \XNorm{ I-P_{BS}B }^{1/(2q+1)}  \\
   &\le& \XNorm{ I+\Omega_2\Omega_1^{+} }^{1/(2q+1)} \sigma_{k+1}^{1/(2q+1)}(B)  \\
   &=& \XNorm{ I+\Omega_2\Omega_1^{+} }^{1/(2q+1)} \sigma_{k+1}(A).
\end{eqnarray*}
So, in particular, the use of the power method drives down the sub-optimality of the additional error exponentially fast as the power $q$ increases.

Finally, let's conclude with a comment about projecting onto extra dimensions and ``keeping'' them, i.e., not filtering them through the rank $k$ space.
Perhaps surprisingly, this is helpful to improve the failure probability.
(This is somewhat different than with Blendenpik in the LS case.)
In particular, it can be used to make the failure probability very small as a function of $p$, the number of extra samples/dimensions.
For Gaussian projections, this holds true, and the result takes a particularly simple form, as is given in the following lemma.
\begin{lemma}
Let $A\in\mathbb{R}^{m \times n}$, fix a rank parameter $k$, and let $p \ge 2$ be an oversampling factor.
Then, 
\begin{eqnarray*}
\Expect{ \FNorm{ \left(I-P_{AS}\right)A } } &\le& \left(1+\frac{k}{p-1}\right)^{1/2} \FNorm{A-A_k}  \\
\Expect{ \TNorm{ \left(I-P_{AS}\right)A } } &\le& \left(1+\frac{k}{p-1}\right)^{1/2} \TNorm{A-A_k} + \frac{e\sqrt{k+p}}{p} \FNorm{A-A_k} .
\end{eqnarray*}
\end{lemma}
\begin{Proof}
The proof is omitted, but it uses the main structural result in an essential manner.
\end{Proof}

Here are some final comments.
\begin{itemize}
\item
The proof for this last result makes use of straightforward result, including the basic structural result, and the analysis is sufficiently fine that we can't get bounds for this by looking at just the top part of the spectrum.
So, both the improved failure probability as well as the improved convergence rate rely on the same structural result that gave improved results for the CSSP.
\item
We can get similar bounds with the expectation removed, and there the failure probability decreases exponentially in $p$ as the extra dimensions increases.
\item
This last result shows that there are several regimes of interest for $p$:
\begin{itemize}
\item
$p=0$: this was in the CSSP
\item
$p=5$ or so, which is a modest oversampling which is used in random projections in practice.
\item
$p \gtrsim k$, in which case the multiplicative factors start to become small
\item
$p = \Theta\left(k\log(k)\right)$ or $p = \Theta\left(k \log(k)/\epsilon^2\right)$, which is the regime where the worst-case analysis is applied.
\end{itemize}
\end{itemize}

\newpage

\section{%
(11/13/2013):  Low-rank Matrix Approximation with Element-wise Sampling}

Today, we will switch gears and start to discuss a different way to construct low-rank matrix approximations: randomly sample elements, rather than rows/columns, from the input matrix.
Here is the reading for today and next time.
\begin{itemize}
\item
Achlioptas and McSherry, ``Fast Computation of Low-Rank Matrix Approximations'' (the JACM version)
\end{itemize}

In particular, today we will cover the following topics.
\begin{itemize}
\item
Review of general approaches to low-rank matrix approximation.
\item
An introduction to the basic ideas of element-wise sampling.
\item
A deterministic structural result that is useful for analyzing element-wise sampling algorithms.
\item
An introduction to a specific element-wise sampling result.
\end{itemize}

\subsection{Review of Some General Themes}

So far, we have been talking about sampling/projection of rows/columns---i.e., we have been working with the actual columns/rows or linear combinations of the columns/rows of an input matrix $A$.
Formally, this means that we are pre- or post-multiplying the input matrix $A$ with a sampling/projection/sketching operator (that itself can be represented as a matrix) to construct another matrix $A^{\prime}$ (with different dimensions) that is similar to $A$ in some way---e.g., the eigenvalues, subspaces, the fraction of norm it captures, etc. are similar to the original matrix $A$.

This approach makes sense; and if access to full columns and/or rows is possible, then it is probably the best approach.
After all, matrices are ``about'' their columns/rows, in the sense that if you have control over the column/row space and the various null spaces, then you basically have control over the entire matrix. 
As we will see, this is reflected in the stronger results that exist for column/row sampling than for element-wise sampling.
In many data applications, e.g., the DNA SNPs, astronomy, etc., as well as data correlation matrices, etc., the actual columns/rows ``mean'' something and/or we have relatively-easy access to most or all of the elements in a given column/row. 
Alternatively, in other applications, e.g., in scientific computing and HPC, one often just wants some basis that is ``good'' in some well-defined sense, and it doesn't matter what that basis ``means'' or how it is constructed, but access to entire columns/rows is relatively-straightforward, e.g., by performing matrix-vector multiplications.

In other cases, however, one might want to access the matrix in different ways, e.g., across groups of columns, which might correspond to a cluster in a graph or in some underlying geometry; or access submatrix blocks, e.g., the intersection of sets of rows and columns, since one might want to do bi-clustering; or access individual elements, since an individual element might be meaningful, which is the case in many internet and social media applications.
For example, in a prototypical recommendation system application, individual elements correspond to the rating that a given user gave to a given movie.

We are now going to switch to talking about random sampling algorithms that access \emph{elements} of the input matrix.
Here are several reasons why this might be of interest.
\begin{itemize}
\item
It is \emph{algorithmically interesting}, since it corresponds to another way to access the data that will have similarities and differences with the column/row sampling algorithms for sampling/projection that we have been discussing.
\item
It is a semi-plausible model of \emph{data access}, e.g., in movie recommendation systems, where individual entries arguably ``mean'' something more than the full columns/rows in terms of how the data are generated and accessed.  
As an idealization, this leads to the so-called ``matrix completion problem,'' which has received a lot of interest recently.
\item
It is a plausible model for \emph{interactive analytics}.
E.g., if large column/row leverage scores correspond to important/interesting/outlying data---and for many applications they are very non-uniform---then it stands to reason (and is true) that often their non-uniformity is not uniform along the other direction.
E.g., high leverage SNPs might not be high leverage uniformly in all subpopulations or in all individuals.
\end{itemize}

As before, there are two quite different ways of thinking about this problem, which parallels the algorithmic-statistical perspective we had before.
\begin{itemize}
\item
\textbf{Algorithmic perspective.}
In this approach, one formulates the problem roughly as follows.
Given an input data matrix that is arbitrary or worst-case, but explicitly or implicitly given as input, we sample a small number of elements from the input matrix, and we try to do it in such a way that we get a good approximation to the best low-rank approximation of that data matrix.
Most of our previous discussion has adopted this algorithmic perspective.
\item
\textbf{Statistical perspective.}
In this approach, one formulates the problem roughly as follows.
Given a model for the unobserved data matrix, we assume that we observe a part of that data matrix according to some rule, and then we try to develop a procedure such that we can compute/predict the original unobserved data matrix exactly or approximately.
Relatedly, one tries to establish sufficient conditions such that this computation is successful.
For example, with rows and columns, if I know that the original matrix is exactly rank $k$, then if I have \emph{any} set of exactly $k$ linearly independent rows, then I can reconstruct the entire matrix.
That last statement is obvious from a linear algebraic perspective, but the assumption of being exactly rank $k$ can viewed as a (very strong) statistical model (in which case one can ask about relaxing it or using a different procedure, e.g., sampling elements, that is less trivial).
\end{itemize}
Importantly, similar ideas appear in both perspectives, but they are handled differently.
In particular, in the algorithmic approach, one needs to identify important or influential or outlying things, whether columns/rows or elements or something else, e.g., by biasing the sample toward high-leverage components; while in the statistical approach, one needs to make some sort of niceness assumption which typically amounts to assuming that there don't exist any very high leverage components.
(That is, some underlying structure is important, but in one case it must be found, while in the other case it must be assumed not to exist.)

\subsection{Introduction to Element-wise Sampling}

We will be covering the AM07 paper, which is one of the earliest element-wise sampling results in the area.
In particular, we will review here their motivation, since it is very nice pedagogically, and since it complements our discussion so far.
The paper was introduced in TCS, and so it adopts the algorithmic approach, but many of the more recent developments in element-wise sampling that adopt a statistical approach can be understood in terms of it.
Although the motivation for their work was less general than these methods have come to be viewed and used, it is good to know their motivation, since it is related to some of the iterative algorithms we discussed, and since it informed some of the design decisions they made in developing the algorithm. 

The motivation for their work was that they were interested in accelerating the computation of good low-rank approximations to an arbitrary matrix $A\in\mathbb{R}^{m \times n}$ when $A$ has strong spectral structure, i.e., when the singular values of interest $\gg$ those of a random matrix of similar size.
In particular, recall that orthogonal iteration and Lanczos iteration, two common algorithms for computing low-rank matrix approximation, operate by performing repeated matrix-vector multiplications.
To get around the memory requirements, etc., of exactly-optimal low-rank approximation, one might find it acceptable to work with a nearly-optimal low-rank approximation.

To have an efficient method for computing near-optimal rank-$k$ approximations with an iterative algorithm like orthogonal/Lanczos iteration, the rough idea they propose is to do the following.
\begin{itemize}
\item
Randomly sample or quantize the entries of the input matrix $A$ to get a matrix $\hat{A}$.
\item
Use Lanczos/orthogonal iteration to get a best rank $k$ approximation $\hat{A}_k$ to $\hat{A}$.
\item
Show that $\|A-\hat{A}_k\| \le \|A-A_k\| + ADDL$, where $\|\cdot\|$ is some matrix norm and $ADDL$ is some additional (additive) error term.
\end{itemize}
So, this approach speeds up the computation of a good low-rank approximation to $A$ (in theory, at least and so far, since their algorithm hasn't been implemented except as a proof of principle) by reducing the number of non-zero entries in the matrix and/or the representation size of those entries.
In particular, recall that iterative algorithms require time that is $O\left(\mbox{nnz}(A)\right)$ multiplied by the number of iterations.
(We will treat the iterative algorithm as a black box, and most of our effort will be to show that the best rank $k$ approximation of the sparsified matrix is not much worse than the best rank $k$ approximation of the original matrix.)
The analysis of this procedure is based on the idea that sampling/quantizing the entries of a matrix can be viewed as adding a random matrix (albeit, a specially-structured and data-dependent random matrix) to the input and then exploiting that the random matrix has weak spectral structure (in which case it only substantially affects the bottom part of the spectrum of the original matrix).

Before getting into the details, here is a thought experiment to make some of these ideas somewhat more precise.
Suppose we want to get $\|A-\hat{A}_k\| \le \|A-A_k\| + ADDL$, and we will use randomization in the following way to do this.
Say that we have an allotment of $ADDL$ and we use it up by adding to $A$ (which is a matrix that is reasonably-well approximated by a low-rank matrix) a matrix $G$ that consists of i.i.d. $N(0,\sigma)$ random variables.
This won't ``help'' computationally, at least with the motivation of speeding up iterative algorithms, since the matrix $A+G$ is no less dense than the matrix $A$, but it shouldn't ``hurt'' us ``too much.'' 
By that, we mean that if the original matrix $A$ had signal in the top part of the spectrum and noise in the bottom part of the spectrum, then we primarily added noise to the bottom part of the spectrum.
More precisely, the reason is that if $\sigma$ is not too large, then $\hat{A}_k = \left(A+G\right)_k$ well-approximates $A$ nearly as well as does $A_k$.
The reason for this latter observation has to do with the stability/robustness w.r.t. Gaussian noise that is well-understood, and it stems from the observation that no low-dimensional subspace ``describes well'' the matrix $G$.
That is, if $k$ is small, then $\|G_k\|$ is small and $\|G-G_k\|$ is large.
(If this latter claim isn't ``obvious'' by now, then recall that, in terms of statistical modeling, low-rank approximations are often used precisely to remove such Gaussian noise in the hypothesized statistical model.)

While more typical in terms of statistical modeling, being Gaussian is \emph{not} essential for the above line of reasoning, and by now it should be clear that this would also hold for any of a wide range of random projection matrices.
That is well-known in random matrix theory, and it was elucidated most clearly in RandNLA by AM07, but it holds more generally (Gaussian, Rademacher, sub-Gaussian, Hadamard, other sparsity-respecting constructions, with appropriate choices of parameters, etc.).
In particular, to get these results to generalize, the following is sufficient for the random variable.
\begin{itemize}
\item
Independence
\item
Mean zero
\item
Small variance
\end{itemize}
If $N$ is any random matrix with entries $N_{ij}$ satisfying these three conditions, then $\|N_k\| \sim \|G_k\|$.
(In that case, we can ask about finding matrices $N$ that have better algorithmic properties, in a manner analogous to how there are a range of different JL-like constructions that have better algorithmic property than the original JL construction while still obtaining similar guarantees.  
Indeed, AM07 shows that $\|N_k\|$ bounds the influence that $N$ has on the optimal rank $k$ approximation to $A+N$, and so if $\|A_k\| \gg \|N_k\|$, then $\left(A+N\right)_k$ will be well-described by $A$.)
In particular, AM07 does the following.
\begin{itemize}
\item
Design a random matrix $N$---\emph{that depends on the input matrix $A$}---but that still satisfies these three conditions.
\item
Choose $N$ such that $A+N$ has better sparsity, etc. properties.
\item
Exploit this phenomenon for computational gain by decreasing the time that each matrix-vector multiplication takes in traditional iterative algorithms.
\end{itemize}

Here is a toy example illustrating this approach.
Let $N$ be a random matrix such that $N_{ij} = \pm A_{ij}$ with equal probability, $\forall i,j$.
Then, $\Expect{N_{ij}}=0$ and $\hat{A}=A+N$ has half the number of non-zero entries as $A$, in expectation.
(This is similar to what we saw before, when we observed that with $\{\pm1\}$ random variables in a random projection matrix, we could set $2/3$ of the entries to zero, in expectation, and still obtain the same concentration results.)
This basic idea can be extended to keeping any $p>0$ fraction of the entries with $ADDL$ error growing as $1/\sqrt{p}$.

It turns out that one can get even better sparsification/variance properties if we choose the probability of keeping an entry to depend on the magnitude of that entry.
(We say sparsification/variance together, since we will be zeroing out entries to sparsify the matrix, and the bottleneck to getting even sparser is typically that the variance in the relevant estimators is not small enough.
So, if we can reduce the variance then we can get sparser.)
In particular, let's keep entries i.i.d. with the following probability.
\begin{itemize}
\item
$\Probab{ \mbox{ keeping } A_{ij} } \sim A_{ij}^{2}  .$
\end{itemize}
If we do this, then we focus attention on the larger entries of $A$ (i.e., those which contribute more to the variance).
This should do particularly well when entries vary a lot in magnitude.
Using the same reasoning, we can also quantize entries to be in $\{-1,1\}$, which has the advantage that we can represent the entry with a single bit.
Alternatively, we can both sample and quantize.

\subsection{A Deterministic Structural Result}

We will start with a \emph{deterministic structural result} formalizing the idea that perturbation matrices that are poorly-approximated in $\mathbb{R}^{k}$ have little influence on the optimal rank-$k$ approximation.
(That is worth thinking about for a minute, as it is a different intuition than what has motivated most of our previous algorithms, but it is a helpful intuition to have.)
We'll use their notation for simplicity of comparison with the paper.

\begin{lemma}
\label{lem:am07-structure}
Let $A,N\in\mathbb{R}^{m \times n}$, and let $\hat{A}=A+N$.
Then, 
\begin{eqnarray*}
\TNorm{ A-\hat{A}_k } &\le& \TNorm{A-A_k} + 2\TNorm{N_k} \\
\FNorm{ A-\hat{A}_k } &\le& \FNorm{A-A_k} + \FNorm{N_k} + 2\sqrt{ \FNorm{N_k}\FNorm{A_k } }
\end{eqnarray*}
\end{lemma}

\textbf{Remark.}
As with our other structural results, there is no randomness here in the statement of this lemma.
That is, it is a deterministic structural result that holds for any worst-case matrix $A$ and any matrix $N$.
In our RandNLA application, we will apply it to the case where $A$ is reasonably well-approximated by a low-rank matrix and $N$ is one of the sparsifying/quantizing matrices we described above.

\textbf{Remark.}
The error in this lemma scales with $\|N_k\|$, and so if $N$ is poorly-approximated in $\mathbb{R}^{k}$, i.e.,if  $\|N_k\|$ is small, then the additional error caused by adding $N$ to $A$ is bounded, compared with the error of the best rank $k$ approximation to $A$.

\begin{Proof}[of Lemma~\ref{lem:am07-structure}]
To do the proof, we'll prove two claims relating $\|A-B_k\|$ to $\|A-A_k\|$ for arbitrary matrices, for the spectral and Frobenius norm, as well as an intermediate claim.

Here is the first claim.
\begin{claim}
\label{claim1-lem:am07-structure}
For all matrices $A$ and $B$, we have that 
\[
\TNorm{A-B_k} \le \TNorm{A-A_k} + 2\TNorm{ \left(A-B\right)_{k} }  .
\]
\end{claim}
\begin{Proof}[of claim]
\begin{eqnarray*}
\TNorm{A-B_k}
   &\le& \TNorm{A-B} + \TNorm{B-B_k} \quad\mbox{(by the triangle inequality)}  \\
   &\le& \TNorm{A-B} + \TNorm{B-A_k} \quad\mbox{(since $B_k$ is the ``best'' rank $k$ approximation to $B$)}  \\
   &\le& \TNorm{A-B} + \TNorm{B-A} + \TNorm{A-A_k} \quad\mbox{(by the triangle inequality)}  \\
   &=& \TNorm{A-A_k} + 2\TNorm{\left(A-B\right)_k} \quad\mbox{(since $\TNorm{B-A}=\TNorm{A-B}=\TNorm{\left(A-B\right)_k} $)}  \\
\end{eqnarray*}
\end{Proof}

Here is the second claim.
\begin{claim}
\label{claim2-lem:am07-structure}
For all matrices $A$ and $B$, we have that 
\[
\FNorm{ P_{B_k}A } \ge \FNorm{ P_{A_k}A } - 2 \FNorm{ \left( A-B \right)_{k} }  .
\]
Here, $P_{B_k}$ is the projection onto the space spanned by the columns of $B_k$.
\end{claim}
\begin{Proof}[of claim]
The idea is: for all matrices $A$ and $B$, if $\FNorm{\left(A-B\right)_{k} }$ is small, then projecting $A$ onto $P_{B_k}$ is almost as good as projecting $A$ onto $P_{A_k}$.
Here are the details.
\begin{eqnarray*}
\FNorm{ P_{B_k}A } 
   &\ge& \FNorm{ P_{B_k}B } - \FNorm{ P_{B_k}\left(A-B\right) }  \quad\mbox{(triangle inequality)} \\
   &\ge& \FNorm{ P_{A_k}B } - \FNorm{ P_{B_k}\left(A-B\right) }  \quad\mbox{(since projecting onto $B_k$ is the ``best'')} \\
   &\ge& \FNorm{ P_{A_k}A } - \FNorm{ P_{A_k}\left(B-A\right) } - \FNorm{ P_{B_k}\left(B-A\right) }   \quad\mbox{(triangle inequality)} \\
   &\ge& \FNorm{ P_{A_k}A } - 2\FNorm{ P_{ \left(B-A\right)_{k} }\left(B-A\right) }  \quad\mbox{(since $\left(B-A\right)_{k}$ is the best)}  \\
   &=& \FNorm{ P_{A_k}A } - 2 \FNorm{ \left( B-A \right)_{k} }
\end{eqnarray*} 
Above we used that $\FNorm{ P\left(B-A\right) } \le \FNorm{ P_{B-A}\left(B-A\right) }$.
\end{Proof}

Now, we will use this intermediate claim to prove that if $\FNorm{ \left( A-B \right)_{k} }$ is small, then $\FNorm{ A-B_k }$ is not much worse than $\FNorm{ A-A_k }$, and so we can use $B_k$ as a surrogate for $A_k$ w.r.t., $\FNorm{\cdot}$---\emph{even if $\FNorm{B-A}$ is large}, as long as $\FNorm{ \left(A-B\right)_{k} }$ is small.

Here is the third claim.
\begin{claim}
\label{claim3-lem:am07-structure}
For all matrices $A$ and $B$, we have that 
\[
\FNorm{A-B_k} \le \FNorm{A-A_k} + 2\sqrt{ \FNorm{\left(A-B\right)_{k}}\FNorm{A_k} } + \FNorm{ \left(A-B\right)_{k} }  .
\]
\end{claim}
\begin{Proof}[of claim]
First, observe the following.
\begin{eqnarray*}
\FNorm{A-B_k}
   &\le& \FNorm{A-P_{B_k}A} + \FNorm{P_{B_k}A-B}  \quad\mbox{(triangle inequality)}  \\
   &\le& \FNorm{A-P_{B_k}A} + \FNorm{P_{B_k}\left(A-B\right)}  \quad\mbox{(since $P_{B_k}B=B_k$)}  \\
   &=& \left( \FNormS{A} - \FNormS{ P_{B_k}A } \right)^{1/2} + \FNorm{ P_{B_k}\left(A-B\right) }  \quad\mbox{(by the Pythagorean theorem)}  \\
   &\le& \left( \FNormS{A} - \FNormS{ P_{B_k}A } \right)^{1/2} + \FNorm{ P_{\left(A-B\right)_k}\left(A-B\right) }  \quad\mbox{(since $\left(A-B\right)_{k}$ is the best)}  \\
   &\le& \left( \FNormS{A} - \FNormS{ P_{B_k}A } \right)^{1/2} + \FNorm{ \left(A-B\right)_k }    .
\end{eqnarray*}
The comment about ``by the Pythagorean theorem'' above is that 
$$\FNormS{A-P_{B_k}A}=\FNormS{A}-\FNormS{P_{B_k}A} ,$$ 
in which case we can apply the Pythagorean theorem to each column of $A$.

So, to establish the claim, we just need to bound the first term.
To do so, use Claim~\ref{claim2-lem:am07-structure}, from which it follows that
\begin{eqnarray*}
\FNormS{ P_{B_k}A }
   &\ge& \FNormS{ P_{A_k}A } + 4 \FNormS{ \left( A-B \right)_{k} } -4 \FNorm{ P_{A_k}A }\FNorm{ \left( A-B \right)_{k} }  \\
   &\ge& \FNormS{ P_{A_k}A } - 4 \FNorm{ P_{A_k}A }\FNorm{ \left( A-B \right)_{k} }  .
\end{eqnarray*}
Thus, it follows that 
\begin{eqnarray*}
\left( \FNormS{A}-\FNormS{P_{B_k}A} \right)^{1/2}
   &\le& \left( \FNormS{A} - \FNormS{P_{A_k}A} + 4 \FNorm{ P_{A_k}A } \FNorm{ \left(A-B\right)_{k} } \right)^{1/2}  \\
   &=& \left( \FNormS{A-A_k} + 4 \FNorm{ P_{A_k}A } \FNorm{ \left(A-B\right)_{k} } \right)^{1/2}  \\ & & \hspace{50mm}  \quad\mbox{(by the Pythagorean result above)}  \\
   &\le& \FNorm{A-A_k} + 2 \sqrt{ \FNorm{ P_{A_k}A } \FNorm{ \left(A-B\right)_{k} } }  .
\end{eqnarray*}
From this the claim follows.
\end{Proof}

From this the lemma follows.
\end{Proof}

\subsection{Introduction to the Element-wise Sampling Algorithm}

To apply this deterministic structural result to develop a provably-good element-wise random sampling algorithm, we will need a result from random matrix theory.
(Actually, we can do it more simply with a matrix Chernoff bound, and this has been developed in subsequent work, but for ease of comparison we will follow AM07 here.).

\textbf{Fact.}
Let $G\in\mathbb{R}^{m \times n}$, with $m < n$, and with entries i.i.d., $N(0,\sigma^2)$ r.v.
Then, w.p. $\ge 1-e^{-\Theta(n})$, we have that
\begin{eqnarray*}
\TNorm{G_k} &\le& 4\sigma\sqrt{n} \\
\FNorm{G_k} &\le& 4\sigma\sqrt{kn}  .
\end{eqnarray*}
The first result is somewhat like Wigner's semicircle law (but the details are importantly different, and in particular it is \emph{not} an asymptotic result); and the second result is since $\FNorm{A} \le \sqrt{k}\TNorm{A}$, for all $A$.

To get a sense of ``scale,'' by which I mean ``how big is big and how small is small,'' note also that the trivial rank $k$ approximation obtained by keeping just the first $k$ rows of $G$.
Call that matrix $D$, i.e., it is just the first $k$ rows from $G$.
This gives w.h.p. that $\FNorm{D} \sim \sigma\sqrt{kn}$. 
Since $\mbox{rank}(D) \le k$, we have also that $\TNorm{D} \ge \FNorm{D}/\sqrt{k}$.

So, the above fact says that the optimal rank-$k$ approximation improves this trivial approximation by only a factor of $\le 4$.
The reason for this is the near orthogonality of the rows of $G$.
By contrast, for a general matrix $A\in\mathbb{R}^{m \times n}$, with $\sigma=|A_{ij}|$ we can have that $\|A_k\|$ can be $\sim \sigma\sqrt{mn}$, in either norm.
The main results below say that the effect of random quantization and random sparsification, as described above, is qualitatively the same as adding Gaussian random noise.

Let's start with a more rigorous statement of the above fact.
A very brief history is the following.
\begin{itemize}
\item
Wigner's semicircle law, which makes a similar claim, but asymptotically in convergence.
\item
Furedi-Komos, which is what AM07 originally used.
\item
Vu's improvement, using a result of Alon that is due to Talagrand.
\end{itemize}
Also, there has been a lot of work in recent years improving these results; many of them are simplified with the matrix concentration results we discussed.

Here is a theorem making precise the above discussion.
\begin{theorem}
Given a matrix $A\in\mathbb{R}^{m \times n}$, with $m \le n$, and fix an $\epsilon > 0$.
Let $\hat{A}$ be a random matrix with entries independent random variables such that for all $i,j$:
\begin{itemize}
\item
$\Expect{\hat{A}_{ij}} = A_{ij}$
\item
$\Varnce{\hat{A}_{ij}}\le\sigma^2$
\item
$\hat{A}_{ij}$ takes values in the interval of length $\kappa$, where $\kappa = \left( \frac{ \log(1+\epsilon) }{ 2 \log(m+n) } \right)^{2}\cdot \sigma \cdot \sqrt{m+n}  $.
\end{itemize}
Then, for all $\theta > 0$, and for all $m+n\ge 152$, we have that 
\[
\Probab{ \TNorm{A-\hat{A} } \ge 2\left(1+\epsilon+\theta\right)\sigma\sqrt{m+n} } < 2 \exp\left( \frac{16\theta^2}{\epsilon^4} \left(\log(n)\right)^4 \right)
\]
\end{theorem}

\textbf{Remark.}
We have stated the above result as it appears in AM07.
Satisfying the range constraint is awkward, but important, and so we will consider more-or-less awkward ways to do it.
In recent years, there have been several improvements to that result which simplify it somewhat, and likely more sophisticated techniques could simplify it even more.
I mention that as an FYI, but we won't have time to go into that in detail.

Given this result, we will state a simple sparsification result and a simple quantization result, as we discussed above, and then we will state a more complicated sparsification result that is more comparable with the previous additive error algorithms via column/row sampling.

\begin{theorem}
Let $A\in\mathbb{R}^{m \times n}$, with $m \le n$, and let $b = \max_{ij} |A_{ij}|$.
(Think of $b$ as analogous to the variance parameter.)
Let $\hat{A}\in\mathbb{R}^{m \times n}$ be a random matrix with entries distributed i.i.d as 
\[
\hat{A}_{ij} = \left\{ \begin{array}{l l}
                    b & \quad \text{with probability $\frac{1}{2}+\frac{A_{ij}}{2b}$}  \\
                    -b & \quad \text{with probability $\frac{1}{2}-\frac{A_{ij}}{2b}$}
                 \end{array}
         \right.  .
\]
Then, $\forall$ sufficiently large $n$, w.p. $\ge 1-\exp\left( -19\left(\log(n)\right)^{4} \right)$, the matrix $\Delta = A-\hat{A}$ satisfies
\begin{eqnarray*}
\TNorm{\Delta_k} &<& 4b\sqrt{n} \\
\FNorm{\Delta_k} &<& 4b\sqrt{kn}    .
\end{eqnarray*} 
\end{theorem}

\begin{theorem}
Let $A\in\mathbb{R}^{m \times n}$, with $76 \le m \le n$, and let $b = \max_{ij} |A_{ij}|$.
(Again, think of $b$ as analogous to the variance parameter.)
For $p\ge\frac{ \left(8\log(n)\right)^{4} }{n}$, 
let $\hat{A}\in\mathbb{R}^{m \times n}$ be a random matrix with entries distributed i.i.d as 
\[
\hat{A}_{ij} = \left\{ \begin{array}{l l}
                    A_{ij}/p & \quad \text{with probability $p$}  \\
                    0 & \quad \text{with probability $1-p$}
                 \end{array}
         \right.  .
\]
Then, w.p. $\ge 1-\exp\left( -19\left(\log(n)\right)^{4} \right)$, the matrix $\Delta = A-\hat{A}$ satisfies
\begin{eqnarray*}
\TNorm{\Delta_k} &<& 4b\sqrt{n/p} \\
\FNorm{\Delta_k} &<& 4b\sqrt{kn/p}    .
\end{eqnarray*} 
\end{theorem}

\textbf{Remark.}
These two results are complementary, and in particular they can be combined.

\textbf{Remark.}
As stated, these results are \emph{not} immediately-comparable to the additive-error or relative-error bounds that we provided before.
We will get to that result in the next class.

\begin{Proof}[of both results]
We just have to fit together all the pieces that we have been discussing.

We can apply the random matrix theorem to $\TNorm{N}$ with $\epsilon=3/10$ and $\theta=1/10$.
Since $\sqrt{m+n} \le \sqrt{2n}$, we have that $2\left(1+3/10+1/10\right)\sqrt{2} < 4$ and also that $$2\exp\left( -\frac{16\theta^2\left(\log(n)\right)^{4}}{\epsilon^4} \right)<\exp\left( -19\left(\log(n)\right)^{4} \right)  .$$
Then, we can use the results that $\TNorm{N_k}=\TNorm{N}$ and $\FNorm{N_k}\le\sqrt{k}\TNorm{N}$ to get the spectral and Frobenius norm bounds, respectively.
To deal with the range constraint, recall that $\kappa = \left( \frac{\log(\left(1+\epsilon\right)}{2\log(m+n)} \right)^{2}\sigma\sqrt{m+n}$. 

For the quantization theorem, using that $\epsilon=3/10$ and that $2b<\kappa$ gives that $m+n > 10^{10}$, which is ``sufficiently large'' in the theorem. 

For the sampling theorem, the lower bound on $p \ge \left( \frac{2\log(m+n)}{\log\left(1+\epsilon\right)} \right)^{4}\frac{1}{m+n}$, which simplifies to $p > \frac{8}{n}\left(\log(n)\right)^{4}$.
\end{Proof}

\newpage

\section{%
(11/18/2013):  Low-rank Matrix Approximation with Element-wise Sampling, Cont.}

We continue with the discussion from last time.
There is no new reading, just the same as last~class.

In particular, today we will cover the following topics.
\begin{itemize}
\item
A more sophisticated version of the element-wise sampling algorithm that is roughly comparable to the additive-error column/row sampling algorithms.
\item
Introductory discussion of the extension of element-wise sampling algorithm to the matrix completion problem.
\end{itemize}

\subsection{A More Sophisticated Version of the Element-wise Sampling Algorithm}

Let's start by describing a more sophisticated version of the element-wise random sampling algorithm we discussed last time.
This is interesting in and of itself as well as since it will enable us to make a connection with the additive-error random sampling algorithm.

Again, we'll follow the motivation from AM07, which describes it nicely.
Observe that in the algorithm from last class that when we sparsify the matrix $A$ by keeping every entry with the same probability $p$, then we get 
\[
\Varnce{ \hat{A}_{ij} } = \frac{1-p}{p}A_{ij} \quad \forall i,j  .
\]
In particular, this means that smaller entries of $A$ lead to random variables with smaller variance.
On the other hand, the bound on $\TNorm{A-\hat{A}}$ depends on the maximum variance.
Thus, to improve the results, one idea is to keep entries of $A_{ij}$ with probability $p_{ij}\le p$, so that all entries in $\hat{A}$ have roughly the same variance.
This will help us to get sparser matrices, while keeping similar quality-of-approximation bounds.

In more detail, if we choose 
\[
\hat{A}_{ij} = \left\{ \begin{array}{l l}
                    A_{ij}/p_{ij} & \quad \text{with probability $p_{ij}$}  \\
                    0 & \quad \text{otherwise}
                 \end{array}
         \right.  ,
\]
then
\begin{eqnarray*}
\Expect{\hat{A}_{ij}} &=& A_{ij}   \\
\Varnce{\hat{A}_{ij}} &=& A_{ij}^{2}\left(\frac{1}{p_{ij}}-1\right)  .
\end{eqnarray*}
In this case, if $p_{ij} = p\frac{A_{ij}^2}{b^2}$, where $p\in(0,1]$ and $b=\max_{ij}|A_{ij}|$, then
\[
\Varnce{\hat{A}_{ij}} = \frac{b^2}{p} - A_{ij} \approx \frac{b^2}{p}  ,
\] 
$\forall i,j$.
With this choice of $p_{ij}$, we have that
\[
\Expect{\mbox{ number of entries kept }} = \sum_{ij}p_{ij} = \sum_{ij} p\frac{A_{ij}^{2}}{b^2} = \frac{p}{b^2}\FNormS{A} = \frac{b}{p^2}mn\mbox{Avg}(A_{ij})^2 \ll pmn ,
\]
where $pmn$ is what is obtained for uniform sampling.

There is the technical issue here that complicates things, but it is a real issue, and that is the so-called range constraint, i.e., the bound on the range that the random variable is allowed to take.
The issue is that if we choose a very low-probability element, then we must rescale it by a lot, and this might violate the range constraint.
(Many in the improvements since AM07 have to do with fixing this issue; we will summarize these below.)
If we let 
\[
p_{ij} = \max\left\{ \tau_{ij} , \sqrt{\tau_{ij}\frac{8\left(\log(n)\right)^{4}}{n}} \right\}  ,
\]
where $\tau = p\left(A_{ij}/b\right)^{2}$, then we have the following theorem.

\begin{theorem}
\label{thm:elementwise-xx1}
Let $A\in\mathbb{R}^{m \times n}$, with $76 \le m \le n$, and let $b = \max_{ij} |A_{ij} |$.
For $p>0$, let
\begin{eqnarray*}
\tau_{ij} &=& \frac{p}{b^2}A_{ij}^{2}  \\
p_{ij} &=& \max_{ij}\left\{ \tau_{ij} , \tau_{ij} \frac{8\left(\log(n)\right)^{4}}{n} \right\}  .
\end{eqnarray*}
Let $\hat{A}\in\mathbb{R}^{m \times n}$ be a random matrix with entries identically distributed as 
\[
\hat{A}_{ij} = \left\{ \begin{array}{l l}
                    A_{ij}/p_{ij} & \quad \text{with probability $p_{ij}$}  \\
                    0 & \quad \text{otherwise}
                 \end{array}
         \right.  .
\]
Then, we have the following.
\begin{itemize}
\item
W.p. $\ge 1-\exp\left( -19\left(\log(n)\right)^{4} \right)$, the matrix $\Delta=A-\hat{A}$ satisfy the following:
\begin{eqnarray*}
\TNorm{ N_k } &<& 4 b \sqrt{n/p } \\
\FNorm{ N_k } &<& 4b\sqrt{kn/p }  .
\end{eqnarray*}
\item
$
\Expect{ \mbox{nnz}(\hat{A}) } 
\le \frac{p}{b^2}\FNormS{A} + m\left(8 \log(n)\right)^{4} 
= pmn \cdot \mbox{Avg}\left[ \left( \frac{A_{ij}}{b} \right)^{2} \right] 
+ m \left( 8 \log(n) \right)^{4}  ,
$
where $\mbox{Avg}\left[ \left( \frac{A_{ij}}{b} \right)^{2} \right] = \frac{\FNormS{A}}{mn}$.
\end{itemize}
\end{theorem}

\begin{Proof}
Choosing $p_{ij} = \max\left\{ \cdot , \cdot \right\}$ ensures that the range constraint is not violated, since no probability is too small.
Since $p_{ij} \le \tau_{ij} + \frac{\left(8\log(n)\right)^{4}}{n}$, using $p_{ij}$ instead of $\tau_{ij}$ adds no more than $mn\frac{\left(8\log(n)\right)^{4}}{n}$ elements of $\hat{A}$ in expectation.
The rest of the proof is similar to what we did before. 
\end{Proof}

\textbf{Remark.}
We have bounded the number of nonzero entries in the sampled matrix in terms of various parameters, but we haven't actually showed that it is small, e.g., in the sense of guaranteeing that it is not larger than a pre-specified value.
We will get to this soon.

Given the result in Theorem~\ref{thm:elementwise-xx1}, let's ask the following questions.
\begin{itemize}
\item
How long does it take to compute this approximation?
\item
How does this approximation compare with the previous additive-error and relative-error bounds that we had with column/row sampling?
\item
Can these results be improved/extended, either in worse-case or under assumptions on the input?
\end{itemize}
 
\paragraph{Running time.}
To answer the question about running time, we will now give an algorithm that, given $n$ and any $s>0$, in one pass over the data produces a matrix $\hat{A}$ with probability $p=\frac{sb^2}{\FNormS{A}}$, s.t. $\Expect{\mbox{nnz}\left(\hat{A}\right)} \le s m \left(8 \log(n) \right)^{4}$.
That is, it is efficient in the Pass Efficient Model, since it uses only $2$ passes over the data and roughly linear in $m+n$ additional space and time.

Note that, given $n$, $b$, and fixed $p$, it is easy to do non-uniform sampling in a single pass over the data using probabilities 
\[
p_{ij} = \max\left\{ \tau_{ij} , \tau_{ij} \frac{ \left(8 \log(n)\right)^{4}}{n} \right\}  ,
\]
with $\tau_{ij} = p \left( A_{ij}/b\right)^2$.
This gives $\Expect{\mbox{nnz}\left(\hat{A}\right)} \le \frac{ p \FNormS{A} }{ b^2 } + m\left(8\log(n)\right)^{4}$.
This is problematic if we want to implement the algorithm since the first term is not known at the outset, and that might correspond to more entries than we want to keep.
To remedy this will involve a slight re-parameterization.
In particular, to make this term $=s$, for a pre-specified $s$, we can let $p=\frac{sb^2}{\FNormS{A}}$.
To compute this takes one full pass over the data (which is acceptable in the pass efficient model).
Here is the entire algorithm to do this.

This \textsc{Sample}($s,n$) algorithm takes as input $s$ and $n$, and it does the following.
\begin{enumerate}
\item
Let $Q$ be an empty priority queue, and let $z=0$
\item
For all $A_{ij}$, do the following.
\begin{enumerate}
\item
$ z \leftarrow z + A_{ij}^{2}$.
\item
Choose a number $r_{ij}$ u.a.r. from $[0,1]$.
\item
Insert $A_{ij}$ into $Q$ with key 
$\kappa_{ij} = \max \left\{ \frac{sA_{ij}^{2}}{r_{ij}} , \frac{sA_{ij}^{2}}{r_{ij}^{2}}\frac{\left(8\log(n)\right)^{4}}{n} \right\}$.
\item
Remove from $Q$ all the elements with key smaller than $z$.
\end{enumerate}
\item
Return the contents of $Q$.
\end{enumerate}

Given this \textsc{Sample}($s,n$) algorithm, here is a lemma regarding its performance.

\begin{lemma}
Let $A\in\mathbb{R}^{m \times n}$, with $76 \le m \le n$.
Then, \textsc{Sample}($s,n$) yields a matrix $\hat{A}$ such that 
\begin{itemize}
\item
w.p. $\ge 1-\exp\left( -19\left(\log(n)\right)^{4} \right)$, we have that the matrix $\Delta = A-\hat{A}$ is s.t.
\begin{eqnarray*}
\TNorm{\Delta_k} &\le& 4\sqrt{\frac{n}{s}}\FNorm{A} \\
\FNorm{\Delta_k} &\le& 4\sqrt{\frac{kn}{s}} \FNorm{A}.
\end{eqnarray*}
\item
$\Expect{ \mbox{nnz}\left(\hat{A}\right) } \le s + m \left(8 \log(n) \right)^{4}$.
\end{itemize}
\end{lemma}

\begin{Proof}
Let $p = \frac{sb^2}{\FNormS{A}}$, and define $\tau_{ij} = p \left(\frac{A_{ij}}{b}\right)^{2} = s \left( \frac{A_{ij}}{\FNorm{A}}\right)^{2}$.
Then, by Theorem~\ref{thm:elementwise-xx1}, it suffices to show that each entry of $A_{ij}$ is kept by \textsc{Sample}($s,n$) w.p. $= p_{ij} = \max\left\{ \tau_{ij} , \tau_{ij} \frac{\left(8\log(n)\right)^{4}}{n} \right\} $.
But an element $A_{ij}$ is in $Q$ when \textsc{Sample}($s,n$)  terminates iff 
\[
\frac{sA_{ij}^{2}}{r_{ij}} \ge \FNorm{A} 
\quad\mbox{or}\quad 
\frac{sA_{ij}^{2}}{r_{ij}^{2}}\frac{\left(8 \log(n)\right)^{4}}{n} \ge \FNorm{A}  .
\]
But this is equivalent to $r_{ij} \le p_{ij}$ and each $r_{ij}$ chosen uniformly over $[0,1]$.
\end{Proof}

\paragraph{Comparison with column/row sampling algorithms.}
To answer the question about comparing this element-wise sampling algorithm with the previous column/row sampling methods, observe the following.
Given a matrix $A$, we can call \textsc{Sample}($s,n$), where $s = \frac{16n}{\epsilon^2}$ to get a matrix $\hat{A} \in \mathbb{R}^{m \times n}$, where $\hat{A}$ has $O\left( n/\epsilon^2 + m \left(\log(n)\right)^{4} \right)$ non-zeros, sampled from the distribution of the main theorem, with $p = \frac{16nb^2}{\epsilon^2\FNormS{A}}$.
This $\tilde{O}\left(m+n\right)$ dependency corresponds roughly to keeping a small number of columns/rows.
Then, by the structural lemma and the main theorem, w.p. $\ge 1-\exp\left( -19\left(\log(n)\right)^{4} \right)$, we have that 
\begin{eqnarray*}
\TNorm{A-\hat{A}_{k}} &\le& \TNorm{A-A_k} + \epsilon \FNorm{A} \\
\FNorm{A-\hat{A}_{k}} &\le& \FNorm{A-A_k} + 3\sqrt{\epsilon}k^{1/4}\FNorm{A}.
\end{eqnarray*}
So, the element-wise sampling algorithm, when the sampling is done with probability distribution that is proportional to the entries-squared, gives additive-error bounds, where the scale of the additive error is $\FNorm{\cdot}$ (just as with the additive-error version of the column/row sampling algorithms).
(Thus, the results are very similar to the additive-error column/row sampling algorithms, but they are not directly comparable, due to some minor differences; see AM07 for the details.)

\paragraph{Improving/extending these results.}
To answer the question about possible extensions and improvements, observe that there are several types of extensions of this basic set up that are interesting.
\begin{itemize}
\item
\textbf{Dealing with small entries.}
We followed AM07, where smaller entries are sampled with higher probability, but there were complexities having to do with the range constraint, etc.
Similar but slightly stronger results can be gotten by dealing with smaller entries in other ways.
For example:
Drineas-Zouzias-11 zero out sufficiently small entries; Achlioptas-Karnin-Liberty-13 sample with respect to $|A_{ij}|$ and get bounds with respect to the stable rank; and Kundu-Namirajan-Drineas-13 sample with respect to a convex combination of a distribution that is proportional to the elements and a distribution that is proportional to the square of the elements.
We are not going to go though these results, although they lead to improvements in some of what we have discussed, and there are likely additional improvements that can be made.
They also lead to somewhat simpler results, due to recent developments in matrix concentration bounds.
\item
\textbf{Getting better bounds.}
We should note something about the structure of the bounds and what one might hope to achieve.
We might hope for bounds on $\TNorm{A-\hat{A}}$ and $\TNorm{A-\hat{A}_k}$; and although we have bounds for $\FNorm{A-\hat{A}_k}$, we can't expect to get interesting bounds for $\FNorm{A-\hat{A}}$ (basically since we have zeroed out most of the elements of the input matrix, which leads to a large Frobenius norm difference between the two matrices).
This is true, even though the low-rank approximations to them are close with respect to the Frobenius norm, which is all that is needed to apply the sparsified matrix in the context of an iterative algorithm, which was the original motivation.

One might wonder what happens if we make additional assumptions on the matrix, e.g., what if the matrix is \emph{exactly} rank $k$?
In that case, i.e., when the matrix is exactly rank $k$, then relative-error approximation algorithms like we have for column/row sampling mean that the error is $\epsilon$ times $0$, meaning that the matrix is reconstructed exactly.
That is, if the matrix is exactly rank $k$, then one can use the leverage-based sampling algorithms or a random projection algorithm to reconstruct the matrix exactly.
(Of course, in that case, the column/row-wise sampling is less interesting, since in that case \emph{any} set of exactly $k$ columns can be used to reconstruct the matrix---numerical and robustness issues aside.)
\item
\textbf{Matrix Completion.}
This leads to the question of whether we can get relative error approximations and/or reconstruct the matrix exactly using element-wise sampling.
The short answer is No, or Probably No, in general, but the answer is Yes under some assumptions (that are rather strong, but that are difficult to remove).
This work has gone on subsequent to but somewhat-independently of the AM07 developments in TCS, and it goes by the name ``matrix completion.'' 
This it is loosely motivated by recommendation systems problems, and several key papers in the area are due to Candes-Recht-09; Candes-Tao-10; Recht-11; and Gross-11.
\end{itemize}

BTW, the question of exactly versus approximately low-rank for element-wise methods is actually quite non-trivial.
Reconstructing or well-approximating a matrix with element-wise sampling is a much harder problem than with column/row sampling/projection methods, and getting relative error or stronger additive error guarantees is challenging.
In general, where there are such results, one needs to make rather strong assumptions to obtain them.
Here are two examples of assumptions that people have used to get better results with element-wise sampling.
\begin{itemize}
\item
\textbf{For Matrix Completion.}
Here, the goal is to impute missing entries from an unobserved matrix.
The assumption on the input is that that the coherence properties are nice, e.g., flat leverage scores, and that data are presented ii.d.; or that the coherence properties are arbitrary but that the data are presented in a way that conform to that; etc.
\item
\textbf{For Laplacian Solvers.}
Here, the goal is to sample entires from an adjacency matrix to sparsity the graph to get faster solvers.
The assumption on the input is that the matrix is not arbitrary but has additional structure such as that it is the adjacency matrix or the Laplacian matrix of a graph.
In this case, one can take advantage of the graph theoretic structure to get interesting results.
The reason is basically that the entries of the Laplacian matrix correspond to rows/columns of the edge-incidence matrix, and thus careful element-wise sampling of the matrix corresponds to column/row sampling of a different matrix.
\end{itemize}
For the next two classes we will deal with the first of these topics, and for the two classes after that we will deal with the second of these two topics.

\subsection{Introduction to matrix completion}

Since we have a bit of time left, we will cover some of the introduction to matrix completion now.
During the next two classes, we will get into more detail.

The so-called matrix completion problem has to do with imputing or predicting the entries of a matrix from a partially observed version of the matrix.
Since the problem is ill-posed, one needs to make assumptions on the matrix.
This need not have anything to do with sampling elements, e.g., one could do other randomized or deterministic things, but---motivated by recommendation systems applications---one popular variant of the problem does consider element-wise sampling.

We will be covering the results in the following two papers.
\begin{itemize}
\item
Recht, ``A Simpler Approach to Matrix Completion''
\item
Chen, Bhojanapalli, Sanghavi, and Ward, ``Coherent Matrix Completion''
\end{itemize}
As I mentioned above, the basic connection between these papers and the AM07 TCS-style results is that the element-wise sampling algorithm we discussed gives additive-error algorithms for arbitrary input (just as did row/column sampling when the sampling was with respect row norms and not leverage scores).
In general, to get relative-error algorithms is much harder, e.g., one needs the matrices to be exactly low rank, one needs very strong incoherence assumptions, one needs to use tools from convex optimization, etc.
But, if one has a relative-error approximation, then under the assumption that the matrix is exactly low-rank, the matrix is reconstructed exactly.

(That is, a relative-error algorithms gives the \emph{exact} answer if the matrix is \emph{exactly} low-rank.
The reason is that the additional error is $\epsilon$ times a scale which equals $0$.
That's not the way most people think about it, but it's true, and it's helpful to know that to understand the connections between RandNLA algorithms and recent work on matrix completion.
Of course, in the particular context of reconstructing exactly rank $k$ matrices, that is not so helpful, since one just needs to choose \emph{any} set of $k$ linearly independent columns.
And the AM07 algorithm is not helpful, since it is an additive-error algorithm.)

So far, we have considered a problem parameterized roughly as follows.
Given a matrix $A$, sample entires to construct a matrix $\hat{A}$ such that $\tilde{A} \approx A$ in some sense (i.e., we have adopted mostly the algorithmic perspective).
To understand the connections with recent work on matrix completion, we need to ask a different question (more like the statistical perspective) with the following parameterization.
Given an unseen matrix $A$ that satisfies some niceness assumptions, and assuming that I see entries from in some way, reconstruct $A$, exactly or approximately.

The simplest thing to do is to make strong assumptions on $A$ and the element presentation.
For example, assume that $A$ is ``nice'' in that the leverage scores, cross leverage scores, etc., are uniform, and assume that the elements of $A$ are presented u.a.r. and also assume that $A$ is exactly rank $k$, then we can call a black box (that involves convex optimization, in the form or $\ell_1$ minimization or some related nuclear norm minimization) and we solve the problem.
There are of course extensions of this basic setup to exactly low-rank plus a few point-wise spikes, to very small additional noise, etc., but that is the basic setup.

We should note that the incoherence assumption is an extremely strong assumption and that the assumption that the entries are sampled i.i.d. is also extremely strong (e.g., with respect to plausible data generative processes in the motivating application domains).
The CBSW paper describes a variant of this approach to matrix completion that is designed to relate to some of what we have been discussing.
\begin{itemize}
\item
$A$ will still need to be exactly rank $k$
\item
$A$ can be arbitrary, i.e., there is no niceness assumption
\item
The elements of $A$ are presented according to a complicated leverage-based distribution.
\end{itemize}
Some comments on this approach.
\begin{itemize}
\item
The point is that this coherence-based or leverage-based approach of CBSW is a strictly stronger result, since the assumption of data presentation essentially takes care of the problem with bad cases with respect to niceness.
In particular, we can assume that $A$ is nice with respect to its coherence properties, in which case the complicated leverage-based distribution becomes the uniform distribution.
\item
This approach is more like the structural approach we have been following, where it highlights the key structural property.
Then, if one wants to make strong assumptions about data presentation and matrix niceness one can, but this approach makes explicit the relationship between the two.
Moreover, this approach makes it easier to relate this more statistical perspective to the more algorithmic perspective we have been following.

\end{itemize}

\newpage

\section{%
(11/20/2013): Element-wise Sampling and Matrix Completion}

Today and in the next class we will discuss element-wise sampling and matrix completion.
Here is the reading for today.
\begin{itemize}
\item
Recht, ``A Simpler Approach to Matrix Completion'' 
\item
Chen, Bhojanapalli, Sanghavi, and Ward, ``Coherent Matrix Completion''
\end{itemize}
The basic connection here is that the element-wise sampling algorithm we discussed in the last class gives additive-error algorithms (just as did row/column sampling when the sampling was with respect row norms and not leverage scores) but to get relative-error algorithms is much harder, e.g., one needs the matrices to be exactly low rank, one needs very strong incoherence assumptions, one needs to use tools from convex optimization, etc.
(Note that a relative-error algorithms gives the exact answer if the matrix is exactly low-rank.)
I'm not going to tex up these notes in detail.

\newpage

\section{%
(11/25/2013): Element-wise Sampling and Matrix Completion, Cont.}

Today, we continue with the discussion of element-wise sampling and matrix completion.
There is no new reading for today.
As with last class, I'm not going to tex up these notes.

\newpage

\section{%
(12/02/2013): Element-wise Sampling of Graphs and Linear Equation Solving}

Today and in the next class we will discuss element-wise sampling and how this relates to recent work on solving linear equations with Laplacian constraint matrices.
Here is the reading for today.
\begin{itemize}
\item
Batson, Spielman, Srivastava, and Teng, ``Spectral Sparsification of Graphs: Theory and Algorithms''
\item
Koutis, Miller, and Peng, ``A fast solver for a class of linear systems''
\item
Spielman and Srivastava, ``Graph Sparsification by Effective Resistances'' 
\item
Drineas and Mahoney, ``Effective Resistances, Statistical Leverage, and Applications to Linear Equation Solving''
\end{itemize}

\subsection{Overview}

We have seen problems that can be written in the form of a system of linear equations with Laplacian constraint matrices, i.e.,  
\[
Lx = b .
\]
For example, we saw this with the various semi-supervised learning methods as well as with the MOV weakly-local spectral method.
In some cases, this arises in slightly modified form, e.g., as an augmented/modified graph and/or if there are additional projections (e.g., the Zhou et al paper on ``Learning with labeled and unlabeled data on a directed graph,'' that is related to the other semi-supervised methods we discussed, does this explicitly).
Today and next time we will discuss how to solve linear equations of this form.

\subsection{Basic statement and outline}

While perhaps not obvious, solving linear equations of this form is a useful \emph{algorithmic primitive}---like divide-and-conquer and other such primitives---much more generally, and thus there has been a lot of work on it in recent years.

Here is a more precise statement of the use of this problem as a primitive.
\begin{definition}
The \emph{Laplacian Primitive} concerns systems of linear equations defined by Laplacian constraint matrices:
\begin{itemize}
\item
\textsc{INPUT}: a Laplacian $L \in \mathbb{R}^{n \times n}$, a vector $b\in\mathbb{R}^{n}$ such that $\sum_{i=1}^{n} b_i = 0$, and a number $\epsilon>0$.
\item
\textsc{OUTPUT}: a vector $\tilde{x}_{opt} \in \mathbb{R}^{n}$ such that $\|\tilde{x}_{opt} - L^{\dagger} v \|_{L} \le \| L^{\dagger} b\|_{L}$, where for a vector $z\in\mathbb{R}^{n}$ the $L$-norm is given by $\|z\|_L=\|z^TLz\|_2$.
\end{itemize}
\end{definition}

While we will focus on linear equations with Laplacian constraint matrices, most of the results in this area hold for a slightly broader class of problems.
In particular, they hold for any linear system $Ax=b$, where $A$ is an SDD (symmetric diagonally dominant) matrix (i.e., that the diagonal entry of each row is larger, or not smaller, than the sum of the absolute values of the off-diagonal entries in that row).
The reason for this is that SDD systems are linear-time reducible to Laplacian linear systems via a construction that only doubles the number of nonzero entries in the matrix.

As mentioned, the main reason for the interest in this topic is that, given a fast, e.g., nearly linear time algorithm, for the Laplacian Primitive, defined above, one can obtain a fast algorithm for all sorts of other basic graph problems.
Here are several examples of such problems.
\begin{itemize}
\item
Approximate Fiedler vectors.
\item
Electrical flows.
\item
Effective resistance computations.
\item
Semi-supervised learning for labeled data.
\item
Cover time of random walks.
\item
Max flow and min cut and other combinatorial problems.
\end{itemize}
Some of these problems we have discussed.
While it might not be surprising that problems like effective resistance 
computations and semi-supervised learning for labeled data can be solved with this primitive, it should be surprising that max flow and min cut and other combinatorial problems can be solved with this primitive.
We won't have time to discuss this in detail, but some of the theoretically fastest algorithms for these problems are based on using this primitive.

Here is a statement of the basic result that led to interest in this area.
\begin{theorem}[ST]
There is a randomized algorithm for the Laplacian Primitive that runs in expected time $O\left( m \log^{O(1)} (n) \log \left(1/\epsilon\right)  \right)$, where $n$ is the number of nodes in $L$, $m$ is the number of nonzero entries in $L$, and $\epsilon$ is the precision parameter.
\end{theorem}
Although the basic algorithm of ST had something like the $50^{th}$ power in the exponent of the logarithm, it was a substantial theoretical breakthrough, and since then it has been improved by KMP to only a single log, leading to algorithms that are practical or almost practical.
Also, although we won't discuss it in detail, many of the local and locally-biased spectral methods we have discussed arose out of this line of work in an effort to develop and/or improve this basic result.

At a high level, the basic algorithm is as follows.
\begin{enumerate}
\item
Compute a sketch of the input by sparsifying the input graph.
\item
Use the sketch to construct a solution, e.g., by solving the subproblem with any black box solver or by using the sketch as a preconditioner for an iterative algorithm on the original~problem.
\end{enumerate}
Thus, the basic idea of these methods is very simple; but to get the methods to work in the allotted time, and in particular to work in nearly-linear time, is very complicated.

Today and next time, we will discuss these methods, including a simple but slow method in more detail and a fast but complicated method in less detail.
\begin{itemize}
\item
\textbf{Today.} 
We will describe a simple, non-iterative, but slow algorithm.
This algorithm provides a very simple version of the two steps of the basic algorithm described above; and, while slow, this algorithm highlights several basic ideas of the more sophisticated versions of these~methods.
\item
\textbf{Next time.}
We will describe a fast algorithm provides a much more sophisticated implementation of the two steps of this basic algorithm.
Importantly, it makes nontrivial use of combinatorial ideas and couples the linear algebra with combinatorial preconditioning in interesting~ways.
\end{itemize}

\subsection{A simple slow algorithm that highlights the basic ideas}

Here, we describe in more detail a very simple algorithm to solve Laplacian-based linear systems.
It will be good to understand before we get to the fast but more complicated versions of the algorithm.

Recall that $L = D-W = B^TWB$ is our Laplacian, where $B$ is the $m \times n$ edge-incidence matrix, and where $W$ is an $m \times m$ edge weight matrix.
In particular, note that $m > n$ (assume the graph is connected to avoid trivial cases), and so the matrix $B$ is a \emph{tall} matrix.

Here is a restatement of the above problem.
\begin{definition}
Given as input a Laplacian matrix $L \in \mathbb{R}^{n \times n}$, a vector $b \in \mathbb{R}^{n}$, compute 
\[
\mbox{argmin}_{x\in\mathbb{R}^{n}} \| Lx - b \|_2  .
\]
The minimal $\ell_2$ norm $x_{opt}$ is given by $x_{opt} = L^{\dagger}b$, where $L^{\dagger}$ is the Moore-Penrose generalized inverse of $L$.
\end{definition}
We have reformulated this as a regression since it makes the proof below, which is based on RLA (Randomized Linear Algebra) methods, cleaner.

The reader familiar with linear algebra might be concerned about the Moore-Penrose generalized inverse since, e.g., it is typically not well-behaved with respect to perturbations in the data matrix.
Here, the situation is particularly simple: although $L$ is rank-deficient, (1) it is invertible if we work with vectors $b\perp\vec{1}$, and (2) because this nullspace is particular simple, the pathologies that typically arise with the Moore-Penrose generalized inverse do \emph{not} arise here.
So, it isn't too far off to think of this as the inverse.  

Here is a simple algorithm to solve this problem.
This algorithm takes as input $L$, $b$, and $\epsilon$; and it returns as output a vector $\tilde{x}_{opt}$.
\begin{enumerate}
\item
Form $B$ and $W$, define $\Phi = W^{1/2}B \in \mathbb{R}^{m \times n}$, let $U_{\Phi}\in\mathbb{R}^{m \times n}$ be an orthogonal matrix spanning the column space of $\Phi$, and let $\left( U_{\Phi} \right)_{(i)}$ denote the $i^{th}$ row of $U_{\Phi}$.
\item
Let $p_i$, for $i \in [n]$ such that $\sum_{i=1}^{n}p_i=1$ be given by 
\begin{equation}
\label{eqn:lev-scores}
p_i \ge \beta\frac{ \| \left( U_{\Phi} \right)_{(i)}\|_2^2  }{ \| U_{\Phi} \|_F^2 } = \frac{\beta}{n} \| \left( U_{\Phi} \right)_{(i)} \|_2^2
\end{equation}
for some value of $\beta\in(0,1]$.
(Think of $\beta=1$, which is a legitimate choice, but the additional flexibility of allowing $\beta\in(0,1)$ will be important in the next class.)
\end{enumerate}

A key aspect of this algorithm is that the sketch is formed by choosing elements of the Laplacian with the probabilities in Eqn.~(\ref{eqn:lev-scores}); these quantities are known as the statistical leverage scores, and they are of central interest in RLA.
Here is a definition of these scores more generally.

\begin{definition}
Given a matrix $A\in\mathbb{R}^{m \times n}$, where $m > n$, the $i^{th}$ leverage score is
\[
\left(P_{A}\right)_{ii} = \left(U_AU_A^T\right)_{ii} = \| \left( U_A \right)_{ii} \|_2^2 ,
\]
i.e., it is equal to the diagonal element of the projection matrix onto the column span of $A$.
\end{definition}

Here is a definition of a seemingly-unrelated notion that we talked about before.

\begin{definition}
Given $G=(V,E)$, a connected, weighted, undirected graph with $n$ nodes, $m$ edges, and corresponding weights $w_e \ge 0$, for all $e \in E$, let $L=B^TWB$.
Then, the effective resistance $R_{e}$ across edge $e \in E$ are given by the diagonal elements of the matrix $R=BL^{\dagger}B$.
\end{definition}

Here is a lemma relating these two quantities.

\begin{lemma}
Let $\Phi = W^{1/2} B$ denote the scaled edge-incidence matrix.
If $\ell_i$ is the leverage score of the $i^{th}$ row of $\Phi$, then $\frac{\ell_i}{w_i}$ is the effective resistance of the $i^{th}$ edge.
\end{lemma}
\begin{Proof}
Consider the matrix 
\[
P = W^{1/2}B \left( B^T W B \right)^{\dagger} B^T W^{1/2} \in \mathbb{R}^{m \times m} ,
\]
and notice that $P = W^{1/2}R W^{1/2}$ is a rescaled version of $R = B L^{\dagger} B$, whose diagonal elements are the effective resistances.
Since $\Phi = W^{1/2}B$, it follows that 
\[
P = \Phi \left( \Phi^T \Phi \right)^{\dagger} \Phi^T .
\]
Let $U_{\Phi}$ be an orthogonal matrix spanning the columns of $\Phi$.
Then, $P=U_{\Phi}U_{\Phi}^T$, and so 
\[
P_{ii} = \left( U_{\Phi} U_{\Phi}^T \right)_{ii} = \| \left( U_{\Phi} \right)_{(i)} \|_2^2 ,
\]
which establishes the lemma.
\end{Proof}

So, informally, we sparsify the graph by biasing our random sampling toward edges that are ``important'' or ``influential'' in the sense that they have large statistical leverage or effective resistance, and then we use the sparsified graph to solve the subproblem.

Here is the main theorem for this algorithm.
\begin{theorem}
With constant probability, $\| x_{opt} - \tilde{x}_{opt} \|_L \le \epsilon \| x_{opt} \|_L$.
\end{theorem}
\begin{Proof}
The main idea of the proof is that we are forming a sketch of the Laplacian by randomly sampling elements, which corresponds to randomly sampling rows of the edge-incidence matrix, and that we need to ensure that the corresponding sketch of the edge-incidence matrix is a so-called subspace-preserving embedding.
If that holds, then the eigenvalues of the edge-incidence matrix and it's sketch are close, and thus the eigenvalues of the Laplacian are close, and thus the original Laplacian and the sparsified Laplacian are ``close,'' in the sense that the quadratic form of one is close to the quadratic form of the other.

Here are the details.

By definition, 
\[
\| x_{opt} - \tilde{x}_{opt} \|_L^2 
   = \left(x_{opt}-\tilde{x}_{opt} \right)^{T} L \left(x_{opt}-\tilde{x}_{opt} \right) .
\]
Recall that $L = B^TWB$, that $x_{opt} = L^{\dagger}b$, and that $\tilde{x}_{opt} = \tilde{L}^{\dagger}b$.
So, 
\begin{eqnarray*}
\| x_{opt} - \tilde{x}_{opt} \|_L^2 
  &=& \left(x_{opt}-\tilde{x}_{opt} \right)^{T} B^TWB \left(x_{opt}-\tilde{x}_{opt} \right)  \\
  &=& \| W^{1/2} B \left( x_{opt}-\tilde{x}_{opt} \right) \|_2^2 
\end{eqnarray*}
Let $\Phi \in \mathbb{R}^{m \times n}$ be defined as $\Phi = W^{1/2}B$, and let its SVD be $\Phi = U_{\Phi} \Sigma_{\Phi} V_{\Phi}^T$.
Then
\[
L = \Phi^T\Phi = V_{\Phi} \Sigma_{\Phi}^{2} V_{\Phi}^T
\] 
and \[
x_{opt} = L^{\dagger} b = V_{\Phi} \Sigma_{\Phi}^{-2} V_{\Phi}^T b .
\]
In addition
\[
\tilde{L} = \Phi^T S^T S \Phi = \left( S\Phi \right)^{T} \left( S\Phi \right) 
\]
and also
\[
\tilde{x}_{opt} 
   = \tilde{L}^{\dagger}b 
   = \left( S\Phi \right)^{\dagger} \left( S\Phi \right)^{T\dagger} b 
   = \left( SU_{\Phi}\Sigma_{\Phi}V_{\Phi}^{T}\right)^{\dagger} \left( SU_{\Phi}\Sigma_{\Phi}V_{\Phi}^{T}\right)^{T\dagger} b  
\]
By combining these expressions, we get that 
\begin{eqnarray*}
\| x_{opt} - \tilde{x}_{opt} \|_L^2
  &=& \| \Phi \left( x_{opt} - \tilde{x}_{opt} \right) \|_2^2 \\
  &=& \| U_{\Phi}\Sigma_{\Phi}V_{\Phi}^{T} \left( V_{\Phi} \Sigma_{\Phi}^{-2} V_{\Phi}^T - \left( SU_{\Phi}\Sigma_{\Phi}V_{\Phi}^{T}\right)^{\dagger} \left( SU_{\Phi}\Sigma_{\Phi}V_{\Phi}^{T}\right)^{T\dagger} \right) b \|_2^2 \\
  &=& \| \Sigma_{\Phi}^{-1} V_{\Phi}^T b - \Sigma_{\Phi} \left( SU_{\Phi}\Sigma_{\Phi}V_{\Phi}^{T}\right)^{\dagger} \left( SU_{\Phi}\Sigma_{\Phi}V_{\Phi}^{T}\right)^{T\dagger} V_{\Phi} b \|_2^2
\end{eqnarray*}

Next, we note the following:
\[
\mathbb{E}\left[ \| U_{\Phi}^T S^T S U_{\Phi} - I  \|_2 \right] \le \sqrt{\epsilon}  ,
\]
where of course the expectation can be removed by standard methods.
This follows from a result of Rudelson-Vershynin, and it can also be obtained as a matrix concentration bound.
This is a key result in RLA, and it holds since we are sampling $O\left(\frac{n}{\epsilon} \log \left( \frac{n}{\epsilon} \right) \right)$ rows from $U$ according to the leverage score sampling probabilities.

From standard matrix perturbation theory, it thus follows that 
\[
\left| \sigma_i \left( U_{\Phi}^TS^TSU_{\Phi} \right) - 1 \right|
   = \left| \sigma_i^2\left(SU_{\Phi}\right)-1 \right| \le \sqrt{\epsilon}  .
\]

So, in particular, the matrix $SU_{\Phi}$ has the same rank as the matrix $U_{\Phi}$.
(This is a so-called subspace embedding, which is a key result in RLA; next time we will interpret it in terms of graphic inequalities that we discussed before.)

In the rest of the proof, let's condition on this random event being true.

Since $SU_{\Phi}$ is full rank, it follows that
\[
\left( SU_{\Phi}\Sigma_{\Phi}\right)^{\dagger} = \Sigma_{\Phi}^{-1} \left( SU_{\Phi} \right)^{\dagger} .
\]
So, we have that 
\begin{eqnarray*}
\| x_{opt} - \tilde{x}_{opt} \|_L^2
  &=& \| \Sigma_{\Phi}^{-1} V_{\Phi}^T b - \left( SU_{\Phi}\right)^{\dagger}\left(SU_{\Phi}\right)^{T\dagger} \Sigma_{\Phi}^{-1}V_{\Phi}^T b  \|_2^2  \\
  &=& \| \Sigma_{\Phi}^{-1} V_{\Phi}^T b - V_{\Omega}\Sigma_{\Omega}^{-2}V_{\Omega}^{T} \Sigma_{\Phi}^{-1}V_{\Phi}^T b  \|_2^2 ,
\end{eqnarray*}
where the second line follows if we define $\Omega = S U_{\Phi}$ and let its SVD be 
\[
\Omega = SU_{\Phi} = U_{\Omega} \Sigma_{\Omega} V_{\Omega}^T   .
\]
Then, let $\Sigma_{\Omega}^{-1} = I+E$, for a diagonal error matrix $E$, and use that $V_{\Omega}^TV_{\Omega} = V_{\Omega}V_{\Omega}^T = I$ to write 
\begin{eqnarray*}
\|x_{opt}-\tilde{x}_{opt} \|_L^2
   &=&   \| \Sigma_{\Phi}^{-1}V_{\Phi}^Tb - V_{\Omega}\left(I+E\right) V_{\Omega}^{T} \Sigma_{\Phi}^{-1} V_{\Phi}^{T} b \|_2^2 \\
   &=&   \| V_{\Omega}E V_{\Omega}^T \Sigma_{\Phi}^{-1} V_{\Phi}^{T} b \|_2^2 \\
   &=&   \| E V_{\Omega}^T \Sigma_{\Phi}^{-1} V_{\Phi}^{T} b \|_2^2 \\
   &\le& \| E V_{\Omega}^T\|_2^2 \| \Sigma_{\Phi}^{-1} V_{\Phi}^{T} b \|_2^2 \\
   &=&   \| E \|_2^2 \| \Sigma_{\Phi}^{-1} V_{\Phi}^{T} b \|_2^2
\end{eqnarray*}
But, since we want to bound $\|E\|$, note that 
\[
\left|E_{ii}\right| = \left| \sigma_i^{-2}\left(\Omega\right) - 1 \right|
                    = \left| \sigma_i^{-1}\left(SU_{\Phi}\right) - 1 \right| .
\]
So, 
\[
\|E\|_2 = \max_i \left| \sigma_i^{-2}\left(SU_{\Phi}\right) - 1 \right| \le \sqrt{\epsilon} .
\]
So, 
\[
\| x_{opt}-\tilde{x}_{opt} \|_L^2 \le \epsilon \| \Sigma_{\Phi}^{-1}V_{\Phi}^T b \|_2^2 .
\]
In addition, we can derive that 
\begin{eqnarray*}
\| x_{opt} \|_L^2 
   &=& x_{opt}^T L x_{opt} \\
   &=& \left( W^{1/2}B x_{opt} \right)^T \left( W^{1/2}B x_{opt} \right) \\
   &=& \| \Phi x_{opt} \|_2^2 \\
   &=& \| U_{\Phi} \Sigma_{\Phi} V_{\Phi}^T V_{\Phi} \Sigma_{\Phi}^{-2} V_{\Phi}^T b \|_2^2 \\
   &=& \| \Sigma_{\Phi}^{-1} V_{\Phi}^{T} b \|_2^2  .
\end{eqnarray*}
So, it follows that 
\[
\|x_{opt}-\tilde{x}_{opt} \|_L^2 \le \epsilon \| x_{opt} \|_L^2 ,
\] 
which establishes the main result.
\end{Proof}

Before concluding, here is where we stand.
This is a very simple algorithm that highlights the basic ideas of Laplacian-based solvers, but it is not fast.
To make it fast, two things need to be done.
\begin{itemize}
\item
We need to compute or approximate the leverage scores quickly.
This step is very nontrivial.
The original algorithm of ST (that had the $\log^{50}(n)$ term) involved using local random walks (such as what we discussed before, and in fact the ACL algorithm was developed to improve this step, relative to the original ST result) to construct well-balanced partitions in nearly-linear time.
Then, it was shown that one could use effective resistances; this was discovered by SS independently of the RLA-based method outlined above, but it was also noted that one could call the nearly linear time solver to approximate them.
Then, it was shown that one could relate it to spanning trees to construct combinatorial preconditioners.
If this step was done very carefully, then one obtains an algorithm that runs in nearly linear time.
In particular, though, one needs to go beyond the linear algebra to map closely to the combinatorial properties of graphs, and in particular find low-stretch spanning trees.
\item
Instead of solving the subproblem on the sketch, we need to use the sketch to create a preconditioner for the original problem and then solve a preconditioned version of the original problem.
This step is relatively straightforward, although it involves applying an iterative algorithm that is less common than popular CG-based methods.
\end{itemize}
We will go through both of these in more detail next time.

\newpage

\section{%
(12/04/2013): Element-wise Sampling of Graphs and Linear Equation Solving, Cont.}

Today, we continue with the discussion of element-wise sampling of graphs and linear equation solving.
There is no new reading for today.

\subsection{Laplacian solvers, cont.}

Last time, we talked about a very simple solver for Laplacian-based systems of linear equations, i.e., systems of linear equations of the form $Ax=b$, where the constraint matrix $A$ is the Laplacian of a graph.
This is not fully-general---Laplacians are SPSD matrices of a particular form---but equations of this form arise in many applications, certain other SPSD problems such as those based on SDD matrices can be reduced to this, and there has been a lot of work recently on this topic since it is a primitive for many other problems.
The solver from last time is very simple, and it highlights the key ideas used in fast solvers, but it is very slow.
Today, we will describe how to take those basic ideas and, by coupling them with certain graph theoretic tools in various ways, obtain a ``fast'' nearly linear time solver for Laplacian-based systems of linear equations.

In particular, today will be based on the Batson-Spielman-Srivastava-Teng and the Koutis-Miller-Peng articles.

\subsection{Review from last time and general comments}

Let's start with a review of what we covered last time.

Here is a very simple algorithm.
Given as input the Laplacian $L$ of a graph $G=(V,E)$ and a right hand side vector $b$, do the following.
\begin{itemize}
\item
Construct a sketch of $G$ by sampling elements of $G$, i.e., rows of the edge-node incidence matrix, with probability proportional to the leverage scores of that row, i.e., the effective resistances of that edge.
\item
Use the sketch to construct a solution, e.g., by solving the subproblem with a black box or using it as a preconditioner to solve the original problem with an iterative method. 
\end{itemize}

The basic result we proved last time is the following.
\begin{theorem}
Given a graph $G$ with Laplacian $L$, let $x_{opt}$ be the optimal solution of $Lx=b$; then the above algorithm returns a vector $\tilde{x}_{opt}$ such that, with constant probability, 
\begin{equation}
\label{eqn:solution-approximation}
\| x_{opt} - \tilde{x}_{opt} \|_{L} \le \epsilon \| x_{opt} \|_{L}  .
\end{equation}
\end{theorem}
The proof of this result boiled down to showing that, by sampling with respect to a judiciously-chosen set of nonuniform importance sampling probabilities, then one obtains a data-dependent subspace embedding of the edge-incidence matrix.
Technically, the main thing to establish was that, if $U$ is an $m \times n$ orthogonal matrix spanning the column space of the weighted edge-incidence matrix, in which case $I = I_n = U^TU$, then 
\begin{equation}
\label{eqn:subspace-embedding}
\| I - \left(SU\right)^{T}\left(SU\right) \|_2 \le \epsilon ,
\end{equation}
where $S$ is a sampling matrix that represents the effect of sampling elements from $L$.

The sampling probabilities that are used to create the sketch are weighted versions of the statistical leverage scores of the edge-incidence matrix, and thus they also are equal to the effective resistance of the corresponding edge in the graph.
Importantly, although we didn't describe it in detail, the theory that provides bounds of the form of Eqn.~(\ref{eqn:subspace-embedding}) is robust to the exact form of the importance sampling probabilities, e.g., bounds of the same form hold if any other probabilities are used that are ``close'' (in a sense that we will discuss) to the statistical leverage scores.

The running time of this simple strawman algorithm consists of two parts, both of which the fast algorithms we will discuss today improve upon.
\begin{itemize}
\item
Compute the leverage scores, exactly or approximately.
A naive computation of the leverage scores takes $O(mn^2)$ time, e.g., with a black box QR decomposition routine.
Since they are related to the effective resistances, one can---theoretically at least compute them with any one of a variety of fast nearly linear time solvers (although one has a chicken-and-egg problem, since the solver itself needs those quantities).
Alternatively, since one does not need the exact leverage scores, one could hope to approximate them in some way---below, we will discuss how this can be done with low-stretch spanning trees.
\item
Solve the subproblem, exactly or approximately.
A naive computation of the solution to the subproblem can be done in $O(n^3)$ time with standard direct methods, or it can be done with an iterative algorithm that requires a number of matrix-vector multiplications that depends on the condition number of $L$ (which in general could be large, e.g., $\Omega(n)$) times $m$, the number of nonzero elements of $L$.
Below, we will see how this can be improved with sophisticated versions of certain preconditioned iterative algorithms.
\end{itemize}

More generally, here are several issues that arise.
\begin{itemize}
\item
Does one use exact or approximate leverage scores?
Approximate leverage scores are sufficient for the worst-case theory, and we will see that this can be accomplished by using LSSTs, i.e., combinatorial techniques. 
\item
How good a sketch is necessary?
Last time, we sampled $\Theta\left( \frac{ n \log(n) }{\epsilon^2} \right)$ elements from $L$ to obtain a $1\pm\epsilon$ subspace embedding, i.e., to satisfy Eqn.~(\ref{eqn:subspace-embedding}), and this leads to an $\epsilon$-approximate solution of the form of Eqn~(\ref{eqn:solution-approximation}).
For an iterative method, this might be overkill, and it might suffice to satisfy Eqn.~(\ref{eqn:subspace-embedding}) for, say, $\epsilon = \frac{1}{2}$.
\item
What is the dependence on $\epsilon$?
Last time, we sampled and then solved the subproblem, and thus the complexity with respect to $\epsilon$ is given by the usual random sampling results.
In particular, since the complexity is a low-degree polynomial in $\frac{1}{\epsilon}$, it will be essentially impossible to obtain a high-precision solution, e.g., with $\epsilon = 10^{-16}$, as is of interest in certain~applications.
\item
What is the dependence on the condition number $\kappa(L)$?
In general, the condition number can be very large, and this will manifest itself in a large number of iterations (certainly in worst case, but also actually quite commonly).
By working with a preconditioned iterative algorithm, one should aim for a condition number of the preconditioned problem that is quite small, e.g., if not constant then $\log(n)$ or less.
In general, there will be a tradeoff between the quality of the preconditioner and the number of iterations needed to solve the preconditioned~problem.
\item
Should one solve the subproblem directly or use it to construct a preconditioned to the original problem?
Several of the results just outlined suggest that an appropriate iterative methods should be used and this is what leads to the best results. 
\end{itemize}

\textbf{Remark.}
Although we are not going to describe it in detail, we should note that the LSSTs will essentially allow us to approximate the large leverage scores, but they won't have anything to say about the small leverage scores.
We saw (in a different context) when we were discussing statistical inference issues that controlling the small leverage scores can be important (for proving statistical claims about unseen data, but not for claims on the empirical data).
Likely similar issues arise here, and likely this issue can be mitigated by using implicitly regularized Laplacians, e.g., as as implicitly computed by certain spectral ranking methods we discussed, but as far as I know no one has explicitly addressed these questions.

\subsection{Solving linear equations with direct and iterative methods}

Let's start with the second step of the above two-level algorithm, i.e., how to use the sketch from the first step to construct an approximate solution, and in particular how to use it to construct a preconditioner for an iterative algorithm.
Then, later we will get back to the first step of how to construct the sketch.

As you probably know, there are a wide range of methods to solve linear systems of the form $Ax=b$, but they fall into two broad categories.
\begin{itemize}
\item
\textbf{Direct methods.}
These include Gaussian elimination, which runs in $O(n^3)$ time; and Strassen-like algorithms, which run in $O(n^{\omega})$ time, where $\omega = 2.87 \ldots 2.37$.
Both require storing the full set of in general $O(n^2)$ entires.
Faster algorithms exist if $A$ is structured.
For example, if $A$ is $n \times n$ PSD with $m$ nonzero, then conjugate gradients, used as a direct solver, takes $O(mn)$ time, which if $m = O(n)$ is just $O(n^2)$.
That is, in this case, the time it takes it proportional to the time it takes just to write down the inverse.
Alternatively, if $A$ is the adjacency matrix of a path graph or any tree, then the running time is $O(m)$; and so on.
\item
\textbf{Iterative methods.}
These methods don't compute an exact answer, but they do compute an $\epsilon$-approximate solution, where $\epsilon$ depends on the structural properties of $A$ and the number of iterations, and where $\epsilon$ can be made smaller with additional iterations.
In general, iterations are performed by doing matrix-vector multiplications.
Advantages of iterative methods include that one only needs to store $A$, these algorithms are sometimes very simple, and they are often faster than running a direct solver.
Disadvantages include that one doesn't obtain an exact answer, it can be hard to predict the number of iterations, and the running time depends on the eigenvalues of $A$, e.g., the condition number $\kappa(A) = \frac{\lambda_{max}(A)}{\lambda_{min}(A)}$.
Examples include the Richardson iteration, various Conjugate Gradient like algorithms, and the Chebyshev~iteration.
\end{itemize}

Since the running time of iterative algorithms depend on the properties of $A$, so-called \emph{preconditioning methods} are a class of methods to transform a given input problem into another problem such that the modified problem has the same or a related solution to the original problem; and such that the modified problem can be solved with an iterative method more quickly.

For example, to solve $Ax=b$, with $A\in\mathbb{R}^{n \times n}$ and with $m=\textbf{nnz}(A)$, if we define $\kappa(A) = \frac{\lambda_{max}(A)}{\lambda_{min}(A)}$, where $\lambda_{max}$ and $\lambda_{min}$ are the maximum and minimum non-zero eigenvalues of $A$, to be the condition number of $A$, then CG runs in 
\[
O\left( \sqrt{\kappa{(A)}}\log\left(1/\epsilon\right) \right)
\]
iterations (each of which involves a matrix-vector multiplication taking $O(m)$ time) to compute and $\epsilon$-accurate solution to $Ax=b$.
By an $\epsilon$-accurate approximation, here we mean the same notion that we used above, i.e., that $$\|\tilde{x}_{opt}-A^{\dagger}b\|_{A}\le\epsilon\|A^{\dagger}b\|_{A},$$ where the so-called $A$-norm is given by $\|y\|_{A}=\sqrt{y^TAy}$.
This $A$-norm is related to the usual Euclidean norm as follows: $\|y\|_{A}\le\kappa(A)\|y\|_2$ and $\|y\|_2\le\kappa(A)\|y\|_{A}$.
While the $A$-norm is perhaps unfamiliar, in the context of iterative algorithms it is not too dissimilar to the usual Euclidean norm, in that, given an $\epsilon$-approximation for the former, we can obtain an $\epsilon$-approximation for the latter with $O\left( \log\left( \kappa(A)/\epsilon \right) \right)$ extra iterations.

In this context, preconditioning typically means solving 
\[
B^{-1}Ax = B^{-1}b ,
\]
where $B$ is chosen such that $\kappa\left(B^{-1}A\right)$ is small; and it is easy to solve problems of the form $Bz=c$. 
The two extreme cases are $B=I$, in which case it is easy to compute and apply but doesn't help solve the original problem, and $B=A^{-1}$, which means that the iterative algorithm would converge after zero steps but which is difficult to compute.
The running time of the preconditioned problem involves $$O\left( \sqrt{\kappa\left( B^{-1}A \right)} \log\left(1/\epsilon \right) \right)$$ matrix vector multiplications.
The quantity $\kappa\left( B^{-1}A \right)$ is sometimes known as the \emph{relative condition number of $A$ with respect to $B$}---in general, finding a $B$ that makes it smaller takes more initial time but leads to fewer iterations.
(This was the basis for the comment above that there is a tradeoff in choosing the quality of the preconditioner, and it is true more generally.)

These ideas apply more generally, but we consider applying them here to Laplacians.
So, in particular, given a graph $G$ and its Laplacian $L_{G}$, one way to precondition it is to look for a different graph $H$ such that $L_H \approx L_G$.
For example, one could use the sparsified graph that we computed with the algorithm from last class.
That sparsified graph is actually an $\epsilon$-good preconditioned, but it is too expensive to compute.
To understand how we can go beyond the linear algebra and exploit graph theoretic ideas to get good approximations to them more quickly, let's discuss different ways in which two graphs can be close to one another.

\subsection{Different ways two graphs can be close}

We have talked formally and informally about different ways graphs can be close, e.g., we used the idea of similar Laplacian quadratic forms when talking about Cheeger's Inequality and the quality of spectral partitioning methods.
We will be interested in that notion, but we will also be interested in other notions, so let's now discuss this topic in more detail.
\begin{itemize}
\item
\textbf{Cut similarity.}
One way to quantify the idea that two graphs are close is to say that they are similar in terms of their cuts or partitions.
The standard result in this area is due to BZ, who developed the notion of \emph{cut similarity} to develop fast algorithms for min cut and max flow and other related combinatorial problems.
This notion of similarity says that two graphs are close if the weights of the cuts, i.e., the sum of edges or edge weights crossing a partition, are close for all cuts.
To define it, recall that, given a graph $G=(V,E,W)$ and a set $S \subset V$, we can define $\mbox{cut}_G = \sum_{u \in S, v \in \bar{S} } W_{(uv)}$.
Here is the definition.
\begin{definition}
Given two graphs, $G=(V,E,W)$ and $\tilde{G} = (V,\tilde{E},\tilde{W})$, on the same vertex set, we say that $G$ and $\tilde{G}$ are \emph{$\sigma$-cut-similar} if, for all $S\subseteq V$, we have that
\[
\frac{1}{\sigma} \mbox{cut}_{\tilde{G}}(S) \le \mbox{cut}_{G}(S) \le \sigma \mbox{cut}_{\tilde{G}}(S)  .
\]
\end{definition}
As an example of a result in this area, the following theorem shows that every graph is cut-similar to a graph with average degree $O(\log(n))$ and that one can compute that cut-similar graph quickly.
\begin{theorem}[BK]
For all $\epsilon > 0$, every graph $G=(V,E,W)$ has a $\left(1+\epsilon\right)$-cut-similar graph $\tilde{G}=(V,\tilde{E},\tilde{V})$ such that $\tilde{E} \subseteq E$ and  $|\tilde{E}| = O\left( n \log(n/\epsilon^2) \right)$.
In addition, the graph $\tilde{G}$ can be computed in  $O\left( m \log^3 (n) + m \log (n/\epsilon^2) \right)$ time.
\end{theorem}
\item
\textbf{Spectral similarity.}
ST introduced the idea of spectral similarity in the context of nearly linear time solvers.
One can view this in two complementary ways.
\begin{itemize}
\item
As a generalization of cut similarity.
\item
As a special case of subspace embeddings, as used in RLA.
\end{itemize}
We will do the former here, but we will point out the latter at an appropriate point.

Given $G=(V,E,W)$, recall that $L:\mathbb{R}^{n}\rightarrow\mathbb{R}$ is a quadratic form associated with $G$ such that 
\[
L(x) = \sum_{(uv) \in E} W_{(uv)} \left( x_u - x_v \right)^2  .
\]
If $S \subset V$ and if $x$ is an indicator/characteristic vector for the set $S$, i.e., it equals $1$ on nodes $u \in S$, and it equals $0$ on nodes $v \in S$, then for those indicator vectors $x$, we have that $L(x) = \mbox{cut}_{G}(x)$.
We can also ask about the values it takes for other vectors $x$.
So, let's define the following.
\begin{definition}
Given two graphs, $G=(V,E,W)$ and $\tilde{G}=(V,\tilde{E},\tilde{W})$, on the same vertex set, we say that $G$ and $\tilde{G}$ are \emph{$\sigma$-spectrally similar} if, for all $x\in\mathbb{R}^{n}$, we have that 
\[
\frac{1}{\sigma} L_{\tilde{G}}(x) \le L_{G}(x) \le \sigma L_{\tilde{G}}(x) .
\]
\end{definition}
That is, two graphs are spectrally similar if their Laplacian quadratic forms are close.

In addition to being a generalization of cut similarity, this also corresponds to a special case of subspace embeddings, restricted from general matrices to edge-incidence matrices and their associated~Laplacians.
\begin{itemize}
\item
To see this, recall that subspace embeddings preserve the geometry of the subspace and that this is quantified by saying that all the singular values of the sampled/sketched version of the edge-incidence matrix are close to $1$, i.e., close to those of the edge-incidence matrix of the original un-sampled graph.
Then, by considering the Laplacian, rather than the edge-incidence matrix, the singular values of the original and sketched Laplacian are also close, up to a quadratic of the approximation factor on the edge-incidence matrix.
\end{itemize}

Here are several other things to note about spectral embeddings.
\begin{itemize}
\item
Two graphs can be cut-similar but not spectrally-similar.
For example, consider $G$ to be an $n$-vertex path and $\tilde{G}$ to be an $n$-vertex cycle.
They are $2$-cut similar but are only $n$-spectrally similar.
\item
Spectral similarity is identical to the notion of relative condition number in NLA that we mentioned above.
Recall, given $A$ and $B$, then $ A \preceq B$ iff $x^TAx \le x^TBx$, for all $x\in\mathbb{R}^{n}$.
Then, $A$ and $B$, if they are Laplacians, are spectrally similar if $\frac{1}{\sigma}B \preceq A \preceq \sigma B$.
In this case, they have similar eigenvalues, since: from the Courant-Fischer results, if $\lambda_1,\ldots,\lambda_n$ are the eigenvalues of $A$ and $\tilde{\lambda}_1,\ldots,\tilde{\lambda}_n$ are the eigenvalues of $B$, then for all $i$ we have that $\frac{1}{\sigma} \tilde{\lambda}_i \le \lambda_i \le \sigma \tilde{\lambda_i}$.
\item
More generally, spectral similarity means that the two graphs will share many spectral or linear algebraic properties, e.g., effective resistances, resistance distances, etc.
\end{itemize}
\item
\textbf{Distance similarity.}
If one assigns a length to every edge $e \in E$, then these lengths induce a shortest path distance between every $u,v \in V$.
Thus, given a graph $G=(V,E,W)$, we can let $d: V \times V \rightarrow \mathbb{R}^{+}$ be the shortest path distance.
Given this, we can define the following notion of similarity.
\begin{definition}
Given two graphs, $G=(V,E,W)$ and $\tilde{G}=(V,\tilde{E},\tilde{W})$, on the same vertex set, we say that $G$ and $\tilde{G}$ are \emph{$\sigma$-distance similar} if, for all pairs of vertices $u,v \in V$, we have that 
\[
\frac{1}{\sigma} \tilde{d}(u,v) \le d(u,v) \le \sigma \tilde{d}(u,v)  .
\]
\end{definition}
Note that if $\tilde{G}$ is a subgraph if $G$, then $d_{G}(u,v) \le d_{\tilde{G}}(u,v)$, since shortest-path distances can only increase.
(Importantly, this does \emph{not} necessarily hold if the edges of the subgraph are re-weighted, as they were done in the simple algorithm from the last class, when the subgraph is constructed; we will get back to this later.)
In this case, a spanner is a subgraph such that distances in the other direction are not changed too much.
\begin{definition}
Given a graph $G=(V,E,W)$, a \emph{$t$-spanner} is a subgraph of $G$ such that for all $u,v \in V$, we have that $d_{\tilde{G}}(u,v) \le t d_{G}(u,v)$.
\end{definition}
There has been a range of work in TCS on spanners (e.g., it is known that every graph has a $2t+1$ spanner with $O\left( n^{1+1/\epsilon} \right)$ edges) that isn't directly relevant to what we are doing. 

We will be most interested in spanners that are trees or nearly trees.

\begin{definition}
Given a graph $G=(V,E,W)$, a \emph{spanning tree} is a tree includes all vertices in $G$ and is a subgraph of $G$.
\end{definition}
There are various related notions that have been studied in different contexts: for example, minimum spanning trees, random spanning trees, and low-stretch spanning trees (LSSTs).
Again, to understand some of the differences, think of a path versus a cycle.
For today, we will be interested in LSSTs.
The most extreme form of a sparse spanner is a LSST, which has only $n-1$ edges but which approximates pairwise distances up to small, e.g., hopefully polylog,~factors.
\end{itemize}

\subsection{Sparsified graphs}

Here is an aside with some more details about sparsified graphs, which is of interest since this is the first step of our Laplacian-based linear equation solver algorithm.
Let's define the following, which is a slight variant of the above.
\begin{definition}
Given a graph $G$, a \emph{$(\sigma,d)$-spectral sparsifier} of $G$ is a graph $\tilde{G}$ such that
\begin{compactenum}
\item
$\tilde{G}$ is $\sigma$-spectrally similar to $G$.
\item
The edges of $\tilde{G}$ are reweighed versions of the edges of $G$.
\item
$\tilde{G}$ has $\le d|V|$ edges.
\end{compactenum}
\end{definition}
 
\textbf{Fact.}
Expanders can be thought of as sparse versions of the complete graph; and, if edges are weighted appropriately, they are spectral sparsifiers of the complete graph.
This holds true more generally for other graphs.
Here are examples of such results.
\begin{itemize}
\item
SS showed that every graph has a $\left(1+\epsilon, O(\left( \frac{\log(n)}{\epsilon^2} \right) \right)$ spectral sparsifier.
This was shown by SS with an effective resistance argument; and it follows from what we discussed last time: last time, we showed that sampling with respect to the leverage scores gives a subspace embedding, which preserves the geometry of the subspace, which preserves the Laplacian quadratic form, which implies the spectral sparsification claim.
\item
BSS showed that every $n$ node graph $G$ has a $\left(  \frac{\sqrt{d}+1}{\sqrt{d}-1},d\right)$-spectral sparsifier (which in general is more expensive to compute than running a nearly linear time solver).
In particular, $G$ has a $\left( 1+2\epsilon, \frac{4}{\epsilon^2} \right)$-spectral sparsifier, for every $\epsilon\in(0,1)$.
\end{itemize}

Finally, there are several ways to speed up the computation of graph sparsification algorithms, relative to the strawman that we presented in the last class.
\begin{itemize}
\item
Given the relationship between the leverage scores and the effective resistances and that the effective resistances can be computed with a nearly linear time solver, one can use the ST or KMP solver to speed up the computation of graph sparsifiers.
\item
One can use local spectral methods, e.g., diffusion-based methods from ST or the push algorithm of ACL, to compute well-balanced partitions in nearly linear time and from them obtain spectral sparsifiers.
\item
Union of random spanning trees.
It is known that, e.g., the union of two random spanning trees is $O(\log(n))$-cut similar to $G$; that the union of $O\left( \log^2(n)/\epsilon^2 \right)$ reweighed random spanning trees is a $1+\epsilon$-cut sparsifier; and so on.  
This suggests looking at spanning trees and other related combinatorial quantities that can be quickly computed to speed up the computation of graph sparsifiers. 
We turn to this next.
\end{itemize}

\subsection{Back to Laplacian-based linear systems}

KMP considered the use of combinatorial preconditions, an idea that traces back to Vaidya.
They coupled this with a fact that has been used extensively in RLA: that only approximate leverage scores are actually needed in the sampling process to create a sparse sketch of $L$.
In particular, they compute upper estimates of the leverage scores or effective resistance of each edge, and they compute these estimates on a modified graph, in which each upper estimate is sufficiently good.
The modified graph is rather simple: take a LSST and increase its weights.
Although the sampling probabilities obtained from the LSST are strictly greater than the effective resistances, they are not too much greater in aggregate.
This, coupled with a rather complicated iterative preconditioning scheme, coupled with careful accounting with careful data structures, will lead to a solver that runs in $O\left( m \log(n)\log(1/\epsilon) \right)$ time, up to $\log\log(n)$ factors.
We will discuss each of these briefly in turn.

\paragraph{Use of approximate leverage scores.}
Recall from last class that an important step in the algorithm was to use nonuniform importance sampling probabilities.  
In particular, if we sampled edges from the edge-incidence matrix with probabilities $\{p_i\}_{i=1}^{m}$, where each $p_i = \ell_i$, where $\ell_i$ is the effective resistance or statistical leverage score of the weighted edge-incidence matrix, then we showed that if we sampled $r=O\left( n\log(n)/\epsilon\right)$ edges, then it follows that 
\[
\| I - \left(SU_{\Phi}\right)^{T}\left(SU_{\Phi}\right) \|_2 \le \epsilon  ,
\]
from which we were able to obtain a good relative-error solution.

Using probabilities exactly equal to the leverage scores is overkill, and the same result holds if we use any probabilities $p_i^{\prime}$ that are ``close'' to $p_i$ in the following sense: if 
\[
p_i^{\prime} \ge \beta \ell_i   ,
\]
for $\beta\in(0,1]$ and $\sum_{i=1}^{m} p_i^{\prime}=1$, then the same result follows if we sample $r=O\left( n\log(n)/(\beta\epsilon)\right)$ edges, i.e., if we oversample by a factor of $1/\beta$.
The key point here is that it is essential not to underestimate the high-leverage edges too much.
It is, however, acceptable if we overestimate and thus oversample some low-leverage edges, as long as we don't do it too much.

In particular, let's say that we have the leverage scores $\{\ell_1\}_{i=1}^{m}$ and overestimation factors $\{\gamma_i\}_{i=1}^{m}$, where each $\gamma_i \ge 1$.
From this, we can consider the probabilities 
\[
p_i^{\prime\prime} = \frac{\gamma_i\ell_i}{\sum_{i=1}^{m}\gamma_i\ell_i}   .
\]
If $\sum_{i=1}^{m}\gamma_i\ell_i$ is not too large, say $O\left(n\log(n)\right)$ or some other factor that is only slightly larger than $n$, then dividing by it (to normalize $\{\gamma_i\ell_i\}_{i=1}^{m}$ to unity to be a probability distribution) does not decrease the probabilities for the high-leverage components too much, and so we can use the probabilities $p_i^{\prime\prime}$ with an extra amount of oversampling that equals $\frac{1}{\beta} = \sum_{i=1}^{m}\gamma_i\ell_i$.

\paragraph{Use of LSSTs as combinatorial preconditioners.}
Here, the idea is to use a LSST, i.e., use a particular form of a ``combinatorial preconditioning,'' to replace $\ell_i=\ell_{(uv)}$ with the stretch of the edge $(uv)$ in the LSST.
Vaidya was the first to suggest the use of spanning trees of $L$ as building blocks as the base for preconditioning matrix $B$.
The idea is then that the linear system, if the constraint matrix is the Laplacian of a tree, can be solved in $O(n)$ time with Gaussian elimination.
(Adding a few edges back into the tree gives a preconditioned that is only better, and it is still easy to solve.)
Boman-Hendrickson used a LSST as a stand-along preconditioner.
ST used a preconditioner that is a LSST plus a small number of extra edges.
KMP had additional extensions that we describe here.

Two question arise with this approach.
\begin{itemize}
\item
Q1: What is the appropriate base tree?
\item
Q2: Which off-tree edges should added into the preconditioner?
\end{itemize}
One idea is to use a tree that concentrates that maximum possible weight from the total weight of the edges in $L$.
This is what Vaidya did; and, while it led to good result, the results weren't good enough for what we are discussing here.
(In particular, note that it doesn't discriminate between different trees in unweighted graphs, and it won't provide a bias toward the middle edge of a dumbbell graph.)
Another idea is to use a tree that concentrates mass on high leverage/influence edges, i.e., edges with the highest leverage in the edge-incidence matrix or effective resistance in the corresponding Laplacian.  

The key idea to make this work is that of \emph{stretch}.
To define this, recall that for every edge $(u,v)\in E$ in the original graph Laplacian $L$, there is a unique ``detour'' path between $u$ and $v$ in the tree $T$.
\begin{definition}
The \emph{stretch} of the edge with respect to $T$ equals the distortion caused by this detour.
\end{definition}
In the unweighted case, this stretch is simply the length of the tree path, i.e., of the path between nodes $u$ and $v$ that were connected by an edge in $G$ in the tree $T$.
Given this, we can define the~following.
\begin{definition}
The \emph{total stretch} of a graph $G$ and its Laplacian $L$ with respect to a tree $T$ is the sum of the stretches of all off-tree edges.
Then, a \emph{low-stretch spanning tree (LSST)} $T$ is a tree such that the total stretch is low.
\end{definition}
Informally, a LSST is one such that it provides a good ``on average'' detours for edges of the graph, i.e., there can be a few pairs of nodes that are stretched a lot, but there can't be too many such~pairs.

There are many algorithms for LSSTs.
For example, here is a result that is particularly relevant for us.
\begin{theorem}
Every graph $G$ has a spanning tree $T$ with total stretch $\tilde{O}\left( m \log(n) \right)$, and this tree can be found in $\tilde{O}\left( m \log(n) \right)$ time.
\end{theorem}

In particular, we can use the stretches of pairs of nodes in the tree $T$ in place of the leverage scores or effective resistances as importance sampling probabilities: they are larger than the leverage scores, and there might be a few that much larger, but the total sum is not much larger than the total sum of the leverage scores (which equals $n-1$).

\paragraph{Paying careful attention to data structures, bookkeeping, and recursive preconditions.}
Basically, to get everything to work in the allotted time, one needs the preconditioner $B$ that is extremely good approximation to $L$ and that can be computed in linear time.
What we did in the last class was to compute a ``one step'' preconditioner, and likely any such ``one step'' preconditioned won't be substantially easier to compute that solving the equation; and so KMP consider recursion in the construction of their preconditioner.
\begin{itemize}
\item
In a recursive preconditioning method, the system in the preconditioned $B$ is not solved exactly but only approximately, via a recursive invocation of the same iterative method.
So, one must find a preconditioned for $B$, a preconditioned for it, and so on.
This gives s multilevel hierarchy of progressively smaller graphs.
To make the total work small, i.e., $O(kn)$, for some constant $k$, one needs the graphs in the hierarchy to get small sufficiently fast.
It is sufficient that the graph on the $(i+1)^{th}$ level is smaller than the graph on the $i^{th}$ level by a factor of $\frac{1}{2k}$.
However, one must converge within $O(kn)$.
So, one can use CG/Chebyshev, which need $O(k)$ iterations to converge, when $B$ is a $k^2$-approximation of $L$ (as opposed to $O(k^2)$ iterations which are needed for something like a Richardson's iteration).
\end{itemize}

So, a LSST is a good base; and a LSST also tells us which off-tree edges, i.e., which additional edges from $G$ that are not in $T$, should go into the preconditioner.
\begin{itemize}
\item
This leads to an $\tilde{O}\left( m \log^2(n)\log(1/\epsilon) \right)$ algorithm.
\end{itemize}
If one keeps sampling based on the same tree and does some other more complicated and careful stuff, then one obtains a hierarchical graph and is able to remove the the second log factor to yield a potentially practical solver.
\begin{itemize}
\item
This leads to an $\tilde{O}\left( m \log (n)\log(1/\epsilon) \right)$ algorithm.
\end{itemize}
See the BSST and KMP papers for all the details.

\newpage

\section{Additional Topics Not Covered This Semester}

There were several topics that I hoped to cover that we didn't have time to cover.
Here is a summary of the most important.
\begin{itemize}
\item
Sparsity-preserving Random Projections.
These are very sparse but structured random projections that run in ``input sparsity'' time, which means proportional to the number of nonzeros plus ``lower order terms,'' which means polynomial in the low dimension (of a rectangular problem) or the rank parameter (if both dimensions of the input are large).
Here are the basic references.
\begin{itemize}
\item
The basic result: Clarkson and Woodruff, ``Low rank approximation and regression in input sparsity time''
\item
A simpler linear algebraic proof is in Section 3 of: Meng and Mahoney, ``Low-distortion Subspace Embeddings in Input-sparsity Time and Applications to Robust Linear Regression''
\item
A generalization of the previous result: Nelson and Nguyen, ``OSNAP: Faster numerical linear algebra algorithms via sparser subspace embeddings'' 
\end{itemize}
\item
Low-rank Approximation and Kernel-based Learning.
The basic issue here is two-fold: whether one is interested in making claims about a given kernel (i.e., SPSD) matrix of using that kernel matrix for some sort of inferential or statistical task; and how to establish that a low-rank approximation preserves the SPSD property, which is in general hard to do unless one uses uniform sampling or has diagonally dominant input.
Here are the basic references.
\begin{itemize}
\item
Gittens and Mahoney, ``Revisiting the Nystrom Method for Improved Large-Scale Machine Learning'' (the arXiv version---the arXiv version is \emph{much} more detailed than the short ICML version)
\item
Bach, ``Sharp analysis of low-rank kernel matrix approximations'' 
\end{itemize}
\item
Least absolute deviations, i.e., $\ell_1$, regression.
This extends the basic ideas beyond $\ell_2$ objectives.
The following is the most comprehensive reference, but it is not easy reading.
\begin{itemize}
\item
Clarkson, Drineas, Magdon-Ismail, Mahoney, Meng, and Woodruff, ``The Fast Cauchy Transform and Faster Robust Linear Regression''
\end{itemize}
\end{itemize}

%
%

%
%

\end{document}